\documentclass[twocolumn, notitlepage, aps, pra, floatfix, superscriptaddress]{revtex4-2}

\usepackage{enumerate}
\usepackage[linesnumbered,ruled,vlined]{algorithm2e}

\usepackage{csvsimple}
\usepackage{datatool}
\usepackage{graphicx}
\usepackage{multirow}
\usepackage{amsmath}
\usepackage{amssymb}
\usepackage{dsfont}
\usepackage{amsthm}
\usepackage{eucal}
\usepackage{float}
\usepackage{qcircuit}
\usepackage{bm}
\usepackage{upgreek}
\usepackage{mathrsfs}
\usepackage{color}
\usepackage{latexsym}
\usepackage[dvipsnames]{xcolor}
\usepackage{float}
\usepackage{enumerate}
\usepackage{stackengine}
\usepackage{braket}
\usepackage{qtree}
\usepackage[titletoc]{appendix}
\usepackage{natbib}
\usepackage{xcolor}
\usepackage{array}
\usepackage{cancel}
\usepackage{tikz}
\usepackage{xpatch}
\usepackage{comment}
\xpatchbibdriver{misc}
  {\printtext[parens]{\usebibmacro{date}}}
  {\iffieldundef{year}
    {}
    {\printtext[parens]{\usebibmacro{date}}}}
  {}
  {\typeout{There was an error patching biblatex-ieee (specifically, ieee.bbx's @online driver)}}

\usepackage [english]{babel}
\usepackage [autostyle, english = american]{csquotes}
\MakeOuterQuote{"}

\theoremstyle{remark}

\usepackage{hyperref}
\hypersetup{
colorlinks,
linkcolor={blue},
citecolor={red},
urlcolor={blue},
linktocpage
}

\usepackage{tikz,lipsum,lmodern}
\usepackage[most]{tcolorbox}

\usepackage[caption=false]{subfig}

\newcommand{\bb}{\begin{equation}}
\newcommand{\ee}{\end{equation}}
\newcommand{\bbb}{\begin{equation*}}
\newcommand{\eee}{\end{equation*}}

\stackMath

\DeclareMathOperator{\Tr}{Tr}

\allowdisplaybreaks 

\begin{document}


\title{Non-CSS color codes on 2D lattices : Models and Topological Properties}

\author{Pramod Padmanabhan}%
\author{Abhishek Chowdhury}
\affiliation{School of Basic Sciences, Indian Institute of Technology, Bhubaneswar, India}%
\email{Email : ppadmana, achowdhury@iitbbs.ac.in}
\author{Fumihiko Sugino}
\affiliation{Research and Education Center for Natural Sciences, Keio University, Hiyoshi, Yokohama, Japan}
\email{Email : fusugino@gmail.com}%
\author{Mrittunjoy Guha Majumdar}
\affiliation{Quantum Optics \& Quantum Information, Department of Instrumentation and Applied Physics, Indian Institute of Science, Bengaluru, India}
\email{mrittunjoyg@iisc.ac.in}
\author{Krishna Kumar Sabapathy}
\affiliation{Xanadu, Toronto, Ontario M5G 2C8, Canada}
\email{krishnakumar.sabapathy@gmail.com}
\begin{abstract}
The two-dimensional color code is an alternative to the toric code that encodes more logical qubits while maintaining crucial features of the $\mathbb{Z}_2\times\mathbb{Z}_2$ toric code in the long wavelength limit. However its short range physics include single qubit Pauli operations that violate either three or six stabilisers as opposed to the toric code where single qubit Pauli operations violate two or four stabilisers. Exploiting this fact we construct several non-CSS versions of the two-dimensional color code falling into two families -  those where either three, four or five stabilisers are violated and those which violate exactly four stabilisers for all the three types of single qubit Pauli operations. These models are not equivalent to the original color code by a local unitary transformation. Nevertheless the code spaces of the CSS and non-CSS versions are related by local unitaries which reflects the fact that their long range physics coincide.  This also implies that the non-CSS versions host transversal Clifford gates and hence support fault-tolerant computations. As a consequence of the non-CSS structure, the logical operators are of a mixed type which in some cases include all the three Pauli operators making them potentially useful for protection against biased Pauli noise.

\end{abstract}

\maketitle 

\tableofcontents

\section{Introduction}
\label{sec:intro}
One of the major challenges in building quantum computers is detecting and correcting errors in the form of noise and decoherence that could occur during information processing. In the quest to move from the `noisy-intermediate-scale-quantum era'  to full-fledged `fault-tolerant application-scale-quantum era', we need to deploy quantum error correcting codes (QECC) \cite{terhal2015}. The role of QECC is to reduce the errors in the qubits that carry the information by adding redundancies.  The way this is achieved is using multiple physical qubits to encode a single logical qubit using a suitable code, such that the logical qubit has lower error rates compared to the underlying physical qubit. 

 A good candidate to overcome this is by using the method of fault tolerant quantum computing, which are realized via many-body systems with topological order, a prominent example being the toric code \cite{Kitaev:1997wr} also known as the surface codes. These models can be thought of as a Hamiltonian realization of planar gauge theories \cite{deWildPropitius:1995hk, Kogut1979AnIT} based on discrete, finite groups and more generally known as the quantum double models \cite{Ferreira:2013oca, Buerschaper2013AHO}. Following Kitaev's work another Hamiltonian realizing the abelian $\mathbb{Z}_2\times\mathbb{Z}_2$ topological phase appeared in the form of the color code \cite{bombin2006} which is not written in the form of a lattice gauge theory. Instead, in these models the degrees of freedom or the physical qubits are located on the vertices of a trivalent and tricolorable lattice. The standard examples of such lattices include the hexagonal lattice (6,6,6),  the square-octagonal lattice (4,8,8) and a (4,6,12) lattice. While these are regular lattices with translational invariance, it is possible to construct several other irregular lattices on which the color code is well defined. A remarkable property of the color code is that it encodes more number of qubits than the toric code on the same space, supports fault tolerant computation as they host transversal gates and is more efficient for quantum error correction \cite{Gottesman1997StabilizerCA,Preskill2004LectureNF}. The simplest example of the color code is the one defined on a triangle with 7-qubits and this coincides with the 7-qubit Steane code \cite{steane1996,salas2004,anderson2014,goto2016minimizing,fowler2011constructing,campbell2017roads}, a well known example of a fault tolerant code. 

The color code has seen several interesting generalizations \cite{da_Silva_2018,katzgraber2010,watson2015,Vuillot2019QuantumPC,Amaro2020ScalableCO, Kargarian2008EntanglementPO,Bombin2010TopologicalSC,Bombin2013GaugeCC, Burton2018SpectraOG, Jones2016GaugeCC,Brown2016FaulttolerantEC} and has also been experimentally realized \cite{Bravyi2015DoubledCC,Nigg2014QuantumCO,PoulsenNautrup2017FaulttolerantIB,Hilder2021FaulttolerantPR,RyanAnderson2021RealizationOR,Bermudez2019FaulttolerantPO,Litinski2017CombiningTH, Bartlett2021,Satzinger2021,Semeghini2021}. Apart from direct generalizations several theoretical works have explored the physics and error correction properties of the color code and related models \cite{Kubica2018TheAO,Vasmer2019ThreedimensionalSC,Scruby2020AHO,Yoshida2011ClassificationOQ,Pastawski2015HolographicQE,Fowler2012SurfaceCT,Yoshida2010FrameworkFC,Bombin2014DimensionalJI,Bombin2013AnIT,Kubica2015UnfoldingTC,Sarvepalli2012TopologicalSC,Sarvepalli2012EfficientDO,Aloshious2021DecodingTC, Brown2015FaultTW,Katzgraber2009ErrorTF,Beverland2014ProtectedGF,Harrington2004AnalysisOQ,Horsman2012SurfaceCQ,Andrist2012UnderstandingTQ,Kubica2018UngaugingQE}. Connections between toric and color codes were presented in \cite{aloshious2018,aloshious2019local,kubica2018ungauging,kubica2015unfolding,Vasmer2021MorphingQC}. 

Both the surface codes and the color codes are examples of Calderbank-Steane-Shor (CSS) codes as each of the stabilizers are entirely composed of either the Pauli, $x$ or $z$ operators. It is of interest to consider non-CSS generalizations as they are thought to be of particular interest in biased noise settings \cite{Stephens2013HighthresholdTQ,tuckett2018ultrahigh,tuckett2019tailoring,tuckett2020fault} where one Pauli error is more likely than the other. In this paper we take a step in this direction by presenting several non-CSS color codes on 2D trivalent, tricolorable lattices. We provide a generic method to construct abelian stabilizer codes on trivalent lattices that are naturally non-CSS. We find that these models are not equivalent to the original color code by a local unitary (LU) transformation. This is despite the fact that the non-CSS versions share the same topological properties as the original color code, that is they continue to realize the $\mathbb{Z}_2\times\mathbb{Z}_2$ topological phase. By this we mean that the code spaces of the two models and the anyonic excitations are LU related. This also implies that the non-CSS versions continue to support fault tolerant quantum computation. However the original color code has more excitations that are local and immobile, that cannot be moved around like the anyons. The non-CSS versions have different kinds of immobile excitations when compared to the original color code. They differ in the fact that their energies are not the same as their CSS counterparts. This serves as the starting point of the generalizations and they naturally lead to non-CSS versions of the color code. 

The paper is organized as follows. We begin with an overview of the original CSS color code in Sec. \ref{sec:preliminaries}. We go over some of its topological properties, including the computation of its ground state degeneracy on genus, $g$ surface and their anyon content. We also look at the immobile excitations of the theory through the application of Pauli operators on an individual qubit and call these the {\it elementary excitations}. The anyons can be obtained by combining these elementary excitations. The construction of the non-CSS stabilizer codes on general trivalent lattice starts in Sec. \ref{sec:construction1}. We take a bottom-up approach here by looking at operator configurations around different types of vertices on a trivalent lattice and at operator configurations along edges. These building blocks are used to construct the full stabilizer codes on hexagonal lattices in Sec. \ref{sec:models666488}. We discuss three types of non-CSS color codes, the $[444]$-color codes, the deconfined $[345]$-color codes and the confined $[345]$-color codes. The notation is explained in the text. In this language the original color code becomes the $[336]$-color codes. We present all their non-CSS versions for completion. The topological properties of each of these models is discussed in Sec. \ref{sec:TOorder}. The finite versions of these codes, namely the triangle codes are studied in Sec. \ref{sec:finitenonCSScodes}. We thoroughly analyze one example of a fully mixed non-CSS code in Sec. \ref{sec:examplemixed[444]} to study the mixed nature of the string logicals in these models. In Sec. \ref{sec:higher-genus} we construct trivalent and tricolorable lattices on higher genus surfaces and build the non-CSS codes on them. We end with an outlook in Sec. \ref{sec:outlook}. There are several appendices exploring more models. We present these for completion.  

\section{Preliminaries}
\label{sec:preliminaries}
To begin with we go over the stabilizer formalism and the topological properties of the canonical color code as originally proposed \cite{bombin2006}. We also study the unitary operations that permute Pauli matrices which will be used later in the paper.

\subsection{Stabilizer formalism and commuting projector codes}
\label{subsec:stabilizerformalism}
Stabilizer codes provide a convenient framework to study QECCs\,\cite{Gottesman1997StabilizerCA}. We briefly recall some of the basic concepts of the stabilizer framework that we will use in the rest of the article. Let $\mathcal{P}_n$ denote the Pauli group on $n$ qubits. The Pauli group is defined as the group generated by the following operators 
\begin{align}
\mathcal{P}_n =     \langle X, Y, Z, \pm 1\!\!1, \pm i 1\!\!1 \rangle^{\otimes n},   
\end{align}
i.e., its the group generated by n-strings of single qubit Pauli operators with $\omega 1\!\!1$, where $\omega^4 = 1$. We will only be concerned with stabilizer codes associated with the Pauli group in this article. A stabilizer code is described by a stabilizer group $\mathcal{S}$ which is an abelian subgroup of the Pauli group that is generated by stabilizers (commuting operators) $s$ where 
\begin{align}
    \mathcal{S} = \langle s \in \mathcal{P}_n\,|\, [s_i,s_j] = 0 \rangle, 
\end{align}
and also $-1\!\!1$ is not in the stabilizer set. Let us denote by the centralizer of $\mathcal{S}$ by $\mathcal{Z}$, which consists of the elements of $\mathcal{P}$ that commute with all the elements of $\mathcal{S}$.  The code space $\mathcal{C}$ is defined as the subspace of the $n$-qubit Hilbert space stabilized by $\mathcal{S}$, where 
\begin{align}
    |\psi\rangle \in \mathcal{C} ~ \text{if } s|\psi\rangle = +1 |\psi\rangle, ~ \forall ~ s \in \mathcal{S}. 
\end{align}
Then the set of logical operators of the code belong to the set $\mathcal{Z} - \mathcal{S}$. Note that these elements leave the code space invariant, however, they would apply some non-trivial  operation on the individual code words, that would be undetected by the syndromes of the stabilizers. 

So we see that the stabilizer code is completely specified by the stabilizers, and the underlying group structure gives rise to the specifics of the code. The stabilizer Hamiltonian is just defined as 
\begin{align}
    H(\mathcal{S}) = -\sum_j^{|\mathcal{S}|} s_j.  
\end{align}
Note that code words are the ground states of $H(\mathcal{S})$ since all the stabilizers account for $-1$ energy in the Hamiltonian.  

There is an alternative representation for stabilizer code Hamiltonians into what are known as commuting projector code (CPC) Hamiltonians. A CPC Hamiltonian has terms made of projectors $\mathbb{P} = \langle P_j \rangle $ such that $[P_i,P_j] =0$, and 
\begin{align}
    H(\mathbb{P}) = -\sum_{j=1}^{|\mathbb{P}|} P_j. 
\end{align}
There are no other constraints of CPC Hamiltonains. It is straightforward to recast any Pauli stabilizer code Hamiltonian as a CPC by defining 
\begin{align}\label{eq:stab2cpc}
    P_j = \frac{1}{2} [1\!\!1 + s_j]. 
\end{align}
Note that in this formulation the projector onto the code space $P_{\mathcal{C}}$ is given by 
\begin{align}\label{eq:codeproj}
    P_{\mathcal{C}} = \prod_{j=1}^{|\mathbb{P}|} P_j.
\end{align}
For topological codes defined on finite lattices, the dimension of the code space is ${\rm dim} \mathcal{C} = 2^{n - |S|}$.  Note that the dimension of the code space is  also known as the ground state degeneracy (GSD) of the corresponding stabilizer code or CPC Hamiltonian. An alternate method to obtain ${\rm dim} \mathcal{C}$ is to compute the trace of $ P_{\mathcal{C}}$. For codes defined on a closed manifold of a given genus, one would need to use the Euler formula for such a counting method to determine ${\rm dim} \mathcal{C}$. We will deploy different methods depending on the context.  


\subsection{Overview of the canonical color code}
\label{subsec:canonicalcc}

\begin{figure}[h]
    \centering
    \includegraphics[width=8cm]{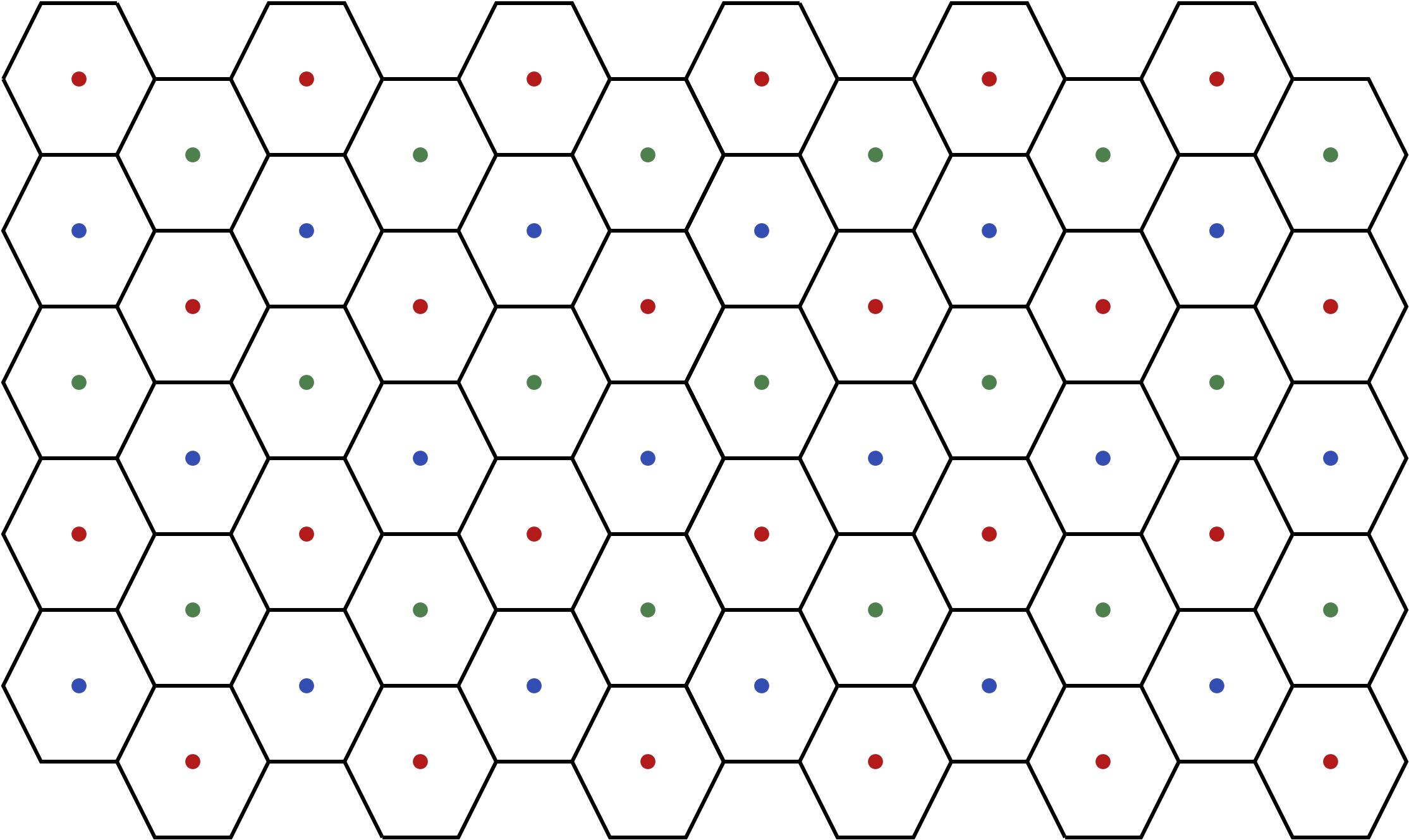}
 \caption{A hexagonal lattice as an example of a trivalent, tricolorable lattice supporting the canonical color code.}
    \label{fig:hexlat}
\end{figure}

Two-dimensional color codes can be cast as a commuting projector Hamiltonians in a manner similar to abelian toric codes or quantum double Hamiltonians~\cite{Kitaev:1997wr,Ferreira:2013oca, Buerschaper2013AHO}. The models are defined on a trivalent and tricolorable lattice discretizing a 
two-dimensional surface. An example is as shown in Fig. \ref{fig:hexlat}, where the physical qubits are placed on the vertices (sites) of the lattice.

   

\begin{figure}[h]
    \centering
    \includegraphics[width=8cm]{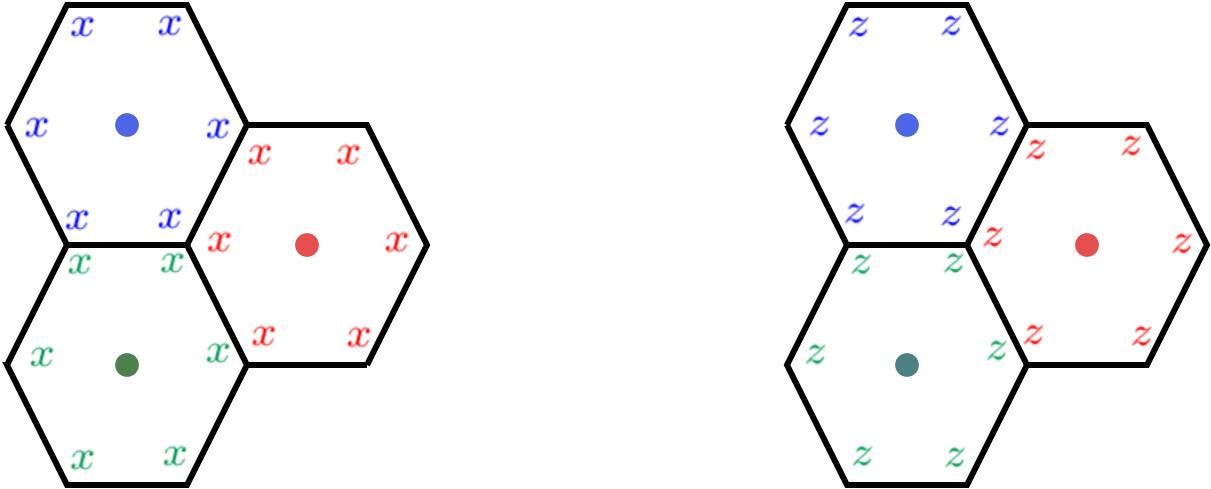}
 \caption{The canonical color code stabilizers depicted in a unit cell of the trivalent and tricolorable lattice. The figures in the left and right panel denote the two types of stabilizers associated with each colored face. }
    \label{fig:stabiliserscc}
\end{figure}

The projectors are defined for the hexagons of each color and are built out of the stabilizers, as
\begin{equation} \label{pxcc}
{\color{red}PX_j}  =  \frac{1}{2}\left(1+{\color{red}X}_{j}\right), 
\end{equation}
 and similar expressions for the other two colors. The operators $X_{j}$, for each color, are the Pauli $X$ stabilizers as depicted in Fig. \ref{fig:stabiliserscc},
with $X_{j}=\prod\limits_{j=1}^6x_{v_j}$, where $v_j$ are the vertices filling up the face ($f$) labelled by the subscript $j$. The operator 1 is shorthand for $\prod\limits_{j=1}^6 1_{v_j}$. Analogous projectors are defined for the $Z$ stabilisers by replacing $x$ with $z$ in Eq.\,(\ref{pxcc}). 

Using these operators we write down the commuting projector Hamiltonian as 
\begin{eqnarray}\label{eq:hcc}
H & = & -\sum\limits_{j=1}^{{\color{red}f}}~\left({\color{red}PX}_{j} + {\color{red}PZ}_{j}\right) \nonumber \\ & - & \sum\limits_{j=1}^{{\color{blue}f}}~\left({\color{blue}PX}_{j} + {\color{blue}PZ}_{j}\right) \nonumber \\ & - & \sum\limits_{j=1}^{{\color{ForestGreen}f}}~\left({\color{ForestGreen}PX}_{j} + {\color{ForestGreen}PZ}_{j}\right),
\end{eqnarray}
where ${\color{red}f}$, ${\color{blue}f}$ and ${\color{ForestGreen}f}$ are the number of red, blue and green faces respectively. For a hexagonal lattice they all equal $\frac{|F|}{3}$, but for an arbitrary trivalent and tricolorable lattice we have, ${\color{red}f}+{\color{blue}f}+{\color{ForestGreen}f}=|F|$.

\subsubsection{Ground State Degeneracy}
The model in Eq.\,\eqref{eq:hcc} is exactly solvable as it is easy to verify that all the terms commute with each other. On a surface with genus $g$, the model has a ground state degeneracy (GSD) given by $2^{4g}$. We compute the GSD as
\begin{equation} \label{eq:trace}
\textrm{GSD} = \Tr\left[\prod\limits_{j=1}^{{\color{red}f}}{\color{red}P_{X_j}}{\color{red}P_{Z_j}}\prod\limits_{j=1}^{{\color{blue}f}}{\color{blue}P_{X_j}}{\color{blue}P_{Z_j}}\prod\limits_{j=1}^{{\color{ForestGreen}f}}{\color{ForestGreen}P_{X_j}}{\color{ForestGreen}P_{Z_j}}\right],    
\end{equation}
which is essentially the trace of the projector to the ground state space of the color code Hamiltonian, Eq.\,(\ref{eq:hcc}). The product consists of several tensor products of Pauli operators having support on different sites, however these terms are traceless and hence do not contribute to the GSD. The only terms contributing to the trace are those of the form $\prod\limits_{j=1}^{|F|}1_{v_j}$ with support on all sites. Thus this GSD computation is an exercise in counting the number of such identity terms. 

Expanding each of the products of projectors in the following way\,: 
\begin{eqnarray}
\prod\limits_{j=1}^{{\color{red}f}}{\color{red}PX}_j & = & 
\frac{1}{2^{\color{red}f}}\left[1 + \prod\limits_{j=1}^{{\color{red}f}}{\color{red}X}_j\right] + \textrm{traceless terms}, \nonumber \\   
\prod\limits_{j=1}^{{\color{red}f}}{\color{red}PZ}_j & = & 
\frac{1}{2^{\color{red}f}}\left[1 + \prod\limits_{j=1}^{{\color{red}f}}{\color{red}Z}_j\right] + \textrm{traceless terms}, 
\end{eqnarray}
and similar expressions for the other colored faces.  

From these expressions we can see that the projector to the ground state manifold has precisely $2^4$ number of identity operator on the entire lattice. These are accounted for by using the following constraints on the stabilisers making up the Hamiltonian defined on a closed manifold of genus $g$,
\begin{eqnarray}\label{constraintscc}
\prod\limits_j{\color{red}X}_j{\color{blue}X}_j & = & \prod\limits_j{\color{blue}X}_j{\color{ForestGreen}X}_j = \prod\limits_j{\color{ForestGreen}X}_j{\color{red}X}_j = 1, \nonumber \\  
\prod\limits_j{\color{red}Z}_j{\color{blue}Z}_j & = & \prod\limits_j{\color{blue}Z}_j{\color{ForestGreen}Z}_j = \prod\limits_j{\color{ForestGreen}Z}_j{\color{red}Z}_j = 1,
\end{eqnarray}

We can thus evaluate the GSD as
\begin{equation}\label{gsdcc}
\textrm{GSD} =  \frac{1}{2^{2{\color{red}f}}}\frac{1}{2^{2{\color{blue}f}}}\frac{1}{2^{2{\color{ForestGreen}f}}}2^{|V|}2^4.
\end{equation}
Using ${\color{red}f}+{\color{blue}f}+{\color{ForestGreen}f}=|F|$ the exponent becomes
\begin{equation}
    |V| - 2|F| + 4 = 4g,
    \label{Euler1}
\end{equation}
 where we have used the Euler formula, $|V|-|E|+|F|=2-2g$ and the fact that $|E|=\frac{3|V|}{2}$ for a trivalent and tricolorable lattice. This completes the first proof for the GSD of the canonical color code on a closed manifold of genus $g$.
 
  The more conventional proof found in the literature involves finding the number of independent stabilisers in the Hamiltonian, Eq.\eqref{eq:hcc}. From the six constraint relations in Eq.\,(\ref{constraintscc}) \footnote{The constraint relations quoted in the literature read as 
 $\prod\limits_j{\color{red}X}_{f_j}  =  \prod\limits_j{\color{blue}X}_{f_j} = \prod\limits_j{\color{ForestGreen}X}_{f_j},$ 
 and $\prod\limits_j{\color{red}Z}_{f_j}  =  \prod\limits_j{\color{blue}Z}_{f_j} = \prod\limits_j{\color{ForestGreen}Z}_{f_j}.$
 This is easily seen to be equivalent to the constraint relations used in Eq.\,(\ref{constraintscc}).}
 we deduce that only four of them are independent, resulting in $2|F|-4$ independent stabilisers for this system. Thus the number of encoded qubits is precisely $|V|-2|F|+4$ which evaluates to $4g$ as shown above. 

We will mostly employ the second method to compute the code space dimension of the non-CSS color codes considered in this paper.

\subsubsection{Code space ${\mathcal{C}}$ and logical operators}
The Hamiltonian in Eq.\,(\ref{eq:hcc}) is a sum of commuting projectors encoding $4g$ logical qubits. One of the ground states can be constructed by applying only the X-type stabilizers on a state left invariant by the Z-type stabilizers,  
\begin{equation}
\ket{\bar{0}_1,\cdots, \bar{0}_{4g}} = N \prod\limits_{j=1}^{{\color{red}f}}{\color{red}PX}_j\prod\limits_{j=1}^{{\color{blue}f}}{\color{blue}PX}_j\prod\limits_{j=1}^{{\color{ForestGreen}f}}{\color{ForestGreen}PX}_j\ket{s},    
\end{equation}
where $g$ denotes the genus of the surface and $\ket{s}$ is a state stabilised by the projectors ${\color{red}PZ}_j, {\color{blue}PZ}_j, {\color{ForestGreen}PZ}_j$ for all $j\in\{1,\cdots, {\color{red} f}\}$, $j\in\{1,\cdots, {\color{blue} f}\}$ and $j\in\{1,\cdots, {\color{ForestGreen} f}\}$ respectively. The $\bar{0}_j$ denotes the $j$th logical zero state and $N$ is the overall normalization of the state. 

\begin{figure}[h]
    \centering
    \includegraphics[width=4cm]{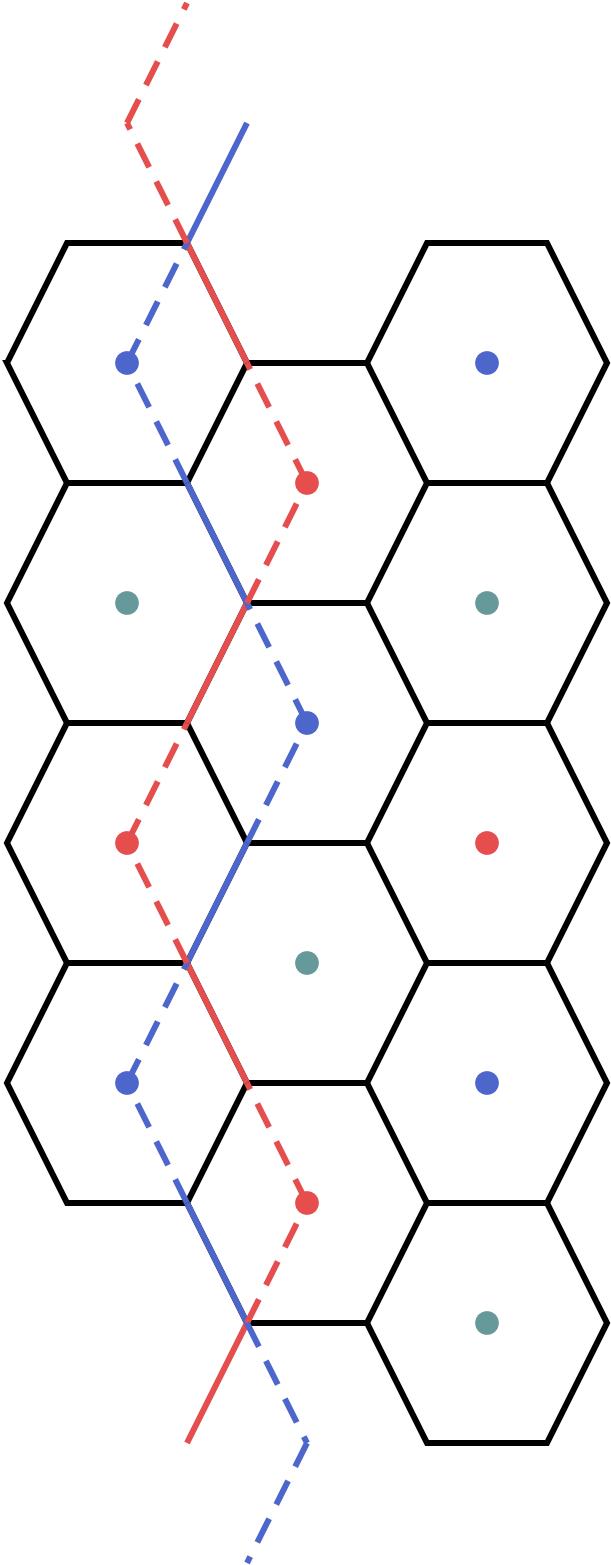}
 \caption{The winding operators on the red and blue shrunk lattices of the color code.}
    \label{fig:windingopcc}
\end{figure}

To construct the other ground states or the code words we identify the winding operators along the non-contractible loops of the genus $g$ manifold. For each non-contractible loop, $C_{\alpha,j}$, where $\alpha\in\{1,2,\cdots, g\}$ enumerates the different holes of the surface and $j\in\{1,2\}$ provides the index for the two loops for a given hole. This is because we have two sets of winding operator along two chosen colors as shown in 
Fig. \ref{fig:windingopcc}. 
We pick the winding operators along the red and blue shrunk lattices as a convention. By a shrunk lattice of a chosen color we mean the sub-lattice containing only the faces of the chosen color and the edges connecting them. Note that this opposes the convention used in the literature.

\begin{figure}[h]
    \centering
    \includegraphics[width=8cm]{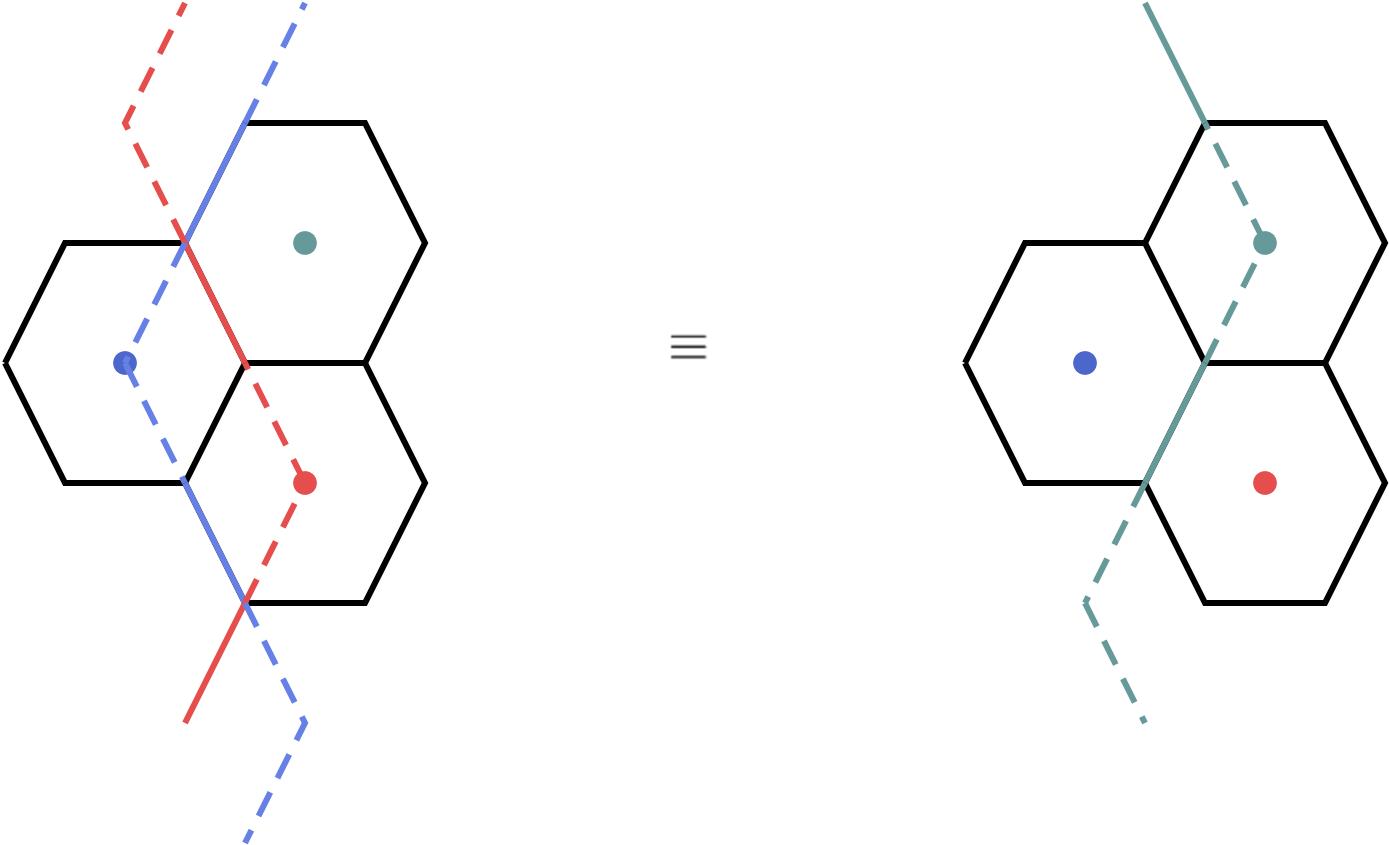}
 \caption{Product of the string operators of the same type on the blue and red shrunk lattice equals a string operator on the green shrunk lattice.}
    \label{fig:stringproductscc}
\end{figure}

There is no loss of generality in this choice as an operator along a given shrunk lattice is obtained as a product of operators along the other two shrunk lattices as shown in Fig. \ref{fig:stringproductscc}. In other words, only two of the three colors are independent with respect to the winding or logical operators. 
The winding operators comprising the logical Pauli $X$ operators are given by
\begin{equation}
{\color{red}\bar{X}}_{\alpha, j} = \prod\limits_{i\in C_{\alpha, j}}x_i;~~{\color{blue}\bar{X}}_{\alpha, j} = \prod\limits_{i\in C_{\alpha, j}}x_i,    
\end{equation}
with $\alpha\in\{1,2,\cdots, g\}$ and $j\in\{1,2\}$ along the red and blue shrunk lattices respectively. 
The logical Pauli $Z$ operators are analogously written as 
\begin{equation}
{\color{red}\bar{Z}}_{\alpha, j} = \prod\limits_{i\in C_{\alpha, j}}z_i;~~{\color{blue}\bar{Z}}_{\alpha, j} = \prod\limits_{i\in C_{\alpha, j}}z_i,    
\end{equation}
along the red and blue shrunk lattices. 

The logical operators for the canonical color code is made up of a single type of Pauli operator as we have just seen. We will see that in some of the non-CSS versions this property changes with the presence of logical operators that are composed of more than one type of Pauli operator. These models can be distinguished using the idea of {\it $P$-distances} that is defined as follows.
In the stabilizer formalism, the {\it code distance} is taken to be the minimum weight of a symmetry operator that cannot be expressed as a product of the stabilizers that make up the commuting projector Hamiltonian. If this minimum weight symmetry operator is composed of a single type of Pauli operator, either $x$, $y$ or $z$, then this is called as a {\it $P$-type logical} and the corresponding weight is denoted the {\it $P$-distance}.

\subsubsection{Elementary excitations}
\begin{figure}[h]
    \centering
    \includegraphics[width=5cm]{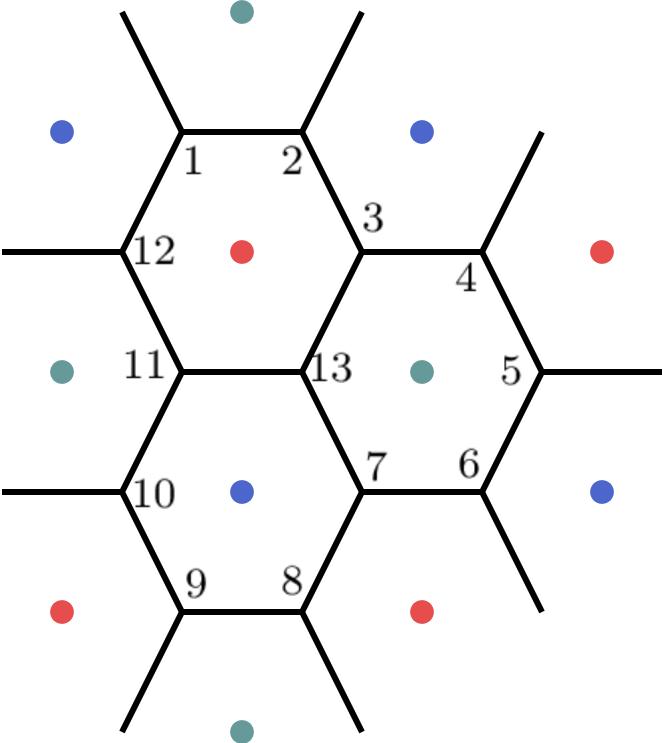}
 \caption{A unit cell in the hexagonal lattice containing a face of each color. The location of the 13 physical qubits are also indicated.}
    \label{fig:lookuplatcc}
\end{figure}

Consider single-qubit bit-flip or phase-flip operations on a vertex of the trivalent lattice. They excites either three or six stabilizers as opposed to the toric code where the same excites two or four stabilizers. More importantly, it is not possible to excite an odd number of stabilisers for the toric code defined on a closed manifold. Before studying the nature of the `triad excitations' we list them for a unit cell of the lattice in Fig.\,\ref{fig:lookuplatcc} in the form of a Table~\ref{luptablecc}.
\begin{table}[ht]
\centering
\begin{tabular}{cccccccc}
\hline\hline
Operation & ${\color{red}X}$ & ${\color{blue}X}$ & ${\color{ForestGreen}X}$ & & ${\color{red}Z}$ & ${\color{blue}Z}$ & ${\color{ForestGreen}Z}$ \\
\hline
$x_j$ & & & & & {\color{red}$\bullet$} & {\color{blue}$\bullet$} & {\color{ForestGreen}$\bullet$} \\ $z_j$ &  {\color{red}$\bullet$} & {\color{blue}$\bullet$} & {\color{ForestGreen}$\bullet$} & & & & \\
$y_j$ &  {\color{red}$\bullet$} & {\color{blue}$\bullet$} & {\color{ForestGreen}$\bullet$} & & {\color{red}$\bullet$} & {\color{blue}$\bullet$} & {\color{ForestGreen}$\bullet$} 
\end{tabular}
\caption{Table of elementary excitations. The index $j$ specifies the physical qubit as shown in Fig. \ref{fig:lookuplatcc} and $j\in\{1,2,\cdots, 13\}$. The columns represent the 6 stabilizers surround each qubit. The colored dots indicate which stabilisers are excited as a result of the single qubit operation. For example, an X operation on the qubit 7 would excite the three Z-type stabilizers that surround the qubit 7. Note that the canonical color  code has a $[336]$ structure for the elementary excitation. This notation will be explained later in the text. }
\label{luptablecc}
\end{table}
Further, the triad excitations are {\it immobile}, i.e.,  they cannot be moved around the lattice using string operators like the deconfined anyons of the abelian toric code. 

\subsubsection{Deconfined anyons}

The topological character of the model warrants the existence of anyonic excitations just as in the case of the abelian toric code. They are obtained from a string of $x$ or $z$ operators acting in the red and blue shrunk lattices. The latter creates the `electric charges' and the former creates the `magnetic fluxes' of the theory. We have
\begin{eqnarray}\label{eq:anyons}
&&{\color{red}e},~~ {\color{red}E_\gamma} = \prod\limits_{j\in{\color{red}\gamma}}x_j;~ {\color{blue}e},~~ {\color{blue}E_\gamma} = \prod\limits_{j\in{\color{blue}\gamma}}x_j\nonumber \\
&&{\color{red}m},~~ {\color{red}M_\gamma} = \prod\limits_{j\in{\color{red}\gamma}}z_j;~ {\color{blue}m},~~ {\color{blue}M_\gamma} = \prod\limits_{j\in{\color{blue}\gamma}}z_j,
\end{eqnarray}
where ${\color{red}\gamma}$ and ${\color{blue}\gamma}$ are paths in the red and blue shrunk lattices respectively as shown in Fig.\,\ref{fig:windingopcc}.
As seen earlier the string operators on the green shrunk lattice can be obtained as a product of the string operators on the red and blue shrunk lattices and hence ${\color{ForestGreen}e}$ and ${\color{ForestGreen}m}$ are not independent excitations. 
The anyons can also be obtained by combining elementary excitations (See Table \ref{luptablecc}) on vertices of a shrunk lattice of a chosen color.


Thus the anyonic content of the color code is isomorphic to that of the $\mathbb{Z}_2\times\mathbb{Z}_2$ toric code as seen from Eq. (\ref{eq:anyons}). However, the spectrum of the color code also contains the immobile `triad' excitations which are not found in the $\mathbb{Z}_2\times\mathbb{Z}_2$ toric code. While the anyons of the theory account for the long range properties of the system, the immobile excitations represent the short range physics. In particular they have no effect on the topological parts of the GSD and the entanglement entropy. 

We will see that the energetics of single qubit excitations around a trivalent vertex can be altered resulting in new stabilizer codes that are non-CSS by construction. Thus the elementary excitations can be thought of as the starting point of the non-CSS color codes.

\subsection{Unitaries permuting the Pauli operators}
\label{subsec:unitaries}
To check the equivalences of the different non-CSS color codes we require the unitary operators that permute the Pauli matrices, $\{x,y,z\}$, and there are precisely 6 of them. We systematically construct these operators in terms of the Hadamard, $H$ and phase gates, $S$. 

\begin{enumerate}
    \item The Hadamard operator, $H=\frac{1}{\sqrt{2}}\left[x+z\right]$, satisfies, $H^2=1$, and has the following action on the Pauli operators under conjugation,
    \begin{equation}
        \left\{\begin{array}{c} x\rightarrow z, \\ y\rightarrow -y, \\ z\rightarrow x\end{array}\right\}.
    \end{equation}
    \item The phase gate, $S=\left(\begin{array}{cc} 1 & 0 \\ 0 & \mathrm{i} \end{array}\right)$, satisfies $S^4=1$, and permutes the Pauli matrices as,
    \begin{equation}
        \left\{\begin{array}{c} x\rightarrow y, \\ y\rightarrow -x, \\ z\rightarrow z\end{array}\right\}.
    \end{equation}
    \item The operator, $t=\frac{1}{\sqrt{2}}\left[y+z\right]= SHS^3$, acts like the `Hadamard' gate interchanging, $y$ and $z$,
    \begin{equation}
        \left\{\begin{array}{c} x\rightarrow -x, \\ y\rightarrow z, \\ z\rightarrow y\end{array}\right\}.
    \end{equation}
    \item The unitaries, $C_1=HS^3$ and $C_2=SH$, cyclically permute the Pauli matrices, 
    \begin{equation}
       C_1: ~\left\{\begin{array}{c} x\rightarrow y, \\ y\rightarrow z, \\ z\rightarrow x\end{array}\right\}.
    \end{equation}
    and
    \begin{equation}
        C_2:~\left\{\begin{array}{c} x\rightarrow z, \\ y\rightarrow x, \\ z\rightarrow y\end{array}\right\}.
    \end{equation}
\item Finally we have the identity transformation which leaves all the Pauli matrices unchanged. 
\end{enumerate}

Note the appearance of the `$-1$''s in the above transformations. It is worth observing that the above unitaries do not form the permutation group of three objects, $S_3$. These are not the only operators that permute the Pauli operators. For example using the transpositions, $\tilde{H}=\frac{1}{\sqrt{2}}\left(z-x\right)$ and $\tilde{t}=\frac{1}{\sqrt{2}}\left(y-z\right)$ interchange $x \leftrightarrow z$, and $y\leftrightarrow z$ respectively, up to a sign. These operators can be expressed in terms of the Hadamard and phase gates as, $\tilde{H}=S^2HS^2$ and $\tilde{t} = -S^3HS$ respectively.


\section{Overview of the main results}
\label{sec:Overview}
This section serves a dual purpose of summarizing our findings and also as a guide to navigate the paper. 

\begin{itemize}
    \item The stabilizer codes to follow are built from the ground up by considering the placement of Pauli operators, $x$, $y$ and $z$ around vertices and along edges of a face appearing in the trivalent lattice. For the vertex configurations we are guided by the local energetics of the Hamiltonian formed out of the stabilizers of the code (dubbed elementary excitations earlier). By considering vertices with unequal energy configurations we are guaranteed inequivalent stabilizer codes that cannot be mapped to each other via LU's. As a result we find {\it two more} color code-like stabilizer codes on trivalent lattices that share the same topological properties as the latter. We label them as the $[444]$- and $[345]$-color codes based on the possible energy excitations around the trivalent vertices of the code. From this perspective the original color code takes the notation, $[336]$-color codes.
    The $P$-distance of the code sets the requirement for the configurations along the edges. We obtain three types of edge configurations, homogeneous ($P$-distance is zero), partially mixed ($P$-distance of some of the string logical operators are 0 and some $\infty$) and finally fully mixed ($P$-distance for the string logical operators is $\infty$). The vertex and edge configurations form the building blocks of the code (See Sec. \ref{sec:construction1}). 
    \item As examples the full non-CSS stabilizer codes are built on the hexagon ($(6,6,6)$) and the square-octagon ($(4,8,8)$) lattices. We show the construction of the $[444]$-color codes on both these lattices and with three different edge configurations, homogeneous, partially and fully mixed. Moreover we provide the codes with and without translational invariance. These can be found in Sec. \ref{sec:models666488}.  
    \item In the case of the $[345]$-color codes, the combinations of vertex and edge configurations result in two types of models : {\it deconfined} and {\it confined} models. In the latter elementary string operators of a given color excite either two or three stabilizers of that color leading to `energetic' strings, that is if the string is extended along certain directions all the stabilizers along the string get excited apart from the end points. Whereas the former case is the more familiar one occurring in the surface codes and the original color code where the strings creating anyons excite only the stabilizers associated to the faces at the endpoints of the string resulting in its deconfined nature. In other words there is no cost for moving anyons in the deconfined $[345]$-color codes and there is an energy cost for moving anyons in the confined counterpart. 
    Furthermore there are three inequivalent $[345]$ vertices that cannot be mapped to each other using a LU (See Fig. \ref{fig:[345]bulkvertexconfigs}).
    An example of each of these different types of $[345]$-color codes on the hexagon and the square-octagon lattice for the three types of edge configurations, with and without translational invariance, can be found in Sec. \ref{sec:models666488}.
    
\end{itemize}

\section{General vertex and edge configurations for trivalent lattices}
\label{sec:construction1}
Our goal is to provide a detailed  construction of non-CSS variants of the color codes on both a finite trivalent lattice tiling a closed two-dimensional surface and on finite trivalent lattices with boundaries. Every such lattice consists of vertices $V$ and edges $E$ that make up the faces $F$ and on each face we consider two stabilizers, just as in the color code. Henceforth, we refer to them as the {\it stabilizer pair}.
As in the color code each face on the trivalent lattice contains an even number of vertices and so we only use squares, hexagons, octagons and so on to tile the surface.  The common regular lattices tiling the plane include 
\begin{itemize}
    \item the hexagon $(6,6,6)$ lattice
    \item the square-octagon $(4,8,8)$ lattice 
    \item  the square-hexagon-dodecagon $(4,6,12)$ lattice. 
\end{itemize} 

We begin the construction of a non-CSS model by considering vertex and edge configurations made out of the three Pauli operators. These are then joined together resulting in stabilizer pairs for each face forming an abelian subgroup of the Pauli group. This construction results in a commuting stabilizer Hamiltonian. Before diving into the configurations we look at the types of vertices possible from the trivalent lattice geometry and the types of edge configurations possible from the desired physical properties of the code. 

\begin{figure}[h]
    \centering
    \includegraphics[width=6cm]{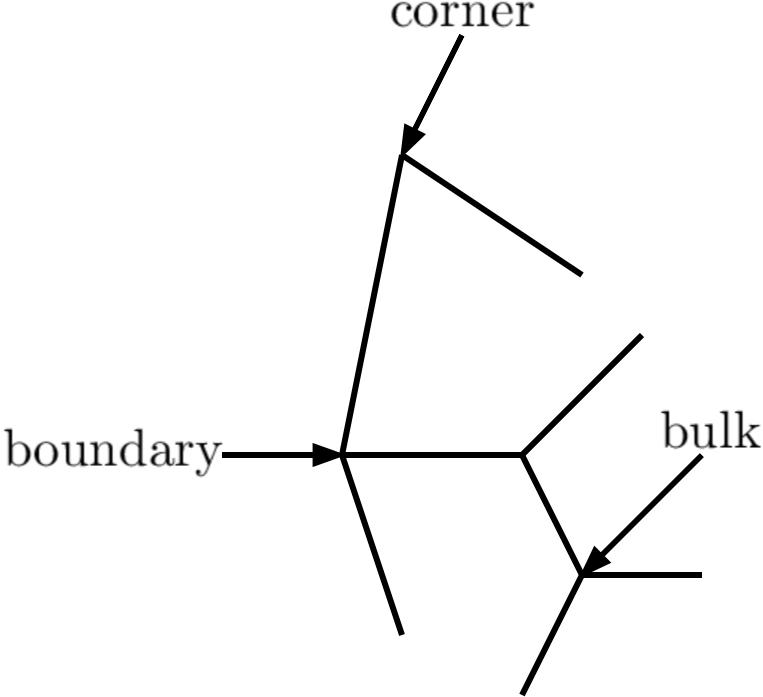}
 \caption{The three types of vertices on a finite trivalent lattice. Only a portion of the entire lattice is shown.}
    \label{fig:3vertextypes}
\end{figure}

Vertices come in three kinds, {\it bulk}, {\it boundary} and {\it corner}. While lattices on closed surfaces consist of only trivalent bulk vertices, finite lattices consist of trivalent vertices in the bulk, vertices that end on the  boundaries of the lattice and those that are located at the corners of the lattice (See Fig. \ref{fig:3vertextypes}). The bulk and boundary type vertices are trivalent whereas the corner type vertices are bivalent. Corner vertices share a single face, whereas boundary and bulk vertices share two and three faces respectively. 

When building edge configurations we keep in mind that the resulting stabilizers have to generate an abelian subgroup of the Pauli group. This is very restrictive, reducing the number of possibilities. Furthermore as the logical operators of the model are built using strings of Pauli operators along edges, we distinguish edge operators according to whether or not they can give rise to P-type operators. This gives us three cases, {\it fully mixed}, {\it partially mixed} and {\it homogeneous} (no mixing) edge configurations. By making these choices for the edge configurations we can either end up in a model where 
\begin{itemize}
    \item the P-distance for the string logicals is infinity when the edge configurations are fully mixed,
    \item a finite distance P-type operator for some of the string logicals when the logicals are partially mixed,
    \item all logicals are of finite P-distance when the logicals are homogeneous or pure.
\end{itemize}

Finally, as a guiding principle that will help classify the space of non-CSS color codes,  we impose another condition on the choice of the Pauli operators on a particular qubit on a given stabilizer pair which we call the {\it anticommuting rule} - {\it The Pauli operators for each qubit for a given stabilizer pair must anticommute.}

It is easy to verify that the stabilizer pairs of the canonical color code satisfy this condition. Imposing this rule picks out the models which share the topological properties of the canonical color code. The reason for this is tied to the {\it quantum double} structure of the algebra of operators appearing in the toric code and the color code \cite{Buerschaper2013AHO, Ferreira:2013oca}. By requiring this condition we preserve this structure for the stabilizers of the non-CSS variants. 

We now turn to systematically exhausting the possible configurations for the three types of vertices and the edges. These can be considered as the building blocks for constructing the full non-CSS color code variants.

\subsection{Bulk vertex configurations} 
\label{subsec:bulk}
A crucial feature of the color codes is the single qubit Pauli operations that result in its elementary excitations (See Table \ref{luptablecc}). In particular, for the canonical color codes,  we see that the $x$, $y$ and $z$ operations violate exactly 3, 6 and 3 stabilizers shared by the trivalent vertex respectively. The fact that we can excite an odd number of stabilizers by local operations distinguishes the color codes from surface codes, such as the toric code. Thinking of the $(363)$ as a local energy configuration of the color code we are led to the possibility of finding other energy configurations that can arise from an alternative choice of stabilizers.

\begin{figure}[h]
    \centering
    \includegraphics[width=5cm]{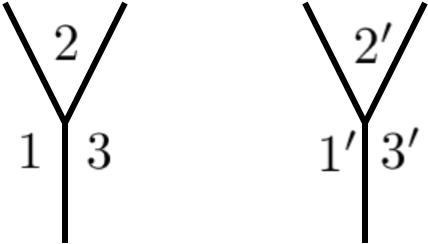}
 \caption{The configuration around a bulk vertex. The integers $n$ and $n'$ with $n\in\{1,2,3\}$ index the positions of the Pauli operators appearing in a configuration. The configurations are chosen such that for a given $n$ the two Pauli operators anticommute, i.e. $\{P_1,P_{1'}\}= \{P_2,P_{2'}\}= \{P_3,P_{3'}\} =0$.}
    \label{fig:YvertexPos}
\end{figure}

To begin a systematic search for the different models we consider a trivalent bulk vertex (See Fig.\ref{fig:YvertexPos}) sharing three faces or six stabilizers, two for each face.
As there are 6 positions to fill with the Pauli $x$, $y$ and $z$ operators we consider the integer solutions of 
$$ a+b+c =6,$$
where $a$, $b$, $c$ are the number of $x$, $y$ and $z$ operators appearing around the vertex. The possible configurations are shown in Table \ref{table:Yconfigs}. 

\begin{table}[ht]
 \centering
 \begin{tabular}{|c|c|cc|cc|c|}
 \hline
 & $a+b+c=6$ & &  Pauli's  & & Energetics & Notation \\
 \hline\hline
 &6+0+0 && $xxxxxx$ && $(0,6,6)$ & \\
 1.& 0+6+0 && $yyyyyy$ && $(6,0,6)$ & $[066]$ \\
 &0+0+6 && $zzzzzz$ && $(6,6,0)$ & \\
 \hline
 &1+5+0/1+0+5 && $xyyyyy$/$xzzzzz$ && $(5,1,6)$/$(5,6,1)$ &  \\
 2.&5+1+0/0+1+5 && $yxxxxx$/$yzzzzz$ && $(1,5,6)$/$(6,5,1)$ & $[156]$ \\
 &5+0+1/0+5+1 && $zxxxxx$/$zyyyyy$ && $(1,6,5)$/$(6,1,5)$ & \\
 \hline
 &2+4+0/2+0+4 && $xxyyyy$/$xxzzzz$ && $(4,2,6)$/$(4,6,2)$ &  \\
 3.&4+2+0/0+2+4 && $yyxxxx$/$yyzzzz$ && $(2,4,6)$/$(6,4,2)$ & $[246]$ \\
 &4+0+2/0+4+2 && $zzxxxx$/$zzyyyy$ && $(2,6,4)$/$(6,2,4)$ & \\
 \hline
 &1+1+4 && $xyzzzz$ && $(5,5,2)$ &  \\
 4.&4+1+1 && $yzxxxx$ && $(2,5,5)$ & $[255]$ \\
 &1+4+1 && $zxyyyy$ && $(5,2,5)$ & \\
 \hline
 &3+3+0 && $xxxyyy$ && $(3,3,6)$ &  \\
 5.&0+3+3 && $yyyzzz$ && $(6,3,3)$ & $[336]$ \\
 &3+0+3 && $xxxzzz$ && $(3,6,3)$ &  \\
 \hline
 &1+2+3/1+3+2 && $xyyzzz$/$xzzyyy$ && $(5,4,3)$/$(5,3,4)$ &  \\
 6.&2+1+3/3+1+2 && $yxxzzz$/$yzzxxx$ && $(4,5,3)$/$(3,5,4)$ & $[345]$ \\
 &2+3+1/3+2+1 && $zxxyyy$/$zyyxxx$ && $(4,3,5)$/$(3,4,5)$ & \\
 \hline
 7.&2+2+2 && $xxyyzz$ && $(4,4,4)$ & $[444]$\\
 \hline
 \end{tabular}
 \caption{Possible configurations at each trivalent vertex. The first column shows the partitions of six which indicate the possible number of the different types of Pauli operators, and the second column indicates the corresponding configuration. The third column specifies the number of stabilizers violated around each vertex upon the application of the Pauli operators, $x$, $y$ and $z$ respectively. The energy configurations are seen as a partition of 12. There is no LU mapping a configuration on some row to a configuration belonging to a different row as the corresponding Hamiltonians have different spectra. 
 The canonical color code and its variants connected by LU's are shown in row 5.}
 \label{table:Yconfigs}
 \end{table}
While the models arising from the possible bulk vertex configurations in Table \ref{table:Yconfigs} can exhibit different features, we are interested in those possessing topological properties of the canonical color code. These are precisely the configurations that obey the {\it anticommuting rule}. This condition drastically reduces the number of possibilities for the number of configurations that we have to consider. 

It is easy to see that only the configurations in rows 5 to 7 of Table. \ref{table:Yconfigs} can possibly satisfy this requirement and so we will only study the models arising from them in the main text. Giving up this condition results in additional lower weight symmetries, for example in the configurations of row 4 of Table \ref{table:Yconfigs}. This results in models with an extensive ground state degeneracy or an enlarged dimension for the code space. We ignore these possibilities in the main text, relegating some of their features to App. \ref{app:[255]cc}. \\

We  denote the models belonging to row 5 of Table \ref{table:Yconfigs} as the $[336]$-color codes. The bulk vertices in this set will be denoted by $[336]$-vertices. Each of these vertices could have configurations from one of the three options shown in row 5. In a similar manner the models from row 6 of Table \ref{table:Yconfigs} are grouped into the set $[345]$-color codes and the corresponding bulk vertices are denoted as the $[345]$-vertices. Finally the models and their bulk vertices in row 7 of Table \ref{table:Yconfigs} are denoted the $[444]$-color codes and the $[444]$-vertices, respectively.

In the following subsections, we will systematically go through operator configurations for the different bulk vertex types mentioned in Table \ref{table:Yconfigs}.

\subsubsection{$[444]$-vertices} 
Each $[444]$ bulk vertex consists of two Pauli, $x$, $y$ and $z$ operators as shown in Table \ref{table:Yconfigs}. There are two types of bulk vertex configurations\,: 

\begin{itemize}
    \item Type\,I - There are $x$, $y$ and $z$ around every vertex for the stabilizers containing that vertex
    \item Type\,II - Two of the Pauli operators are the same and differ from the third on each of the associated stabilizers containing that vertex.
\end{itemize}

\begin{figure}[h]
    \centering
    \includegraphics[width=5cm]{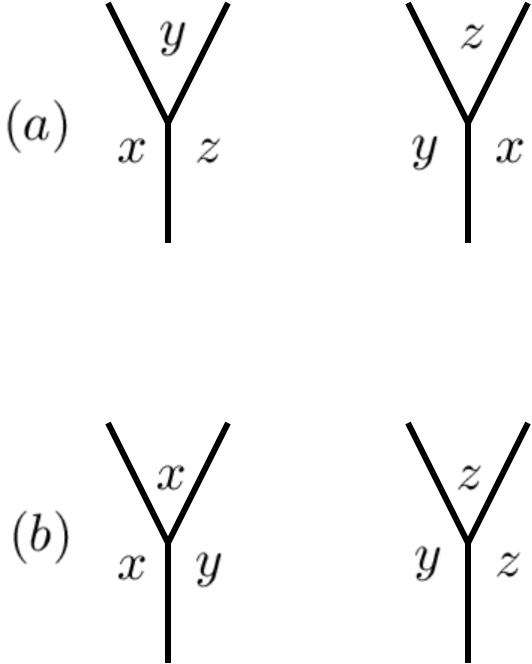}
 \caption{A representative configuration from each set of Type I and Type II $[444]$ bulk vertices. }
    \label{fig:[444]typeI-IIbulkvertexconfigs}
\end{figure}

An example of each of these types is illustrated in Fig. \ref{fig:[444]typeI-IIbulkvertexconfigs}.
Our task now is to find the different configurations corresponding to these two possibilities and group them into different inequivalent classes, that is those that are not related to each other by an LU.\\

\paragraph*{\bf Type I -} We begin with the cyclic configurations around the bulk vertex for the two sets of stabilizers. There are precisely six sets of such cyclic configurations corresponding to the six permutations of $x$, $y$ and $z$. These are $\{(xyz), (xzy), (yxz), (yzx), (zxy), (zyx)\}$, read clockwise around the bulk vertex beginning from position 1 (See Fig. \ref{fig:YvertexPos}). Using the requirement of anticommuting operators we obtain two sets for the pairs of bulk vertex operators, $A =  \{(xyz), (yzx), (zxy)\}$ and $B = \{(xzy),(yxz), (zyx)\}$. In each set we obtain a valid vertex pair by picking any two of the three configurations, for example 
\begin{align} \nonumber
 \left\{\left(\begin{array}{c}xyz \\ zxy\end{array}\right) \right\},    
\end{align}
denotes a vertex pair from set A.
We look for the inequivalent configurations among the three allowed pairs in each set. 

\begin{figure}[h]
    \centering
    \includegraphics[width=9cm]{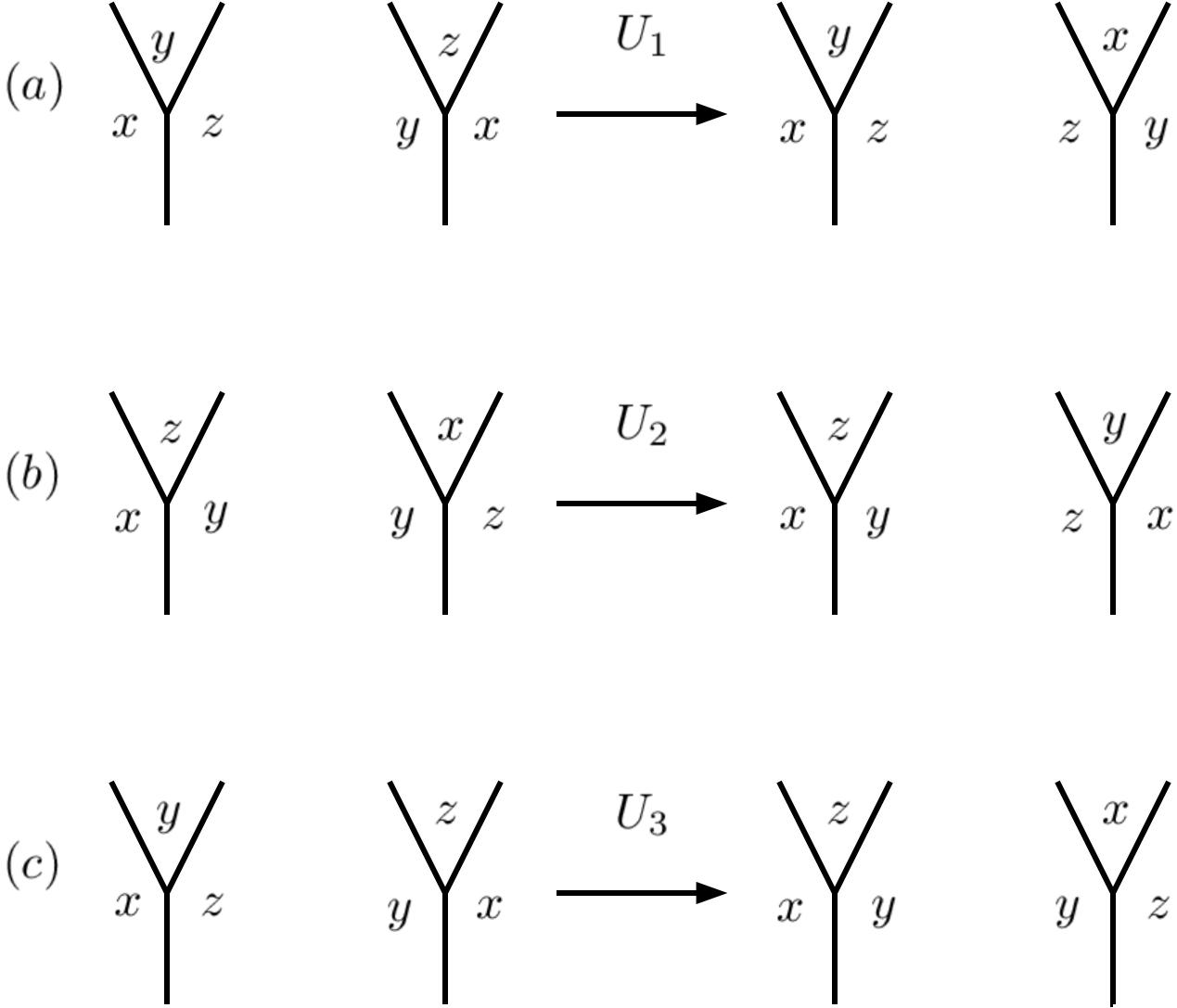}
 \caption{The LU's between Type I $[444]$ bulk vertices. (a)$U_1$ corresponds to a LU between two bulk vertex pairs from set $A$, (b) $U_2$ corresponds to two bulk vertex pairs from set $B$, and (c) $U_3$ corresponds to a LU between a vertex pair from set $A$ and one from set $B$.}
    \label{fig:[444]type1bulkvertexLU}
\end{figure}

Consider the first set $A$ and let the first bulk vertex pair be $(xyz)$ and the second vertex contain the operators from  any of the remaining two choices in the set $A$, namely, $(yzx)$ or $(zxy)$. These two choices are mapped to each other by an LU. In a similar manner we see that just one inequivalent pair of stabilizers is obtained from the second set $B$. Furthermore, we see that there is an LU not only among the different choices within the set, but also between two pairs of bulk vertex operators when one pair of operators is taken from the set $A$ and the other pair from set $B$. The LU's between the different types of $[444]$ bulk vertices are shown in Fig. \ref{fig:[444]type1bulkvertexLU}. The unitary $U_1=C_2$  effects the transformation $\{x \rightarrow z, y\rightarrow x, z\rightarrow y\}$ whereas the unitary $U_2=S$ interchanges $x$ and $y$.
Thus we are left with a lone class of $[444]$ bulk vertex operators of Type I which is given by 
\begin{align} \label{eq:lonetype1}
 \left\{\left(\begin{array}{c}xyz \\ yzx\end{array}\right)\right\}.    
\end{align}

\paragraph*{\bf Type II -} Next we consider the configurations where two operators are the same and the third is different around a single bulk vertex of the pair. We have three sets of such operators, that is when either $x$, $y$ or $z$ is the operator that repeats itself. The stabilizer pairs satisfying the anticommuting rule are given by
\begin{eqnarray}
A = \left\{\begin{array}{ccc} \left(\begin{array}{c}xxy \\ yzz\end{array}\right) & \left(\begin{array}{c}xyx \\ yzz\end{array}\right) & \left(\begin{array}{c}yxx \\ zyz\end{array}\right) \\
\left(\begin{array}{c}xxy \\ zyz\end{array}\right) & \left(\begin{array}{c}xyx \\ zzy\end{array}\right) & \left(\begin{array}{c}yxx \\ zzy\end{array}\right) 
\end{array}\right\} \label{eq:444type21}\\
B = \left\{\begin{array}{ccc} \left(\begin{array}{c}xxz \\ zyy\end{array}\right) & \left(\begin{array}{c}xzx \\ zyy\end{array}\right) & \left(\begin{array}{c}zxx \\ yzy\end{array}\right) \\
\left(\begin{array}{c}xxz \\ yzy\end{array}\right) & \left(\begin{array}{c}xzx \\ yyz\end{array}\right) & \left(\begin{array}{c}zxx \\ yyz\end{array}\right)  
\end{array}\right\} \label{eq:444type22}\\
C = \left\{\begin{array}{ccc} \left(\begin{array}{c}yyx \\ xzz\end{array}\right) & \left(\begin{array}{c}yxy \\ xzz\end{array}\right) & \left(\begin{array}{c}xyy \\ zxz\end{array}\right) \\
\left(\begin{array}{c}yyx \\ zxz\end{array}\right) & \left(\begin{array}{c}yxy \\ zzx\end{array}\right) & \left(\begin{array}{c}xyy \\ zzx\end{array}\right)  
\end{array}\right\}. \label{eq:444type23}
\end{eqnarray}
Before we look for the inequivalent pairs among these three sets we notice that each of these sets can be mapped into each other by a permutation of the $x$, $y$ and $z$ operators and hence it is enough to consider any one of these sets. Take for example a set in Eq. \eqref{eq:444type21} mapping to the set in Eq. \eqref{eq:444type22}, by interchanging $y$ and $z$, and to the set in Eq. \eqref{eq:444type23}, by interchanging $x$ and $y$. Hence we look for the inequivalent pairs of stabilizers in the set in Eq. \eqref{eq:444type21} and we find three of them,
\begin{equation}\label{eq:444type2}
\left\{\begin{array}{ccc}
\left(\begin{array}{c}xxy \\ yzz\end{array}\right) & \left(\begin{array}{c}xyx \\ yzz\end{array}\right) & \left(\begin{array}{c}xxy \\ zyz\end{array}\right) 
\end{array}\right\}.
\end{equation}

\begin{figure}[h]
    \centering
    \includegraphics[width=8cm]{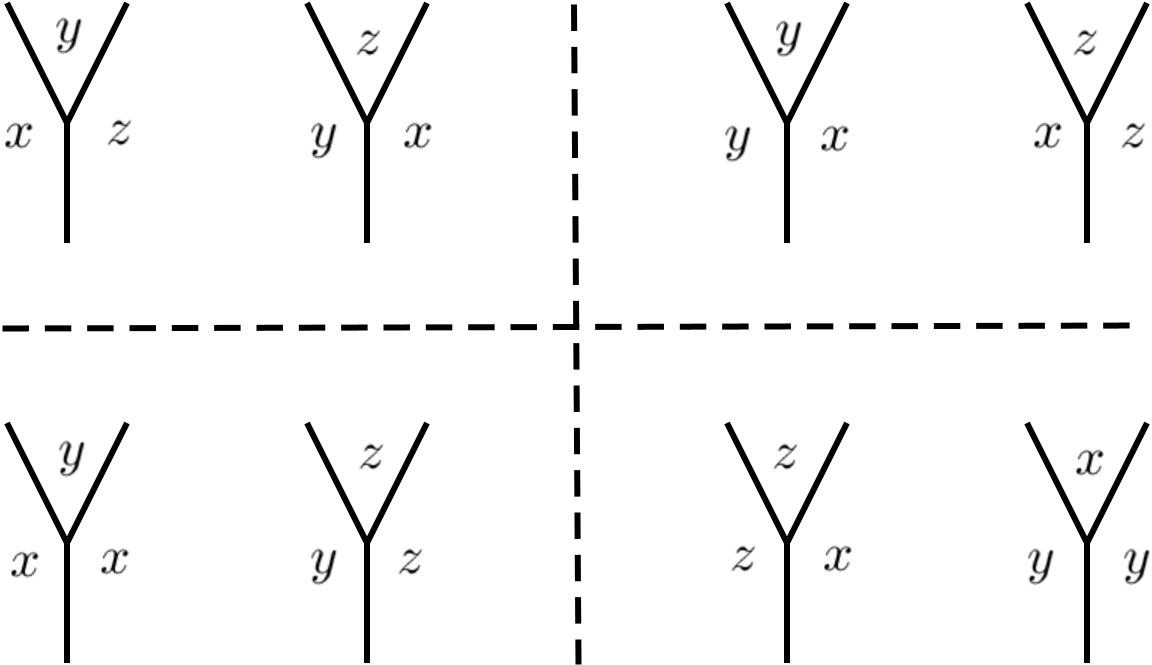}
 \caption{The four inequivalent $[444]$ bulk vertex pair operator configurations from Eqs.\,\eqref{eq:lonetype1} and \eqref{eq:444type2}.}
    \label{fig:[444]bulkvertexconfigs}
\end{figure}

Along with the lone cyclic configuration in Eq.\,\eqref{eq:lonetype1}, we obtain in total four classes of inequivalent vertex configurations as shown in Fig. \ref{fig:[444]bulkvertexconfigs}.\\

\begin{figure}[h]
    \centering
    \includegraphics[width=8cm]{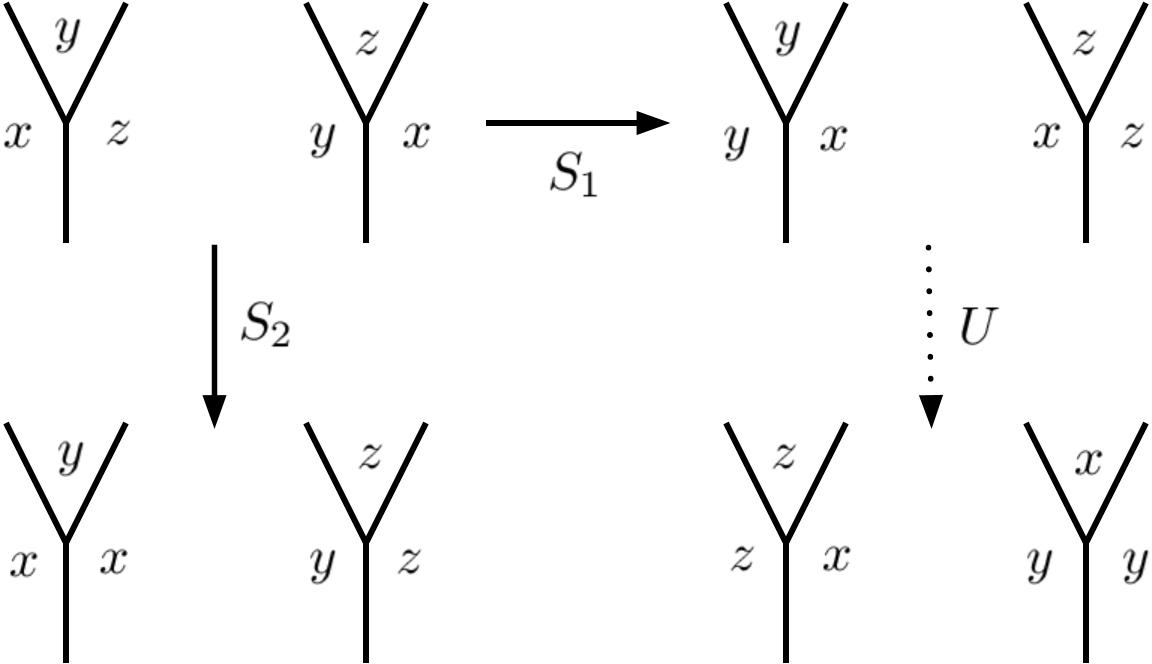}
 \caption{The equivalences between the four $[444]$ bulk vertex configurations. $S_1$ and $S_2$ represent shuffling of stabilizers and $U$ a unitary transformation. In this case the $x$ and $z$ operators in position 3 are shuffled and the unitary effecting the transformation, $\{x\rightarrow y, y\rightarrow z, z\rightarrow x\}$ or $C_1$, is applied.}
    \label{fig:[444]bulkvertexconfigLU}
\end{figure}

\paragraph*{\bf Shuffling stabilizers -} Each bulk vertex shares three stabilizers corresponding to the three colors adjacent to this vertex. We can interchange the stabilizers of the same color and this amounts to an interchange of the Pauli operators in the bulk vertex configuration for the pair. This can also be thought of as relabelling the indices I and II.
This operation leaves the Hamiltonians built out of these stabilizers invariant as it translates to a shuffling of stabilizers in the Hamiltonian. Taking this into account, we see that each of the four classes of the $[444]$ bulk vertices  can be mapped into each other as shown in Fig. \ref{fig:[444]bulkvertexconfigLU}.
Repeating this for the other classes we conclude that there is just a single type of bulk vertex in the $[444]$ case. Any other type of $[444]$ bulk vertex can be mapped into this one by a LU.

\subsubsection{$[345]$-vertices} 
\label{sec:[345]vertices}
Now we find the inequivalent pairs of $[345]$ bulk vertices by looking at the configurations in row 6 of Table \ref{table:Yconfigs}. Among the six possible configurations in row 6, namely, $\{xyyzzz, xyyyzz, xxyzzz, xxxyzz, xxyyyz, xxxyyz\}$, we need to consider just one of them as the rest can be obtained by permutations of the Pauli operators and are thus LU equivalent to each other. For example $xxyyyz$ can be obtained from $xyyzzz$ by replacing 
$x,y,z$ with $z,x,y$, respectively.
Thus we consider permutations of $xyyzzz$ to build the stabilizers around the $[345]$ bulk vertices.

\begin{figure}[h]
    \centering
    \includegraphics[width=4cm]{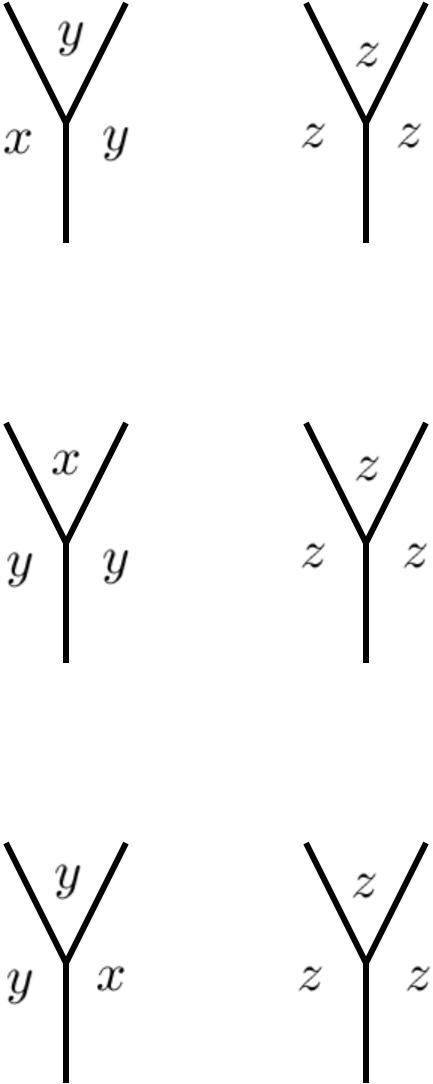}
 \caption{The three inequivalent $[345]$ bulk vertex configurations.}
    \label{fig:[345]bulkvertexconfigs}
\end{figure}

We have three sets of operator pairs that satisfy the anticommuting rule for the $xyyzzz$ configurations,
\begin{align}
A &= \left\{\begin{array}{cccc}
\left(\begin{array}{c}xyy \\ zzz\end{array}\right) & \left(\begin{array}{c}xzy \\ zyz\end{array}\right) & \left(\begin{array}{c}xyz \\ zzy\end{array}\right) & \left(\begin{array}{c}xzz \\ zyy\end{array}\right)
\end{array}\right\}\label{eq:345Yvertex1} \\ 
B &= \left\{\begin{array}{cccc}
\left(\begin{array}{c}yxy \\ zzz\end{array}\right) & \left(\begin{array}{c}zxy \\ yzz\end{array}\right) & \left(\begin{array}{c}yxz \\ zzy\end{array}\right) & \left(\begin{array}{c}zxz \\ yzy\end{array}\right) 
\end{array}\right\}\label{eq:345Yvertex2} \\
C &= \left\{\begin{array}{cccc}
\left(\begin{array}{c}yyx \\ zzz\end{array}\right) & \left(\begin{array}{c}yzx \\ zyz\end{array}\right) & \left(\begin{array}{c}zyx \\ yzz\end{array}\right) & \left(\begin{array}{c}zzx \\ yyz\end{array}\right) 
\end{array}\right\}.\label{eq:345Yvertex3}
\end{align}
 Notice that in each set given by Eqs. \eqref{eq:345Yvertex1}, \eqref{eq:345Yvertex2}, \eqref{eq:345Yvertex3} the operator pairs can be obtained from each other by shuffling. Thus each set can be represented by a single pair of operators. On the other hand to relate the operator pairs belonging to different sets we need to rotate the bulk vertex. This feature is different from what we saw for the $[444]$-vertices and it leads to three inequivalent models on an arbitrary trivalent lattice. The three inequivalent $[345]$ bulk vertices are shown in Fig. \ref{fig:[345]bulkvertexconfigs}.

\subsubsection{$[336]$-vertices} 
Bulk vertices corresponding to the canonical color code belong to the $[336]$ type and are listed in row 5 of table \ref{table:Yconfigs}. Analogous to the arguments in the earlier cases it is easy to see that it is enough to consider the pairs of stabilizers arising from considering bulk vertices that include $\{xxxzzz\}$ as other configurations are obtained by permuting the Pauli matrices. The possible configurations obtained by rearranging the three $x$ and $z$ operators around the bulk vertex obeying the anticommuting rule are
\begin{equation}
\left\{\begin{array}{cccc}
\left(\begin{array}{c}xxx \\ zzz\end{array}\right) & \left(\begin{array}{c}zxx \\ xzz\end{array}\right) & \left(\begin{array}{c}xzx \\ zxz\end{array}\right) & \left(\begin{array}{c}xxz \\ zzx\end{array}\right)
\end{array}\right\}.
\end{equation}
However, it is rather easy to check that there is just one inequivalent class in this set as the four pairs of operators can be mapped into each other by a simple rearrangement of the stabilizers. Thus we conclude that there is just one $[336]$ bulk vertex possible and every other configuration is LU equivalent to this one. \\~\\

\noindent\fbox{%
    \parbox{0.47\textwidth}{%
        There is one equivalence class of bulk vertex operator configuration for the [444] and [336] vertex types and three inequivalent ones for the [345] bulk vertex type.
    }%
}
   
\subsection{Boundary vertex configurations}
\begin{figure}[h]
    \centering
    \includegraphics[width=5cm]{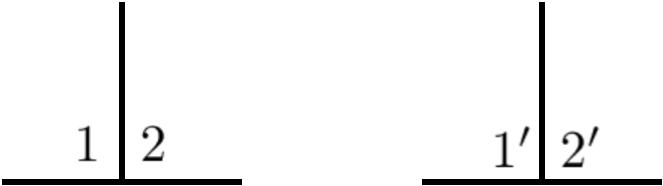}
 \caption{The four positions of a boundary vertex for a stabilizer pair are indexed by $n$ and $n'$ with $n\in\{1,2\}$. The configurations are such that the Pauli operators anticommute for a given $n$, i.e. $\{P_1,P_{1'}\} = \{P_2,P_{2'}\}=0$.}
    \label{fig:boundaryvertexconfigsPos}
\end{figure}

For the vertices ending on the boundaries of the finite lattice (See Fig. \ref{fig:boundaryvertexconfigsPos}), there are just four places to fill with the Pauli operators as opposed to the six spots around the bulk trivalent vertices. Thus possible configurations of the Pauli operators in this case are the integer solutions of 
$$a+b+c =4.$$

\begin{table}[ht]
\centering
\begin{tabular}{|c|c|cc|cc|c| }
\hline
& $a+b+c=4$ && Pauli's && Energetics & Notation \\
\hline \hline
  &4+0+0 && $xxxx$ && $(0,4,4)$ &\\
 1.&0+4+0 && $yyyy$ && $(4,0,4)$ & [044] \\
 &0+0+4 && $zzzz$ && $(4,4,0)$ & \\
 \hline
 &3+1+0/1+3+0 && $xxxy$/$xyyy$ && $(1,3,4)$/$(3,1,4)$ & \\
 2.&3+0+1/1+0+3 && $xxxz$/$xzzz$ && $(1,4,3)$/$(3,4,1)$ & [134] \\
 &0+1+3/0+3+1 && $yzzz$/$yyyz$ && $(4,3,1)$/$(4,1,3)$ &  \\
 \hline
 &2+1+1 && $xxyz$ && $(2,3,3)$ &  \\
 3.&1+2+1 && $xyyz$ && $(3,2,3)$ & [233] \\
 &1+1+2 && $xyzz$ && $(3,3,2)$ & \\
 \hline
 &2+2+0 && $xxyy$ && $(2,2,4)$ & \\
 4.&2+0+2 && $xxzz$ && $(2,4,2)$ & [224] \\
 &0+2+2 && $yyzz$ && $(4,2,2)$ & \\
 \hline
 \end{tabular}
 \caption{Possible configurations at each boundary vertex. The first column shows the partitions of four indicating the number of the different types of Pauli operators, whereas the second column indicates the corresponding configuration. The third column specifies the number of stabilizers violated around each vertex upon the application of the Pauli operators, $x$, $y$ and $z$ respectively. The energy configurations are seen as a partition of 8. There is no LU mapping a configuration on some row to a configuration belonging to a different row as the Hamiltonians built out of them have different spectra.}
 \label{table:Bconfigs}
 \end{table}

However, if we take into the account the anticommuting rule we are left with just two sets of solutions, namely, the rows 3 and 4 in Table.\,\ref{table:Bconfigs}. While there are three possibilities for the operators in each case, they can be mapped into each other using LU's. So we take the $\{(2,1,1), (2,2,0)\}$ configurations as the representative elements in this class. Upon the application of Pauli operators on the boundary vertex qubits, leads to an energy configuration of either $\{[233],~ \rm{or}~ [224]\}$ for the $\{(2,1,1), (2,2,0)\}$ boundary vertices respectively as seen from Table \ref{table:Bconfigs}). We label these boundary vertices by its energetics just as in the case of the bulk vertex configurations. We explore further details of these two vertex types in the following subsections. 

\subsubsection{$[233]$-vertices}
\begin{figure}[h]
    \centering
    \includegraphics[width=6cm]{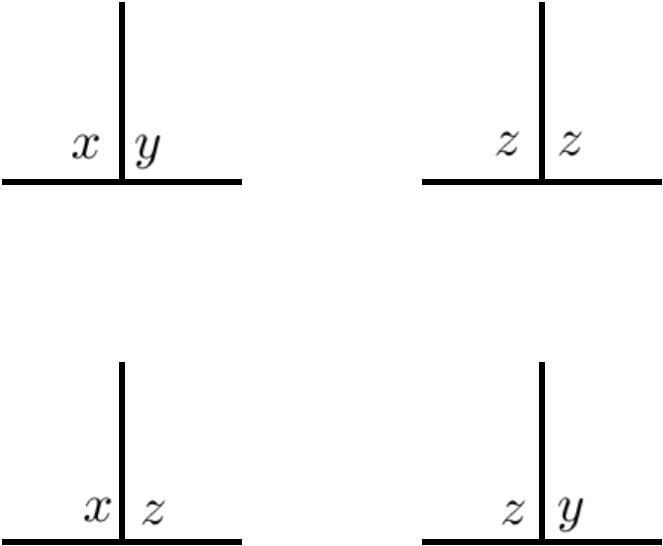}
 \caption{Boundary vertex configurations of the $[233]$ type. They are inequivalent up to an interchange of the stabilizers.}
    \label{fig:[233]boundaryvertexconfigs}
\end{figure}

There are three choices for the vertices, namely $xyzz$, $xyyz$ and $xxyz$ which are LU equivalent to each other.  Considering $xyzz$, we find two sets of stabilizer pairs that are compatible with the anticommuting rule,
\begin{align}
A &= \left\{\begin{array}{cc}
\left(\begin{array}{c}xy \\ zz\end{array}\right) &\left(\begin{array}{c}yx \\ zz\end{array}\right) \\
\end{array}\right\}\label{eq:boundvertconfig1} \\
B &= \left\{\begin{array}{cc}
\left(\begin{array}{c}xz \\ zy\end{array}\right) &\left(\begin{array}{c}zx \\ yz\end{array}\right)
\end{array}\right\}\label{eq:boundvertconfig2}.
\end{align}
The operator pairs in Eq. (\ref{eq:boundvertconfig1}) are mapped into each other with a LU and so are the operator pairs in Eq. (\ref{eq:boundvertconfig2}). However, there is no LU between these two sets implying there are two inequivalent classes of $[233]$ boundary vertices as shown in Fig. \ref{fig:[233]boundaryvertexconfigs}. Nevertheless, they can be transformed into each other by a shuffling of the stabilizers.

\subsubsection{$[224]$-vertices}
\begin{figure}[h]
    \centering
    \includegraphics[width=6cm]{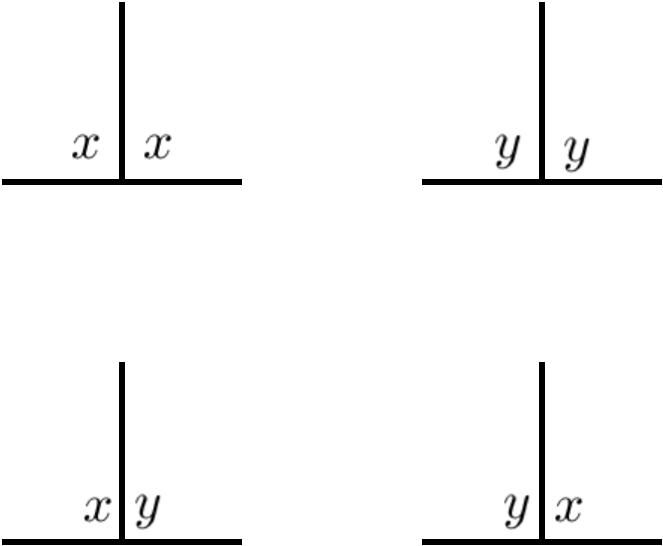}
 \caption{Boundary vertex configurations of the $[224]$ type. They are inequivalent up to an interchange of the stabilizers.}
    \label{fig:[224]boundaryvertexconfigs}
\end{figure}

We have three choices for the configurations, $xxyy$, $yyzz$, $zzxx$ that are LU equivalent to each other. There are two sets of stabilizer pairs resulting from $xxyy$,
\begin{equation}
\left\{\begin{array}{cc}
\left(\begin{array}{c}xx \\ yy\end{array}\right) & \left(\begin{array}{c}xy \\ yx\end{array}\right)  
\end{array}\right\}.
\end{equation}
These are mapped into each other by a shuffling of the stabilizers as in the earlier case. These configurations are shown in Fig. \ref{fig:[224]boundaryvertexconfigs}.

\subsection{Corner vertex configurations}
\begin{figure}[h]
    \centering
    \includegraphics[width=5cm]{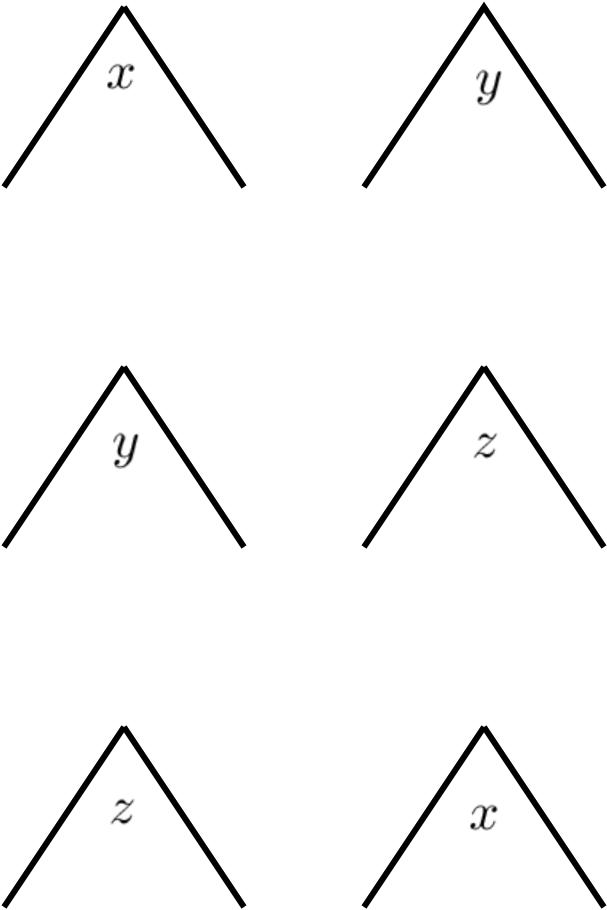}
 \caption{The three possible corner vertex configurations obeying the anticommuting rule. Note that the corner vertices are bivalent.}
    \label{fig:cornervertexconfigs}
\end{figure}

At the corners of a finite lattice we have just two positions for the opeator pair to be filled with the Pauli operators. With the anticommuting property we find that there are precisely three possibilities for the operator pairs at the corner, namely, $x$ and $y$, or $y$ and $z$, or $z$ and $x$ (See Fig. \ref{fig:cornervertexconfigs}).

\subsection{Edge configurations} 
\label{subsec:edge}
\begin{figure}[h]
    \centering
    \includegraphics[width=4cm]{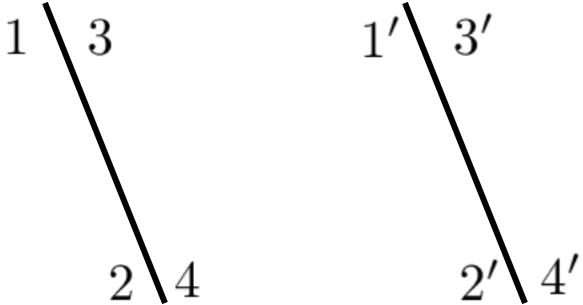}
 \caption{The edges for an operator pair. The numbers index the physical qubit and hence the position of the Pauli operators acting on them. To satisfy the anticommuting rule the Pauli operators at $n$ and $n'$ with $n\in\{1,2,3,4\}$, must anticommute.}
    \label{fig:econfigPos}
\end{figure}

As mentioned earlier there are three types of edge configurations that could lead to models with varying P-distances for the logical operators. For each edge on a stabilizer pair (See Fig. \ref{fig:econfigPos})
we have four possible positions, with each position containing two neighbouring qubits. Thus for each position we require two Pauli operators and we choose them such that they satisfy the anticommuting rule and they lead to commuting stabilizers.

\begin{figure}[h]
    \centering
    \includegraphics[width=5cm]{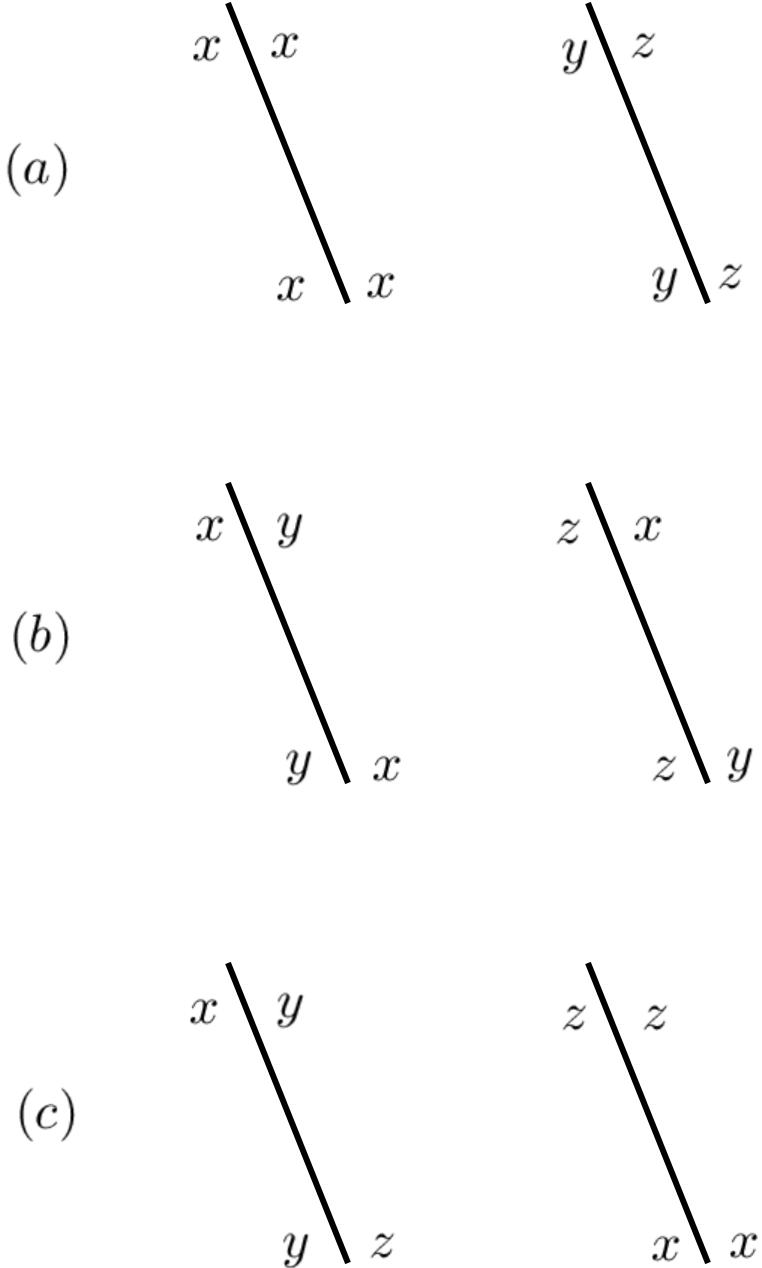}
 \caption{An example for each type of edge configuration leading to logical operators with different P-distances. Note that the stabilizers formed out of these edge configurations will commute with each other. (a) Homogeneous configuration where there is no mixing of the Pauli operators. (b) Partially mixed edge configuration. (c) Fully mixed edge configuration.}
    \label{fig:3econfigs}
\end{figure}

The fully mixed logicals that result in models with infinite P-distance are obtained from the edge configurations,
\begin{equation}\label{eq:edgemix}
\{(xy), (yz), (zx)\}~\textrm{or}~\{(yx), (zy), (xz)\}.    
\end{equation}
The four positions for an edge in the stabilizer pair can be filled using the configurations in either of the sets.
The configurations giving partially mixed logicals are either 
\begin{equation}\label{eq:edgepartmix}
\{(xx), (yz), (zy)\}~\textrm{or}~\{(xz), (yy), (zx)\}~\textrm{or}~\{(xy), (yx), (zz)\}.    
\end{equation}
Finally the choice leading to no mixing or an homogeneous edge configuration is obtained using
\begin{equation}\label{eq:edgehom}
\{(xx), (yy), (zz)\}.    
\end{equation}
An example for each case is illustrated in Fig. \ref{fig:3econfigs}.
\\~\\
\noindent\fbox{%
    \parbox{0.47\textwidth}{%
    There are three types of edges depending on the \textit{mixing} of the four operator positions with respect to the stabilizer content, namely, homogeneous, partially mixed, and fully mixed. 
        }%
}

\section{Non-CSS color codes on hexagonal lattices}
\label{sec:models666488}
We are now in a position to build the full stabilizer codes by combining the inequivalent vertex and the different edge configurations according to what type of code we require. The construction works for both regular and irregular trivalent lattices, i.e., with and without translational invariance. We can work on both finite lattices with boundaries and on lattices that discretize a closed surface. We begin with the hexagonal lattice on a torus.

The construction of the stabilizer pairs for the hexagonal lattice proceeds by consistently joining the vertex and edge configurations keeping in mind the requirements of the code and this is best illustrated through figures. We will carry out the procedure for each vertex type separately in what follows. We will first show the method locally in each case and then give examples on a full lattice. By local construction we mean that the construction is carried out on a small part of the lattice, mostly on two or three edges that do not form a closed figure. While it is always possible to do this for the different vertex types and edge configurations the method fails in some instances when we try to close the figures. In this section we will only provide the definitions of the various non-CSS color codes and leave the analysis of their properties in sections following this. 

\subsection{Construction of $[444]$-color codes}
\begin{figure*}[h]
    \centering
    \includegraphics[width=16cm]{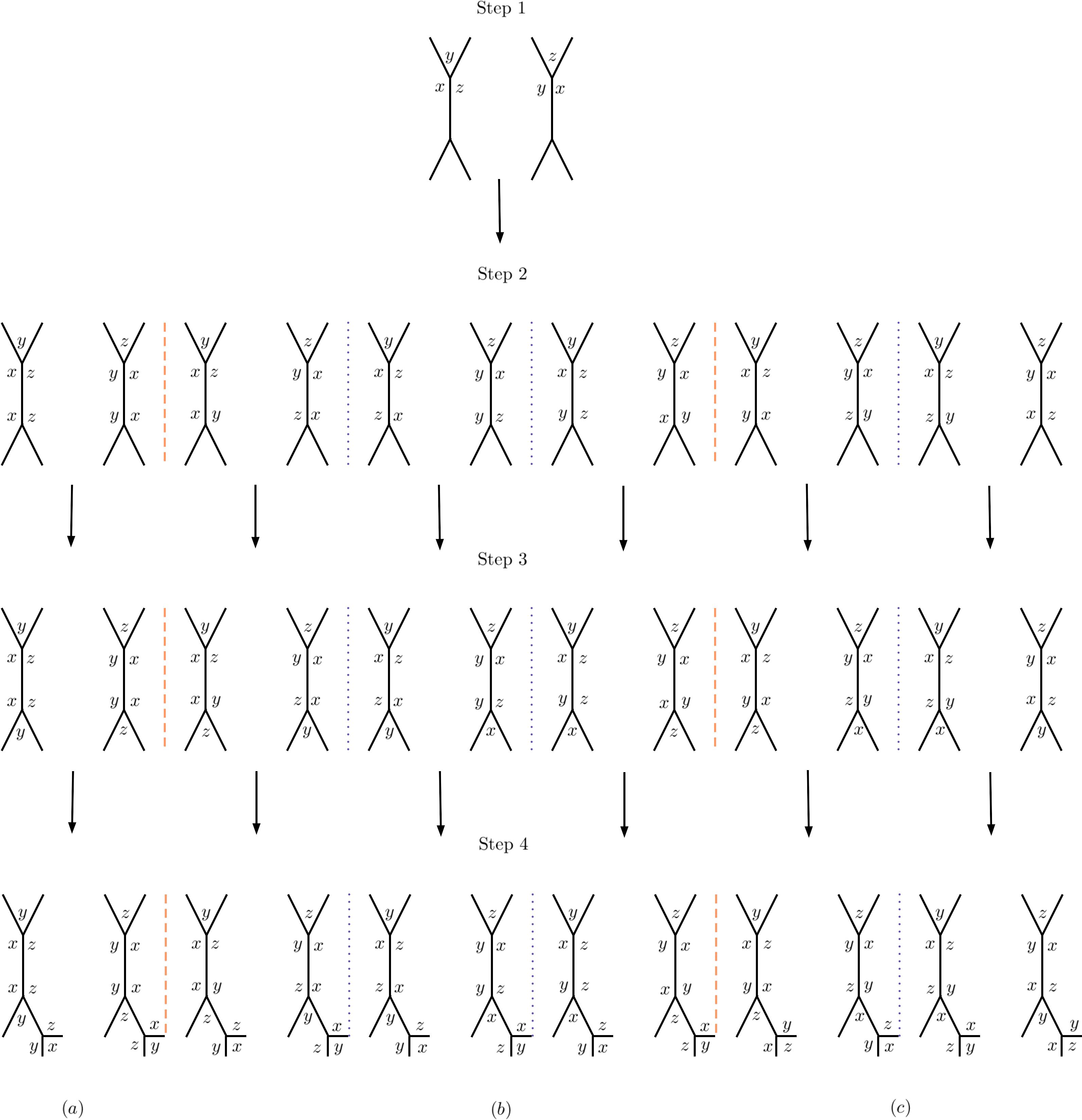}
 \caption{The different steps involved in the construction of the $[444]$-color code in a local part of a trivalent lattice. In step 2 (a) homogeneous, (b) partially mixed and (c) fully mixed edge configurations are shown. There are more possibilities at step 4 and we show just one of them.}
    \label{fig:[444]cclocalconstruction}
\end{figure*}

The local construction proceeds using the following steps :
\begin{enumerate}
\item Start with a pair of two adjacent trivalent $[444]$-vertices connected by an edge 
and choose a $[444]$-vertex for one of the trivalent vertices. This choice is arbitrary and there is no loss of generality for a different choice.
\item Fill the edge with an allowed edge configuration. This can be done in three ways; by placing either a fully mixed edge configuration (there are two choices for this) or a partially mixed edge configuration (there are three choices for this) or an homogeneous edge configuration. 
\item Complete the configuration on the second vertex such that it is also of the $[444]$-vertex type. 
\item Using the second $[444]$-vertex we can continue the above steps to further extend the code on the lattice.
\end{enumerate}
The procedure for the construction is summarized in Fig. \ref{fig:[444]cclocalconstruction}.

While this procedure can be continued in a consistent way for an arbitrary trivalent lattice with or without boundaries, we will look at examples of trivalent lattices such as the $(6,6,6)$- and $(4,8,8)$-lattices.\\ 

\begin{figure}[h]
    \centering
    \includegraphics[width=8.5cm]{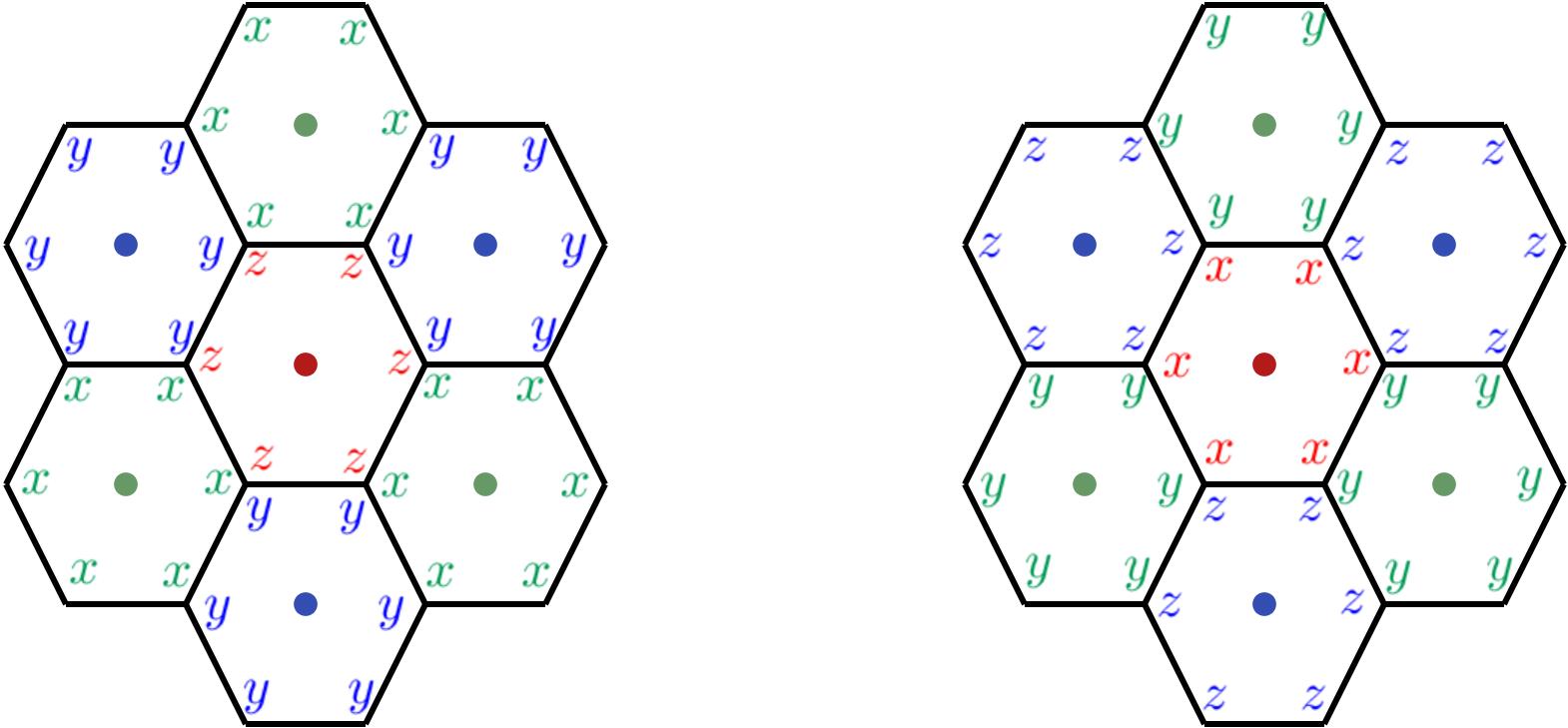}
 \caption{The canonical $[444]$-color code. There is no mixing in each edge for this case. }
    \label{fig:[444]ccnomixing}
\end{figure}

\paragraph{\bf Hexagonal lattice -}\, We can construct consistent $[444]$-color codes both with and without translational invariance on the hexagon lattice. Starting with the former, the stabilizers for the simplest $[444]$-color code that we can write down on the hexagon lattice are shown in Fig. \ref{fig:[444]ccnomixing}.
This model contains edge configurations of the homogeneous type and around every vertex we have all the three Pauli operators. 
We will denote this model as the {\it canonical $[444]$-color code}. Every red face has the same set of stabilizers as the original color code, whereas on the blue (green) faces we have $Y$ and $Z$ ($X$ and $Y$) type of stabilizers. It is clear that these stabilizers can also be defined on an arbitrary trivalent, tricolorable lattice. 

\begin{figure}[h]
    \centering
    \includegraphics[width=8cm]{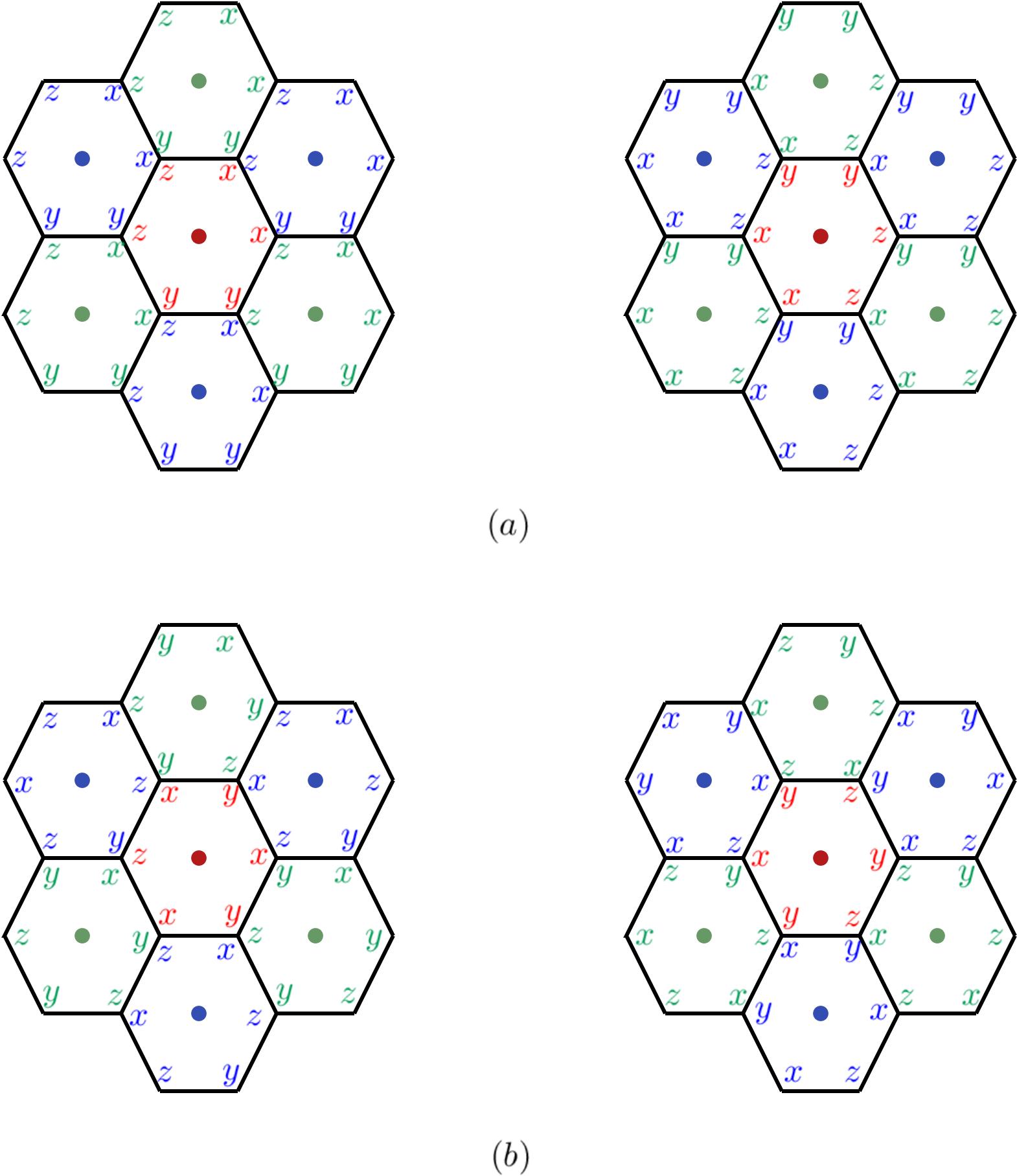}
 \caption{The $[444]$-color code with edges that are  partially mixed (a) and fully mixed (b).}
    \label{fig:[444]ccpartialfullmixing}
\end{figure}

\begin{figure}[h]
    \centering
    \includegraphics[width=8cm]{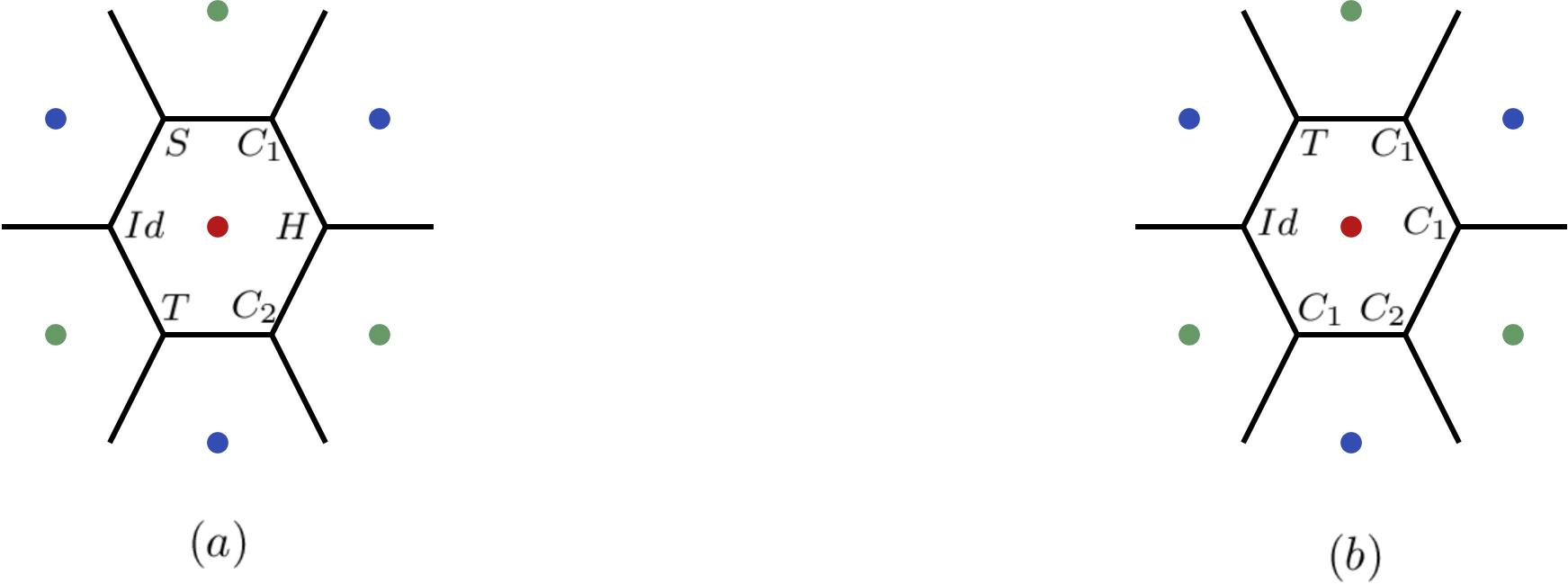}
 \caption{(a) The LU between the translationally invariant $[444]$-color code with (a) partially mixed, (b) fully mixed edge configurations and the canonical $[444]$-color code.}
    \label{fig:[444]ccLUtocanonicalwti}
\end{figure}

We can now construct the $[444]$-color codes where the edge configurations are partially or fully mixed as shown in Fig. \ref{fig:[444]ccpartialfullmixing}.
The models with partially and fully mixed edge configurations are equivalent to the canonical $[444]$-color code by an LU. This is demonstrated explicitly in Fig. \ref{fig:[444]ccLUtocanonicalwti}.

The non-translationally invariant $[444]$-color codes with partially and fully mixed edge configurations are shown in App. \ref{app:NTIccs[444]}. The codes on the square-octagon lattice, with and without translational invariance, can be found in App. \ref{app:488latticemodels}.

\subsection{Construction of $[345]$-color codes}
The construction here goes along the lines of the $[444]$-color codes. However there are two crucial differences from the $[444]$ case. First, there are three types of inequivalent $[345]$ bulk vertices to choose from, Eqs. (\ref{eq:345Yvertex1}), (\ref{eq:345Yvertex2}) and (\ref{eq:345Yvertex3}). We will begin our construction from the first configuration in each of these inequivalent classes guaranteeing three inequivalent stabilizer codes. Second, we have more possibilities for combining the chosen bulk vertex with an edge configuration. We find two types of {\it combined vertex-edge} configurations. 
\begin{enumerate}
\item The combined vertex-edge configurations admit deconfined anyons as in the color codes. That is either two or four stabilizers are excited by a local string configuration exciting the end points of the string as shown in Fig. \ref{fig:[345]deconfinededgeconfigs}.
\begin{figure}[h]
    \centering
    \includegraphics[width=8cm]{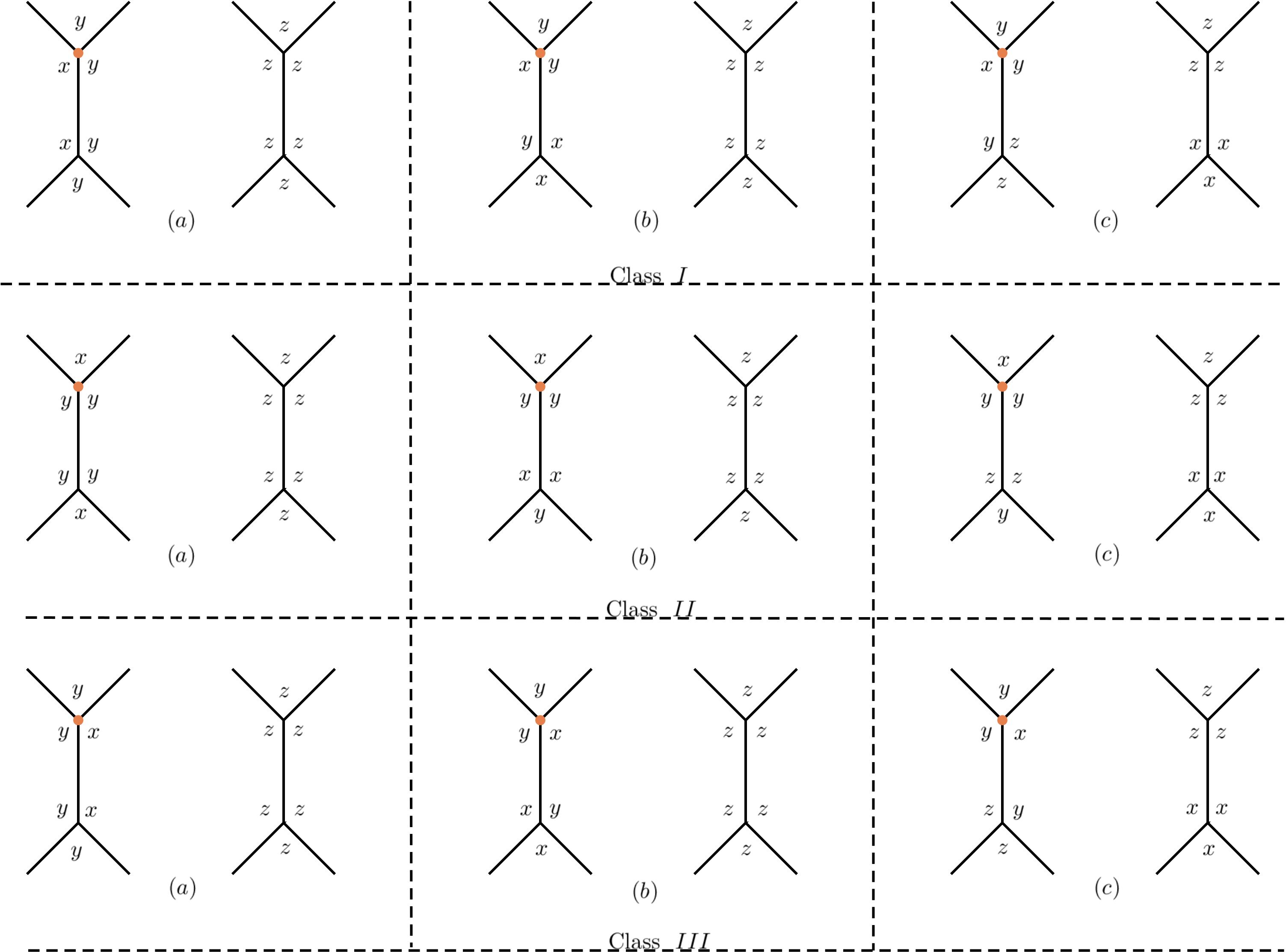}
 \caption{The deconfined edge configurations possible for three inequivalent $[345]$-color codes (See Fig. \ref{fig:[345]bulkvertexconfigs}), indicated by an orange dot. The edge configurations can be either (a) homogeneous, (b) partially or (c) fully mixed.} 
    \label{fig:[345]deconfinededgeconfigs}
\end{figure}
String operators formed out of the edge configurations in each of the figures in Fig. \ref{fig:[345]deconfinededgeconfigs} lead to deconfined anyons. For example in Class I (b), the string operators, $(yx)$, $(zz)$ excite two of the four stabilizers at the end points of the string, and the operator $(xy)$ excites all the four stabilizers. This is similar to what occurs in the $[336]$ color code and the $[444]$-color codes introduced earlier.
We will denote the models built out of these as the {\it deconfined $[345]$-color codes}.
\item The vertex-edge combines lead to confined anyons along the direction defined by the edge. This follows when the local string configurations excite either two or three of the four stabilizers located at the end points of the string (See Fig. \ref{fig:[345]confinededgeconfigs}). 
\begin{figure}[h]
    \centering
    \includegraphics[width=8cm]{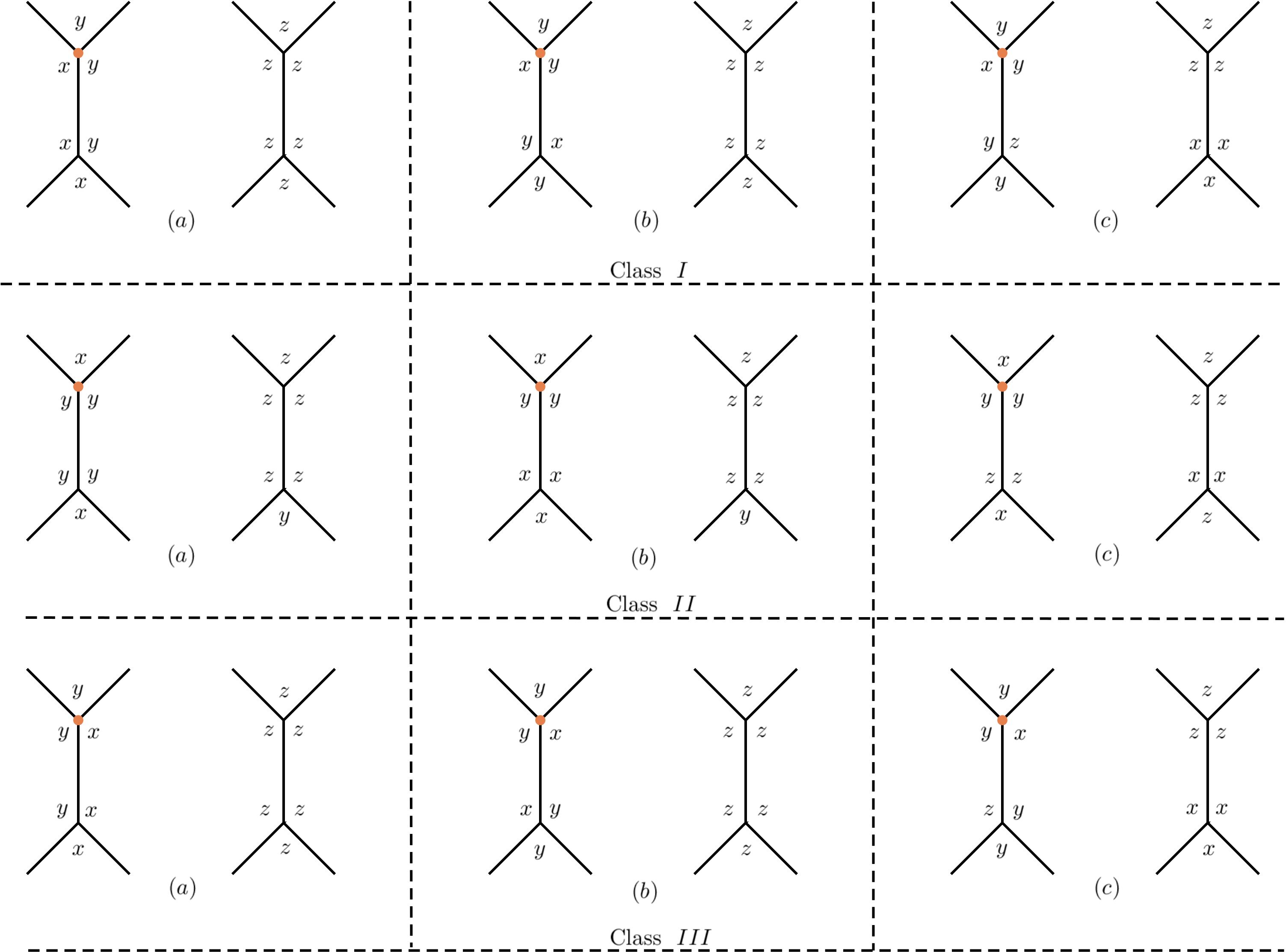}
 \caption{The confined edge configurations possible for the three inequivalent classes of a $[345]$-color code  (See Fig. \ref{fig:[345]bulkvertexconfigs}), indicated by an orange dot. The edge configurations can be either (a) homogeneous, (b) partially or (c) fully mixed.}
    \label{fig:[345]confinededgeconfigs}
\end{figure}
String operators formed out of the edge configurations in each of the figures can lead to confined anyons. For example in Class I (b), the string operators, $(yx)$, $(xy)$ excite three of the four stabilizers at the end points of the string, and the operator $(zz)$ excites two of the four stabilizers. Extending this string along a particular direction can lead to an {\it energetic string} that not only excites the stabilizers at the end points of the string but also the stabilizers along the path of the string.  Depending on the rest of the code it is possible to create deconfined and confined anyons with appropriate choices for the string operators.
We will call the models built out of these configurations as the {\it confined $[345]$-color codes}.
\end{enumerate}

\paragraph{\bf Deconfined $[345]$-color codes -}\, The steps for the local construction in the deconfined setting are illustrated in  Fig. \ref{fig:[345]ccdeconfinedlocalconstruction}.
As in the $[444]$-color codes we can construct stabilizer codes with or without translational invariance. The construction also holds for an arbitrary trivalent, tricolorable lattice as it is local. We will call the codes where none of the edge configurations are mixed as the {\it canonical $[345]$-deconfined color codes}. 

\begin{figure}[h]
    \centering
    \includegraphics[width=8cm]{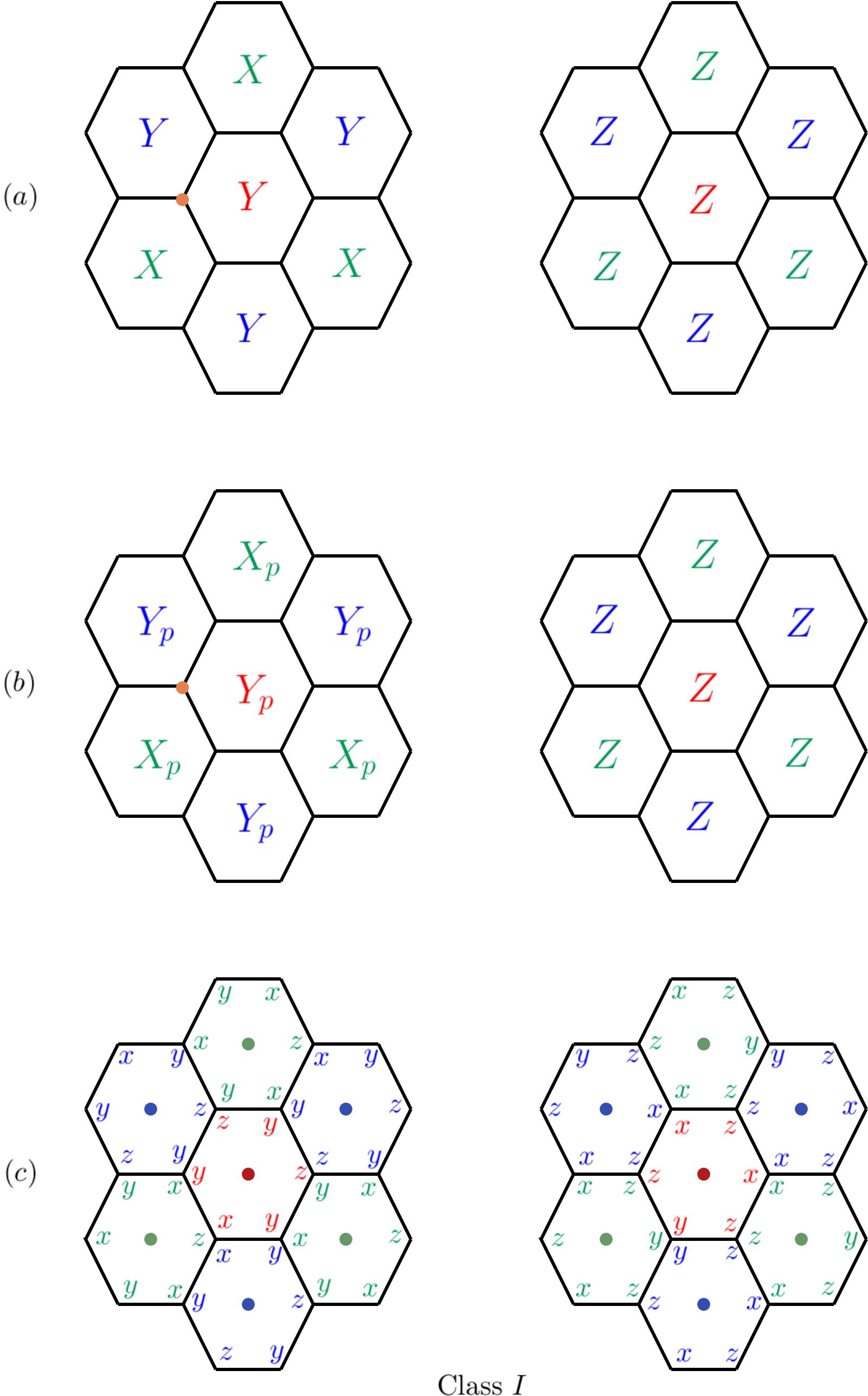}
 \caption{The translationally invariant deconfined $[345]$-color codes in the Class $I$ type for (a) homogeneous, (b) partially and (c) fully mixed edge configurations.}
    \label{fig:[345]ccdeconfinedClassI}
\end{figure}

We can have three inequivalent translational invariant codes on the hexagon lattice corresponding to the $[345]$-bulk vertices shown in Fig. \ref{fig:[345]bulkvertexconfigs}. The edge configurations are such that the vertex-edge combines are of the deconfined type as shown in Fig. \ref{fig:[345]deconfinededgeconfigs}.
Furthermore the edge configurations can be either homogeneous, partially or fully mixed 
as shown in Fig. \ref{fig:[345]ccdeconfinedClassI}.

The non-translationally invariant versions of the deconfined $[345]$-color codes are written down in App. \ref{app:NTIccs[345]deconfined} and the codes on the square-octagon lattice in App. \ref{app:488latticemodels}.

\begin{figure}[h]
    \centering
    \includegraphics[width=8cm]{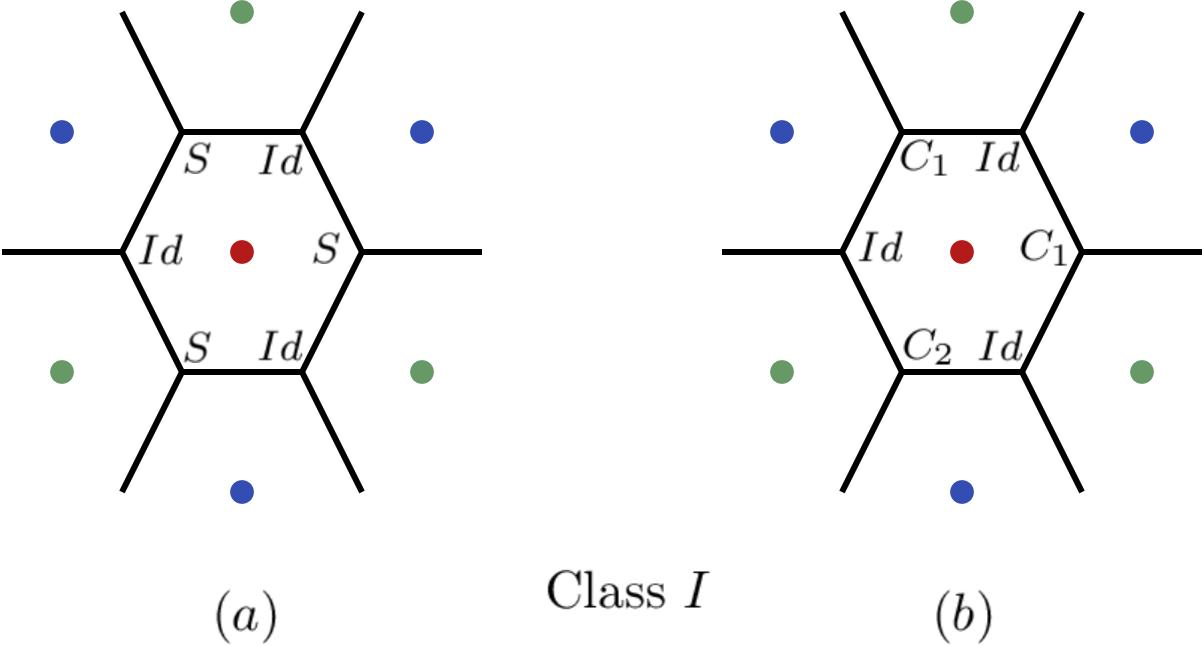}
 \caption{The LU's between the Class I canonical deconfined $[345]$-color code on the hexagon lattice and the (a) partially mixed, (b) fully mixed translationally invariant deconfined $[345]$-color codes.}
    \label{fig:[345]ccdeconfinedLUtocanonicalClassI}
\end{figure}

The translationally invariant $[345]$-color codes with partially and fully mixed edge configurations are equivalent to the canonical $[345]$-deconfined color code by an LU as shown in Fig. \ref{fig:[345]ccdeconfinedLUtocanonicalClassI}. \\ 

\paragraph{\bf Confined $[345]$-color codes - }\, 
Fig. \ref{fig:[345]ccconfinedlocalconstruction} shows the steps for the local construction of the confined $[345]$-color codes. Using this we can proceed to construct the confined $[345]$-color codes on different lattices. However, we do not obtain consistent color codes for any even sided face in the lattice. By this we mean that either not all the vertex-edge combines are confined, i.e. some come out to be deconfined, or not all the vertices are of the $[345]$ bulk vertex type. For example while it is possible to define a confined $[345]$-color code on the square and hexagon it is not possible to obtain an octagon or a decagon (See App. \ref{app:NTIccs[345]confined}). 

\begin{figure}[h]
    \centering
    \includegraphics[width=8cm]{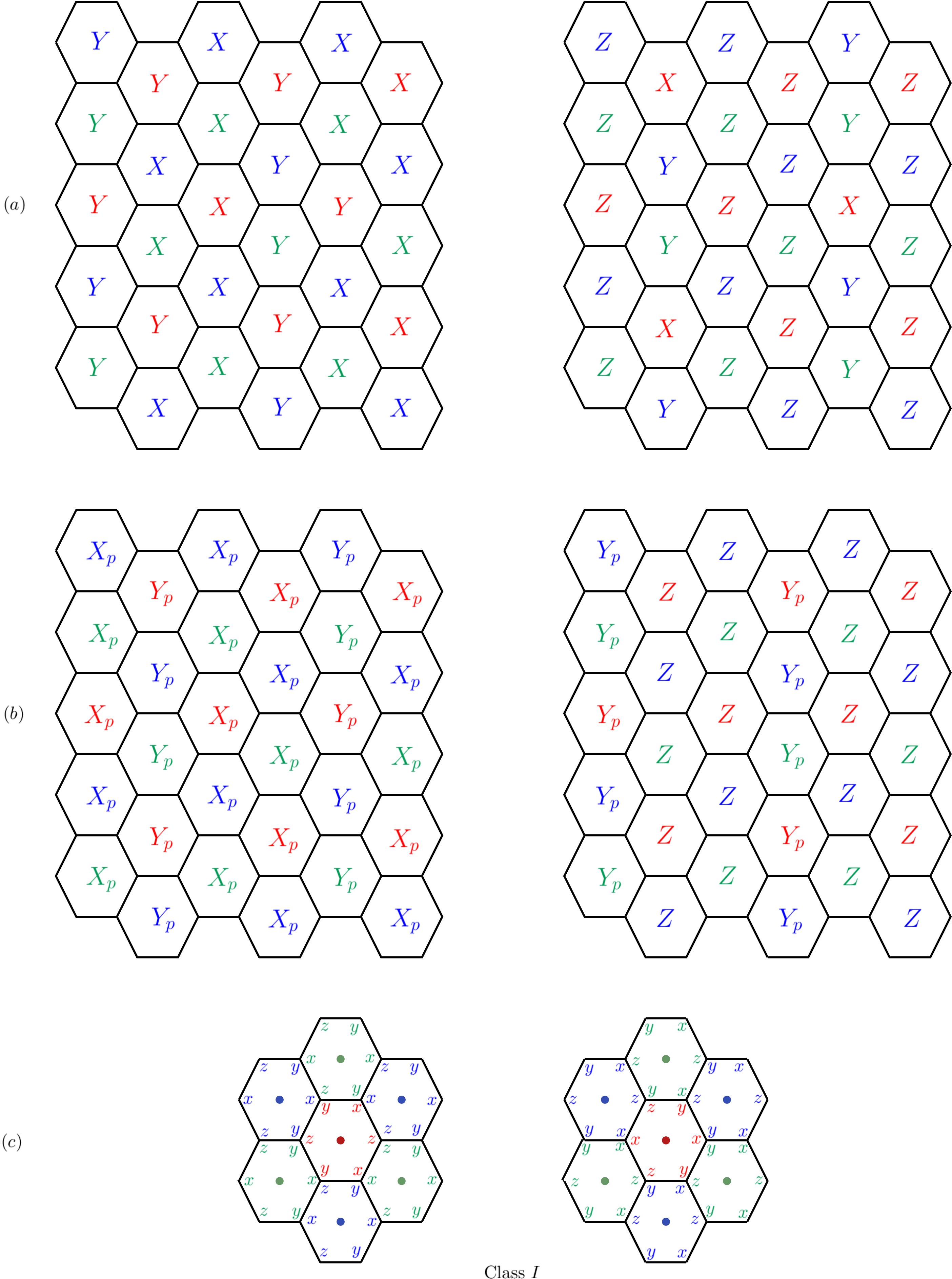}
 \caption{The translatioanlly invariant Class I canonical confined $[345]$-color codes on the hexagonal lattice with (a) homogeneous, (b) partially mixed and (c) fully mixed edge configurations.}
    \label{fig:[345]ccconfinedClassIwti}
\end{figure}

The consistent models on the hexagonal lattice, with translational invariance for Class I, is shown in Figs. \ref{fig:[345]ccconfinedClassIwti}, 
models in the classes II and III with translational invariance are explained in App. \ref{app:[345]classIIclassIII} and models without translational invariance are shown in Figs. \ref{fig:[345]ccconfinedpartialmixingwoti} and \ref{fig:[345]ccconfinedfullmixingwoti} of App. \ref{app:NTIccs[345]confined}

\subsection{Construction of $[336]$-color codes}
For completion we show how this construction procedure works for the $[336]$-color codes that are equivalent to the original color code by an LU. As a result we find color codes that have mixed edge configurations. We begin with the local construction of this type in Fig. 
\ref{fig:[336]cclocalconstruction} 
in App. \ref{app:localconstruction}.

\begin{figure}[h]
    \centering
    \includegraphics[width=8cm]{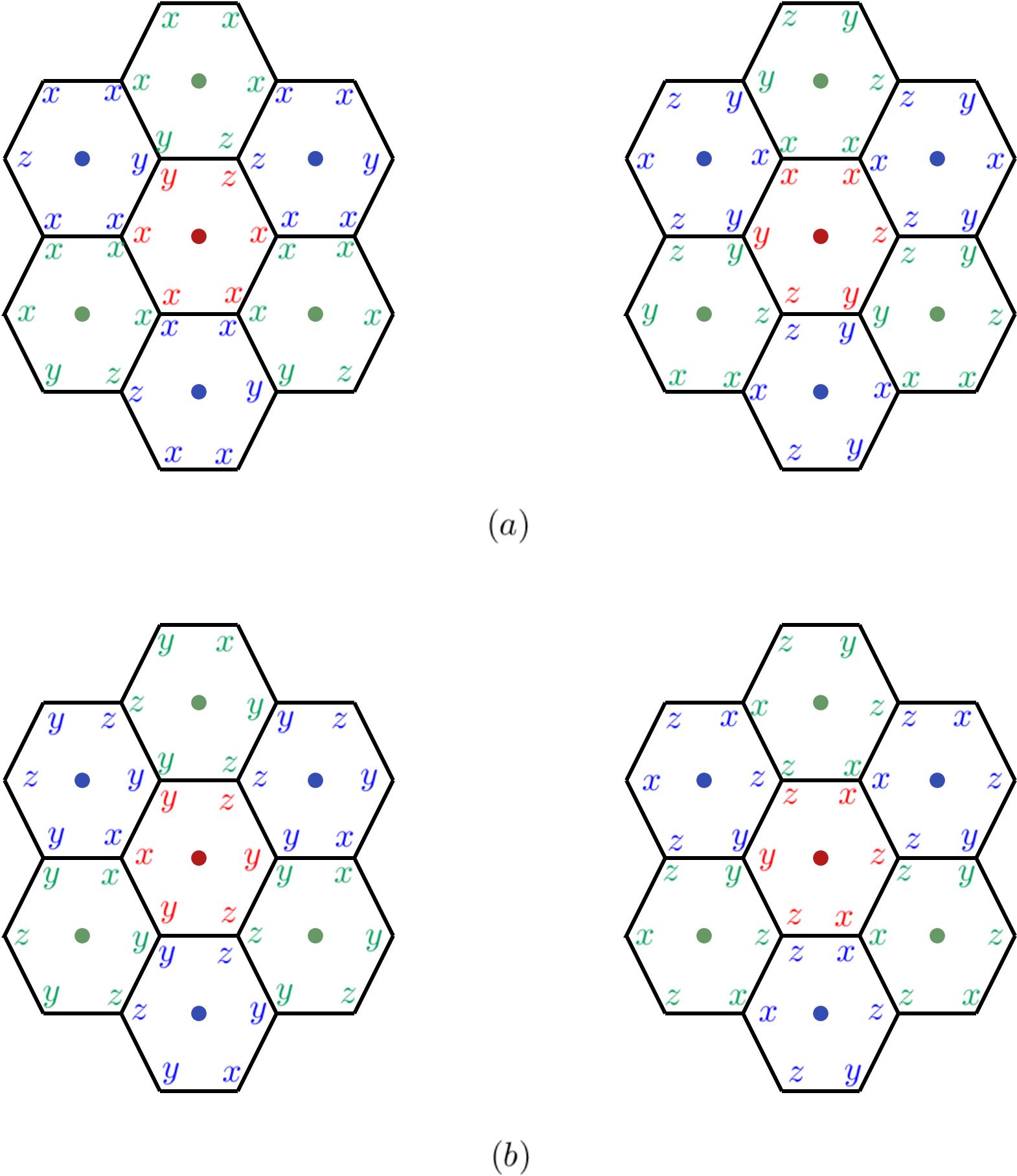}
    \caption{The $[336]$-color codes on the hexagonal lattice with (a) partially and (b) fully mixed edge configurations. These codes are equivalent to the original $[336]$-color code by an LU.}
    \label{fig:[336]ccpartialfullmixing}
\end{figure}

Using this the non-CSS $[336]$-color codes on the hexagonal lattice for partially and fully mixed edge configurations, with  translational invariance is shown in Figs. \ref{fig:[336]ccpartialfullmixing} and the non-translationally invariant analogue is shown in Fig. \ref{fig:[336]ccpartialfullmixingwoti} of App. \ref{app:NTIccs[336]}.


\section{$\mathbb{Z}_2\times\mathbb{Z}_2$ topological order of non-CSS color codes} 
\label{sec:TOorder}
So far we have studied the general construction of stabilizer codes on trivalent lattices and explored how to apply this construction to generate  different types of color codes in two-dimensional lattices. An important feature we used was the bulk vertex energetics to make a broad classification of the non-CSS color codes. In other words, we exploited the local `short-range' physics to obtain various models.  

In this Section, we study the long-range properties of the canonical $[444]$ and $[345]$-color codes on the hexagonal lattice and show that they realize the $\mathbb{Z}_2\times\mathbb{Z}_2$ topological phase just as the $[336]$-color codes that contain, in particular, the canonical CSS color code. We provide detailed derivations to establish this aspect of non-CSS color codes. 

Due to the non-CSS nature of these codes we will denote the two stabilizers on each face as 
\begin{align}
 &\text{Set I\,:} \{{\color{red}r_{I}}, {\color{blue}b_{I}}, {\color{ForestGreen}g_{I}}\},\\~~&\text{Set II\,:} \{ {\color{red}r_{II}}, {\color{blue}b_{II}}, {\color{ForestGreen}g_{II}} \}.    
\end{align}
While this notation is useful to describe the stabilizers on each of the colored faces in the non-CSS color codes with partially mixed and fully mixed edge configurations, it is not needed for the codes with homogeneous edge configurations. In the latter case we will denote the stabilizers as either of the Pauli operators, $X$, $Y$ or $Z$ with the appropriate color.

The Hamiltonian, is given by
\begin{align}\label{ncsscc}
H & =  -\sum\limits_{j=1}^{{\color{red}f}}~\left({\color{red}Pr_I}_j + {\color{red}Pr_{II}}_j\right)  -  \sum\limits_{j=1}^{{\color{blue}f}}~\left({\color{blue}Pb_I}_j + {\color{blue}Pb_{II}}_j\right) \nonumber \\ & -  \sum\limits_{j=1}^{{\color{ForestGreen}f}}~\left({\color{ForestGreen}Pg_I}_j + {\color{ForestGreen}Pg_{II}}_j\right),
\end{align}
with $j$ acting as the index for the faces and $\color{red}f$, $\color{blue}f$ and $\color{ForestGreen}f$ denoting the number of red, blue and green faces respectively. Note that for a general trivalent, tricolorable lattice ${\color{red}f}\neq{\color{blue}f}\neq\color{ForestGreen}f$. They are equal in the case of a hexagonal lattice.

The GSD can be computed either by evaluating the trace of the projector to the ground state manifold,
or by identifying the constraints satisfied by the stabilizers to obtain the number of independent stabilizers. We will find that there are precisely $2|F|-4$ independent stabilizers for both the $[444]$-color code and the deconfined $[345]$-color code. This implies that the number of encoded qubits is the same as the original CSS color code, $|V|-2|F|+4=4g$, with $g$ being the genus.  
We will also write down the logical Pauli's in each case and account for the GSD. Along this process we also provide a description of the anyon content for the different codes that indicates the type of topological phase.

We primarily work on the hexagonal lattice and for the case of the translationally invariant stabilizer codes. We believe the topological properties continue to hold for the non-translational versions of these codes as well.

\subsection{The $[444]$-color codes}
The stabilizers of the canonical $[444]$-color codes is shown in Fig. \ref{fig:[444]ccnomixing}. As all the edge configurations are homogeneous, we can think of the code simply as green faces made up of the $X$ and $Y$ stabilizers, the blue faces with $Y$ and $Z$ stabilizers and the red with $Z$ and $X$ stabilizers. 
This makes contrast to the original color code where the $Z$ and $X$ stabilizers make up all three colors.  

\subsubsection{Ground State Degeneracy}
The constraints among the stabilizers of this model are easily read off from the code in Fig. \ref{fig:[444]ccnomixing} as, 
\begin{eqnarray}
\prod\limits_{j=1}^{{\color{red}f}}{\color{red}r_I}\prod\limits_{j=1}^{{\color{blue}f}}{\color{blue}b_{II}} & = & \prod\limits_{j=1}^{{\color{red}f}}{\color{red}r_I}\prod\limits_{j=1}^{{\color{ForestGreen}f}}{\color{ForestGreen}g_I}\prod\limits_{j=1}^{{\color{ForestGreen}f}}{\color{ForestGreen}g_{II}} = 1, \label{eq:[444]constraint1}\\
\prod\limits_{j=1}^{{\color{blue}f}}{\color{blue}b_I}\prod\limits_{j=1}^{{\color{ForestGreen}f}}{\color{ForestGreen}g_{II}} & = & \prod\limits_{j=1}^{{\color{blue}f}}{\color{blue}b_I}\prod\limits_{j=1}^{{\color{red}f}}{\color{red}r_I}\prod\limits_{j=1}^{{\color{red}f}}{\color{red}r_{II}} = 1, \label{eq:[444]constraint2}\\
\prod\limits_{j=1}^{{\color{ForestGreen}f}}{\color{ForestGreen}g_I}\prod\limits_{j=1}^{{\color{red}f}}{\color{red}r_{II}} & = & \prod\limits_{j=1}^{{\color{ForestGreen}f}}{\color{ForestGreen}g_I}\prod\limits_{j=1}^{{\color{blue}f}}{\color{blue}b_I}\prod\limits_{j=1}^{{\color{blue}f}}{\color{blue}b_{II}} = 1, \label{eq:[444]constraint3}\\
\prod\limits_{j=1}^{{\color{red}f}}{\color{red}r_{II}}\prod\limits_{j=1}^{{\color{blue}f}}{\color{blue}b_{I}} &\prod\limits_{j=1}^{{\color{blue}f}}& {\color{blue}b_{II}}  =  1, \label{eq:[444]constraint4}\\
\prod\limits_{j=1}^{{\color{blue}f}}{\color{blue}b_{II}}\prod\limits_{j=1}^{{\color{ForestGreen}f}}{\color{ForestGreen}g_{I}} &\prod\limits_{j=1}^{{\color{ForestGreen}f}}& {\color{ForestGreen}g_{II}}  =  1, \label{eq:[444]constraint5}\\
\prod\limits_{j=1}^{{\color{ForestGreen}f}}{\color{ForestGreen}g_{II}}\prod\limits_{j=1}^{{\color{red}f}}{\color{red}r_{I}} & \prod\limits_{j=1}^{{\color{red}f}}& {\color{red}r_{II}}  =  1, \label{eq:[444]constraint6}\\
\prod\limits_{j=1}^{{\color{red}f}}{\color{red}r_{I}}\prod\limits_{j=1}^{{\color{blue}f}}{\color{blue}b_{I}}&\prod\limits_{j=1}^{{\color{ForestGreen}f}}&{\color{ForestGreen}g_{I}} = \prod\limits_{j=1}^{{\color{red}f}}{\color{red}r_{II}}\prod\limits_{j=1}^{{\color{blue}f}}{\color{blue}b_{II}}\prod\limits_{j=1}^{{\color{ForestGreen}f}}{\color{ForestGreen}g_{II}} = 1. \label{eq:[444]constraint7}
\end{eqnarray}
Out of these eleven constraints we can verify that only four are independent just as in the case of the original $[336]$-color code.
We make a choice to pick the following operators 
\begin{eqnarray}
\beta_1 & = & \prod\limits_{j=1}^{{\color{red}f}}{\color{red}r_I}\prod\limits_{j=1}^{{\color{blue}f}}{\color{blue}b_{II}},~ \beta_2 = \prod\limits_{j=1}^{{\color{blue}f}}{\color{blue}b_I}\prod\limits_{j=1}^{{\color{ForestGreen}f}}{\color{ForestGreen}g_{II}}, \nonumber \\  \beta_3  & = & \prod\limits_{j=1}^{{\color{ForestGreen}f}}{\color{ForestGreen}g_I}\prod\limits_{j=1}^{{\color{red}f}}{\color{red}r_{II}},~
\beta_4  =  \prod\limits_{j=1}^{{\color{red}f}}{\color{red}r_I}\prod\limits_{j=1}^{{\color{ForestGreen}f}}{\color{ForestGreen}g_I}\prod\limits_{j=1}^{{\color{ForestGreen}f}}{\color{ForestGreen}g_{II}},
\end{eqnarray}
with the constraints given by 
\begin{align}
    \beta_i = 1, ~ \forall~ i = 1 ~\text{to} ~4.
\end{align}
We observe that the second constraints of Eqs. (\ref{eq:[444]constraint2}) and (\ref{eq:[444]constraint3}) are obtained as $\beta_2\beta_3\beta_4 =1 $ and $\beta_1\beta_2\beta_4=1$ respectively. Similarly, the constraints in Eqs. (\ref{eq:[444]constraint4})-(\ref{eq:[444]constraint6}) are equivalent to $\beta_1\beta_2\beta_3\beta_4=1$, $\beta_1\beta_4=1$ and $\beta_3\beta_4=1$, respectively. The two constraints in Eq. (\ref{eq:[444]constraint7}) are $\beta_2\beta_4=1$ and $\beta_1\beta_3\beta_4=1$ respectively.  

So  the number of encoded qubits is $|V|-2|F|+4$ which is 4 for the hexagonal lattice tiling on a torus. On more general trivalent lattices tiling of a genus-$g$ surface, $|V|-2|F|+4$ reduces to $4g$ as seen earlier. Thus, we conclude that the GSD of the $[444]$-color codes coincides with the original $[336]$-color code. 

\subsubsection{The Code Space}
Now we construct the ground states or the encoded qubits for $[444]$-color code with homogeneous edge configurations on the hexagonal lattice. The construction illustrated here goes through for a more general trivalent, tricolorable lattice discretizing surfaces of arbitrary genus with minor modifications. We work with the stabilizers, $\color{ForestGreen}X$, $\color{ForestGreen}Y$ for the green faces, $\color{blue}Y$, $\color{blue}Z$ for the blue faces and $\color{red}Z$, $\color{red}X$ for the red faces. 

As in the construction of the encoded qubits for the original CSS $[336]$-color code, we project a state diagonalizing one set of the stabilizers (the seed state), say the $Z$ stabilizers, with the remaining stabilizers onto the code space. This procedure applies for the $[444]$-color code as well albeit a crucial difference coming from the choice of the state to be projected. It can be checked that choosing a homogeneous configuration for the seed state, namely a state containing either $\ket{0_z}$, $\ket{0_x}$ or $\ket{0_y}$ on all the vertices is annihilated by the projectors built out of the stabilizers supported on the hexagonal faces. For example ${\color{ForestGreen}PXPY}\ket{s_Z}$, with $\ket{s_Z}$ being the state with $\ket{0_z}$ on the vertices of the hexagon, is 0. Notice that this is non-zero for a square, octagonal or any face whose number of vertices is a multiple of four. 

\begin{figure}[h]
    \centering
    \includegraphics[width=8cm]{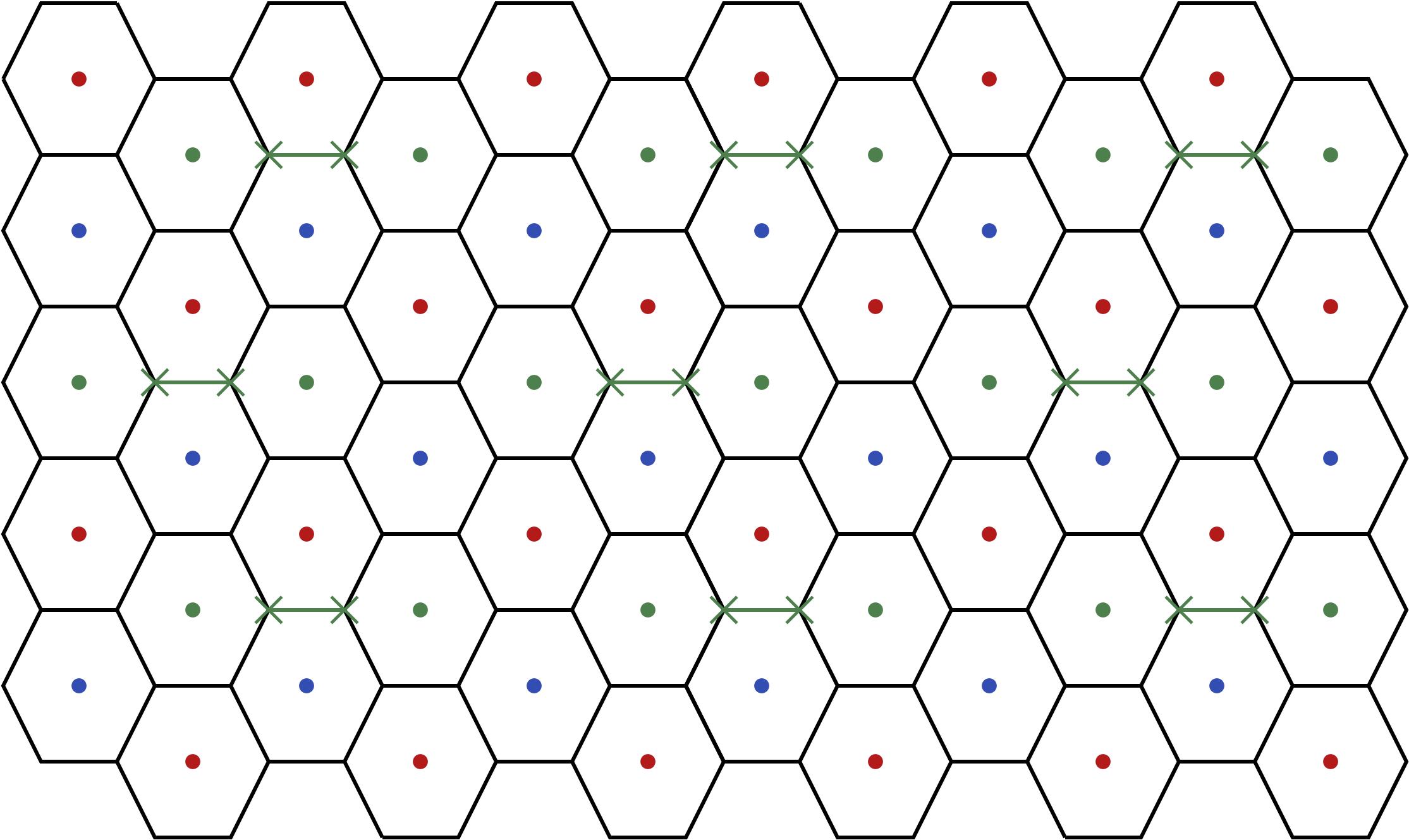}
    \caption{The seed state configuration for the $[444]$-color code with homogeneous edge configurations. Only a part of the hexagonal lattice is shown. The vertices of the green edges are filled with $\ket{1_z}$ states (green crosses) and the remaining vertices are filled with the $\ket{0_z}$ state. }
    \label{fig:[444]ccnomixingseedstate}
\end{figure}

As a result we choose a seed state such that there is one $\ket{1_z}$ state on every green face, with the remaining vertices filled by $\ket{0_z}$. This is attained by a seed state, $\ket{s_{[444]}}$ shown in Fig. \ref{fig:[444]ccnomixingseedstate}.
This choice ensures that some (See Fig. \ref{fig:[444]ccnomixingseedstate}) of the blue and red faces have two $\ket{1_z}$'s and the remaining vertices filled by $\ket{0_z}$. These configurations are not annihilated by the projectors built out of the stabilizers on the red and blue faces as these continue to be +1 eigenstates of the $Z$ stabilizers. Thus the seed state in Fig. \ref{fig:[444]ccnomixingseedstate} is appropriate for constructing the ground states of the $[444]$-color code on the hexagonal lattice. Furthermore we would like to relate the ground state of the $[444]$-color code to the $[336]$-color code by LU's. In this regard we notice that the seed state in Fig. \ref{fig:[444]ccnomixingseedstate}, $\ket{s_{[444]}}$ is obtained from $\ket{s_Z}$, the seed state of the CSS $[336]$-color code by flipping the $\ket{0_z}$'s to $\ket{1_z}$'s along alternating pairs of green faces (shown as green edges in Fig. \ref{fig:[444]ccnomixingseedstate}). We denote these edges as $\color{ForestGreen}e$ in the formulas below. That is,
\begin{equation}
    \ket{s_{[444]}} = \prod\limits_{v_i, v_j\in {\color{ForestGreen} e}}~x_{v_i}x_{v_j}\ket{s_Z}.
\end{equation}
The encoded qubit is obtained from 
\begin{eqnarray}\label{eq:[444]cccanonicalgroundstate}
\ket{\bar{0}_1, \bar{0}_2, \bar{0}_3, \bar{0}_4}_{[444]}  & = &   \prod\limits_{j=1}^{{\color{blue}f}}{\color{blue}PY}_j\prod\limits_{j=1}^{{\color{red}f}}{\color{red}PX}_j\nonumber \\
& &\prod\limits_{j=1}^{{\color{ForestGreen}f}}{\color{ForestGreen}PX}_j{\color{ForestGreen}PY}_j\ket{s_{[444]}}, \nonumber \\
& = &  \prod\limits_{v_i, v_j\in {\color{ForestGreen} e}}~x_{v_i}x_{v_j}\prod\limits_{j=1}^{{\color{blue}f}}{\color{blue}PY}_j\prod\limits_{j=1}^{{\color{red}f}}{\color{red}PX}_j\nonumber \\
& &\prod\limits_{j=1}^{{\color{ForestGreen}f}}{\color{ForestGreen}PX}_j{\color{ForestGreen}PY^\perp}_j\ket{s_Z} \nonumber \\
& = & \prod\limits_{v_i, v_j\in {\color{ForestGreen} e}}~x_{v_i}x_{v_j}\prod\limits_{j=1}^{{\color{blue}f}}{\color{blue}PX^\perp}_j\prod\limits_{j=1}^{{\color{red}f}}{\color{red}PX}_j\nonumber \\
& & \prod\limits_{j=1}^{{\color{ForestGreen}f}}{\color{ForestGreen}PX}_j{\color{ForestGreen}PX}_j\ket{s_Z} \nonumber \\
& = & \prod\limits_{v_i, v_j\in {\color{blue} e}}~z_{v_i}z_{v_j}\prod\limits_{v_i, v_j\in {\color{ForestGreen} e}}~x_{v_i}x_{v_j}\prod\limits_{j=1}^{{\color{blue}f}}{\color{blue}PX}_j \nonumber \\
& & \prod\limits_{j=1}^{{\color{red}f}}{\color{red}PX}_j\prod\limits_{j=1}^{{\color{ForestGreen}f}}{\color{ForestGreen}PX}_j\ket{s_Z} \nonumber \\
& = & U \ket{\bar{0}_1, \bar{0}_2, \bar{0}_3, \bar{0}_4}_{[336]}, \nonumber \\
\end{eqnarray}
where 
\begin{align}
    U=\left(\prod\limits_{v_i, v_j\in {\color{blue} e}}~z_{v_i}z_{v_j}\right)\left(\prod\limits_{v_i, v_j\in {\color{ForestGreen} e}}~x_{v_i}x_{v_j}\right).
\end{align} Here ${\color{blue}e}$ are horizontal edges connecting blue faces. However we use only alternating horizontal edges while taking the product of the operators of the form $z_{v_i}z_{v_j}$. This results in the identity
\begin{equation}
   \prod\limits_{j=1}^{{\color{blue}f}}{\color{blue}PX}_j \prod\limits_{v_i, v_j\in {\color{blue} e}}~z_{v_i}z_{v_j}\ket{s_Z} = \prod\limits_{v_i, v_j\in {\color{blue} e}}~z_{v_i}z_{v_j} \prod\limits_{j=1}^{{\color{blue}f}}{\color{blue}PX^\perp}_j \ket{s_Z},
\end{equation}
which is required in the derivation of Eq. (\ref{eq:[444]cccanonicalgroundstate}). We have used the following identities,
\begin{eqnarray}
{\color{ForestGreen}PY^\perp}\ket{s_Z} & \equiv & {\color{ForestGreen}PX}\ket{s_Z}, \\
{\color{blue}PY}\ket{s_Z} & \equiv & {\color{blue}PX^\perp}\ket{s_Z},
\end{eqnarray}
on hexagonal faces while deriving Eq. (\ref{eq:[444]cccanonicalgroundstate}). The operators $PX^\perp$ denote the projectors orthogonal to $PX$.  Thus we find that the ground state or the encoded qubit of the $[444]$-color code with homogeneous edge configurations are equivalent to those of the original CSS $[336]$-color code by an LU. The form of the LU is such that the ground state of the former code is an excited state for the latter. Due to this LU the logical Pauli operators of the $[444]$-color code are obtained by conjugating the logical Pauli's of the $[336]$-color code. In particular the logical $X$'s along two colored lattices can be used to construct the other encoded qubit states spanning the entire code space.
In particular this implies that the logical Pauli's of the canonical $[444]$-color code coincide with those of the CSS $[336]$-color codes. 
Even though the canonical 444-color code is a non-CSS code, its logical Pauli operators are of the homogenous type. 

It is also worth noting that though the stabilizers of the two codes are not LU equivalent to each other, their code spaces are still LU equivalent. Furthermore the LU equivalence to the CSS $[336]$-color code guarantees fault tolerant nature of the code space of the canonical $[444]$-color codes as the logical Clifford operators continue to remain transversal. These statements are not too surprising as the long-range properties of the $[444]$- and $[336]$-color codes are expected to coincide. The difference between the two is attributed to the short range spectrum of the two models. 

\subsubsection{Anyon Content}
In the CSS $[336]$-color code electric charges, $e$ (magnetic fluxes, $m$) are those states violating the $X$ ($Z$) stabilizers, obtained using local string operators acting on the shrunk lattices of a chosen color. The non-CSS nature of the $[444]$-color codes makes way for a similar narrative with `electric charges' (`magnetic fluxes') being those states obtained when the $I$ ($II$) stabilizer invariance conditions are not satisfied. The `dyonic' excitations are obtained when both the $I$ and the $II$ stabilizer invariance conditions are simultaneously not satisfied, such as for a $Y$ error on a bulk vertex for the CSS $[336]$-color code. The choice of what we label $I$ and $II$ kind of stabilizer is a convention and the picture remains the same when we make the alternative choice. 

Adopting this convention we see that $\color{red}e$ and $\color{red}m$ excitations are created by acting with string operators made of the Pauli $x$ and Pauli $z$ operators on the red shrunk lattice respectively. In a similar manner the $\color{blue}e$ and $\color{blue}m$ excitations are created by string operators made of Pauli $z$ and Pauli $y$ operators on the blue shrunk lattice. These string operators either commute or anticommute with the LU, 
\begin{align}
U=\left(\prod\limits_{v_i, v_j\in {\color{blue} e}}~z_{v_i}z_{v_j}\right)\left(\prod\limits_{v_i, v_j\in {\color{ForestGreen} e}}~x_{v_i}x_{v_j}\right),    
\end{align}
mapping the code space of the CSS $[336]$-color code to 
the code space of the canonical $[444]$-color code, and vice versa. Thus up to a sign the anyonic excitations of the canonical $[444]$-color code are LU related to the anyons of the CSS $[336]$-color code. Thus we expect them to behave as the anyons of a $\mathbb{Z}_2\times\mathbb{Z}_2$ topologically ordered system.

As in the original $[336]$-color code the operators creating the green anyons are obtained by taking a product of the blue and red strings. However there is a difference in which operators are chosen for the $[444]$-color code. The $\color{ForestGreen}e$ is obtained by combining the strings creating $\color{red}\epsilon$ (red dyons) and $\color{blue}m$, whereas $\color{ForestGreen}m$ is found by combining the strings creating $\color{red}e$ and $\color{blue}\epsilon$ (blue dyons). This is seen as a consequence of the non-CSS nature of the $[444]$-color codes. 

\subsection{Deconfined $[345]$-color codes}
There are three inequivalent $[345]$-color codes corresponding to the inequivalent bulk vertex configurations in Fig. \ref{fig:[345]bulkvertexconfigs}. We analyze the topological properties of the canonical $[345]$-color code defined in class $I$ (See Fig. \ref{fig:[345]ccdeconfinedClassI}). Here the green faces have the same set of stabilizers as the original ($[336]$)-color code and both the blue and red faces have the $Y$ and $Z$ stabilizers.

\subsubsection{Ground State Degeneracy}
As in the previous cases we identify the independent constraints among the stabilizers to compute the GSD or the dimension of the code space. A set of four independent constraints can be read off from the stabilizers of the code in Fig. \ref{fig:[345]ccdeconfinedClassI},
\begin{eqnarray}
\beta_1 & = & \prod\limits_{j=1}^{{\color{red}f}}{\color{red}r_I}_j\prod\limits_{j=1}^{{\color{blue}f}}{\color{blue}b_{I}}_j,~ \beta_2 = \prod\limits_{j=1}^{{\color{blue}f}}{\color{blue}b_{II}}_j\prod\limits_{j=1}^{{\color{ForestGreen}f}}{\color{ForestGreen}g_{II}}_j, \nonumber \\  \beta_3  & = & \prod\limits_{j=1}^{{\color{ForestGreen}f}}{\color{ForestGreen}g_{II}}_j\prod\limits_{j=1}^{{\color{red}f}}{\color{red}r_{II}}_j,~
\beta_4  =  \prod\limits_{j=1}^{{\color{red}f}}{\color{red}r_I}_j\prod\limits_{j=1}^{{\color{ForestGreen}f}}{\color{ForestGreen}g_I}_j\prod\limits_{j=1}^{{\color{ForestGreen}f}}{\color{ForestGreen}g_{II}}_j. \nonumber \\
\end{eqnarray}
This leads to $|V|-2|F|+4$ encoded qubits which reduces to $4g$ for a trivalent, tricolorable lattice discretizing a surface of genus $g$.

\subsubsection{The Code Space}
\begin{figure}[h]
    \centering
    \includegraphics[width=8cm]{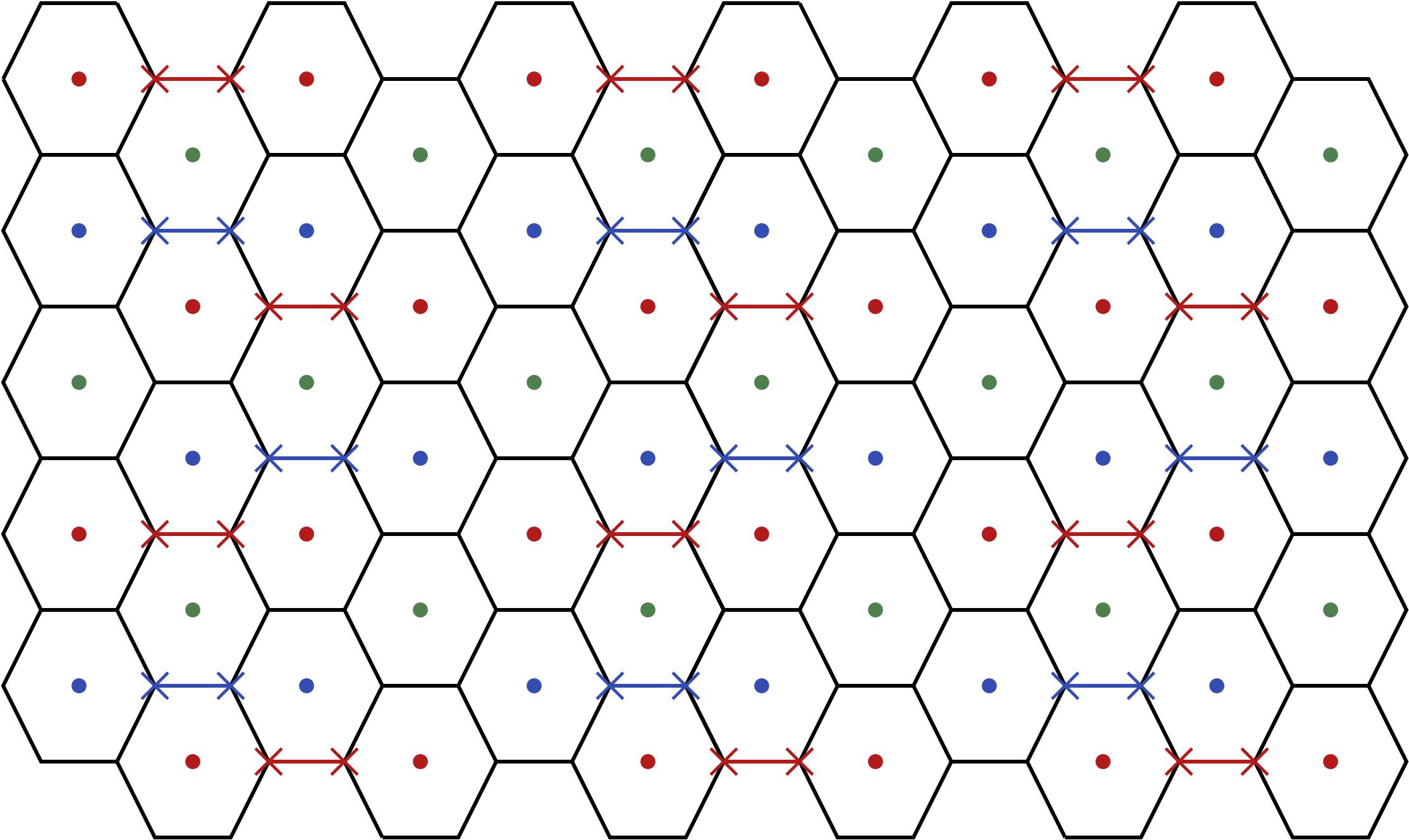}
    \caption{The supports of the $z_{v_i}z_{v_j}$ operators on the ${\color{blue}e}$ and ${\color{red}e}$ edges. }
    \label{fig:[345]ccdeconfinedgroundstate}
\end{figure}

In this case we can choose the product state $\ket{s_Z}$ as the seed state that can be projected to the code space.  A ground state is constructed as,
\begin{eqnarray}\label{eq:345cc0000state}
\ket{\bar{0}_1\bar{0}_2\bar{0}_3\bar{0}_4}_{[345]} & = &  
\prod\limits_{j=1}^{{\color{red}f}}{\color{red}PY}_j  \prod\limits_{j=1}^{{\color{blue}f}}{\color{blue}PY}_j  \prod\limits_{j=1}^{{\color{ForestGreen}f}}{\color{ForestGreen}PX}_j\ket{s_Z}, \nonumber \\
& = & \prod\limits_{j=1}^{{\color{red}f}}{\color{red}PX^\perp}_j  \prod\limits_{j=1}^{{\color{blue}f}}{\color{blue}PX^\perp}_j  \prod\limits_{j=1}^{{\color{ForestGreen}f}}{\color{ForestGreen}PX}_j\ket{s_Z}, \nonumber \\
& = & \prod\limits_{v_i, v_j\in {\color{red}e}}z_{v_i}z_{v_j}\prod\limits_{v_i, v_j\in {\color{blue}e}}z_{v_i}z_{v_j} \nonumber \\
& & \prod\limits_{j=1}^{{\color{red}f}}{\color{red}PX}_j  \prod\limits_{j=1}^{{\color{blue}f}}{\color{blue}PX}_j  \prod\limits_{j=1}^{{\color{ForestGreen}f}}{\color{ForestGreen}PX}_j\ket{s_Z}, \nonumber \\
& = & U \ket{\bar{0}_1\bar{0}_2\bar{0}_3\bar{0}_4}_{[336]}, \nonumber \\
\end{eqnarray}
where 
\begin{align}
    U=\prod\limits_{v_i, v_j\in {\color{red}e}}z_{v_i}z_{v_j}\prod\limits_{v_i, v_j\in {\color{blue}e}}z_{v_i}z_{v_j}
\end{align} is the LU that maps the ground state of the deconfined $[345]$-color code with homogeneous edge configurations 
to that of the CSS $[336]$-color code, and vice versa. Here ${\color{red}e}$ and ${\color{blue}e}$ are alternating horizontal edges connecting red and blue faces respectively (See Fig. \ref{fig:[345]ccdeconfinedgroundstate}).
This LU equivalence to the CSS $[336]$-color code ensures that the deconfined $[345]$-color code supports fault-tolerant computing as well. The remaining ground states or encoded qubits are constructed with the help of the logical Pauli $X$ operators which are the same as the ones for the CSS $[336]$-color codes. 

\subsubsection{Anyon Content}
As in the cases of the original $[336]$-color code and the $[444]$-color codes, the deconfined $[345]$-color codes have the same set of anyons as a $\mathbb{Z}_2\times\mathbb{Z}_2$ topologically ordered phase. We construct these anyons in the red and blue shrunk lattices without loss of generality. The anyons on the green shrunk lattice are not independent and can be obtained from the latter. The $\color{red}e$ and $\color{blue}e$ electric charges are obtained from string operators made of the Pauli $z$ operators. The $\color{red}m$ and $\color{blue}m$ magnetic fluxes are obtained from string operators made of the Pauli $y$ operators. These string operators act on the ground state 
$\ket{\bar{0}_1\bar{0}_2\bar{0}_3\bar{0}_4}_{[345]}$ of Eq. (\ref{eq:345cc0000state}) to create the appropriate excitaitions. As they either commute or anticommute with the LU,
\begin{align}
U=\prod\limits_{v_i, v_j\in {\color{red}e}}z_{v_i}z_{v_j}\prod\limits_{v_i, v_j\in {\color{blue}e}}z_{v_i}z_{v_j},    
\end{align}
 the resulting excitations are LU equivalent to the excitations of the CSS $[336]$-color code. In particular, the electric charges of the deconfined $[345]$-color codes are LU equivalent to the electric charges of the CSS $[336]$-color code whereas the magnetic fluxes are LU equivalent to the dyons of the CSS $[336]$-color code.

\subsection{Confined $[345]$-color codes}
\begin{figure}[h]
    \centering
    \includegraphics[width=8cm]{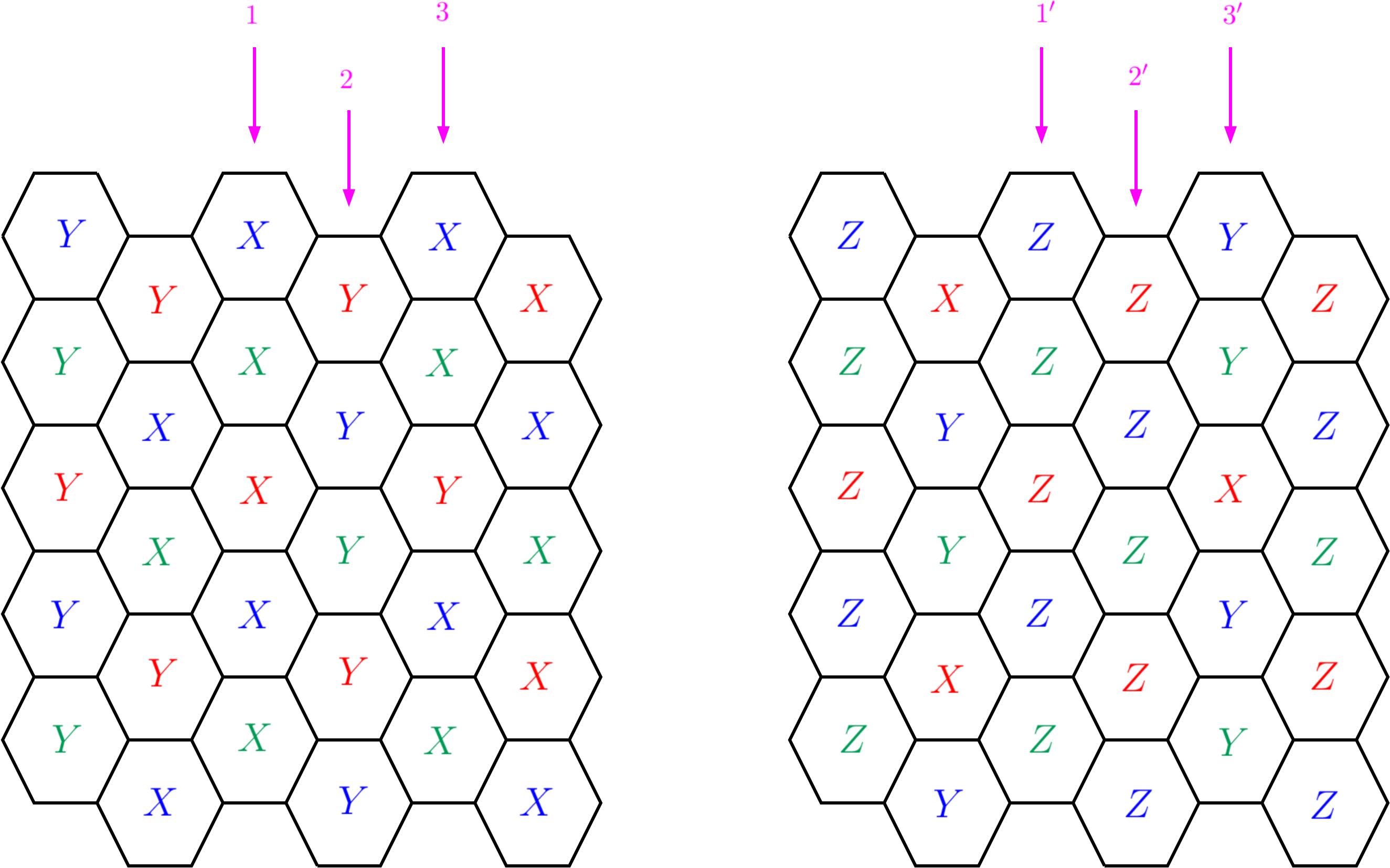}
    \caption{The translationally invariant confined $[345]$-color code with the column pairs, $j-j'$ with $j\in\{1,2,3\}$, indicated.}
    \label{fig:[345]ccconfinedClassIwithcolumns}
\end{figure}

There are three types of inequivalent $[345]$ bulk vertices (See Fig. \ref{fig:[345]bulkvertexconfigs}) that can be used to build the confined $[345]$-color codes that are translationally invariant and have homogeneous edge configurations (See Fig. \ref{fig:[345]ccconfinedClassIwti}). We analyze the model belonging to Class $I$ without loss of any of the features. Unlike the previous models the translational invariance in this system manifests itself in a different manner. Thinking of the hexagonal lattice in terms of columns we see that in the class $I$ model of the confined $[345]$-color code (See Fig. \ref{fig:[345]ccconfinedClassIwti}), there is an $X$ and $Z$ column in the stabilizer pair, followed by a $Y$ and $Z$ column which is continued by a column pair that is a mixture of $X$ and $Y$'s. In the third column we notice that for a face of a given color there are one $X$ and one $Y$ operators similar to what we saw in the canonical $[444]$-color code. We indicate these column pairs as column 1-1', column 2-2' and column 3-3' to refer to them easily in what follows. The class $I$ confined $[345]$-color code is shown again in Fig. \ref{fig:[345]ccconfinedClassIwithcolumns} but this time with the column pairs clearly indicated.
\subsubsection{Ground State Degeneracy}
The dimension of the code space is much harder to compute in this case due to the peculiar nature in which the translational invariance is implemented in this system. Nevertheless by inspecting the stabilizers in Fig. \ref{fig:[345]ccconfinedClassIwithcolumns} we see that we can pick four independent constraints among the set of stabilizers just by looking at the colors : these are ${\color{red}r_I}{\color{blue}b_I}$, ${\color{red}r_{II}}{\color{blue}b_{II}}$, ${\color{blue}b_{I}}{\color{ForestGreen}g_{I}}$ and ${\color{blue}b_{II}}{\color{ForestGreen}g_{II}}$.  Following this color scheme it is quite complicated to write down the stabilizer constraints for the case of the confined $[345]$-color code and so we request the reader to convince themselves that this is indeed the case by looking at the stabilizer pairs for this system in Fig. \ref{fig:[345]ccconfinedClassIwithcolumns}. We will see indirect ways of affirming this fact by studying the LU equivalence between the code spaces of the confined $[345]$- and the CSS $[336]$-color codes next.  

\subsubsection{The Code Space}
\begin{figure}[h]
    \centering
    \includegraphics[width=8cm]{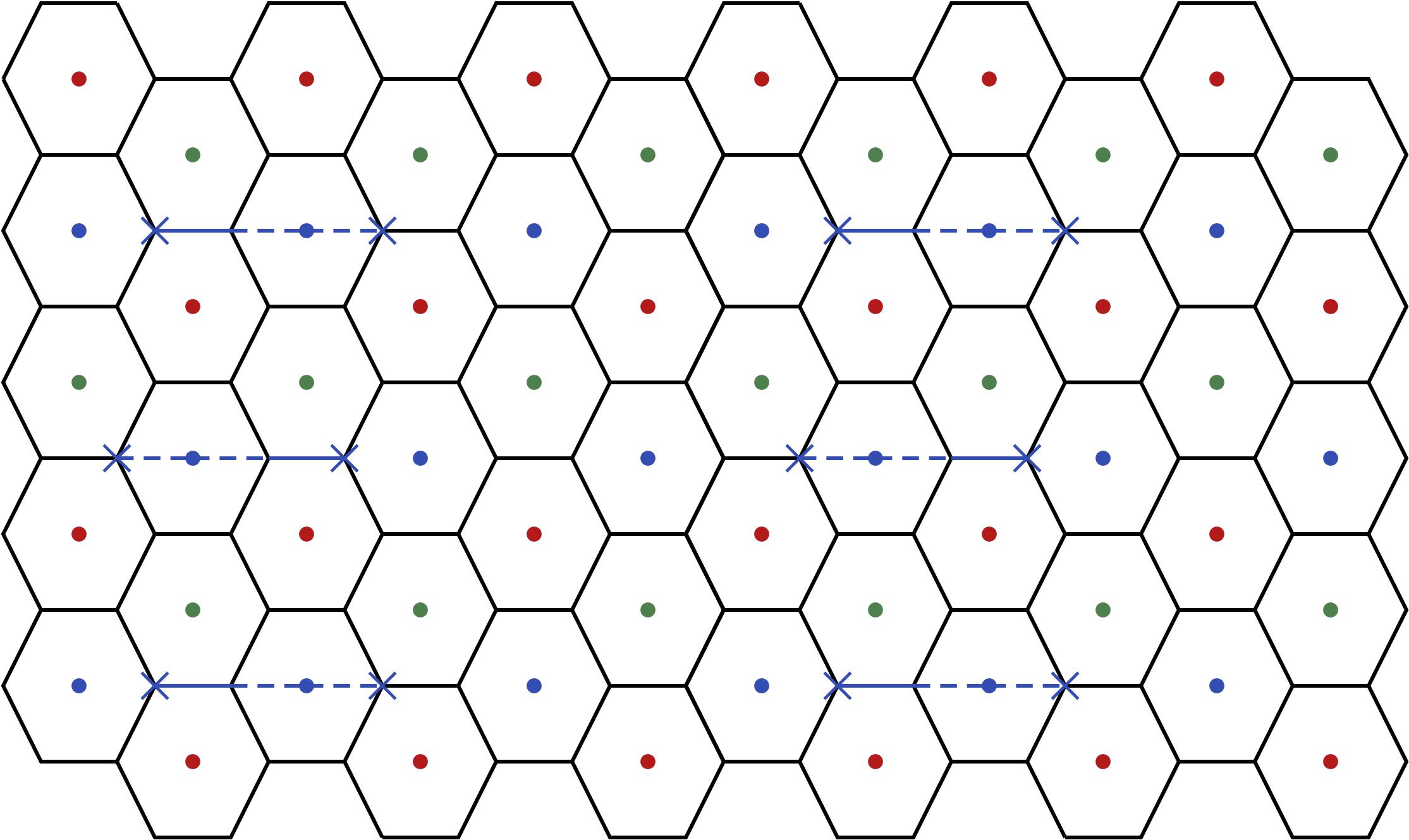}
    \caption{The support of the edges along the blue shrunk lattice, connecting column 3's. The sites marked with blue crosses are filled with the $\ket{1_z}$ state in the tensor product with the remaining sites filled with $\ket{0_z}$.}
    \label{fig:[345]ccconfinednomixingClassIseedstate}
\end{figure}

The operators occurring in the stabilizer pair in this case is similar to what we saw for the canonical $[444]$-color code and due to this a homogeneous seed state cannot be projected onto the code space for this system. To rectify this we choose a seed state by taking tensor products of $\ket{0_z}$'s and $\ket{1_z}$'s as we did while analyzing the code space of the canonical $[444]$-color code. The operators in the column 3 of the stabilizer pairs contain the $X$ and $Y$ operators for each of the colored faces. We choose the seed state such that these faces contain exactly one $\ket{1_z}$ while the remaining are made of $\ket{0_z}$'s. This is achieved by flipping the $\ket{0_z}$'s of the $\ket{s_Z}$ state along the edges in the blue shrunk lattice as shown in Fig. \ref{fig:[345]ccconfinednomixingClassIseedstate}.
Thus the seed state, $\ket{s}_{[345], \textrm{confined}}$ is 
\begin{equation}
    \ket{s}_{[345], \textrm{confined}} = \prod\limits_{v_i, v_j, v_k\in {\color{blue}e}}x_{v_i}x_{v_j}x_{v_k}\ket{s_Z}.
\end{equation}
This state continues to be a +1 eigenstate of the $Z$ stabilizers. The encoded qubit of the confined $[345]$-color code of class $I$ can now be evaluated as,
\begin{eqnarray}\label{eq:[345]ccconfinedI0000state}
\ket{\bar{0}_1\bar{0}_2\bar{0}_3\bar{0}_4}_{[345], \textrm{confined}} & = &  \prod\limits_{j=1}^{{\color{red}f_1}}{\color{red}PX}_j\prod\limits_{j=1}^{{\color{blue}f_1}}{\color{blue}PX}_j\prod\limits_{j=1}^{{\color{ForestGreen}f_1}}{\color{ForestGreen}PX}_j \nonumber \\
& & \prod\limits_{j=1}^{{\color{red}f_2}}{\color{red}PY}_j\prod\limits_{j=1}^{{\color{blue}f_2}}{\color{blue}PY}_j\prod\limits_{j=1}^{{\color{ForestGreen}f_2}}{\color{ForestGreen}PY}_j \nonumber \\
& & \prod\limits_{j=1}^{{\color{red}f_3}}{\color{red}PX}_j{\color{red}PY}_j \nonumber \\
& & \prod\limits_{j=1}^{{\color{blue}f_3}}{\color{blue}PX}_j{\color{blue}PY}_j \nonumber \\
& & \prod\limits_{j=1}^{{\color{ForestGreen}f_3}}{\color{ForestGreen}PX}_j{\color{ForestGreen}PY}_j \nonumber \\
& & \prod\limits_{v_i, v_j, v_k\in {\color{blue}e}}x_{v_i}x_{v_j}x_{v_k}\ket{s_Z} \nonumber \\
& = & \prod\limits_{v_i, v_j, v_k\in {\color{blue}e}}x_{v_i}x_{v_j}x_{v_k} \nonumber \\
& & \prod\limits_{j=1}^{{\color{red}f_1}}{\color{red}PX}_j\prod\limits_{j=1}^{{\color{blue}f_1}}{\color{blue}PX}_j\prod\limits_{j=1}^{{\color{ForestGreen}f_1}}{\color{ForestGreen}PX}_j \nonumber \\
 & & \prod\limits_{j=1}^{{\color{red}f_2}}{\color{red}PX^\perp}_j\prod\limits_{j=1}^{{\color{blue}f_2}}{\color{blue}PX^\perp}_j\prod\limits_{j=1}^{{\color{ForestGreen}f_2}}{\color{ForestGreen}PX^\perp}_j \nonumber \\
& &\prod\limits_{j=1}^{{\color{red}f_3}}{\color{red}PX}_j\prod\limits_{j=}^{{\color{blue}f_3}}{\color{blue}PX}_j\prod\limits_{j=1}^{{\color{ForestGreen}f_3}}{\color{ForestGreen}PX}_j \ket{s_Z} \nonumber \\
& = & U\ket{\bar{0}_1\bar{0}_2\bar{0}_3\bar{0}_4}_{[336]},
\end{eqnarray}
where 
\begin{align}\label{eq:[345]ccconfinedILU}
    U=\left(\prod\limits_{v_i, v_j, v_k\in {\color{blue}e}}x_{v_i}x_{v_j}x_{v_k}\right)\left(\prod\limits_{v_i, v_j, v_k\in {\color{red}e}}z_{v_i}z_{v_j}z_{v_k}\right).
\end{align} Here the ${\color{blue}e}$ are horizontal edges in the blue shrunk lattice connecting two column 3's and the ${\color{red}e}$ are horizontal edges in the red shrunk lattice connecting two column 2's. The ${\color{red}f_i}$, ${\color{blue}f_i}$ and ${\color{ForestGreen}f_i}$ for $i\in\{1,2,3\}$ are the total number of red, blue and green faces in columns 1,2,3 respectively.
The state in Eq. (\ref{eq:[345]ccconfinedI0000state}) can also be viewed as an excitation of the $[336]$-color code. However unlike the previous cases this excitation is an instance where sets of three stabilizers of the $[336]$-color code are excited and hence are not deconfined anyons where either two or four stabilizers of a given color are excited. This is not surprising as we are dealing with a confined $[345]$-color code here. The remaining encoded qubits are obtained by acting the winding operators on the $\ket{\bar{0}_1\bar{0}_2\bar{0}_3\bar{0}_4}_{[345], \textrm{confined}}$ state. Notice that the winding operators either commute or anticommute with the LU in Eq. \ref{eq:[345]ccconfinedILU} and hence all the encoded qubits remain LU equivalent to the code space of the CSS $[336]$-color code up to a negative sign.

\subsubsection{Anyon Content}
In this case the elementary strings supported on two sites of a chosen shrunk lattice, either $x_{v_i}x_{v_j}$, $y_{v_i}y_{v_j}$ or $z_{v_i}z_{v_j}$, excite either three or two of the stabilizers in the stabilizer pair. As a result of this extending strings along certain directions can increase or decrease the energy. This is fundamentally different from the deconfined anyons ecountered in the earlier cases including the deconfined $[345]$-color code. However as the string operators creating these excitations by acting on the $\ket{\bar{0}_1\bar{0}_2\bar{0}_3\bar{0}_4}_{[345], \textrm{confined}}$ state given by Eq. (\ref{eq:[345]ccconfinedI0000state}) either commute or anticommute with the LU, $U$ in Eq. (\ref{eq:[345]ccconfinedILU}), the resulting excited states are still equivalent to the anyons of the CSS $[336]$-color code. 

\section{Triangular non-CSS color codes } 
\label{sec:finitenonCSScodes}
We will now discuss the non-CSS color codes on finite lattices that include boundary and corner vertices apart from the bulk vertices. The 7-qubit Steane code is the simplest example of a color code on a finite lattice, where the seven physical qubits are conveniently placed in the geometry of a triangle. The framework introduced here allows the construction of 7-qubit non-CSS codes that are not connected to the Steane code by 
an LU.

There is a single bulk vertex at the center of the 7-qubit triangle implying that there are exactly two more inequivalent Steane codes, one for the $[444]$ bulk vertex and the other for the $[345]$ bulk vertex. The three inequivalent $[345]$ bulk vertex configurations in Fig. \ref{fig:[345]bulkvertexconfigs} are mapped to each other by a global rotation. We also obtain non-CSS $[336]$-Steane codes that are equivalent to the canonical CSS Steane code by 
an LU. We include these for completion. 

Due to the non-CSS nature of the code we denote the pair of stabilizers for every face as $I$ and $II$ along with the color of the face. It is easily seen that each of these operators square to one, generating a $\mathbb{Z}_2$ and together they form a subgroup of the 7-qubit Pauli group. A little bit of work also shows that -1 is not generated from these stabilizers. Using the six stabilizers we construct projectors, for example ${\color{red} Pr_I} = \frac{1+{\color{red} r_I}}{2}$, where 1 is the identity operator acting on the 7-qubits. The code space is spanned by the ground states of the Hamiltonian formed out of these projectors as in the case of the original color code, 
\begin{equation}\label{eq:444scH}
H = -\left({\color{red} Pr_I} + {\color{red} Pr_{II}} + {\color{blue} Pb_I} + {\color{blue} Pb_{II}} + {\color{ForestGreen} Pg_I} + {\color{ForestGreen} Pg_{II}}\right).
\end{equation}
As the six stabilizers of this code are independent the Hamiltonian in Eq. (\ref{eq:444scH}) has two ground states or it encodes a single logical qubit, $\{\ket{0_L}, \ket{1_L}\}$.

\subsection{$[444]$-Steane code}
\label{subsec:[444]sc}

\begin{figure}[h]
    \centering
    \includegraphics[width=8cm]{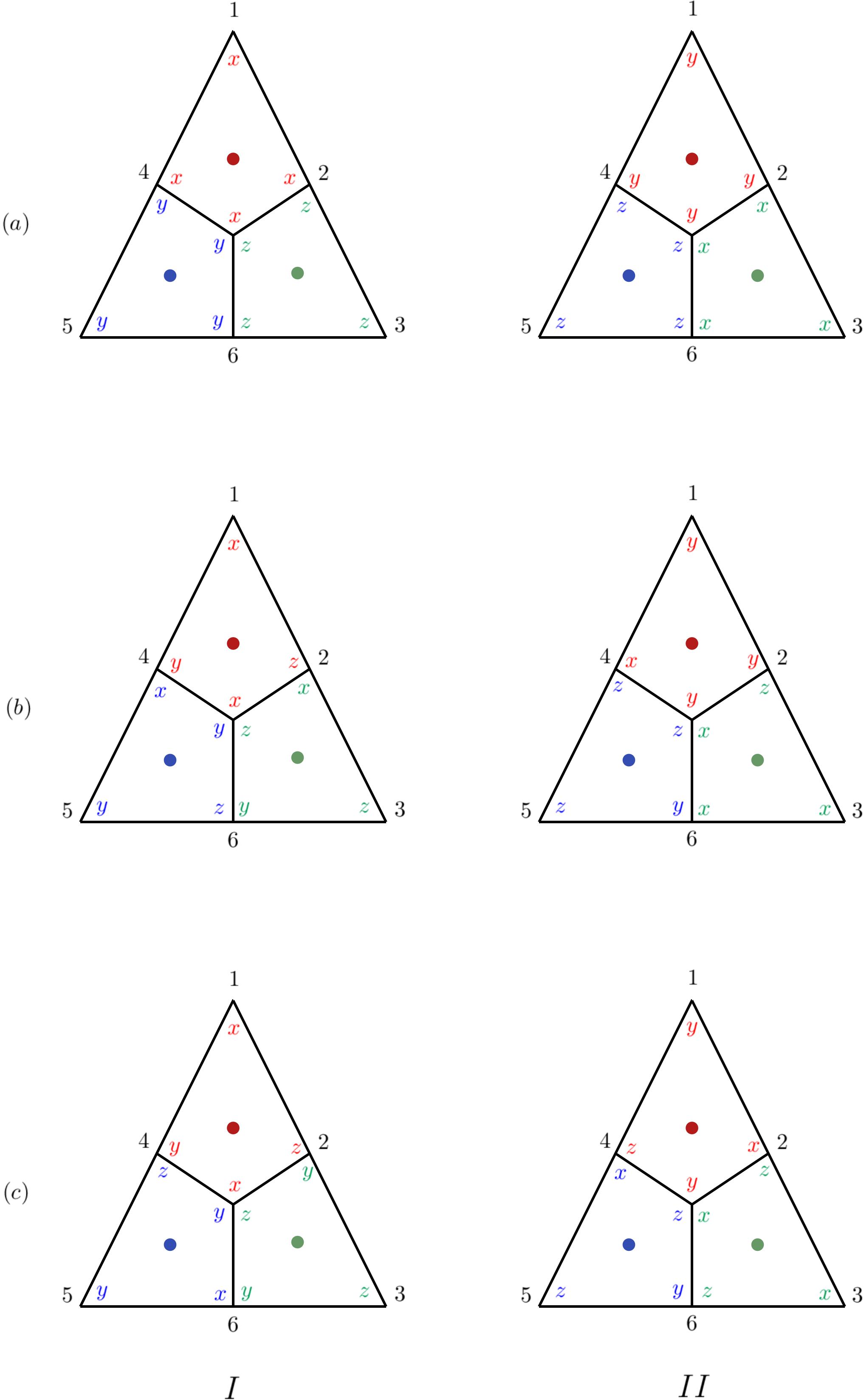}
 \caption{ The stabilizers of the $[444]$-Steane code. The indices of the physical qubits are shown with the 7th qubit indexing the lone bulk vertex. The stabilizers of the left triangle pick up the suffix $I$ and those in the right, $II$. The edge configurations are either (a) homogeneous (unmixed), (b) partially or (c) fully mixed.}
 \label{fig:[444]scstabilizers}
\end{figure}

\paragraph{\bf Homogeneous edge configuration -} The stabilizers for the canonical $[444]$- 7-qubit Steane code  are shown in Fig. \ref{fig:[444]scstabilizers} (a).
By inspection we find that the logical Pauli's for this code remain the same as the original $[336]$-Steane code, 
\begin{eqnarray}\label{eq:[444]sclogicalPaulis}
X_L & = & x_1x_2x_3x_4x_5x_6x_7, \nonumber \\
Z_L & = & z_1z_2z_3z_4z_5z_6z_7, \nonumber \\
Y_L & = & -y_1y_2y_3y_4y_5y_6y_7.
\end{eqnarray}
A more systematic way of arriving at these expressions is as follows. Consider the symmetries of the stabilizers of the $[444]$-Steane code, that are those operators that cannot be written as a product of the stabilizers. We find distance-three operators that have support along the boundaries of the triangle and another set of distance-three operators that have support along the diagonals of the triangle. Explicitly,
\begin{equation}
    \{x_1x_2x_3,~x_3x_6x_5,~x_5x_4x_1\},
\end{equation}
and
\begin{equation}
    \{x_1x_7x_6,~x_3x_7x_4,~x_5x_7x_2\},
\end{equation}
are two sets of distance-three symmetries along the boundaries and diagonals respectively. It is easy to see that any operator in these two sets are equivalent to the 
logical Pauli, $X_L$ (See Eq. (\ref{eq:[444]sclogicalPaulis})) up to a multiplication by a stabilizer of the $[444]$-Steane code with unmixed edge configurations. The logical Pauli's in Eq. (\ref{eq:[444]sclogicalPaulis}) are chosen in such a way that they have support on all the seven qubits. They are nevertheless stabilizer equivalent to appropriate weight-three symmetries. A similar method can be used to find the logical Pauli, $Z_L$ and hence $Y_L$. 

Having obtained the logical Pauli's the ground state can now be constructed by projecting the eigenstate of the $Z_L$ operator using the stabilizers, $\color{red}r_{I}$, $\color{red}r_{II}$, $\color{blue}b_{I}$, $\color{ForestGreen}g_{II}$. A simple calculation shows that this state coincides with that of the original Steane code,
\begin{eqnarray}
\ket{0_L} & = & \frac{1}{8}\left[ \ket{0_10_20_30_40_50_60_7} + \ket{0_10_20_31_41_51_61_7} \right. \nonumber \\ & & \left. + \ket{0_11_21_30_40_51_61_7} + \ket{0_11_21_31_41_50_60_7} \right. \nonumber \\
& & \left. + \ket{1_11_20_31_40_50_61_7} + \ket{1_11_20_30_41_51_60_7} \right. \nonumber \\ & & \left. + \ket{1_10_21_31_40_51_60_7} + \ket{1_10_21_30_41_50_61_7} \right] \label{eq:0Lsc}.
\end{eqnarray}
An easy way to understand this is by observing that the action of the $Y$ stabilizer on the eigenstate of $Z_L$ is the same as the $X$ stabilizer as each of them has the weight four. However this is not true when the stabilizers have the weight six as then we have $\mathrm{i}^6=-1$ and so the $Y$ stabilizer on a hexagon acts as $-X$ stabilizer on the eigenstate of $Z_L$. This will lead to important differences when we consider triangle codes of larger sizes. However for the simplest case of the 7-qubit triangle the $Y$ and $X$ stabilizers have the same action on the eigenstates of $Z_L$.

The other ground state for the $[444]$-Steane code is obtained as $\ket{1_L} = X_L\ket{0_L}$. Thus the codewords of the $[444]$-Steane code coincide with that of the original Steane code or the $[336]$-Steane code with unmixed edge configurations even though the two sets of stabilizers are not equivalent to each other by an LU.
As a consequence the logical Hadamard, phase gate and CNOT are unchanged as well. Thus the $[444]$-Steane code provides a set of stabilizers that supports fault-tolerant computation. \\

\paragraph{\bf Mixed edge configurations -}\,The situation is modified for the $[444]$-Steane codes with partially and fully mixed edge configurations (See Figs. \ref{fig:[444]scstabilizers} (b), (c)). The code space, the logical Pauli's and Cliffords see a change now but can be mapped to the canonical $[444]$-Steane code by the LU's, $H_2S_4(SHS^3)_6$ and $(SH)_2(HS^3)_4(SH)_6$ respectively. 

It is still illustrative to construct the logical operators for the mixed $[444]$-Steane codes from the ground up regardless of their equivalence to the canonical $[444]$-Steane code. This helps in seeing the equivalence class of the logical Pauli's and inspecting their mixed nature.

\subsubsection{The logical operators and the code space} 
 We proceed by identifying the symmetries of this Hamiltonian that are not generated by the stabilizers. As there is just a single encoded logical qubit we expect to find just a single set of the logical Pauli's $\{X_L, Z_L, Y_L\}$, supported on all the seven qubits.

\paragraph{\bf Symmetries - } From the geometry of the seven qubit system the symmetries with support exhausting all the seven qubits can be obtained by considering the symmetries along the boundaries (edges of the triangle) and the diagonals of the triangle (symmetries that include support on the qubit 7). For the $[444]$-Steane code with partially mixed edge configurations (See Fig. \ref{fig:[444]scstabilizers} (b)), the boundary symmetries are given by
\begin{eqnarray}
\left\{ x_1z_2x_3,~x_5x_6x_3,~x_1y_4x_5\right\} \label{eq:444scpartialmixingboundsym1} \\
\left\{ y_1y_2y_3,~y_5z_6y_3,~y_1x_4y_5\right\} \label{eq:444scpartialmixingboundsym2} \\
\left\{ z_1x_2z_3,~z_5y_6z_3,~z_1z_4z_5\right\}, \label{eq:444scpartialmixingboundsym3}
\end{eqnarray}
and the diagonal symmetries are given by 
\begin{eqnarray}
\left\{x_1x_7x_6,~x_5x_7z_2,~x_3x_7y_4\right\}\label{eq:444scpartialmixingdiagsym1} \\
\left\{y_1y_7z_6,~y_5y_7y_2,~y_3y_7x_4\right\}\label{eq:444scpartialmixingdiagsym2} \\
\left\{z_1z_7y_6,~z_5z_7x_2,~z_3z_7z_4\right\}.\label{eq:444scpartialmixingdiagsym3} 
\end{eqnarray}
Notice that each of the symmetries are of weight three and all the operators inside a given set commute with each other. On the other hand operators belonging to different sets anticommute with each other. We also note that the operators in Eqs. (\ref{eq:444scpartialmixingboundsym1}) and (\ref{eq:444scpartialmixingdiagsym1}) commute with each other and so do the operators in Eqs. (\ref{eq:444scpartialmixingboundsym2}) and (\ref{eq:444scpartialmixingdiagsym2}) and the operators in Eqs. (\ref{eq:444scpartialmixingboundsym3}) and (\ref{eq:444scpartialmixingdiagsym3}). Furthermore, we can check that the operators within a set are related to each other, up to a fourth root of unity, by the stabilizers of the model. For example,
\begin{eqnarray}
x_1z_2x_3 & \sim & x_1z_2x_3({\color{red}r_I}) \sim x_3x_7y_4, \\
x_1z_2x_3 & \sim & x_1z_2x_3({\color{red}r_I}{\color{blue}b_{I}}{\color{blue}b_{II}}) \sim x_5x_6x_3.
\end{eqnarray}
The logical Paulis, $X_L$ and $Z_L$, can be written as
\begin{eqnarray}
X'_L & = & x_1z_2x_3y_4x_5x_6x_7 \label{eq:444scpartialmixinglogicalX} \\
Z'_L & = & z_1x_2z_3z_4z_5y_6z_7. \label{eq:444scpartialmixinglogicalZ}
\end{eqnarray}
This choice is arbitrary and other choices just lead to rotated code states. 

\paragraph{\bf Code Space - } To construct the encoded qubit consider the +1 eigenstate of the logical Pauli $Z'_L$, $\ket{s_{Z'}}=\ket{0_{z_1},0_{x_2}, 0_{z_3}, 0_{z_4}, 0_{z_5}, 0_{y_6}, 0_{z_7}}$. We have used the +1 eigenstates of the Pauli operators, $x$, $y$ and $z$ as $\ket{0_x}=\frac{1}{\sqrt{2}}\left[\ket{0}+\ket{1}\right]$, $\ket{0_y}=\frac{1}{\sqrt{2}}\left[\ket{0}+\mathrm{i}\ket{1}\right]$ and $\ket{0_z}\equiv\ket{0}$ respectively. The encoded qubit is then 
\begin{eqnarray}
\ket{0'_L} & = & {\color{red}Pr_I}{\color{red}Pr_{II}}{\color{blue}Pb_{I}}{\color{ForestGreen}Pg_{II}}~\ket{s_{Z'}}, \\
\ket{1'_L} & = & X'_L~\ket{0'_L}.
\end{eqnarray}
Alternatively we can use the LU to the canonical $[444]$-Steane code to construct the logical qubit for the partially mixed model,
\begin{equation}
\ket{0'_L} = H_2S^3_4(SHS^3)_6~\ket{0_L}, ~~\ket{1'_L} = H_2S^3_4(SHS^3)_6~\ket{1_L},
\end{equation}
where $\ket{0_L}$ in Eq. (\ref{eq:0Lsc}), and $\ket{1_L}$ are the encoded qubits of the original $[336]$-Steane code.

\paragraph{\bf Logical Cliffords - } Using the LU, $H_2S_4(SHS^3)_6$ between the canonical $[444]$-Steane code and the partially mixed $[444]$-Steane code we obtain the logical Cliffords as
\begin{eqnarray}
H'_L & = & H_1H_2H_3\left(SHS^3\right)_4H_5\left(SHS^3HSHS^3\right)_6H_7, \\
S'_L & = & S_1\left(HSH\right)_2S_3S_4S_5\left(SHSHS^3\right)_6S_7,
\end{eqnarray}
and 
\begin{eqnarray}
CNOT'_{11'}\cdots CNOT'_{77'}  = & \nonumber \\  \left(H_2H_{2'}S_4S_{4'}t_6t_{6'}\right)&CNOT_L\left(H_2H_{2'}S_4S_{4'}t^\dag_6t^\dag_{6'}\right), \nonumber \\
\end{eqnarray}
where $t=SHS^3$ and $CNOT_L = CNOT_{11'}\cdots CNOT_{77'} $. The primed indices denote another copy of the Steane code. Note that the transversal nature of the Pauli operators is not affected across the various models. 

\subsubsection{The fully mixed $[444]$-Steane code} 
The stabilizers of the $[444]$-Steane code with fully mixed edge configurations (See Fig. \ref{fig:[444]scstabilizers} (c)) are equivalent to the canonical $[444]$-Steane code by the LU, $(SH)_2(HS^3)_4(SH)_6$. 
The boundary symmetries in this case are 
\begin{eqnarray}
\left\{ x_1z_2x_3,~x_5z_6x_3,~x_1y_4x_5\right\} \label{eq:444scfullmixingboundsym1} \\
\left\{ y_1x_2y_3,~y_5x_6y_3,~y_1z_4y_5\right\} \label{eq:444scfullmixingboundsym2} \\
\left\{ z_1y_2z_3,~z_5y_6z_3,~z_1x_4z_5\right\}, \label{eq:444scfullmixingboundsym3}
\end{eqnarray}
and the diagonal symmetries are given by 
\begin{eqnarray}
\left\{x_1x_7z_6,~x_5x_7z_2,~x_3x_7y_4\right\}\label{eq:444scfullmixingdiagsym1} \\
\left\{y_1y_7x_6,~y_5y_7x_2,~y_3y_7z_4\right\}\label{eq:444scfullmixingdiagsym2} \\
\left\{z_1z_7y_6,~z_5z_7y_2,~z_3z_7x_4\right\}.\label{eq:444scfullmixingdiagsym3} 
\end{eqnarray}
As in the previous case the operators within a set commute and those belonging to different sets anticommute. The operators within a single set are related to each other by the action of the stabilizers of the code in Fig. \ref{fig:[444]scstabilizers} (c).

Using the symmetries we build the logical Pauli operators with support on all seven qubits as,
\begin{eqnarray}
X^{''}_L & = & x_1z_2x_3y_4x_5z_6x_7 \label{eq:444scfullmixinglogicalX} \\
Z^{''}_L & = & z_1y_2z_3x_4z_5y_6z_7. \label{eq:444scfullmixinglogicalZ}
\end{eqnarray}
The ground states can be constructed by projecting the eigenstate of $Z_L^{''}$ using the stabilizers of the code in Fig. \ref{fig:[444]scstabilizers} (c). Using the LU the logical qubit in this case is written as 
\begin{eqnarray}
 \ket{0^{''}_L} & = & (HS^3)_2(SH)_4(HS^3)_6\ket{0_L}, \\
 \ket{1^{''}_L} & = & (HS^3)_2(SH)_4(HS^3)_6\ket{1_L},      
\end{eqnarray}
where $\ket{0_L}$ is the original Steane code, Eq. (\ref{eq:0Lsc}) and this coincides with the alternate construction of acting the projectors built out of the stabilizers of the $[444]$-Steane code with fully mixed edge configurations on the +1 eigenstate of $Z_L^{''}$.
The logical Cliffords are 
\begin{eqnarray}
H^{''}_L & = & H_1\left(SHS^3\right)_2H_3\left(HS^3HSH\right)_4H_5\left(SHS^3\right)_6H_7, \nonumber \\ \\
S^{''}_L & = & S_1\left(SHSHS^3\right)_2S_3\left(HSH\right)_4S_5\left(SHSHS^3\right)_6S_7, \nonumber \\
\end{eqnarray}
and the new $CNOT^{''}_L$ gate is obtained by conjugating the old one with $(SH)_2(HS^3)_4(SH)_6(SH)_{2'}(HS^3)_{4'}(SH)_{6'}$.

\subsection{$[345]$-Steane code}
\label{subsec:[345]sc}
\begin{figure}[h]
    \centering
    \includegraphics[width=8cm]{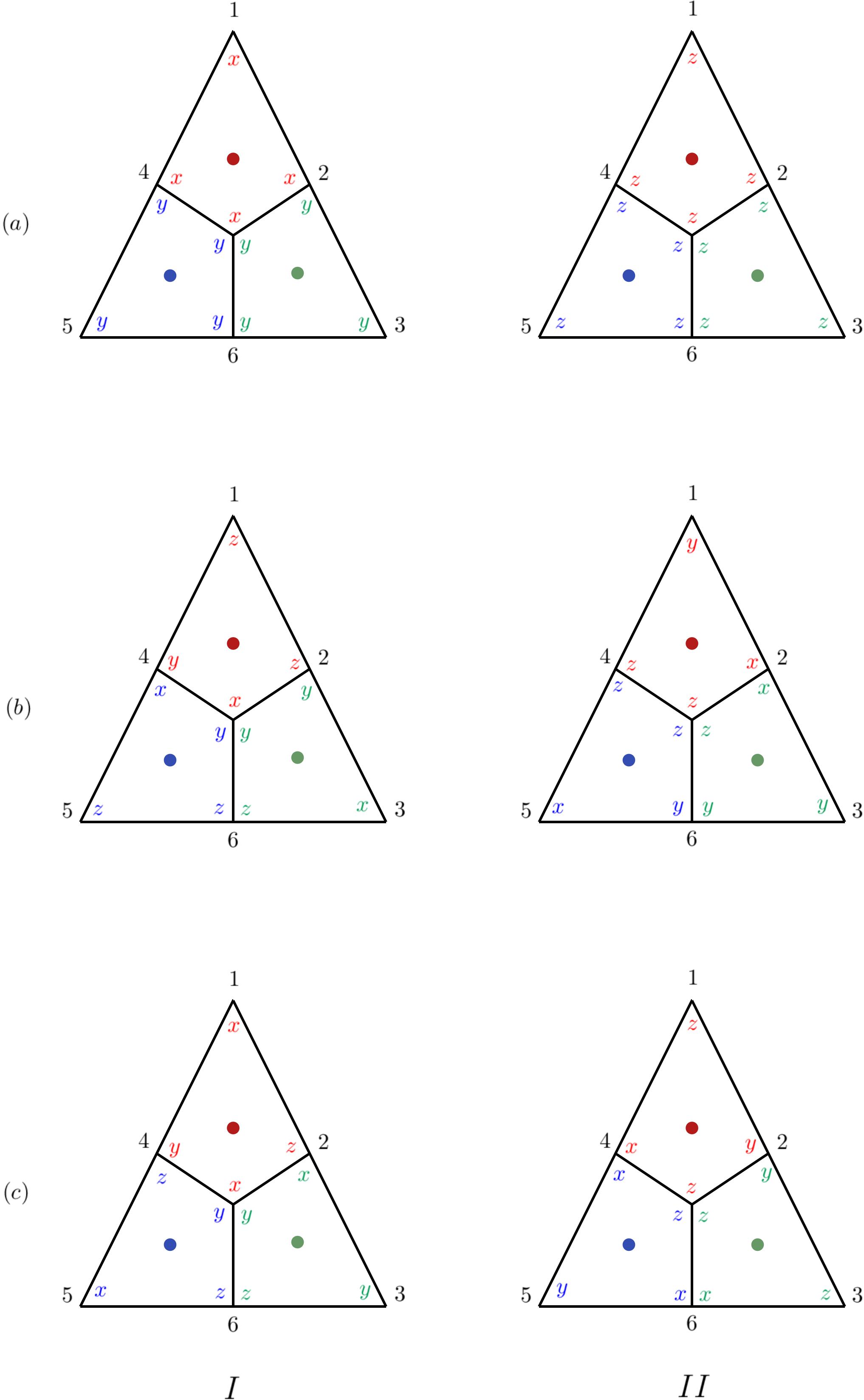}
 \caption{ The stabilizers of the $[345]$-Steane code. The indices of the physical qubits are shown with the 7th qubit indexing the lone bulk vertex. The stabilizers of the left triangle pick up the suffix $I$ and those in the right, $II$. The edge configurations are either (a) homogeneous, (b) partially mixed or (c) fully mixed.}
 \label{fig:[345]scstabilizers}
\end{figure}

The stabilizers for the canonical, partially and fully mixed, $[345]$- 7-qubit Steane code are shown in Fig. \ref{fig:[345]scstabilizers}.
Though there are three inequivalent $[345]$ bulk vertices (See Fig. \ref{fig:[345]bulkvertexconfigs}), they can be mapped into each other by a global rotation of all the stabilizers of the code. This is a consequence of the fact there is just a single bulk vertex to contend with in the Steane code. Hence we pick one of them for further consideration without a loss of generality.
The Hamiltonian is built out of the projectors corresponding to these stabilizers as in the $[444]$-Steane code. The dimension of the code space is once again 2 as all the stabilizers are independent. To construct the logical qubit we once again identify the symmetries and the logical Paulis. We note that the for canonical $[345]$-Steane code in Fig. \ref{fig:[345]scstabilizers} (a), the logical Pauli operators are the same as the original $[336]$-Steane code. The logical qubit is also unchanged and is given by $\ket{0_L}$ and $\ket{1_L}$ in Eq. (\ref{eq:0Lsc}). We find a change in the expressions for the logical operators and the logical qubits when we use partially and fully mixed edge configurations as shown in Figs. \ref{fig:[345]scstabilizers} (b), (c). However these codes are equivalent to the canonical $[345]$-Steane code by LU's, $(SH)_1H_2(SH)_3S_4(HS^3)_5(SHS^3)_6$ and $(SH)_2(HS^3)_4(SH)_5(HS^3)_6$ respectively.

\subsubsection{The partially mixed $[345]$-Steane code} 
The boundary and the diagonal symmetries, whose combined support covers all the seven physical qubits, are given by,
\begin{eqnarray}
\left\{ z_1z_2z_3,~y_5x_6z_3,~z_1y_4y_5\right\} \label{eq:345scpartialmixingboundsym1} \\
\left\{ y_1x_2y_3,~x_5y_6y_3,~y_1z_4x_5\right\} \label{eq:345scpartialmixingboundsym2} \\
\left\{ x_1y_2x_3,~z_5z_6x_3,~x_1x_4z_5\right\}, \label{eq:345scpartialmixingboundsym3}
\end{eqnarray}
and 
\begin{eqnarray}
\left\{z_1x_7x_6,~y_5x_7z_2,~z_3x_7y_4\right\}\label{eq:345scpartialmixingdiagsym1} \\
\left\{y_1z_7y_6,~x_5z_7x_2,~y_3z_7z_4\right\}\label{eq:345scpartialmixingdiagsym2} \\
\left\{x_1y_7z_6,~z_5y_7y_2,~x_3y_7x_4\right\}.\label{eq:345scpartialmixingdiagsym3} 
\end{eqnarray}
As in the case of the $[444]$-Steane code, the symmetries within each $\{\}$ mutually commute 
and anticommute with the symmetries in the other $\{\}$. Also the symmetries in Eqs. (\ref{eq:345scpartialmixingboundsym1}) and (\ref{eq:345scpartialmixingdiagsym1}), those in Eqs. (\ref{eq:345scpartialmixingboundsym2}) and (\ref{eq:345scpartialmixingdiagsym2}), and those in Eqs. (\ref{eq:345scpartialmixingboundsym3}) and (\ref{eq:345scpartialmixingdiagsym3}) mutually commute. 
This results in the logical Pauli's that read as
\begin{eqnarray}
X'_L & = & z_1z_2z_3y_4y_5x_6x_7 \label{eq:345scpartialmixinglogicalX} \\
Z'_L & = & y_1x_2y_3z_4x_5y_6z_7. \label{eq:345scpartialmixinglogicalZ}
\end{eqnarray}
The logical qubit is constructed using the LU as
\begin{eqnarray}
\ket{0'_L} & = & (HS^3)_1H_2(HS^3)_3S^3_4(SH)_5(SHS^3)_6\ket{0_L}, \label{eq:345scpartialmixing0L} \\
\ket{1'_L} & = & (HS^3)_1H_2(HS^3)_3S^3_4(SH)_5(SHS^3)_6\ket{1_L}. \label{eq:345scpartialmixing1L}
\end{eqnarray}
The logical Cliffords are obtained by conjugating the transversal Clifford gates of the canonical $[345]$-Steane code by the LU, $(SH)_1H_2(SH)_3S_4(HS^3)_5(SHS^3)_6$.
 
\subsubsection{The fully mixed $[345]$-Steane code} 
The boundary and the diagonal symmetries, whose combined support covers all the seven physical qubits, are given by,
\begin{eqnarray}
\left\{ x_1z_2x_3,~z_5y_6x_3,~x_1y_4z_5\right\} \label{eq:345scfullmixingboundsym1} \\
\left\{ z_1y_2z_3,~y_5x_6z_3,~z_1x_4y_5\right\} \label{eq:345scfullmixingboundsym2} \\
\left\{ y_1x_2y_3,~x_5z_6y_3,~y_1z_4x_5\right\}, \label{eq:345scfullmixingboundsym3}
\end{eqnarray}
and 
\begin{eqnarray}
\left\{x_1x_7y_6,~z_5x_7z_2,~x_3x_7y_4\right\}\label{eq:345scfullmixingdiagsym1} \\
\left\{z_1z_7x_6,~y_5z_7y_2,~z_3z_7x_4\right\}\label{eq:345scfullmixingdiagsym2} \\
\left\{y_1y_7z_6,~x_5y_7x_2,~y_3y_7z_4\right\}.\label{eq:345scfullmixingdiagsym3} 
\end{eqnarray}
The logical Pauli's are then,
\begin{eqnarray}
X^{''}_L & = & x_1z_2x_3y_4z_5y_6x_7 \label{eq:345scfullmixinglogicalX} \\
Z^{''}_L & = & z_1y_2z_3x_4y_5x_6z_7. \label{eq:345scfullmixinglogicalZ}
\end{eqnarray}
The logical qubit is constructed using the LU as
\begin{eqnarray}
\ket{0^{''}_L} & = & (HS^3)_2(SH)_4(HS^3)_5(SH)_6~\ket{0_L}, \label{eq:345scfullmixing0L} \\
\ket{1^{''}_L} & = & (HS^3)_2(SH)_4(HS^3)_5(SH)_6~\ket{1_L}. \label{eq:345scfullmixing1L}
\end{eqnarray}
The logical Cliffords are obtained by conjugating the transversal Clifford gates of the canonical $[345]$-Steane code by the LU, $(SH)_2(HS^3)_4(SH)_5(HS^3)_6$.
 
\subsection{$[336]$-Steane code}
\label{subsec:[336]sc}
\begin{figure}[h]
    \centering
    \includegraphics[width=8cm]{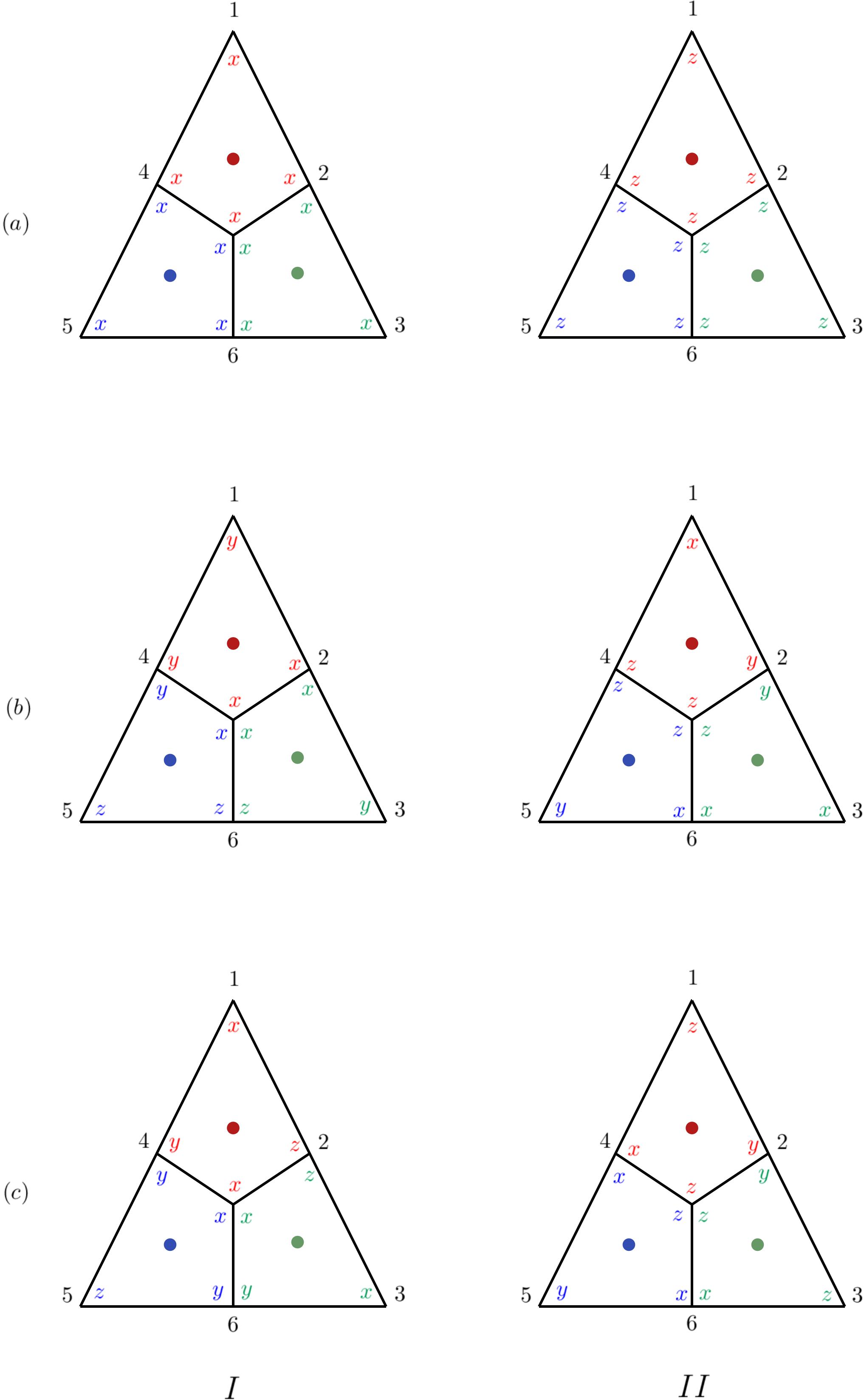}
 \caption{ The stabilizers of the $[336]$-Steane code. The indices of the physical qubits are shown with the 7th qubit indexing the lone bulk vertex. The stabilizers of the left triangle pick up the suffix $I$ and those in the right, $II$. The edge configurations are either (a) homogeneous, (b) partially mixed or (c) fully mixed.}
 \label{fig:[336]scstabilizers}
\end{figure}

For completion we include the canonical, partially and fully mixed $[336]$-Steane codes in Fig. \ref{fig:[336]scstabilizers}.
The canonical $[336]$-Steane code in Fig. \ref{fig:[336]scstabilizers} (a) is the original Steane code and its logical qubit and operators are well known. We will discuss the $[336]$-Steane codes with partially and fully mixed edge configurations that are equivalent to the canonical $[336]$-Steane code by the LU's, $(HS^3)_1(SHS^3)_2(HS^3)_3S_4(SH)_5H_6$ and $(SH)_2(HS^3)_4(SH)_5(HS^3)_6$ respectively.

\subsubsection{The partially mixed $[336]$-Steane code} 
The boundary and the diagonal symmetries, whose combined support covers all the seven physical qubits, are given by,
\begin{eqnarray}
\left\{ y_1x_2y_3,~z_5z_6y_3,~y_1y_4z_5\right\} \label{eq:336scpartialmixingboundsym1} \\
\left\{ x_1y_2x_3,~y_5x_6x_3,~x_1z_4y_5\right\} \label{eq:336scpartialmixingboundsym2} \\
\left\{ z_1z_2z_3,~x_5y_6z_3,~z_1x_4x_5\right\}, \label{eq:336scpartialmixingboundsym3}
\end{eqnarray}
and 
\begin{eqnarray}
\left\{y_1x_7z_6,~z_5x_7x_2,~y_3x_7y_4\right\}\label{eq:336scpartialmixingdiagsym1} \\
\left\{x_1z_7x_6,~y_5z_7y_2,~x_3z_7z_4\right\}\label{eq:336scpartialmixingdiagsym2} \\
\left\{z_1y_7y_6,~x_5y_7z_2,~z_3y_7x_4\right\}.\label{eq:336scpartialmixingdiagsym3} 
\end{eqnarray}
As in the earlier cases, the symmetries within each $\{\}$ mutually commute 
and anticommute with the symmetries in the other $\{\}$. Also the symmetries in Eqs. (\ref{eq:336scpartialmixingboundsym1}) and (\ref{eq:336scpartialmixingdiagsym1}), those in Eqs. (\ref{eq:336scpartialmixingboundsym2}) and (\ref{eq:336scpartialmixingdiagsym2}) and those in Eqs. (\ref{eq:336scpartialmixingboundsym3}) and (\ref{eq:336scpartialmixingdiagsym3}) mutually commute. 
This results in the logical Pauli's that read as
\begin{eqnarray}
X'_L & = & y_1x_2y_3y_4z_5z_6x_7 \label{eq:336scpartialmixinglogicalX} \\
Z'_L & = & x_1y_2x_3z_4y_5x_6z_7. \label{eq:336scpartialmixinglogicalZ}
\end{eqnarray}
The logical qubit is constructed using the LU as
\begin{eqnarray}
\ket{0'_L} & = & (SH)_1(SHS^3)_2(SH)_3S^3_4(HS^3)_5H_6~\ket{0_L},~~~~ \label{eq:336scpartialmixing0L} \\
\ket{1'_L} & = & (SH)_1(SHS^3)_2(SH)_3S^3_4(HS^3)_5H_6~\ket{1_L}.~~~~ \label{eq:336scpartialmixing1L}
\end{eqnarray}
The logical Cliffords are obtained by conjugating the transversal Clifford gates of the canonical $[336]$-Steane code by the LU, $(HS^3)_1(SHS^3)_2(HS^3)_3S_4(SH)_5H_6$.

\subsubsection{The fully mixed $[336]$-Steane code} 
The boundary and the diagonal symmetries, whose combined support covers all the seven physical qubits, are given by,
\begin{eqnarray}
\left\{ x_1z_2x_3,~z_5y_6x_3,~x_1y_4z_5\right\} \label{eq:336scfullmixingboundsym1} \\
\left\{ z_1y_2z_3,~y_5x_6z_3,~z_1x_4y_5\right\} \label{eq:336scfullmixingboundsym2} \\
\left\{ y_1x_2y_3,~x_5z_6y_3,~y_1z_4x_5\right\}, \label{eq:336scfullmixingboundsym3}
\end{eqnarray}
and 
\begin{eqnarray}
\left\{x_1x_7y_6,~z_5x_7z_2,~x_3x_7y_4\right\}\label{eq:336scfullmixingdiagsym1} \\
\left\{z_1z_7x_6,~y_5z_7y_2,~z_3z_7x_4\right\}\label{eq:336scfullmixingdiagsym2} \\
\left\{y_1y_7z_6,~x_5y_7x_2,~y_3y_7z_4\right\}.\label{eq:336scfullmixingdiagsym3} 
\end{eqnarray}
As in the earlier cases, the symmetries within each $\{\}$ mutually commute  
and anticommute with the symmetries in the other $\{\}$. Also the symmetries in Eqs. (\ref{eq:336scfullmixingboundsym1}) and (\ref{eq:336scfullmixingdiagsym1}), those in Eqs. (\ref{eq:336scfullmixingboundsym2}) and (\ref{eq:336scfullmixingdiagsym2}), and those in Eqs. (\ref{eq:336scfullmixingboundsym3}) and (\ref{eq:336scfullmixingdiagsym3}) mutually commute. 
This results in the logical Pauli's that read as
\begin{eqnarray}
X^{''}_L & = & x_1z_2x_3y_4z_5y_6x_7 \label{eq:336scfullmixinglogicalX} \\
Z^{''}_L & = & z_1y_2z_3x_4y_5x_6z_7. \label{eq:336scfullmixinglogicalZ}
\end{eqnarray}
The logical qubit is constructed using the LU as
\begin{eqnarray}
\ket{0^{''}_L} & = & (HS^3)_2(SH)_4(HS^3)_5(SH)_6~\ket{0_L}, \label{eq:336scfullmixing0L} \\
\ket{1^{''}_L} & = & (HS^3)_2(SH)_4(HS^3)_5(SH)_6~\ket{1_L}. \label{eq:336scfullmixing1L}
\end{eqnarray}
The logical Cliffords are obtained by conjugating the transversal Clifford gates of the canonical $[336]$-Steane code by the LU, $(SH)_2(HS^3)_4(SH)_5(HS^3)_6$.

\subsection{The 19-qubit code}
\label{subsec:19qtc}
\begin{figure}[h]
    \centering
    \includegraphics[width=9cm]{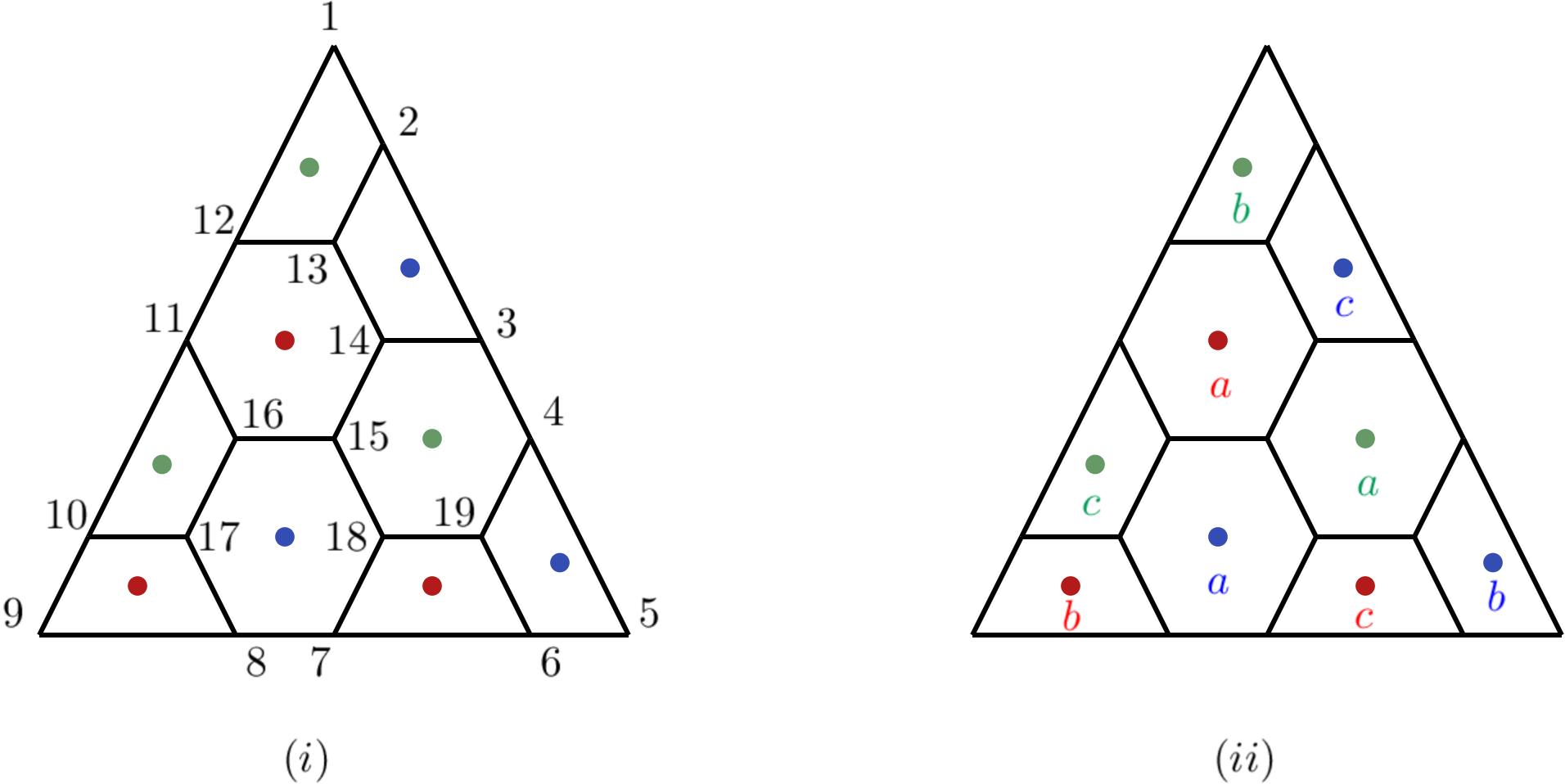}
 \caption{The labels of the (i) 19 qubits and (ii) the faces of different colors for the 19-qubit code are indexed by $\{a, b, c\}$. }
 \label{fig:19qtcconfiguration}
\end{figure}

The triangle geometry of the 7-qubit Steane code can be cut out from the hexagon lattice such that we do not include any full hexagon. A natural generalization of this is cutting out a larger triangle such that we include exactly one full hexagon of each color resulting in the {\it 19-qubit code} (See Fig. \ref{fig:19qtcconfiguration}). The 7-qubit and 19-qubit codes are examples of distance-three and distance-five codes in the triangle code family. 

There are three faces for each color and thus six stabilizers for each color summing up to 18 independent stabilizers that results in a single encoded qubit. The stabilizers are denoted $\color{red}(r_i)_j, \color{blue}(b_i)_j, \color{ForestGreen}(g_i)_j$ for $i\in\{a, b, c\}$ and $j\in\{I, II\}$. The stabilizers generate an abelian subgroup of the 19-qubit Pauli group and the projectors built out of them is used to construct the commuting projector Hamiltonian just as in the previous cases. Before we analyze the $[345]$- and $[444]$-19 qubit codes we first review the case of the original CSS $[336]$-19 qubit codes. We construct the code words or the logical qubit and the appropriate logical Pauli and Clifford operations.

\subsubsection{$[336]$-19 qubit code}
\label{subsubsec:[336]19qtc}
The CSS version of the 19 qubit codes are stabilized by the $X$ and $Z$ operators for each face of the 19 qubit triangle. The logical Pauli's are given by 
\begin{equation}
    X_L = x_1\cdots x_{19},~~Z_L=z_1\cdots z_{19}.
\end{equation}
As there is just one encoded qubit we expect to find just a single set of logical Pauli operators. These can be systematically obtained by identifying the symmetries of the stabilizers just as in the case of the 7 qubit Steane codes. In this case we consider the symmetries along the boundaries of the triangle and this covers the sites 1 through 12 (See Fig. \ref{fig:19qtcconfiguration}). Next we consider symmetries that are stabilizer equivalent to the boundary symmetries by beginning at a corner vertex and then we cut through the triangle spanning bulk vertices and end on a boundary vertex opposing the chosen corner vertex. For example in the green lattice of Fig. \ref{fig:19qtcconfiguration}, three such symmetries exhausting all the bulk vertices are given by, $x_1x_{13}x_{14}x_{19}x_6$, $x_1x_{13}x_{14}x_{18}x_7$ and $x_1x_{13}x_{14}x_{15}x_{16}x_{17}x_8$. Note that these symmetries are at least of distance five. Along with the boundary symmetries, each of which are 
distance-five symmetries, we cover all the sites of the 19 qubit triangle. 

The encoded qubit is constructed by projecting, $\ket{s_Z}$ the eigenstate of the $Z_L$ logical with the $X$ stabilizers,
\begin{equation}
    \ket{0_L} = \prod\limits_{j\in\{a,b,c\}}~{\color{red}PX_j}{\color{blue}PX_j}{\color{ForestGreen}PX_j}~\ket{s_Z},~~\ket{1_L}=X_L~\ket{0_L}.
\end{equation}

\paragraph{\bf Logical Hadamard, $H_L$ - } The transversal logical Hadamard in this case is given by 
\begin{equation}
    H_L = H_1\cdots H_{19}.
\end{equation}
This is verified to be the right expression by checking its action on the logical qubit as,
\begin{eqnarray}
H_L\ket{0_L} & = & H_L\prod\limits_{j\in\{a,b,c\}}~{\color{red}PX_j}{\color{blue}PX_j}{\color{ForestGreen}PX_j}~\ket{s_Z}, \nonumber \\
& = & \prod\limits_{j\in\{a,b,c\}}~{\color{red}PZ_j}{\color{blue}PZ_j}{\color{ForestGreen}PZ_j}~\ket{s_X},
\end{eqnarray}
where $\ket{s_X}$ is the +1 eigenstate of $X_L$. This state is precisely the dual state and is given by $\frac{\ket{0_L}+\ket{1_L}}{\sqrt{2}}$. It is easily seen that $H_L\ket{1_L}=\frac{\ket{0_L}-\ket{1_L}}{\sqrt{2}}$.
Moreover the logical Hadamard consistently acts on the logical Pauli's by interchanging $X_L$ and $Z_L$ under conjugation. 

\paragraph{\bf Logical Phase Gate, $S_L$ - } Obtaining the transversal logical phase gate is a bit more subtle as a naive choice of $S_1\cdots S_{19}$ does not have the right action on the logical qubit. This is seen as follows,
\begin{eqnarray}
S_L\ket{0_L} & = & S_L \prod\limits_{j\in\{a,b,c\}}~{\color{red}PX_j}{\color{blue}PX_j}{\color{ForestGreen}PX_j}~\ket{s_Z}, \nonumber \\
& = & \prod\limits_{j\in\{a,b,c\}}~{\color{red}PY_j}{\color{blue}PY_j}{\color{ForestGreen}PY_j}~\ket{s_Z},
\end{eqnarray}
using $SXS^3=Y$. However the resulting state is not the same as $\ket{0_L}$ as $PY$ supported on a hexagon acts on $\ket{s_Z}$ as $PX^{\perp}$ or the orthogonal complement of $PX$. This shows that the naive choice for the transversal logical phase gate is not the right one. To fix this we use $S^3$ on half of the sites for each hexagon such that the rest of the stabilizers supported on the quadrilaterals transform as before. This will ensure that the $PX$ operators on each hexagon transforms to $PY^\perp$ which acts on $\ket{s_Z}$ in the same way as $PX$. Thus the consistent choice for the logical phase gate is given by 
\begin{equation}
    S_L = \prod\limits_{j\in\mathcal{B}}S^3 \prod\limits_{j\notin\mathcal{B}}S,
\end{equation}
where the set comprises of the sites $\mathcal{B}=\{3,4,7,8,11,12,13,17,19\}$.
This operator has the right action on $X_L$ transforming it to $Y_L$ under conjugation and commutes with $Z_L$ leaving it invariant. 

\paragraph{\bf Logical CNOT Gate, $CNOT_L$ - } The $CNOT$ gate is a two qubit gate which performs the NOT operation with the first qubit as a control qubit. Explicitly it acts as a Pauli $X$ on the second qubit if the first qubit is in the $\ket{1}$ state. As an operator this translates to 
\begin{equation}
    CNOT_{12} = \frac{1}{2}\left[1+Z_1+X_2-Z_1X_2\right], 
\end{equation}
where 1 and 2 index the two qubits. Under conjugation by $CNOT$ the Pauli operators on the two qubits transform as 
\begin{eqnarray}
X_1 \rightarrow X_1X_2, && ~ X_2\rightarrow X_2, \nonumber \\
Z_1 \rightarrow Z_1, && ~ Z_2 \rightarrow Z_1Z_2.
\end{eqnarray}
These expressions will be useful in constructing the logical $CNOT_L$ on the encoded qubits of the $[336]$-19 qubit code. We pick the naive choice that is,
\begin{equation}
    CNOT_L = \prod\limits_{j=1}^{19}~\left(CNOT\right)_{jj'},
\end{equation}
the unprimed and the primed indices denote the two copies of the $[336]$-19 qubit codes
To verify that this choice indeed does the job, we verify its action on two copies of the encoded qubits. Explicitly,
\begin{eqnarray}
CNOT_L\ket{0_L, 0_L} & = & \nonumber \\ & & CNOT_L~\prod\limits_{j\in\{a,b,c\}}~{\color{red}PX_j}{\color{blue}PX_j}{\color{ForestGreen}PX_j} \nonumber \\ & & \prod\limits_{j\in\{a,b,c\}}~{\color{red}PX_{j'}}{\color{blue}PX_{j'}}{\color{ForestGreen}PX_{j'}}\ket{s_Z}\otimes \ket{s_Z} \nonumber \\
& = &\prod\limits_{j,j'\in\{a,b,c\}}~{\color{red}PX_jX_{j'}}{\color{blue}PX_jX_{j'}}{\color{ForestGreen}PX_jX_{j'}} \nonumber \\
& & \prod\limits_{j\in\{a,b,c\}}~{\color{red}PX_{j'}}{\color{blue}PX_{j'}}{\color{ForestGreen}PX_{j'}}\ket{s_Z}\otimes \ket{s_Z} \nonumber\\
\end{eqnarray}
which coincides with $\ket{0_L, 0_L}$ as $PX_jX_{j'}PX_{j'} = PX_jPX_{j'}$. Clearly it leaves $\ket{0_L, 1_L}$ also unchanged as $\ket{0_L, 1_L} = X'_L\ket{0_L, 0_L}$ and $CNOT_L$ does not transform the Pauli $X$ operator acting on the second copy. On the other hand when the control qubit is $\ket{1_L}$, the $CNOT_L$ operator changes the second encoded qubit. For example,
\begin{eqnarray}
CNOT_L \ket{1_L, 0_L} & = & CNOT_LX_L\ket{0_L, 0_L}, \nonumber \\
& = & X_LX'_LCNOT_L\ket{0_L, 0_L}, \nonumber \\
& = & \ket{1_L, 1_L}.
\end{eqnarray}
It is easy to see that $CNOT_L\ket{1_L, 1_L} = \ket{1_L, 0_L}$ as $CNOT_L^2=1$. Thus the naive choice for the transversal $CNOT_L$ gate acts in the desired manner on both the logical qubit and the logical Pauli operators. 

We have thus established the well known fact that the CSS $[336]$- 19 qubit code supports fault-tolerant computation. Next we proceed to the construction of the logical Pauli's, the logical qubit and the logical Cliffords for both the $[345]$- and $[444]$-19 qubit codes. We will see that the code words in each of these cases can be mapped to the code words of the CSS $[336]$-19 qubit code by an LU. 

\subsubsection{$[345]$-19 qubit code}
\label{subsubsec:[345]19qtc}
We begin with the $[345]$-19 qubit codes. Unlike the 7-qubit triangles there are more bulk vertices in the 19-qubit triangles which implies there are more possibilities for inequivalent $[345]$-triangle codes. More precisely there are three inequivalent $[345]$-triangle codes corresponding to the inequivalent bulk vertex configurations in Fig. \ref{fig:[345]bulkvertexconfigs}. And for each of these configurations we can build either a deconfined or a confined $[345]$-19 qubit code. 

\paragraph{\bf Deconfined $[345]$-19 qubit code - } Among the three possible inequivalent $[345]$-vertex types (See Fig. \ref{fig:[345]bulkvertexconfigs}), we discuss just one of them with homogeneous edge configurations. We pick the case where the every red face has one $Y$ and one $Z$ stabilizers whereas each of the blue and green faces have one $X$ and one $Z$ stabilizers. To construct the code words we begin with the logical Pauli operators. These operators coincide with those of the CSS $[336]$-19 qubit codes, that is $X_L=x_1\cdots x_{19}$ and $Z_L=z_1\cdots z_{19}$. Moving on to the construction of the logical qubit we project the $\ket{s_Z}$, the +1 eigenstate of the $Z_L$ operator,
\begin{equation}
   \ket{\tilde{0}_L} =  \prod\limits_{j\in\{a,b,c\}}~{\color{red}PY_j}{\color{blue}PX_j}{\color{ForestGreen}PX_j}~\ket{s_Z}.
\end{equation}
 The projectors on the red faces differ from the projectors used in the construction of the logical qubit in the $[336]$-19 qubit code. As seen earlier the ${\color{red}PY_a}$ projector on the red hexagon acts as ${\color{red}PX^\perp_a}$ on the state $\ket{s_Z}$, whereas the ${\color{red}PY_b}$ and ${\color{red}PY_c}$ on the quadrilaterals act as ${\color{red}PX_b}$ and ${\color{red}PX_c}$ on the state $\ket{s_Z}$. With the $Y$ projectors reducing in this manner the code word becomes,
 \begin{eqnarray}
 \ket{\tilde{0}_L} & = & {\color{red}PX^\perp_a}{\color{red}PX_b}{\color{red}PX_c} \prod\limits_{j\in\{a,b,c\}}~{\color{blue}PX_j}{\color{ForestGreen}PX_j}~\ket{s_Z}, \nonumber \\
& = & z_1z_{12} \prod\limits_{j\in\{a,b,c\}}~{\color{red}PX_j}{\color{blue}PX_j}{\color{ForestGreen}PX_j}~\ket{s_Z}, \nonumber \\
& = & z_1z_{12}~\ket{0_L},
\end{eqnarray}
with $\ket{0_L}$ being the logical qubit of the CSS $[336]$-19 qubit code. Note that $z_1z_{12}$ is not the unique choice that flips ${\color{red}PX_a}$ to ${\color{red}PX^\perp_a}$. Any operator that is equivalent to $z_1z_{12}$ up to a multiplication by a logical Pauli $Z$ also has the same effect. The state $\ket{\tilde{0}_L}$ can also be viewed as an excited state in the spectrum of the CSS $[336]$-19 qubit Hamiltonian. 

The LU equivalence of the $[336]$- and the deconfined $[345]$-19 qubit codes implies that the latter supports fault-tolerant computation as well. The $[345]$-19 qubit codes with partially and fully mixed edge configurations are LU equivalent to the one just discussed with homogeneous edge configurations. Hence they are naturally fault-tolerant in nature. 

\begin{figure}[h]
    \centering
    \scalebox{0.8}{\includegraphics[width=10cm]{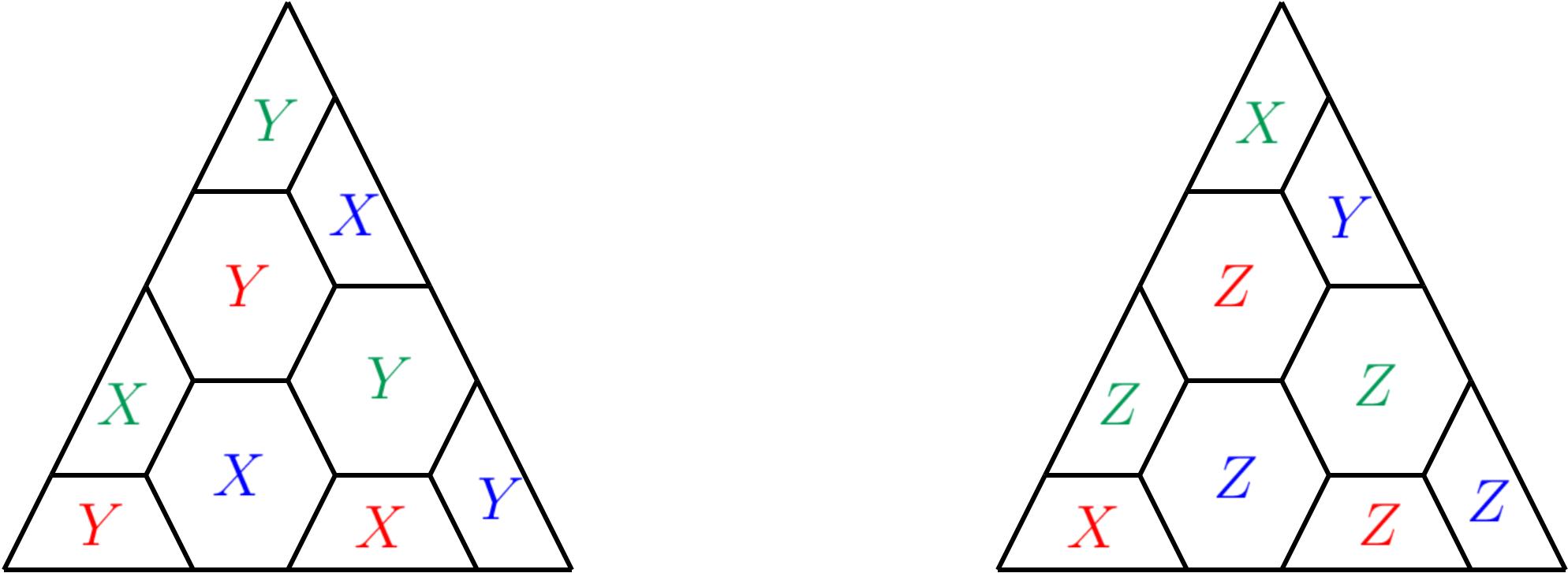}}
 \caption{The stabilizers for the confined $[345]$-19 qubit code with homogeneous edge configurations.}
 \label{fig:[345]19qtcconfined}
\end{figure}

\paragraph{\bf Confined $[345]$-19 qubit code - } The stabilizer pair for a confined $[345]$-19 qubit code is shown in Fig. \ref{fig:[345]19qtcconfined}.
The logical Pauli operators continue to be the same as in the previous cases namely, $X_L=x_1\cdots x_{19}$ and $Z_L=z_1\cdots z_{19}$. The encoded qubit is constructed as 
\begin{eqnarray}
\ket{\tilde{\tilde{0}}_L} & = & {\color{red}PY_a}{\color{ForestGreen}PY_a}{\color{blue}PX_a}{\color{red}PY_b}{\color{red}PX_b}{\color{red}PX_c} \nonumber \\
& & {\color{blue}PY_b}{\color{blue}PY_c}{\color{blue}PX_c}{\color{ForestGreen}PX_b}{\color{ForestGreen}PY_b}{\color{ForestGreen}PX_c}~\ket{s_Z}.
\end{eqnarray}
As we saw earlier the $Y$ projectors on the quadrilaterals act in the same way as the $X$ projectors on the $\ket{s_Z}$ state. The key difference arises in the action of the $Y$ projectors on the red and green hexagons on the $\ket{s_Z}$ state. We have seen that they act as $PX^\perp$ on $\ket{s_Z}$. As a result the $\ket{\tilde{\tilde{0}}_L}$ reduces to 
\begin{equation}
    \ket{\tilde{\tilde{0}}_L} = \left(z_1z_{12}\right)\left(z_4z_5\right)~\ket{0_L},
\end{equation}
with $\ket{0_L}$ being the code word for the $[336]$-19 qubit code, and thereby establishing the LU equivalence between the logical qubit of the confined $[345]$-19 qubit code and the $[336]$-19 qubit code. 

\subsubsection{$[444]$-19 qubit code}
\label{subsubsec:[444]19qtc}
Consider the canonical $[444]$-19 qubit code where the edge configurations are homogeneous. The red stabilizers are made of the $\color{red}X$ and $\color{red}Y$ operators, the blue faces with the $\color{blue}Y$ and $\color{blue}Z$ operators and finally the green faces with the $\color{ForestGreen}Z$ and $\color{ForestGreen}X$ operators. 

It is easy to see that all the symmetries, whose combined support spans all the 19 qubit locations, are of the homogeneous type, that is they are composed of the same Pauli operator. This results in logical Pauli's of the homogeneous type, namely the $X_L$ and the $Z_L$ operators.
However if we construct the logical qubit, by projecting the +1 eigenstate of either of the logical Pauli's using the projectors in the Hamiltonian, we obtain the 0 state as these eigenstates are orthogonal to the code space. This can be checked explicitly by evaluating ${\color{red}(PX_a)_I(PY_a)_{II}}\ket{s_Z}$, where $\ket{s_Z}$ is the +1 eigenstate of $Z_L$. Hence we need to make a different choice for the seed state that generates the ground state or the logical qubit. 
Instead the state
\begin{eqnarray}
\ket{s} & = & \ket{1_{z_1}0_{z_2}\cdots 0_{z_{11}}1_{z_{12}}0_{z_{13}}\cdots 0_{z_{19}}}, \nonumber \\
& = & x_1x_{12}~\ket{s_Z}
\end{eqnarray}
has a non-zero projection to the code space and can be used to construct the the logical qubit, $\ket{0^{''}_L}$, with the projectors, $\prod\limits_{i\in\{a,b,c\}}~{\color{red}(PX_i)}{\color{red}(PY_i)}{\color{blue}(PY_i)}{\color{ForestGreen}(PX_i)}$. The logical qubit, $\ket{0^{''}_L}$ then reduces to,
\begin{eqnarray}
\ket{0^{''}_L} & = & \prod\limits_{i\in\{a,b,c\}}~{\color{red}(PX_i)}{\color{red}(PY_i)}{\color{blue}(PY_i)}{\color{ForestGreen}(PX_i)}x_1x_{12}~\ket{s_Z}, \nonumber \\
& = & x_1x_{12} {\color{red}(PX_a)}{\color{red}(PY^\perp_a)}{\color{blue}(PY_a)}{\color{ForestGreen}(PX_a)} \nonumber \\
& &\prod\limits_{i\in\{b,c\}}~{\color{red}(PX_i)}{\color{red}(PY_i)}{\color{blue}(PY_i)}{\color{ForestGreen}(PX_i)}~\ket{s_Z} \nonumber \\
& = & {\color{red}(PX_a)}{\color{blue}(PX^\perp_a)}{\color{ForestGreen}(PX_a)} \nonumber \\
& &\prod\limits_{i\in\{b,c\}}~{\color{red}(PX_i)}{\color{red}(PX_i)}{\color{blue}(PX_i)}{\color{ForestGreen}(PX_i)}~\ket{s_Z} \nonumber \\
& = & \left(x_1x_{12}\right)\left(z_8z_9\right)\ket{0_L},
\end{eqnarray}
establishing the LU equivalence between the code word of the $[444]$-19 qubit code and the $[336]$-19 qubit code, $\ket{0_L}$. The logical , $\ket{1^{''}_L}$ is obtained as, $\ket{1^{''}_L}=X_L\ket{0^{''}_L}$. The LU equivalence ensures that the $[444]$-19 qubit supports fault-tolerant computation.

The $[444]$-19 qubit codes with partially and fully mixed edge configurations are equivalent to the canonical $[444]$-19 qubit code by LU's. The construction of these codes follow the algorithm developed earlier or alternatively they can be cut out from the corresponding hexagon codes in Fig. \ref{fig:[444]ccpartialfullmixing}.

\section{Example : A $[444]$-color code with fully mixed edge configurations - Mixed logicals}
\label{sec:examplemixed[444]}
One of the motivations for studying non-CSS color codes is the potential presence of logical Pauli's of the mixed type. Though this is not possible in the canonical $[444]$-color codes with homogeneous edge configurations it can be seen to appear in the $[444]$-color codes with either partially or fully mixed edge configurations. We show this for the latter case in this section.

\begin{figure}[h]
    \centering
    \includegraphics[width=8cm]{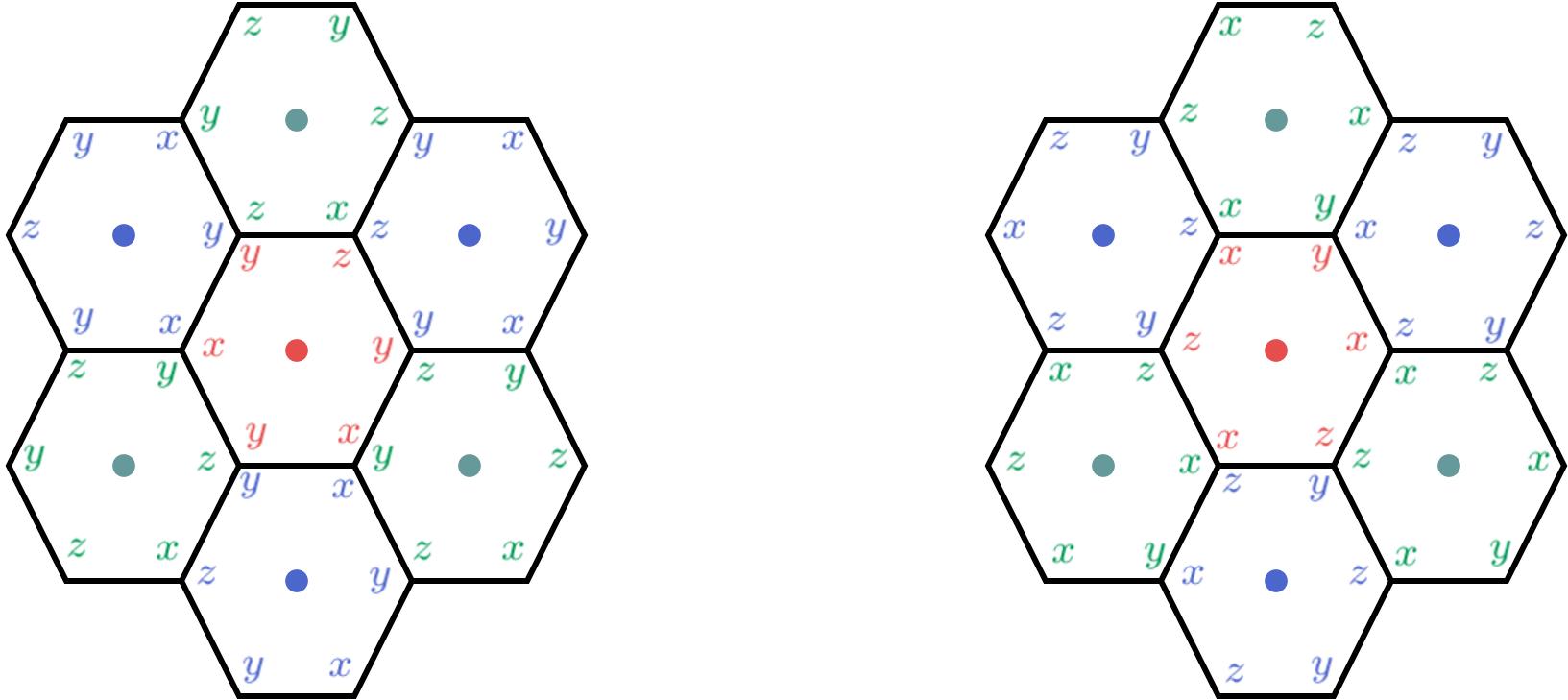}
 \caption{Stabilisers of a $[444]$-color code with fully mixed edge configurations.}
    \label{fig:4ccstabilisers}
\end{figure}

Consider a translationally invariant $[444]$-color code shown in Fig. \ref{fig:4ccstabilisers}.
Upon shuffling the $\color{red}r_I$ and $\color{red}r_{II}$ stabilizers we can find the LU between the $[444]$-color code in Fig. \ref{fig:4ccstabilisers} and the canonical $[444]$-color code in Fig. \ref{fig:[444]ccnomixing}.
Due to the LU we do not expect to see any difference in the fundamental topological properties of the canonical and the fully mixed $[444]$-color code of Fig. \ref{fig:4ccstabilisers}, nevertheless we will now analyze the mixed model in full detail to see explicitly how the logical Pauli's of the mixed type appear.
 
\subsubsection{Ground State Degeneracy}
 The GSD or the code space dimension of this model is the same as that of the canonical $[444]$-color code due to the LU between them. Hence we need to identify the four independent constraints among the stabilizers in Fig. \ref{fig:4ccstabilisers}
 \begin{eqnarray}\label{eq:4ccstabiliserconstraints}
\beta_1 &=& \prod\limits_{j=1}^{{\color{red}f}}{\color{red}r_I}_j\prod\limits_{j=1}^{{\color{blue}f}}{\color{blue}b_I}_j,~ \beta_2 = \prod\limits_{j=1}^{{\color{red}f}}{\color{red}r_{II}}_j\prod\limits_{j=1}^{{\color{ForestGreen}f}}{\color{ForestGreen}g_{II}}_j  \nonumber \\
\beta_3 & = & \prod\limits_{j=1}^{{\color{ForestGreen}f}}{\color{ForestGreen}g_I}_j\prod\limits_{j=1}^{{\color{blue}f}}{\color{blue}b_{II}}_j, ~ \beta_4 = \prod\limits_{j=1}^{{\color{blue}f}}{\color{blue}b_I}_j\prod\limits_{j=1}^{{\color{ForestGreen}f}}{\color{ForestGreen}g_I}_j\prod\limits_{j=1}^{{\color{ForestGreen}f}}{\color{ForestGreen}g_{II}}_j. \nonumber \\
\end{eqnarray}

\subsubsection{The Code space}
The LU between the homogeneous $[444]$-color code and the fully mixed case helps us determine the code space. Though we do not write down the ground state of the fully mixed case explicitly we are still interested in identifying the logical Pauli $X$'s, $\bar{X}_1, \bar{X}_2, \bar{X}_3, \bar{X}_4$, to show their mixed nature.
 While these were relatively easy to construct in the canonical $[444]$-color code, it requires a bit more thought when the code is made up of mixed edge configurations. We systematically do this by considering the string operators that create the anyonic excitations in this code.

\begin{figure}[h]
    \centering
    \includegraphics[width=4cm]{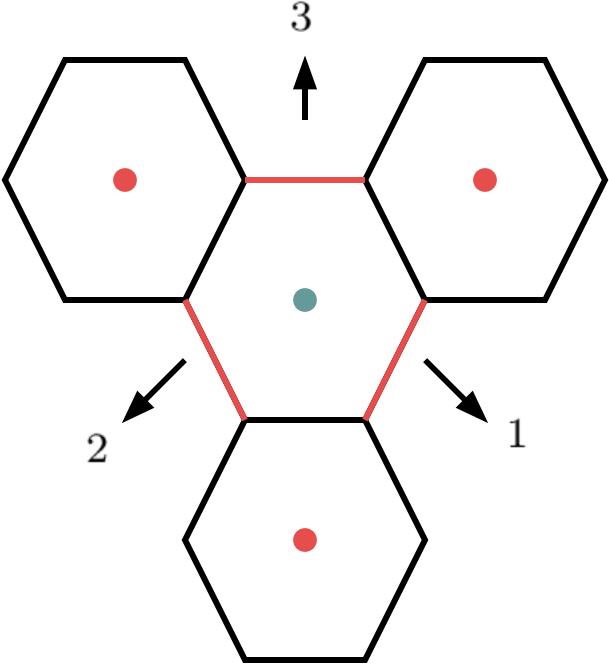}
 \caption{The three independent directions on the red shrunk lattice.}
    \label{fig:elemstringred}
\end{figure}

\begin{figure}[h]
    \centering
    \includegraphics[width=8.5cm]{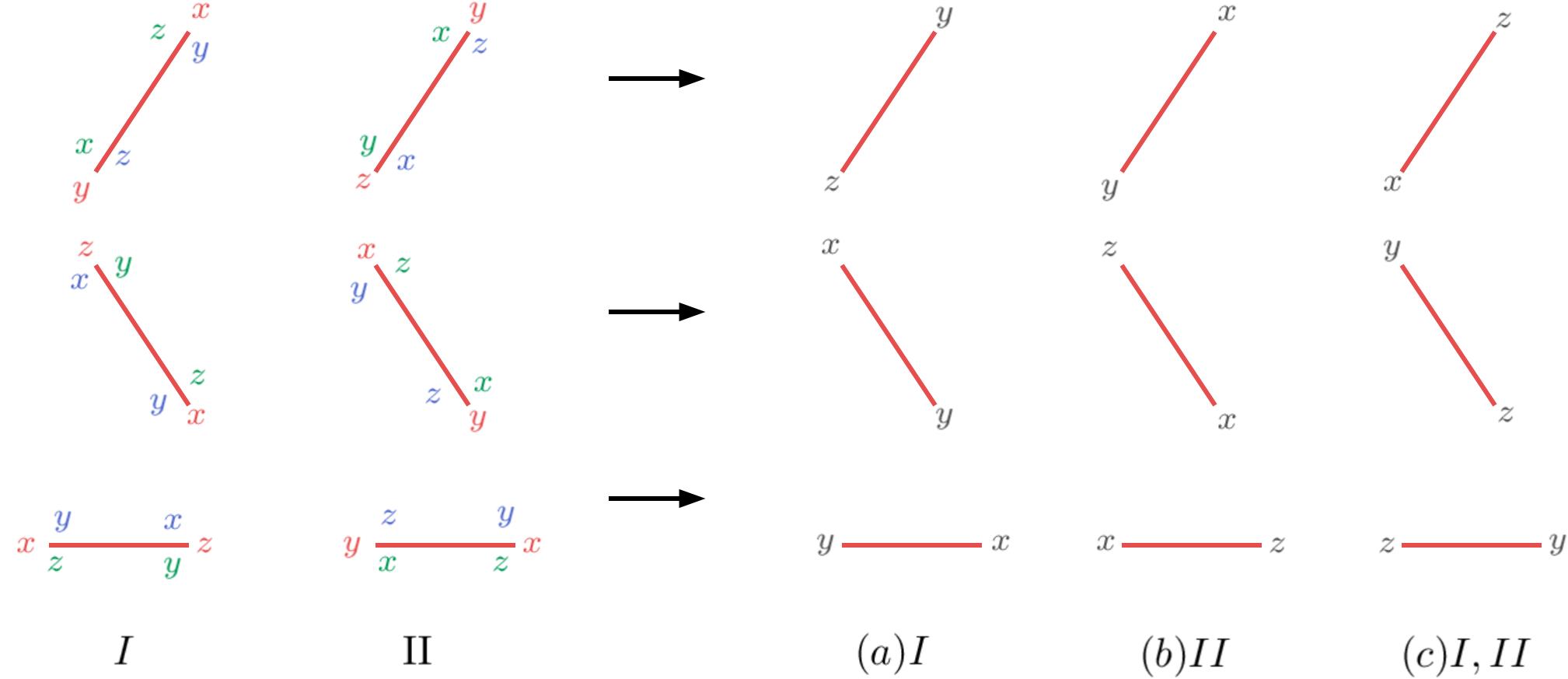}
 \caption{The elementary strings along the three independent directions on the red shrunk lattice for the stabilisers in Fig. \ref{fig:4ccstabilisers}. (a) This elementary string acting on the ground state excites a pair of ${\color{red}r_I}$ stabilizers. (b) This excites a pair of ${\color{red}r_{II}}$ stabilizers. (c) This excites a pair of ${\color{red}r_I}$ and ${\color{red}r_{II}}$ stabilizers.} 
    \label{fig:4ccredelemstrings}
\end{figure}

\begin{figure}[h]
    \centering
    \includegraphics[width=8cm]{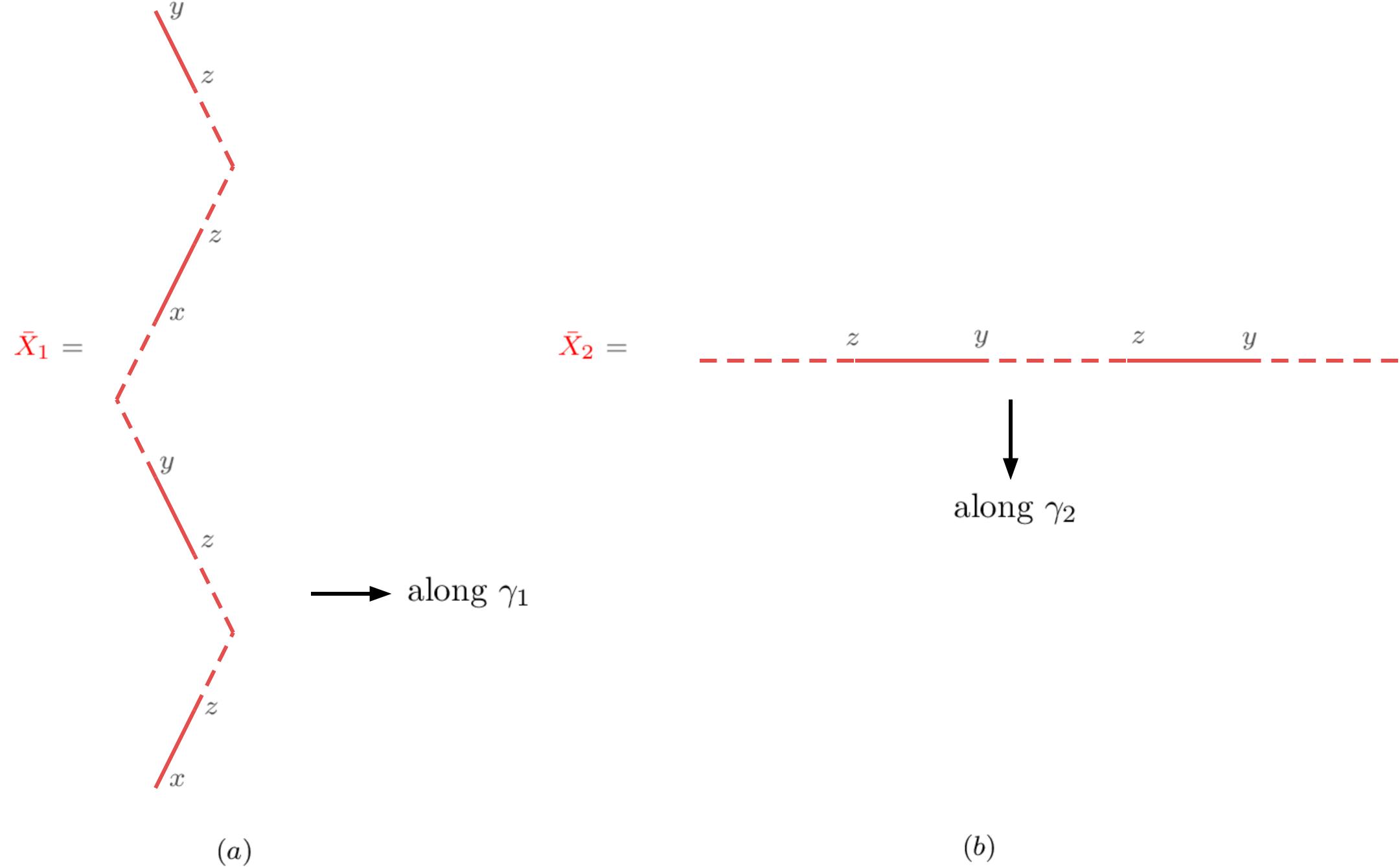}
 \caption{The logical Pauli $X$ operators along the red shrunk lattice. (a) This operator winds along the non-contractible loop $\gamma_1$. (b) This operator winds along the non-contractible loop $\gamma_2$.}
    \label{fig:4ccredlogicalX}
\end{figure}

\begin{figure}[h]
    \centering
    \includegraphics[width=4cm]{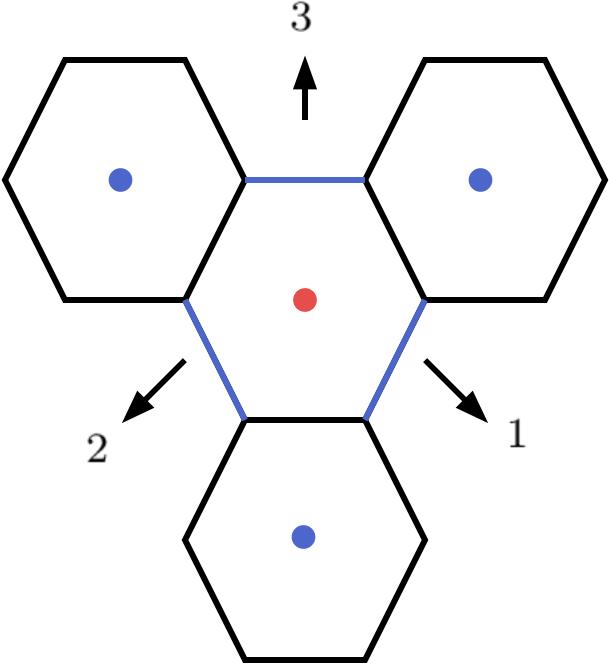}
 \caption{The elementary strings along the three independent directions on the blue shrunk lattice.}
    \label{fig:elemstringblue}
\end{figure}

\begin{figure}[h]
    \centering
    \includegraphics[width=8.5cm]{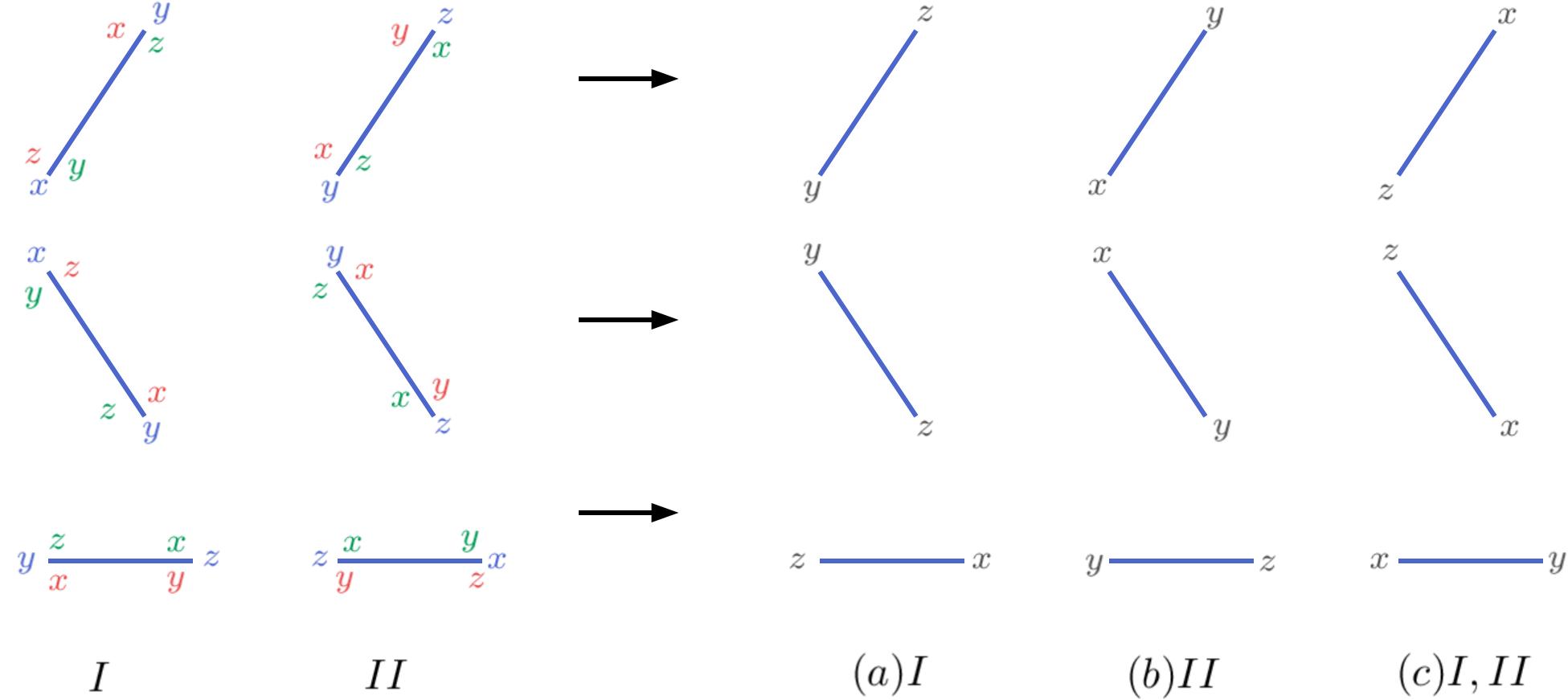}
 \caption{The elementary strings along the three independent directions on the blue shrunk lattice. In (a), (b), (c) we have shown the strings that excite the following pairs of stabilizers, ${\color{blue}b_I}$, ${\color{blue}b_{II}}$ and both the types of blue stabilizers respectively.}
    \label{fig:4ccblueelemstrings}
\end{figure}

\begin{figure}[h]
    \centering
    \includegraphics[width=8cm]{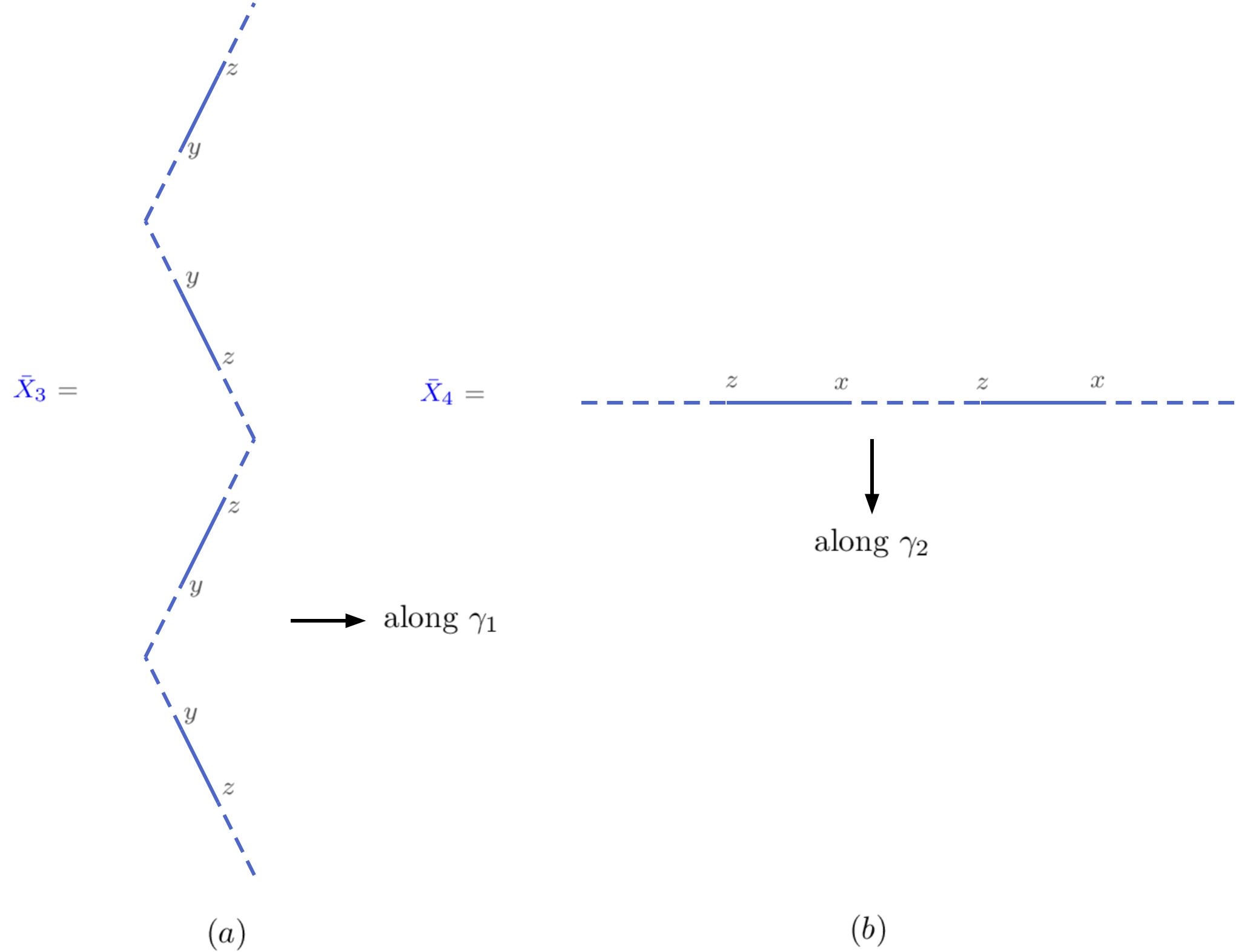}
 \caption{The logical Pauli $X$ operators along the blue shrunk lattice. (a) This operator winds along the non-contractible loop $\gamma_1$. (b) This operator winds along the non-contractible loop $\gamma_2$.}
    \label{fig:4ccbluelogicalX}
\end{figure}

\begin{figure}[h]
    \centering
    \includegraphics[width=8cm]{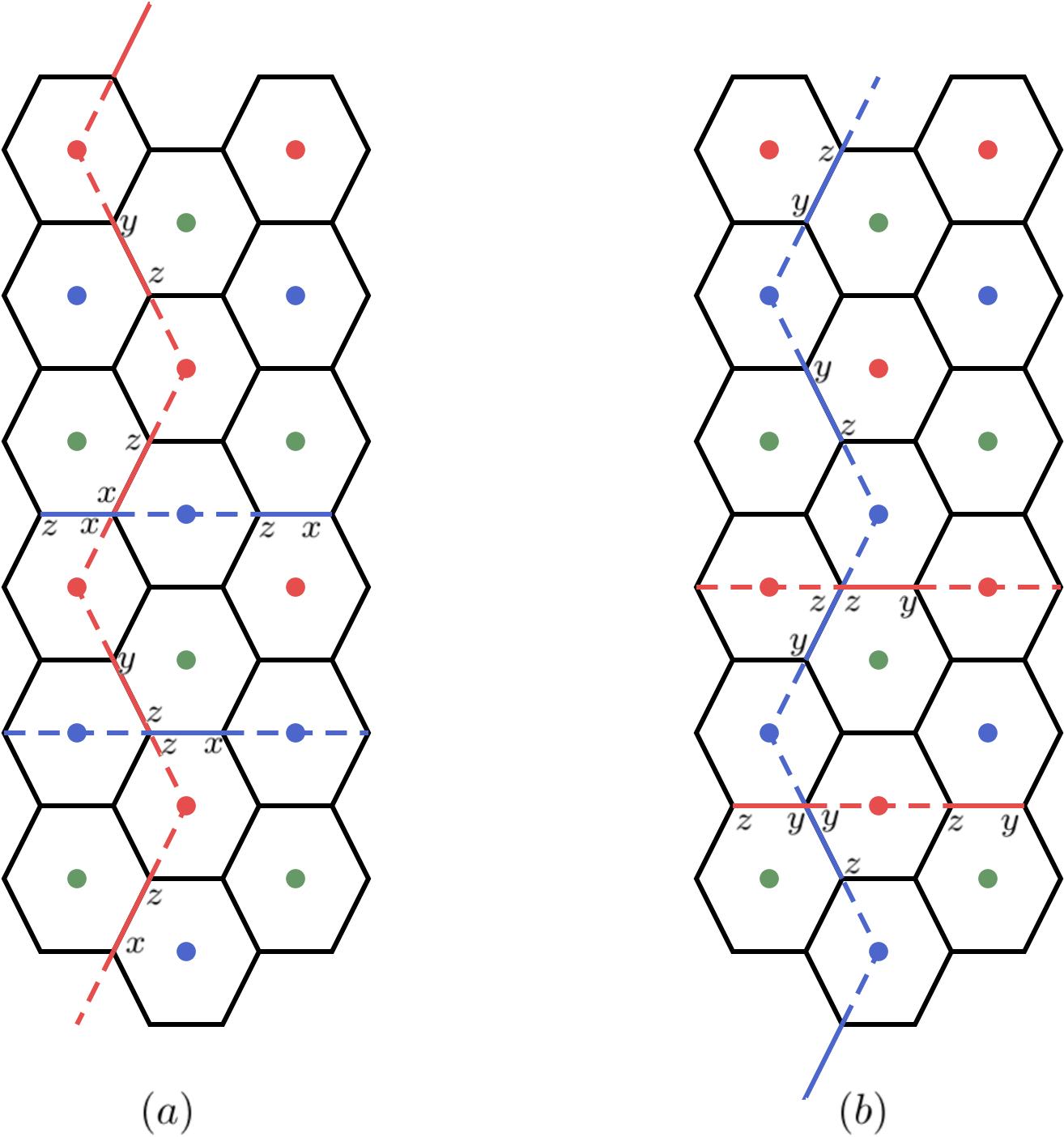}
 \caption{The logical Pauli $X$ operators commute with each other, ($a$) $[{\color{red}\bar{X}_1}, {\color{blue}\bar{X}_4}]=0$, ($b$) $[{\color{red}\bar{X}_2}, {\color{blue}\bar{X}_3}]=0$.}
    \label{fig:4cclogicalXcommuA}
\end{figure}

\begin{figure}[h]
    \centering
    \includegraphics[width=8cm]{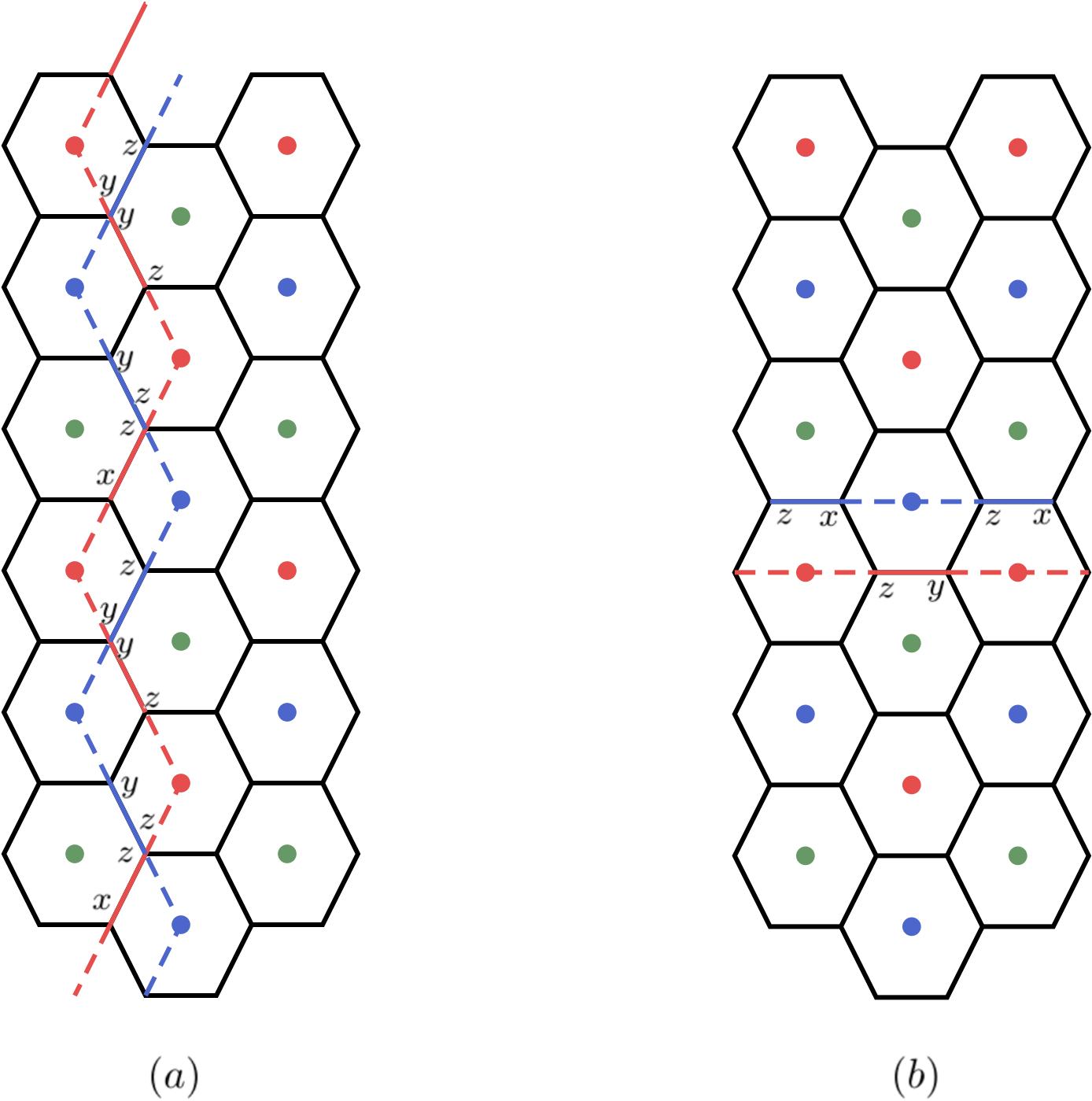}
 \caption{The logical Pauli $X$ operators commute with each other, ($a$) $[{\color{red}\bar{X}_1}, {\color{blue}\bar{X}_3}]=0$, ($b$) $[{\color{red}\bar{X}_2}, {\color{blue}\bar{X}_4}]=0$.}
    \label{fig:4cclogicalXcommuB}
\end{figure}

\paragraph{\bf Logical Pauli $X$'s -} Any path on a chosen shrunk lattice is made of three types of {\it elementary strings} that lie along the three independent directions on the hexagonal lattice as shown in Fig. \ref{fig:elemstringred}.
For the stabilizers in Fig. \ref{fig:4ccstabilisers} the elementary strings along the red shrunk lattice are shown in Fig. \ref{fig:4ccredelemstrings}. Extending any of these strings along the indicated directions excites either two or four of the red stabilizers at the end points of the string.
The logical Pauli's corresponding to the red shrunk lattice can now be constructed as shown in Fig. \ref{fig:4ccredlogicalX}.
 We can follow the same steps to obtain the remaining two logical Pauli $X$ operators by finding the winding operators along the non-contractible loops drawn along the blue shrunk lattice. Consider the elementary strings along the three independent directions on the blue shrunk lattice as shown in Fig. \ref{fig:elemstringblue}.
 We determine the elementary string operators along the three independent directions on the blue shrunk lattice as shown in Fig. \ref{fig:4ccblueelemstrings} by considering the stabilisers in Fig. \ref{fig:4ccstabilisers}.
Using these elementary string operators we can construct the remaining logical Pauli $X$'s for the two non-contractible loops along the blue shrunk lattice as shown in Fig. \ref{fig:4ccbluelogicalX}.
Next we require that the four logical Pauli $X$'s commute with each other. While this is obvious in the homogeneous $[444]$-color code, it needs to be checked for the fully mixed case. This is shown diagrammatically in Figs. \ref{fig:4cclogicalXcommuA} and \ref{fig:4cclogicalXcommuB}.
With all logical Pauli $X$'s in place we can write down all the ground states or encoded qubits of the $[444]$-color code with mixed edge configurations as
\begin{equation}\label{eq:4ccaaaastate}
   \ket{\bar{a}_1\bar{a}_2\bar{a}_3\bar{a}_4}_{[444]}=  {\color{red}\bar{X}_1}^{a_1}{\color{red}\bar{X}_2}^{a_2}{\color{blue}\bar{X}_3}^{a_3}{\color{blue}\bar{X}_4}^{a_4}\ket{\bar{0}_1\bar{0}_2\bar{0}_3\bar{0}_4}_{[444]},
\end{equation}
for $a_1,a_2,a_3,a_4\in\{0,1\}$, exhausting the $2^4$ ground states of the code space for the system on a torus.

\begin{figure}[h]
    \centering
    \includegraphics[width=8cm]{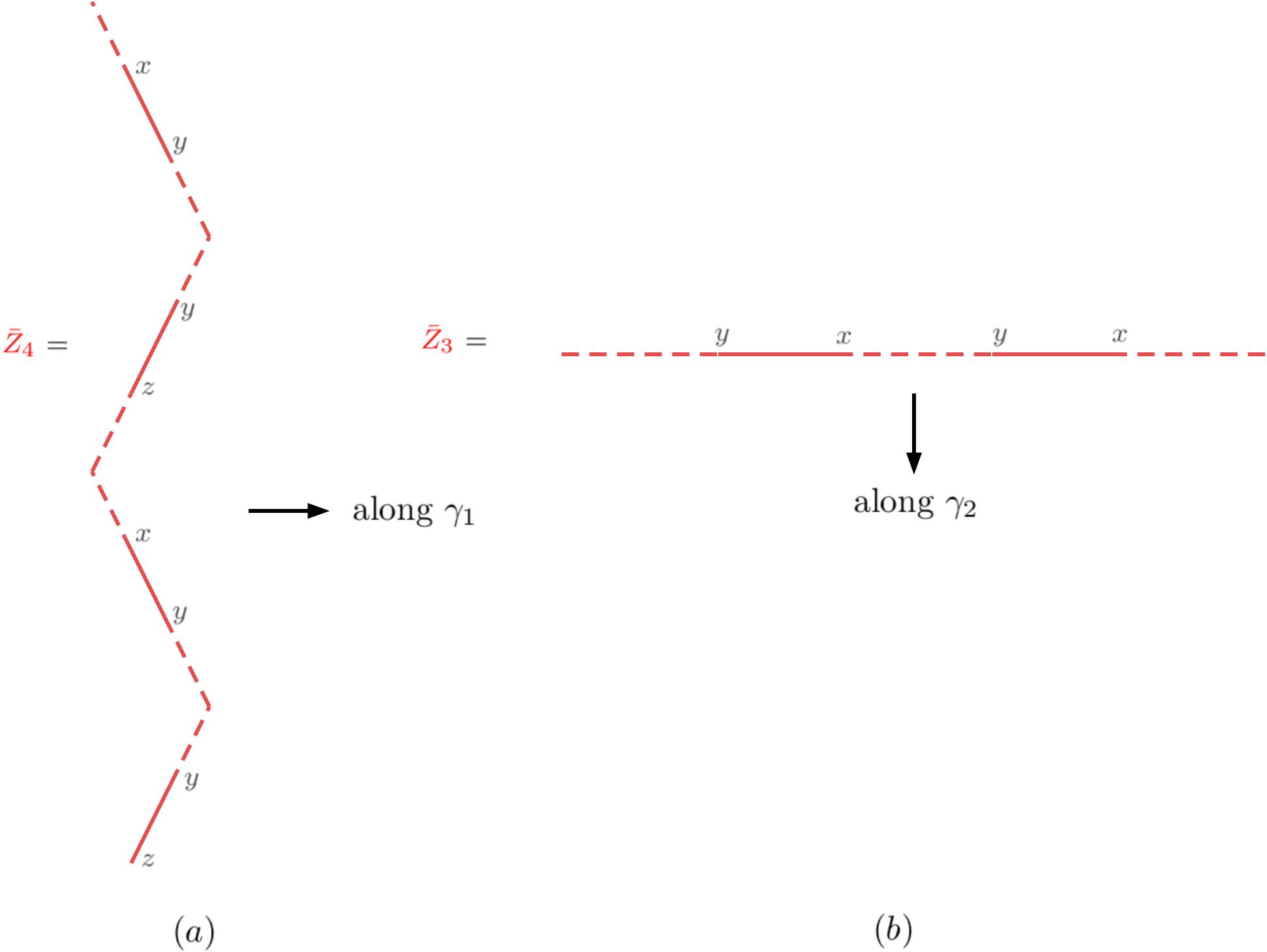}
 \caption{The logical Pauli $Z$ operators along the red shrunk lattice. (a) This operator winds along the non-contractible loop $\gamma_1$. (b) This operator winds along the non-contractible loop $\gamma_2$.}
    \label{fig:4ccredlogicalZ}
\end{figure}

\begin{figure}[h]
    \centering
    \includegraphics[width=8cm]{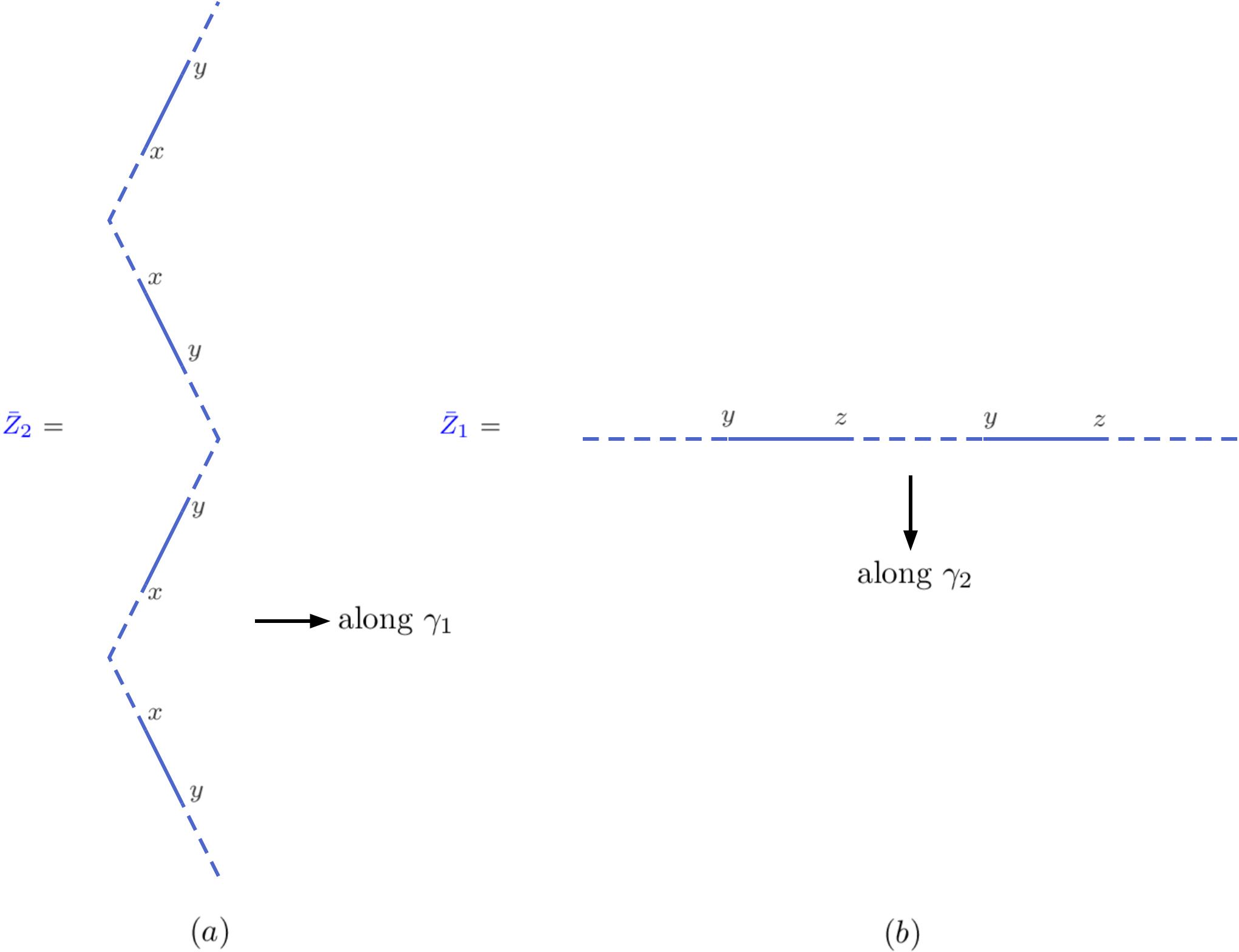}
 \caption{The logical Pauli $Z$ operators along the blue shrunk lattice. (a) This operator winds along the non-contractible loop $\gamma_1$. (b) This operator winds along the non-contractible loop $\gamma_2$.}
    \label{fig:4ccbluelogicalZ}
\end{figure}

\begin{figure}[h]
    \centering
    \includegraphics[width=7cm]{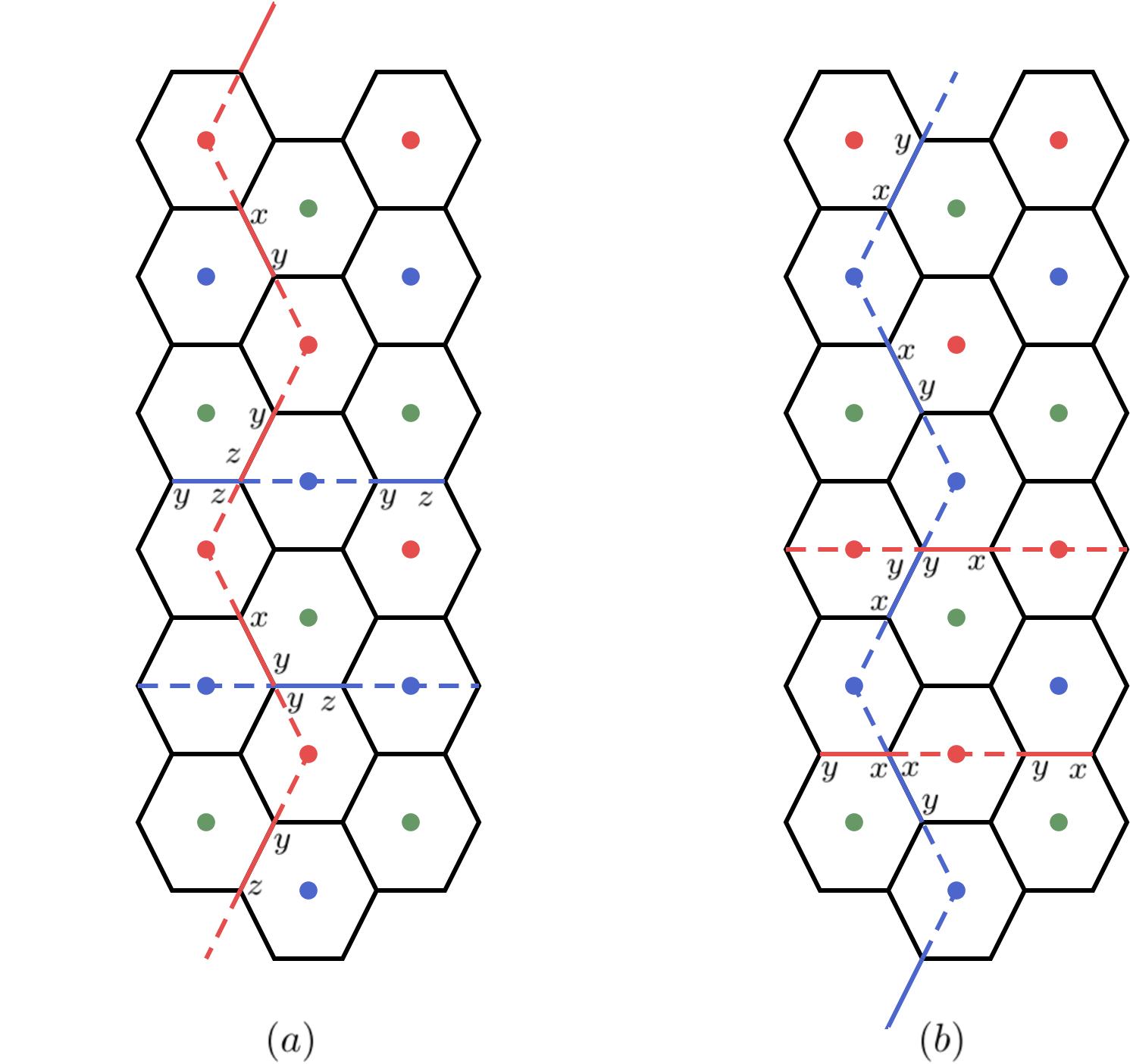}
 \caption{The logical Pauli $Z$ operators commute with each other, ($a$) $[{\color{blue}\bar{Z}_1}, {\color{red}\bar{Z}_4}]=0$, ($b$) $[{\color{blue}\bar{Z}_2},  {\color{red}\bar{Z}_3}]=0$.}
    \label{fig:4cclogicalZcommuA}
\end{figure}

\begin{figure}[h]
    \centering
    \includegraphics[width=7cm]{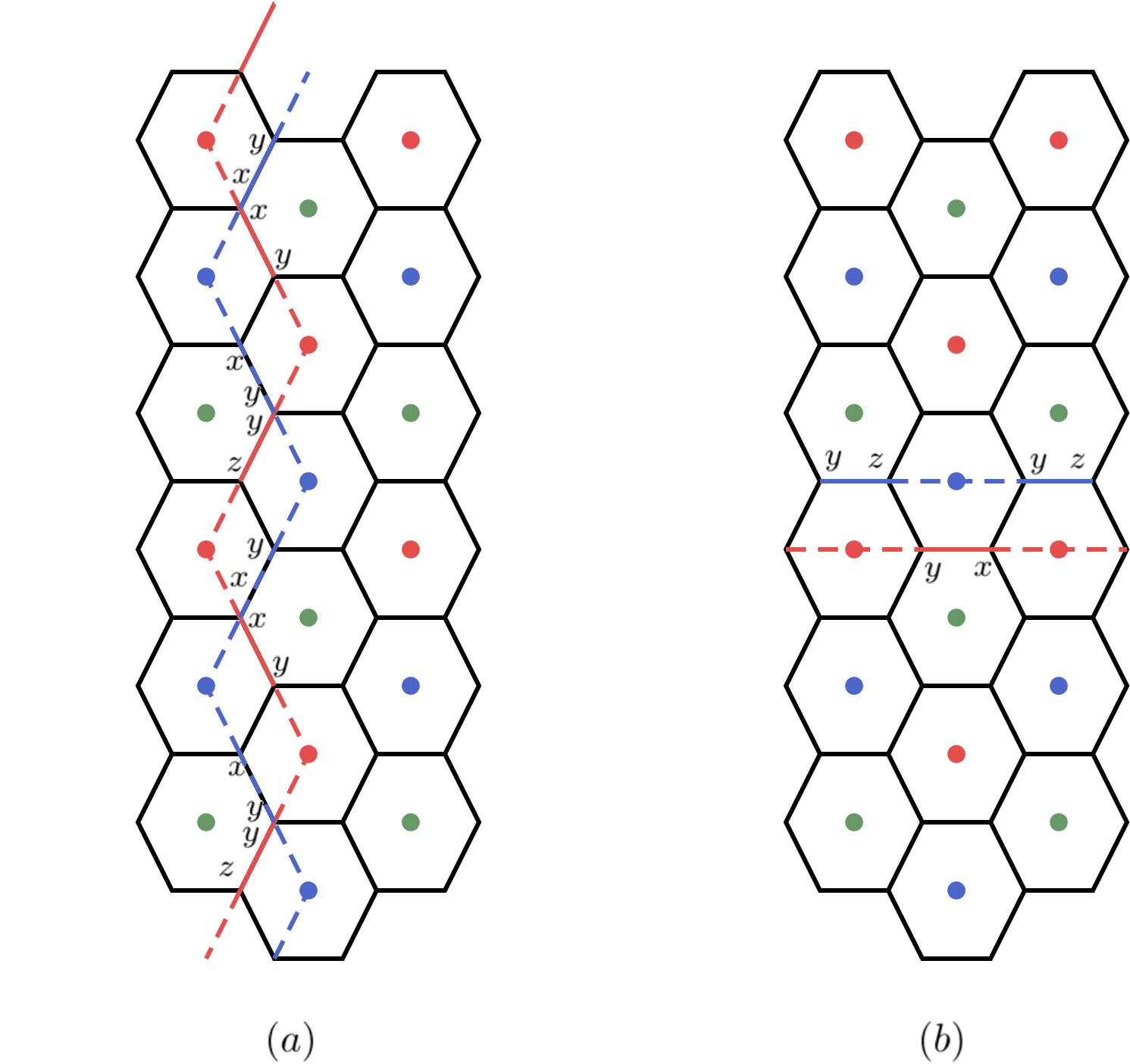}
 \caption{The logical Pauli $Z$ operators commute with each other, ($a$) $[{\color{blue}\bar{Z}_2}, {\color{red}\bar{Z}_4}]=0$, ($b$) $[{\color{blue}\bar{Z}_1}, {\color{red}\bar{Z}_3}]=0$.}
    \label{fig:4cclogicalZcommuB}
\end{figure}

\paragraph{\bf Logical Pauli $Z$'s -} There are four logical $Z$'s, two along the red shrunk lattice and two along the blue shrunk lattice as shown in Figs. \ref{fig:4ccredlogicalZ} and \ref{fig:4ccbluelogicalZ}.
Furthermore the logical $Z$'s commute with each other as shown in Figs. \ref{fig:4cclogicalZcommuA} and \ref{fig:4cclogicalZcommuB}
The logical $Z$'s measure the global fluxes and can be used to distinguish the states in the code space as
\begin{eqnarray}
 & & {\color{blue}\bar{Z}_1}{\color{blue}\bar{Z}_2}{\color{red}\bar{Z}_3}{\color{red}\bar{Z}_4}\ket{\bar{a}_1\bar{a}_2\bar{a}_3\bar{a}_4}_{[444]}   =  \nonumber \\ & &(-1)^{a_1+a_2+a_3+a_4}\ket{\bar{a}_1\bar{a}_2\bar{a}_3\bar{a}_4}_{[444]}.\nonumber \\
\end{eqnarray}

\begin{figure}[h]
    \centering
    \includegraphics[width=7cm]{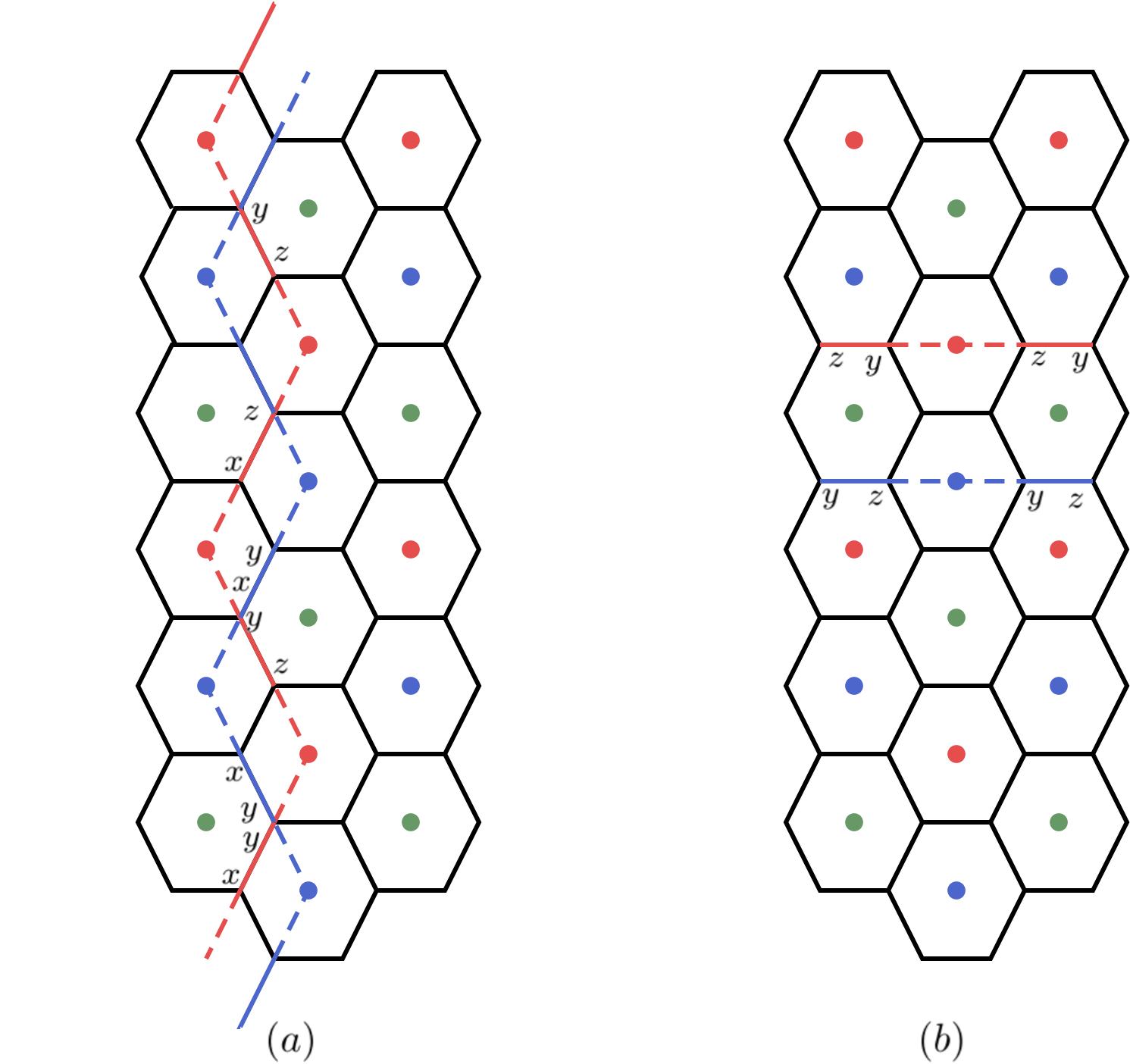}
 \caption{The logical Pauli $X$ and $Z$ operators commute with each other, ($a$) $[{\color{red}\bar{X}_1}, {\color{blue}\bar{Z}_2}]=0$, ($b$) $[{\color{red}\bar{X}_2}, {\color{blue}\bar{Z}_1}]=0$.}
    \label{fig:4cclogicalXZcommuA}
\end{figure}

\begin{figure}[h]
    \centering
    \includegraphics[width=7cm]{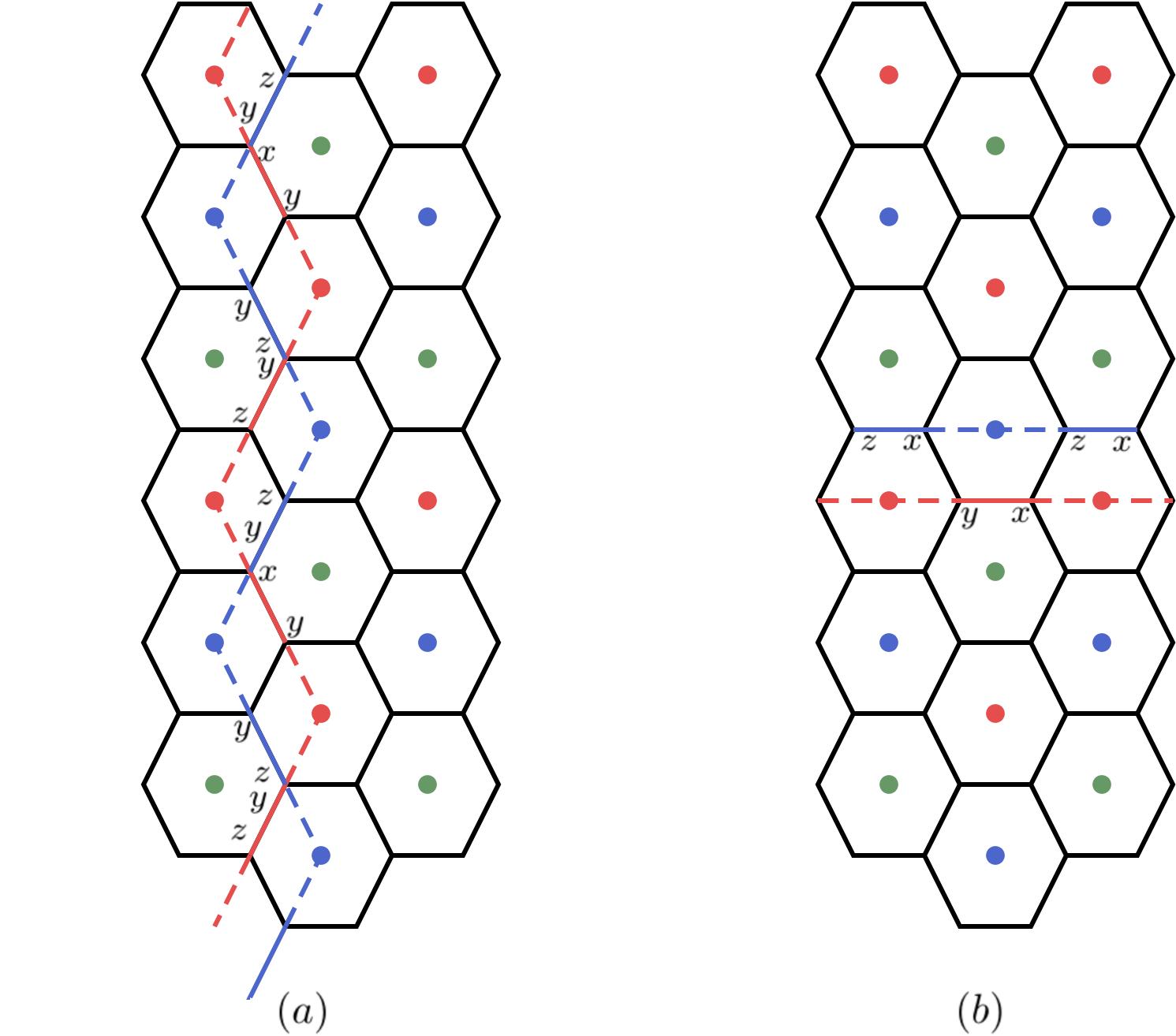}
 \caption{The logical Pauli $X$ and $Z$ operators commute with each other, ($a$) $[{\color{blue}\bar{X}_3}, {\color{red}\bar{Z}_4}]=0$, ($b$) $[{\color{blue}\bar{X}_4}, {\color{red}\bar{Z}_3}]=0$.}
    \label{fig:4cclogicalXZcommuB}
\end{figure}

\begin{figure}[h]
    \centering
    \includegraphics[width=7cm]{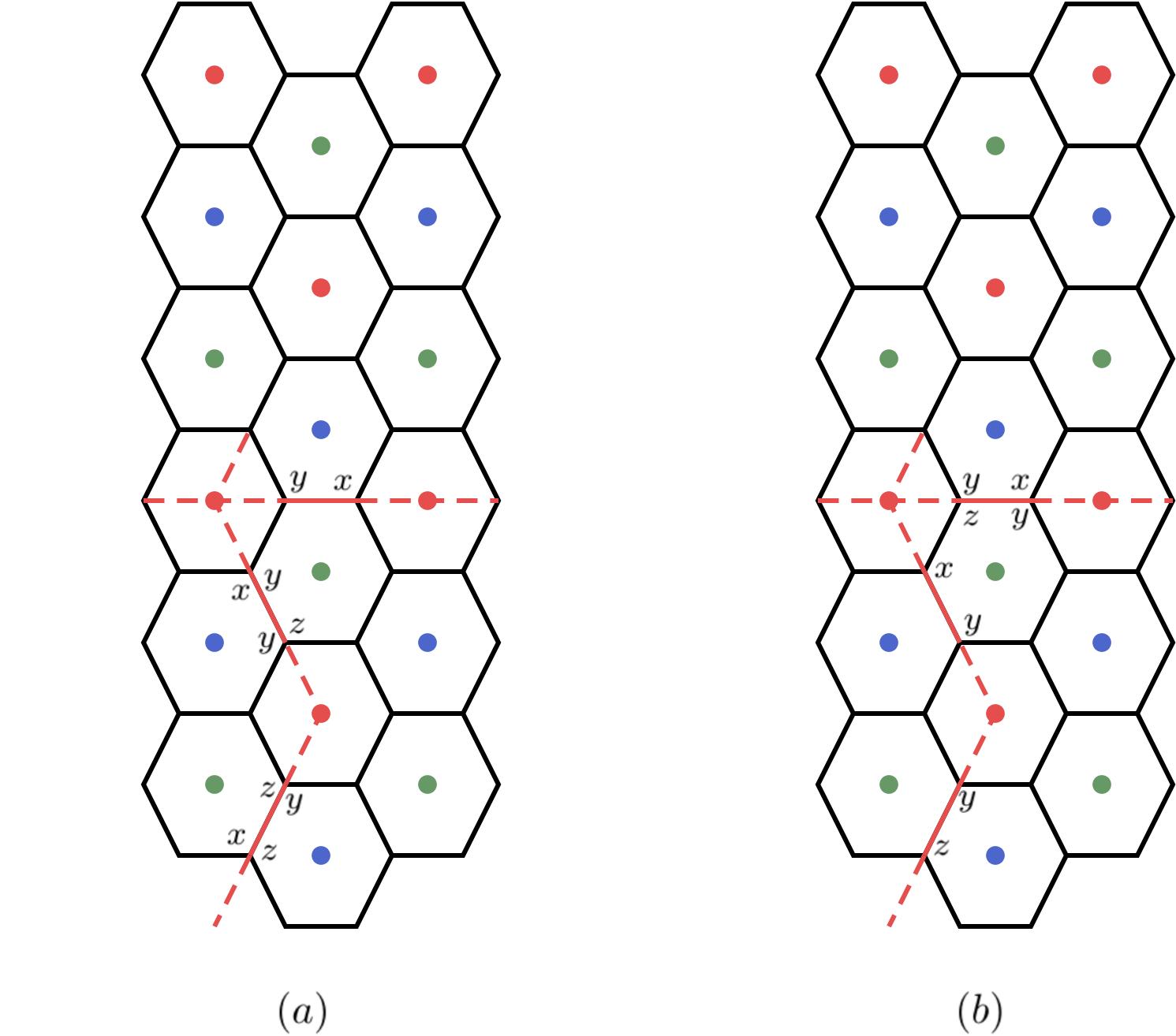}
 \caption{The logical Pauli $X$ and $Z$ operators commute with each other, ($a$) $[{\color{red}\bar{X}_1}, {\color{red}\bar{Z}_4}]=0$, $[{\color{red}\bar{X}_1}, {\color{red}\bar{Z}_3}]=0$, ($b$) $[{\color{red}\bar{X}_2}, {\color{red}\bar{Z}_4}]=0$, $[{\color{red}\bar{X}_2}, {\color{red}\bar{Z}_3}]=0$.}
    \label{fig:4cclogicalXZcommuC}
\end{figure}

\begin{figure}[h]
    \centering
    \includegraphics[width=7cm]{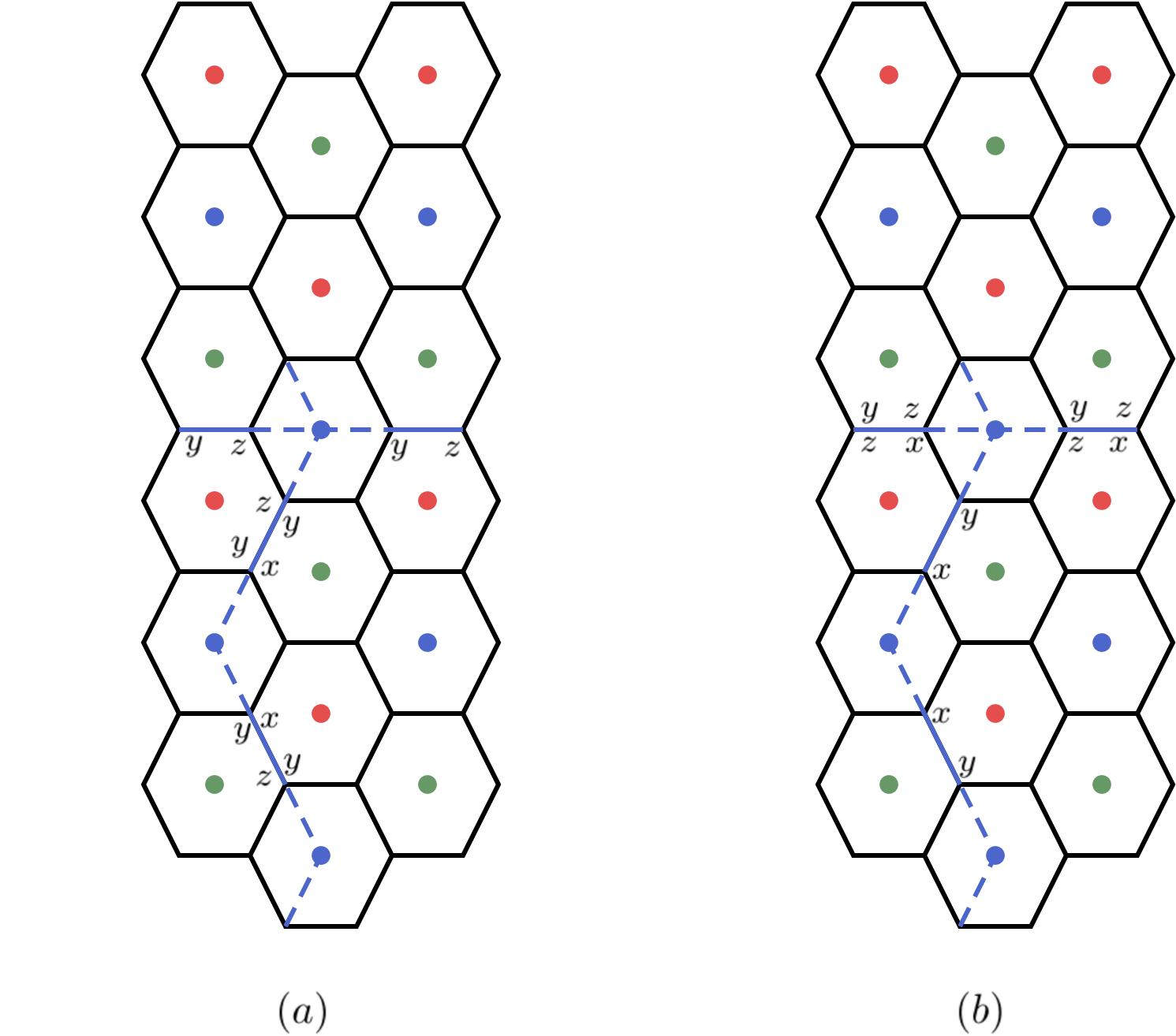}
 \caption{The logical Pauli $X$ and $Z$ operators commute with each other, ($a$) $[{\color{blue}\bar{X}_3}, {\color{blue}\bar{Z}_1}]=0$, $[{\color{blue}\bar{X}_3}, {\color{blue}\bar{Z}_2}]=0$, ($b$) $[{\color{blue}\bar{X}_4}, {\color{blue}\bar{Z}_1}]=0$, $[{\color{blue}\bar{X}_4}, {\color{blue}\bar{Z}_2}]=0$.}=
    \label{fig:4cclogicalXZcommuD}
\end{figure}

\paragraph{\bf Relations between the logical $X$'s and $Z$'s -} Next we check the commutation and anti-commutation relations between the logical Pauli's on the code space as in the CSS  color codes. The relations between $\{{\color{red}\bar{X}_1}, {\color{red}\bar{X}_2}, {\color{blue}\bar{X}_3}, {\color{blue}\bar{X}_4}\}$, $\{{\color{blue}\bar{Z}_1}, {\color{blue}\bar{Z}_2}, {\color{red}\bar{Z}_3}, {\color{red}\bar{Z}_4}\}$ can be summarized as 
\begin{equation}
\{\bar{X}_i, \bar{Z}_j\}=0,~ i=j;~~ [\bar{X}_i, \bar{Z}_j]=0,~i\neq j.   
\end{equation}
We check these relations diagrammatically in Figs. \ref{fig:4cclogicalXZcommuA}, \ref{fig:4cclogicalXZcommuB}, \ref{fig:4cclogicalXZcommuC} and  \ref{fig:4cclogicalXZcommuD}.

\begin{figure}[h]
    \centering
    \includegraphics[width=7cm]{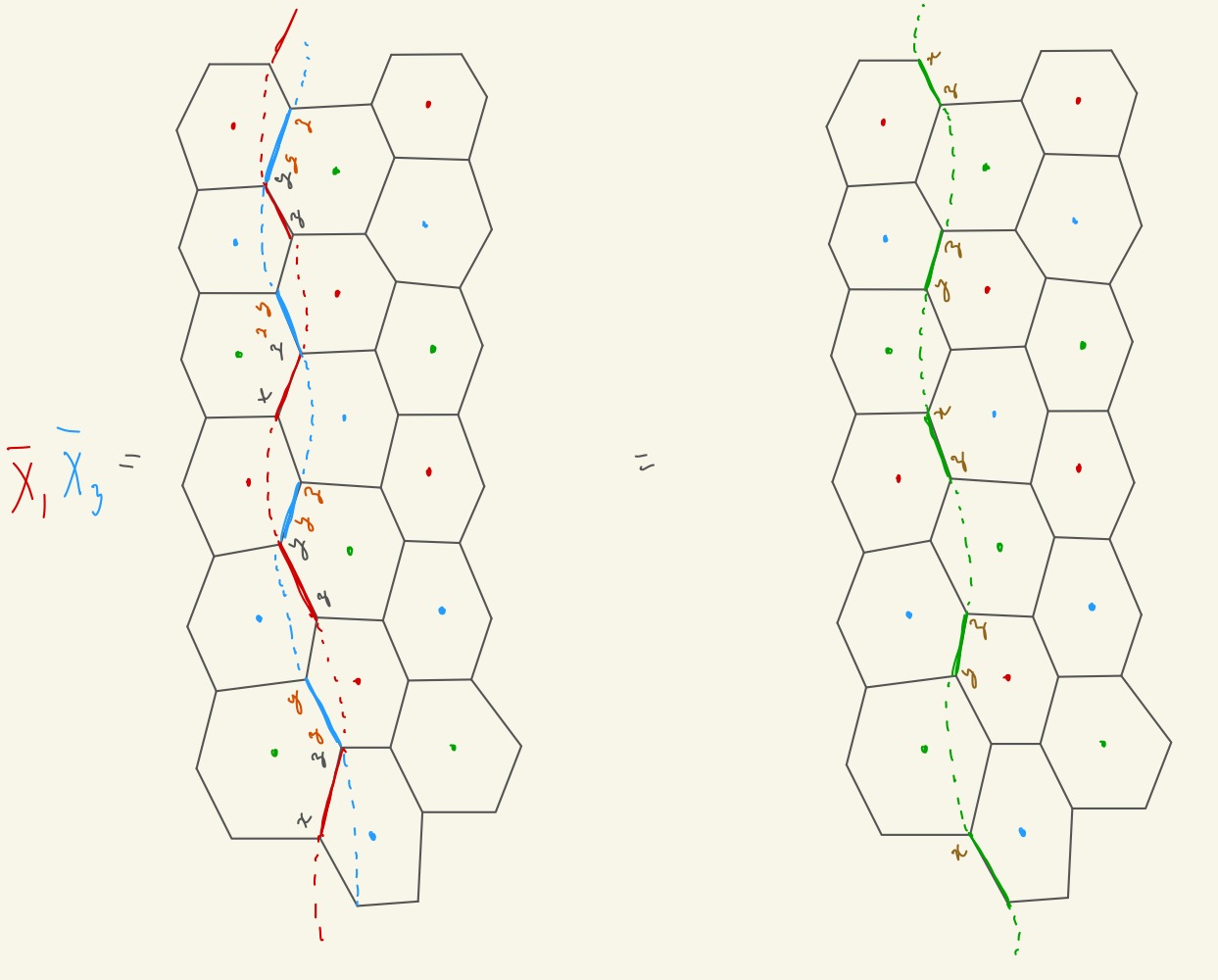}
 \caption{A green logical Pauli $X$ obtained as a product of ${\color{red}\bar{X}_1}{\color{blue}\bar{X}_3}$.}
    \label{fig:4ccgreenlogicalXA}
\end{figure}

\begin{figure}[h]
    \centering
    \includegraphics[width=8cm]{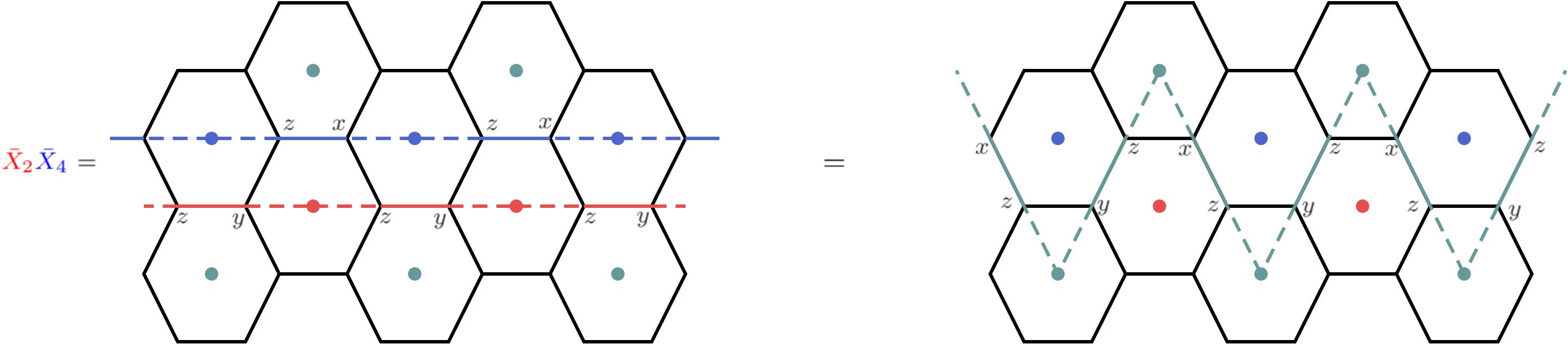}
 \caption{A green logical Pauli $X$ obtained as a product of ${\color{red}\bar{X}_2}{\color{blue}\bar{X}_4}$.}
    \label{fig:4ccgreenlogicalXB}
\end{figure}

\begin{figure}[h]
    \centering
    \includegraphics[width=8cm]{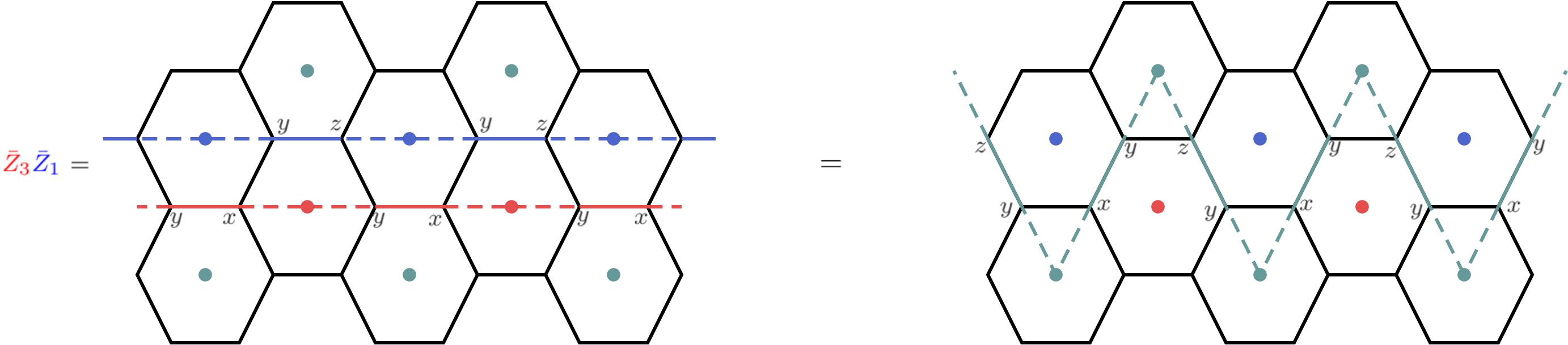}
 \caption{A green logical Pauli $Z$ obtained as a product of ${\color{red}\bar{Z}_3}{\color{blue}\bar{Z}_1}$.}
    \label{fig:4ccgreenlogicalZA}
\end{figure}

\begin{figure}[h]
    \centering
    \includegraphics[width=7cm]{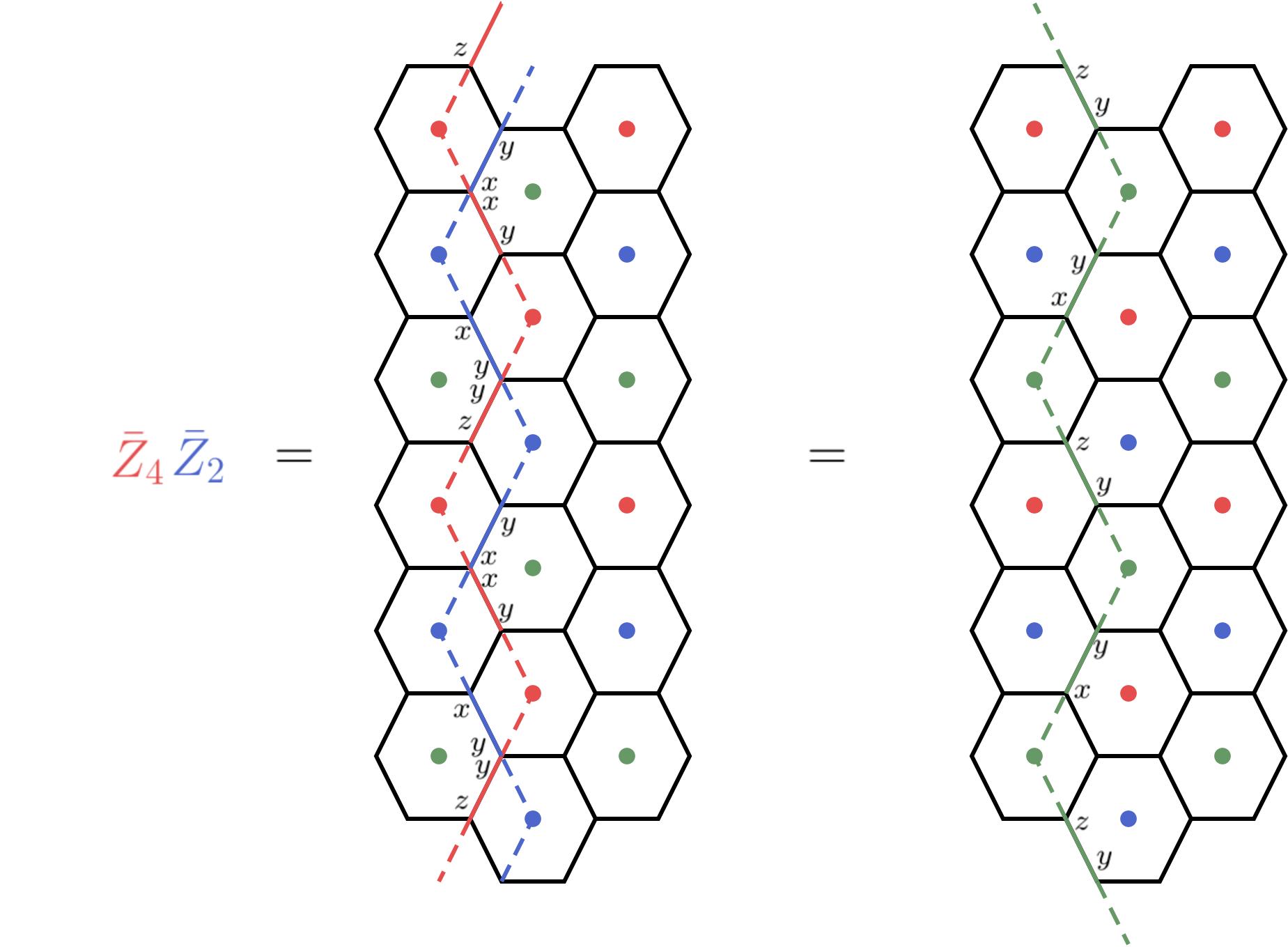}
 \caption{A green logical Pauli $Z$ obtained as a product of ${\color{red}\bar{Z}_4}{\color{blue}\bar{Z}_2}$.}
    \label{fig:4ccgreenlogicalZB}
\end{figure}

\begin{figure}[h]
    \centering
    \includegraphics[width=8.5cm]{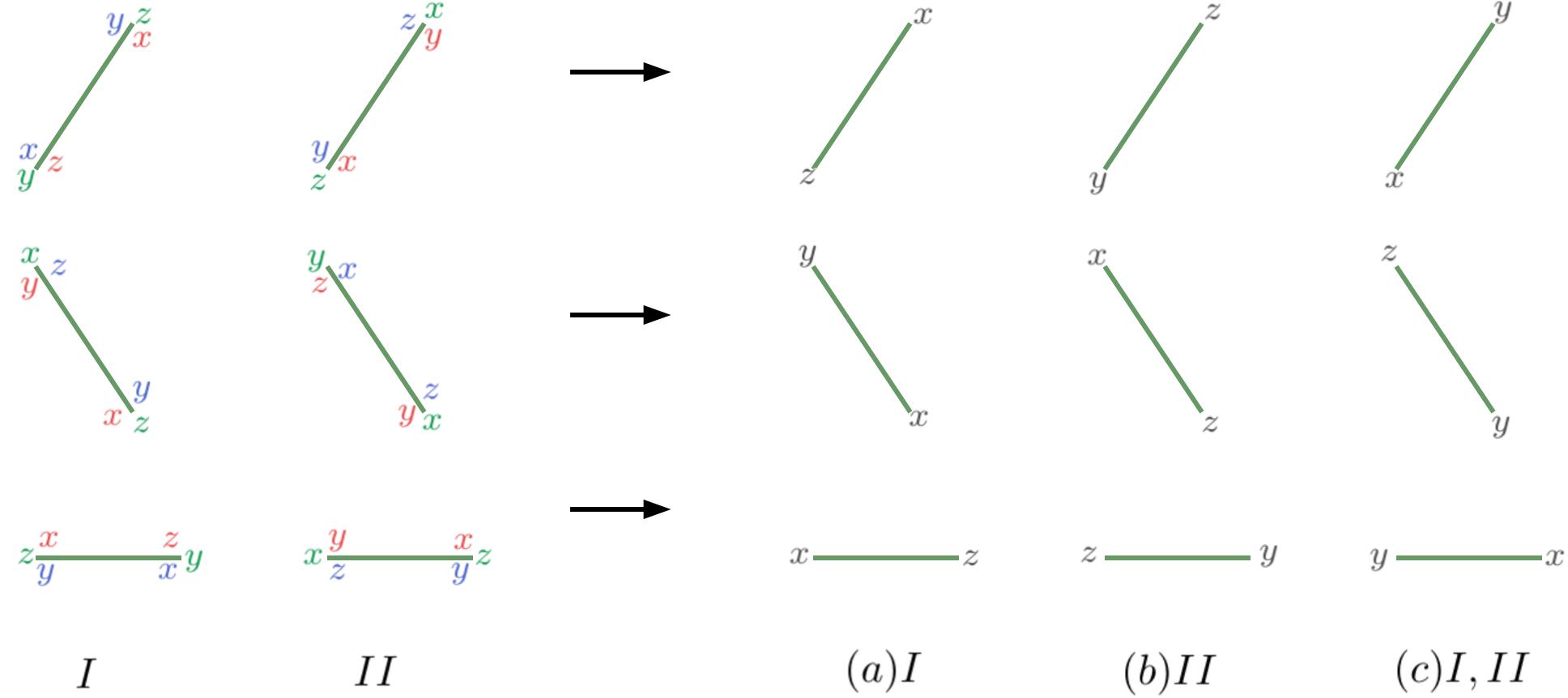}
 \caption{The elementary strings along the three independent directions on the green shrunk lattice for the stabilisers in Fig. \ref{fig:4ccstabilisers}. (a) This elementary string acting on the ground state excites a pair of ${\color{ForestGreen}g_I}$ stabilizers. (b) This excites a pair of ${\color{ForestGreen}g_{II}}$ stabilizers. (c) This excites a pair of ${\color{ForestGreen}g_I}$ and ${\color{ForestGreen}g_{II}}$ stabilizers. Extending any of these strings along the indicated directions excites either two or four of the green stabilizers at the end points of the string.} 
    \label{fig:4ccgreenelemstrings}
\end{figure}

\paragraph{\bf Logicals on the green shrunk lattice -} As in the case of the canonical color code, we can construct logical $X$ and $Z$ operators winding along the non-contractible loops drawn along the green shrunk lattice. However we do not expect them to generate new ground states as these operators can be obtained as a product of the logical operators winding along the red and blue shrunk lattices. This is shown in Figs. \ref{fig:4ccgreenlogicalXA}, \ref{fig:4ccgreenlogicalXB}, \ref{fig:4ccgreenlogicalZA} and  \ref{fig:4ccgreenlogicalZB}.
We can check that these are indeed the right logicals on the green shrunk lattice by considering the set of elementary string excitations along the three independent directions of the green shrunk lattice as shown in Fig. \ref{fig:4ccgreenelemstrings}.
These considerations imply that it is enough to consider just the red and blue logical operators in analyzing the logical space of this model just as in the original ($[336]$)-color code.

\paragraph{\bf Mixed nature of the logical Pauli's -} We claim that all of the logical Pauli operators in this code are of the {\it mixed} type, meaning that each of the logical Pauli's, ${\color{red}\bar{X}_1}, {\color{red}\bar{X}_2}, {\color{blue}\bar{X}_3}, {\color{blue}\bar{X}_4}$ and ${\color{blue}\bar{Z}_1}, {\color{blue}\bar{Z}_2}, {\color{red}\bar{Z}_3}, {\color{red}\bar{Z}_4}$, comprise of more than one type of Pauli operator. This is evident from their expressions in Figs. \ref{fig:4ccredlogicalX}, \ref{fig:4ccredlogicalZ}, \ref{fig:4ccbluelogicalX} and \ref{fig:4ccbluelogicalZ}. However these are not the only choices for non-contractible loops around a torus. We need to show that the logical operators remain mixed for any choice of non-contractible loop. In other words we need to make sure that the entire equivalence class of logical operators that are related to each other via multiplication by the local stabilizers in Fig. \ref{fig:4ccstabilisers} are also mixed. 

\begin{figure}[h]
    \centering
    \includegraphics[width=8cm]{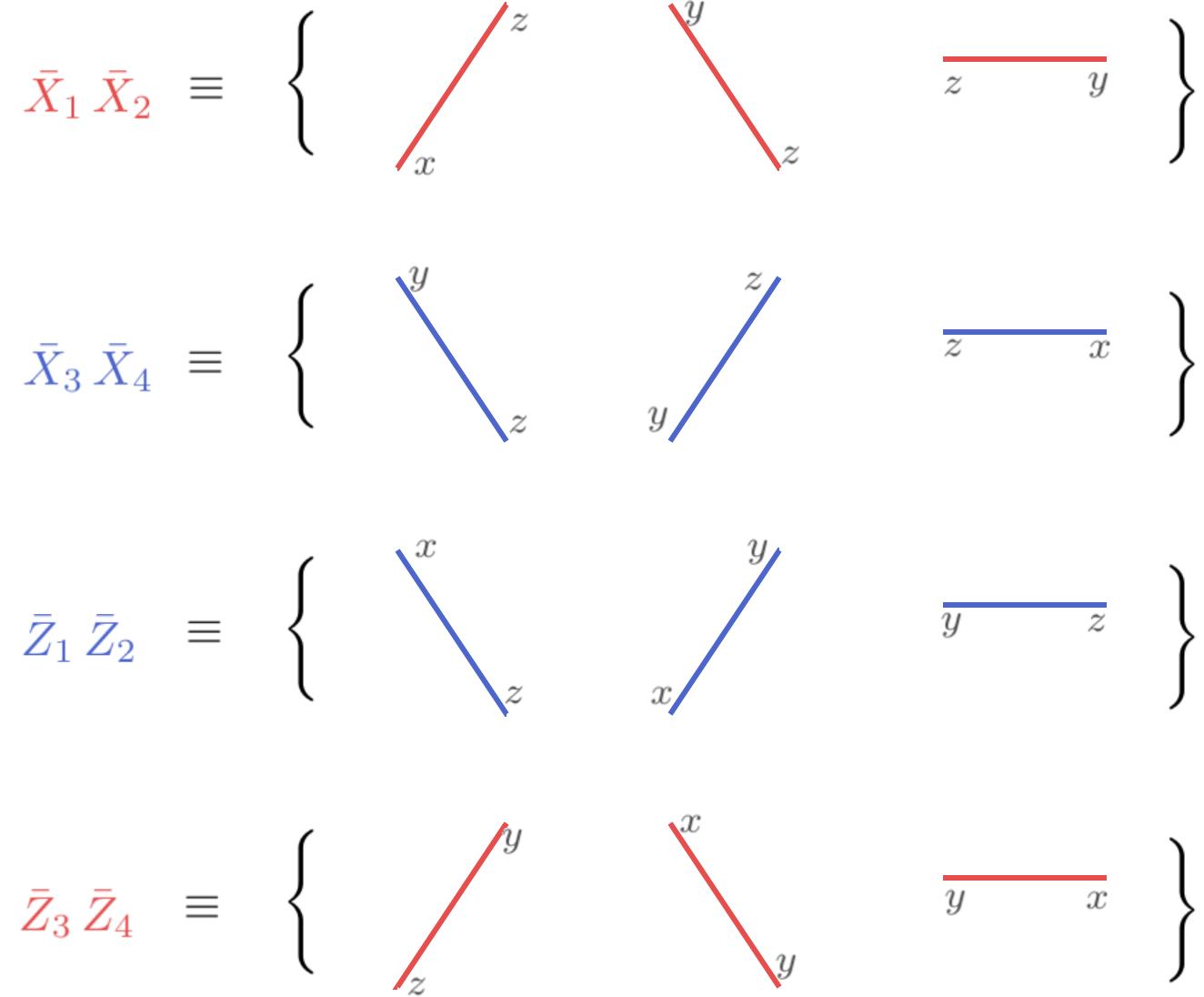}
 \caption{The equivalence class of the red and blue logical Pauli $X$ and $Z$ operators are generated by the elementary strings shown here. Using these we can write down the logical for an arbitrary non-contractible loop around the torus. }
 \label{fig:4ccmixedlogicals}
\end{figure}

\begin{figure}[h]
    \centering
    \includegraphics[width=8cm]{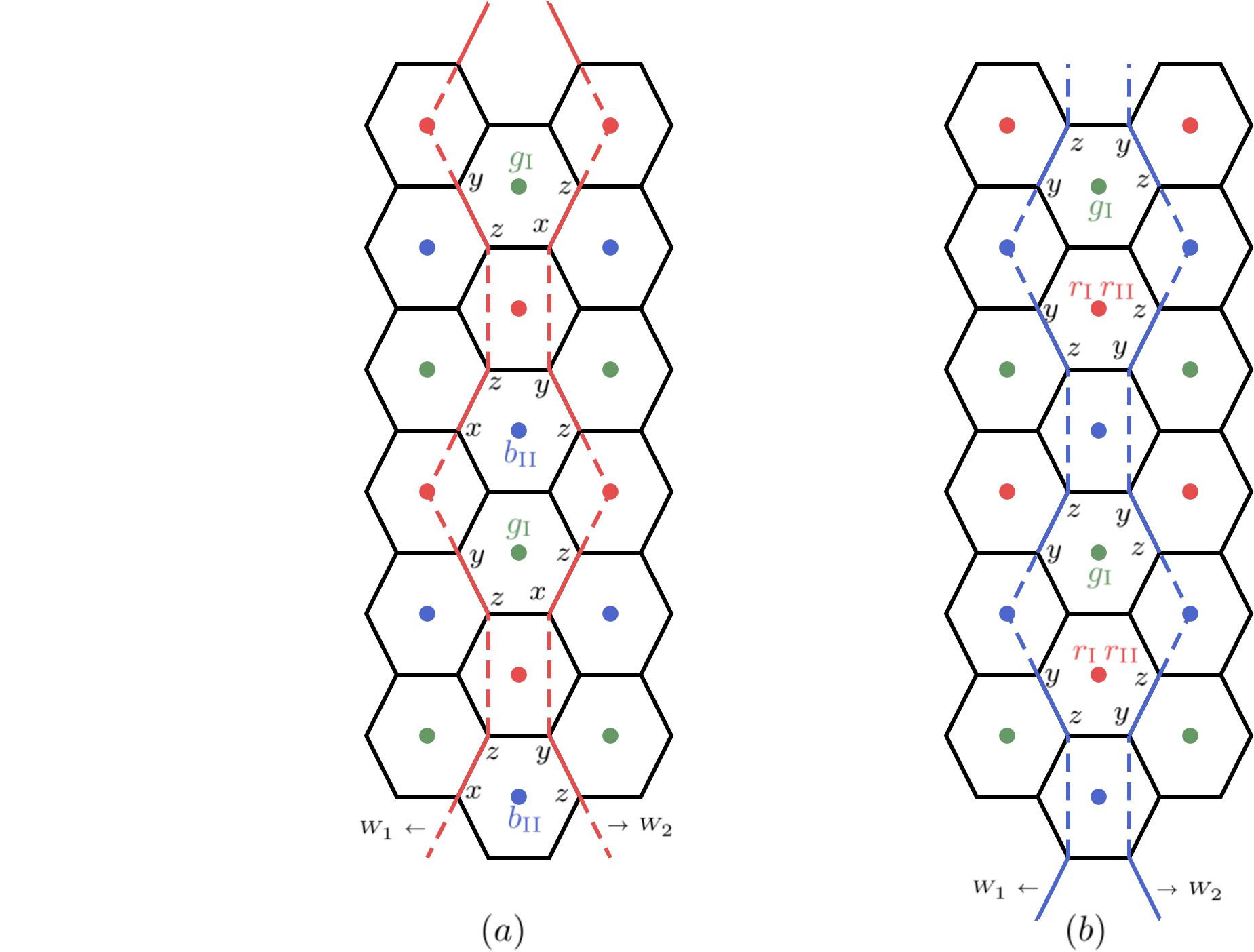}
 \caption{Equivalence of the red and blue logical Pauli $X$ operators. (a) The two winding operators $W_1$ and $W_2$ can be mapped to each other using ${\color{ForestGreen}g_I}{\color{blue}b_{II}}$ between the them. (b) In this case they are mapped to each other using ${\color{ForestGreen}g_I}{\color{red}r_Ir_{II}}$ between them.}
 \label{fig:4cclogicalequivC1}
\end{figure}

 To see this we note that an arbitrary path winding around the torus is traversed on the hexagonal lattice via three independent directions. And for such an arbitrary path winding around the torus the logical $X$ or $Z$ is constructed from the elementary red or blue strings shown in Figs. \ref{fig:4ccredelemstrings} and \ref{fig:4ccblueelemstrings} and since each of these strings along the three different independent directions are mixed we are guaranteed that the entire operator will remain mixed while winding around the torus. The equivalence class of the different logicals are summarized in Fig. \ref{fig:4ccmixedlogicals}
The winding operators for different choices of the non-contractible loop are mapped to each other via the stabilizers of the code, Fig. \ref{fig:4ccstabilisers}. We illustrate this for two examples in Fig. \ref{fig:4cclogicalequivC1}.  
\subsubsection{Anyonic content}
\begin{figure}[h]
    \centering
    \includegraphics[width=9cm]{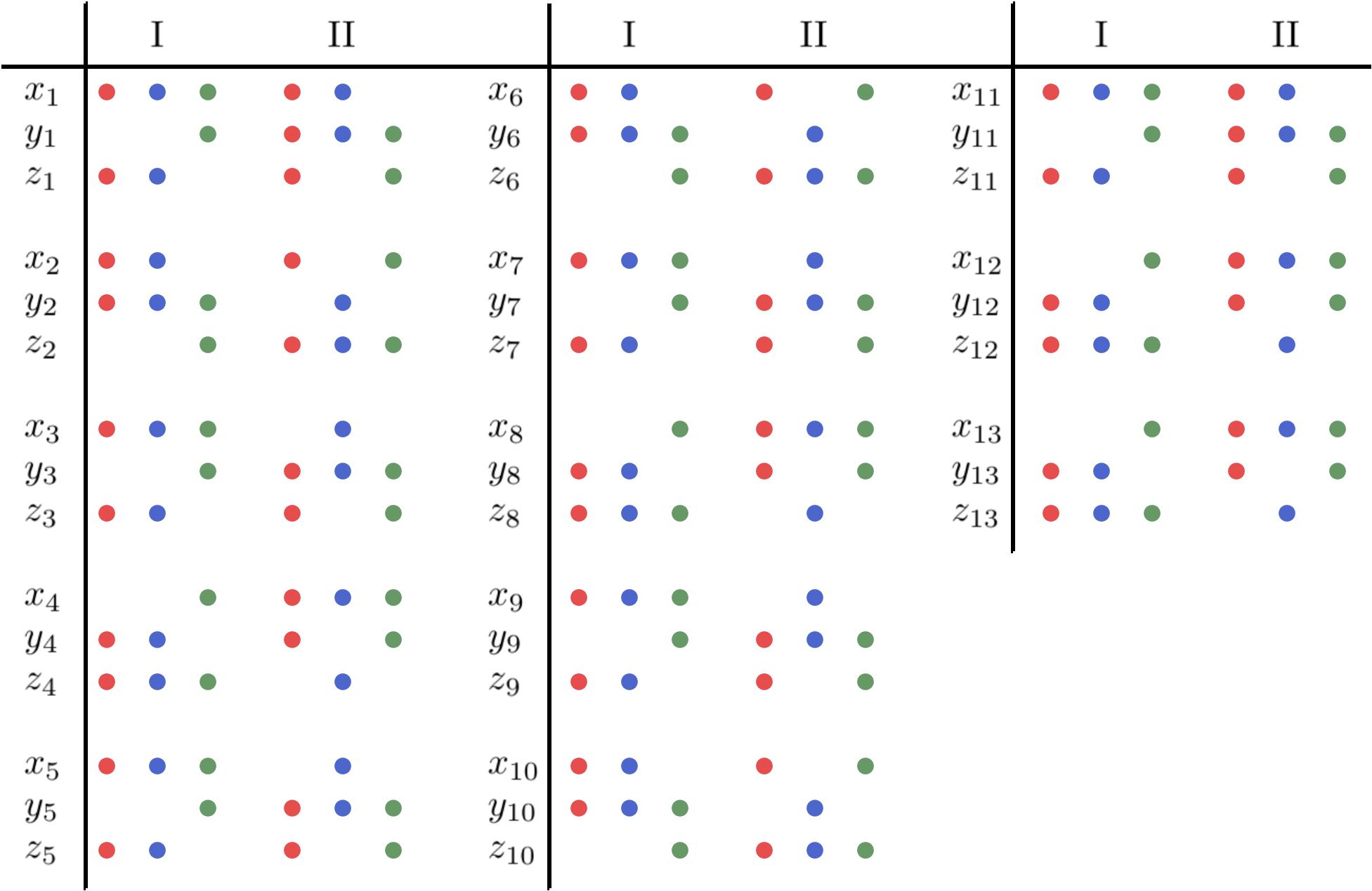}
 \caption{Effect of single qubit operations on the physical qubits of the $[444]$-color code. See Fig. \ref{fig:lookuplatcc} for the location of the physical qubits on the unit cell comprising the three different colors. Notice that for each of the operations precisely four stabilizers are excited.} 
 \label{fig:4cclookuptable}
\end{figure}

While the $[444]$-color code shares the same long range features with the original $[336]$-color code, they have different spectrum resulting from single qubit operations as shown in the table in Fig. \ref{fig:4cclookuptable}. 
Most of the excitations resulting from the single qubit operations are immobile, implying that they cannot be moved around the lattice using string operators with zero energy cost. These excitations account for the short-range physics of the model. In particular they do not contribute to the GSD of the model which is a purely topological feature.

\begin{figure}[h]
    \centering
    \includegraphics[width=7cm]{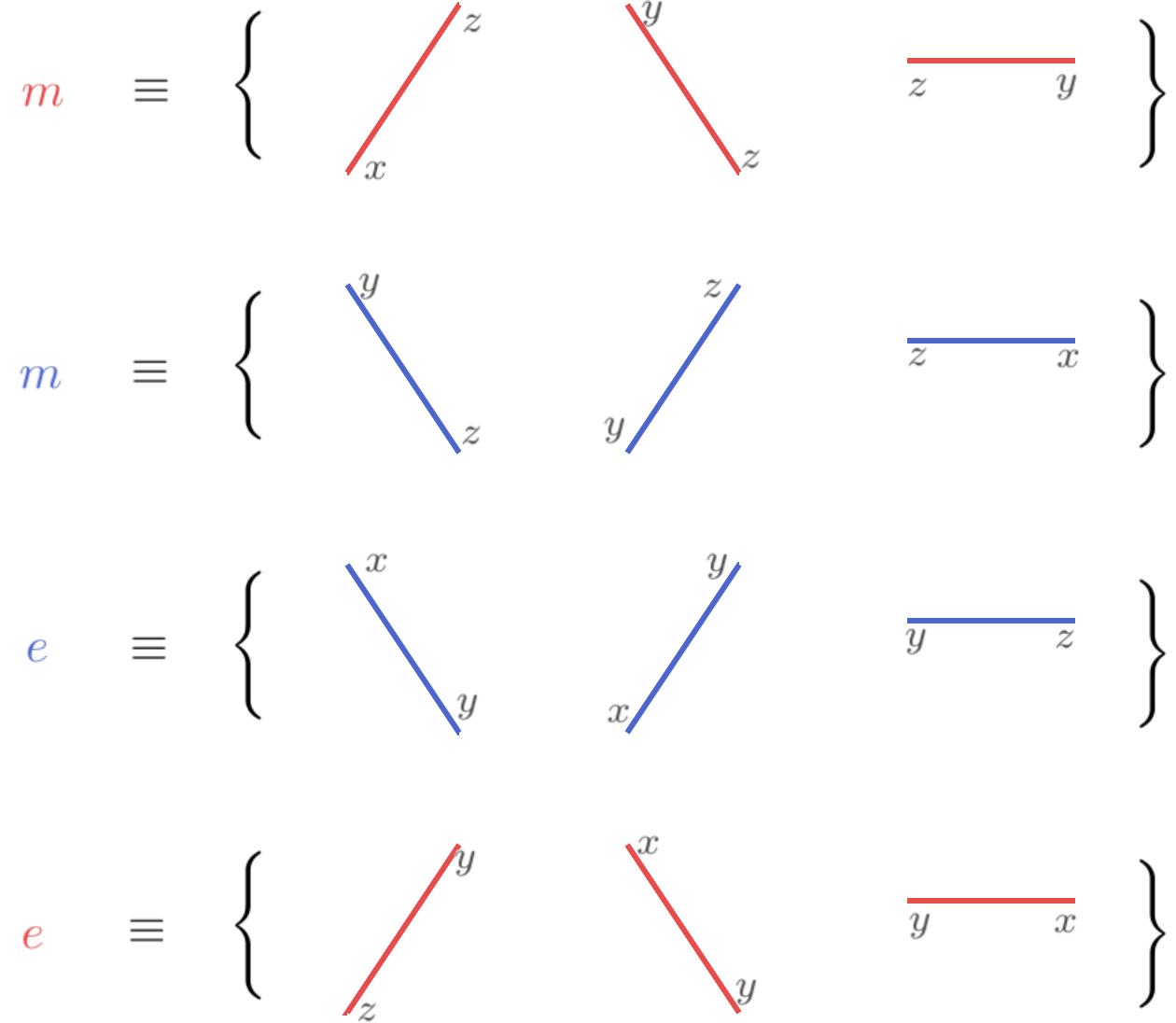}
 \caption{The string operators creating the fundamental anyons, $\{{\color{blue}e}, {\color{red}m}\}; \{{\color{red}e}, {\color{blue}m}\}$, of the $[444]$-color code defined in Fig. \ref{fig:4ccstabilisers}. The figure shows the operators that have to be applied to the three independent directions of the hexagonal lattice to create anyons at the endpoints of the string. }
 \label{fig:4ccanyonstrings}
\end{figure}

On the other hand the excitations which explain the topological nature of the GSD of the model are the deconfined anyons. For the ground states chosen in Eq.\,(\ref{eq:4ccaaaastate}) the two sets of $\mathbb{Z}_2$ toric code anyons,
\begin{equation}
    \{{\color{blue}e}, {\color{red}m}\};~~ \{{\color{red}e}, {\color{blue}m}\},  
\end{equation}
are created by the string operators shown in Fig. \ref{fig:4ccanyonstrings}. 
These string operators can also be obtained by cutting the logical operators such that it no longer winds around the torus. 

To ensure that we have indeed obtained the anyons of the $\mathbb{Z}_2\times\mathbb{Z}_2$ toric code we check their braiding and fusion properties as shown in App. \ref{app:fusionbraiding[444]fm}. 

\section{Examples of trivalent and tricolorable lattices of higher-genus topology}
\label{sec:higher-genus}
In this section we present some examples of trivalent and tricolorable lattices of higher-genus topology with genus two or more. 

\begin{figure}[h]
    \centering
    \includegraphics[width=8cm]{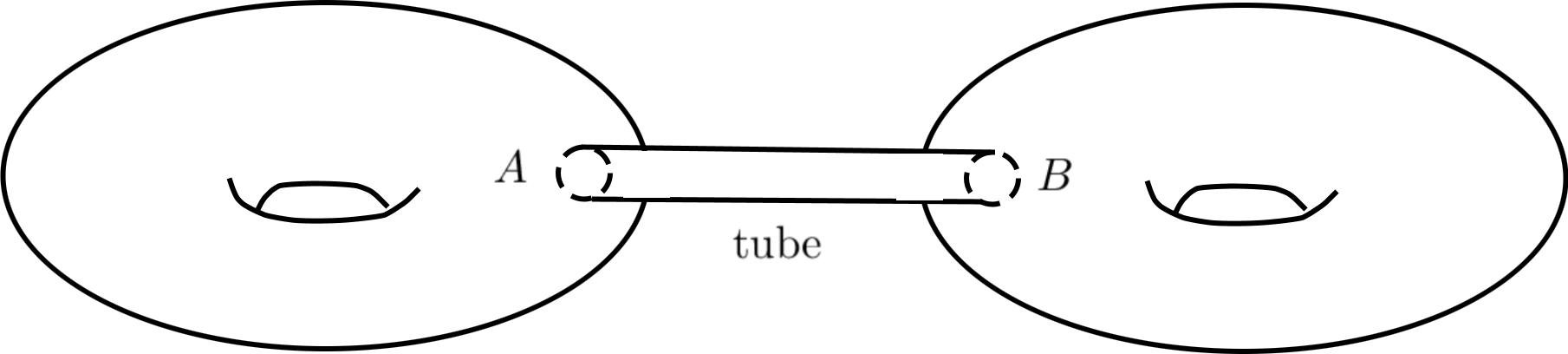}
 \caption{A genus-two surface constructed by connecting two tori with one hole by a tube. 
 A and B denote the holes of the tori at which the tube is connected.}
    \label{fig:genus-two}
\end{figure}

\begin{figure}[h]
    \centering
    \includegraphics[width=8cm]{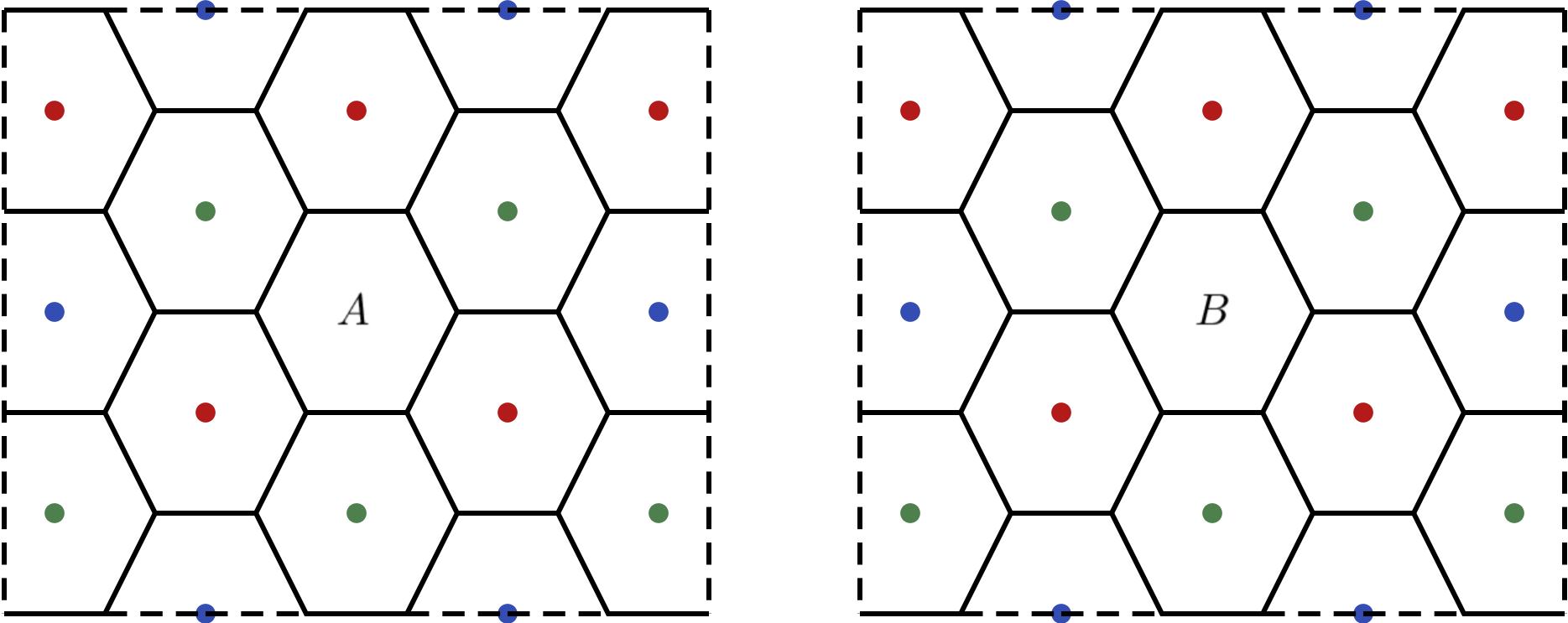}
   \caption{Example of two copies of a hexagonal lattice of tori.with holes A and B. 
 For each of the two, the upper and lower horizontal dashed lines are identified, 
 and the left and right vertical dashed lines are also.}
    \label{fig:two-hexagonal-tori}
\end{figure}

First let us consider a genus-two surface which can be constructed by preparing two tori, with one hole each, connected by a tube as in Fig.~\ref{fig:genus-two}. 
In order to construct trivalent and tricolorable lattices on this topology, we prepare a regular hexagonal lattice of torus topology, remove a face to create a hole, and make two copies of it. Fig.~\ref{fig:two-hexagonal-tori} depicts the hexagonal lattices around the holes A and B. 
%
%
%
%

We divide each edge of the holes into three pieces. Then hexagons adjacent to the holes become octagons. 
Lattice structure of a tube which connects to the tori at the holes is presented below. 
In Fig.~\ref{fig:two-hexagonal-tori2}, colors of the boundaries of the tube are also indicated. 
%
\begin{figure}[h]
    \centering
    \includegraphics[width=8cm]{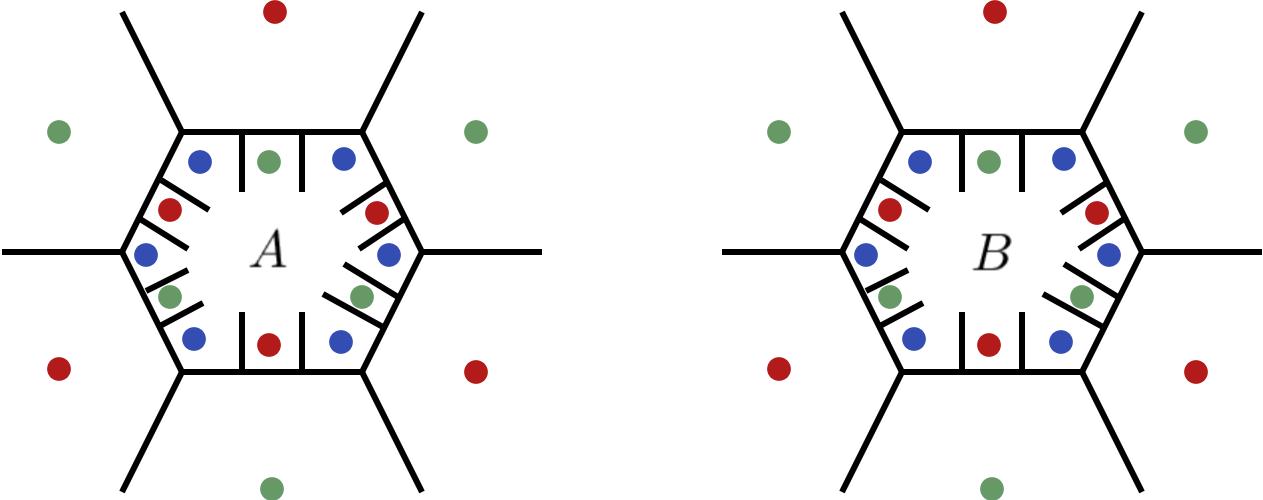}
   \caption{Each edge of the holes in Fig.~\ref{fig:two-hexagonal-tori} is divided into three pieces. 
 Then hexagons adjacent to the holes become octagons. 
 Colors around A and B indicate the colors of the boundaries of the tube.}
    \label{fig:two-hexagonal-tori2}
\end{figure}
%

\begin{figure}[h]
    \centering
    \includegraphics[width=9cm]{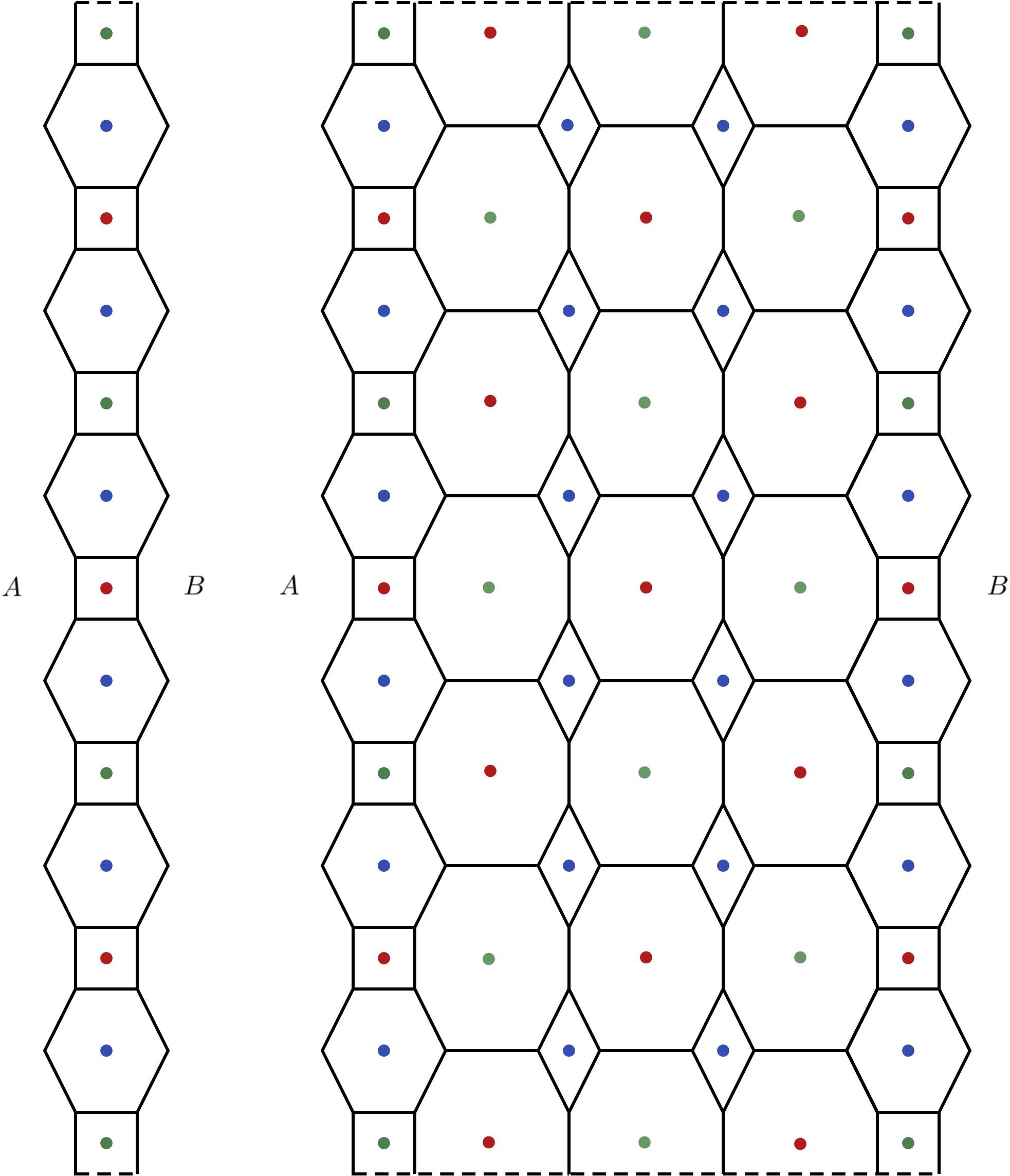}
 \caption{The shortest tube (the left panel) and a longer tube (the right panel).
 In each figure, the upper and lower horizontal dashed lines are identified. 
 }
    \label{fig:tubes}
\end{figure}

The shortest tube is constructed by squares and hexagons as in the left panel of Fig.~\ref{fig:tubes}. 
For longer tubes, their bulk consists of squares and octagons whereas their boundaries consist of 
squares and hexagons as in the right panel of Fig.~\ref{fig:tubes}. 
%
%
 
Connecting either of the tubes to the holed tori at the boundaries A and B, we can construct trivalent and tricolorable lattices realizing genus-two surfaces. 
We can explicitly confirm the Euler formula for these lattices. 
Let $k_f$ denote the number of the vertices of the face $f$. Then $3|V| = \sum_f k_f$ holds for any trivalent lattice, where the sum is taken for all the faces of the lattice.
Together with this, the formula (\ref{Euler1}) leads to 
\begin{equation}
2-2g=\frac16\sum_f(6-k_f).
\label{Euler2}
\end{equation}
Since the above lattices consist only of squares, hexagons and octagons, (\ref{Euler2}) becomes
\begin{equation}
2-2g=\frac13(|F_4|-|F_8|)
\label{Euler3}
\end{equation} 
with $|F_4|$ and $|F_8|$ being the number of squares and the number of octagons, respectively.   
In Fig.~\ref{fig:two-hexagonal-tori2}, we can see that there are six octagons around each of the holes A and B. 
In the left (right) panel of Fig. \ref{fig:tubes} there are six squares (24 squares and 18 octagons).
From these, we can easily confirm that  (\ref{Euler3}) gives $g=2$. 

Repeating a similar procedure we can construct trivalent and tricolorable lattices realizing surfaces with genus being more than two.

\subsection{Example of a non-CSS color code in higher-genus topology}
The trivalent and tricolorable nature of the lattice naturally ensures that the $[444]$- and $[345]$- color codes with homogeneous edge configurations are well defined on the higher-genus surfaces. This implies that the red, blue and green faces have the $X$ and $Y$, $Y$ and $Z$, and the $Z$ and $X$ stabilizers for the former case, whereas on the latter code we have $X$ and $Z$, $Y$ and $Z$, and the $Y$ and $Z$ stabilizers, respectively. The other inequivalent classes in the $[345]$-color codes are obtained by rotations of the colors as seen earlier. For the reason elaborated in Fig. \ref{fig:[345]ccconfinedpossiblepolygons} it is not possible to construct a $[345]$-color code where all the vertex-edge combines are confined as the lattice involves square and octagonal faces along with the hexagonal lattice. Nevertheless it is possible to construct $[345]$-color codes with a mixture of confined and deconfined vertex-edge combines, but we will not consider them here. 

\begin{figure}[h]
    \centering
    \includegraphics[width=9cm]{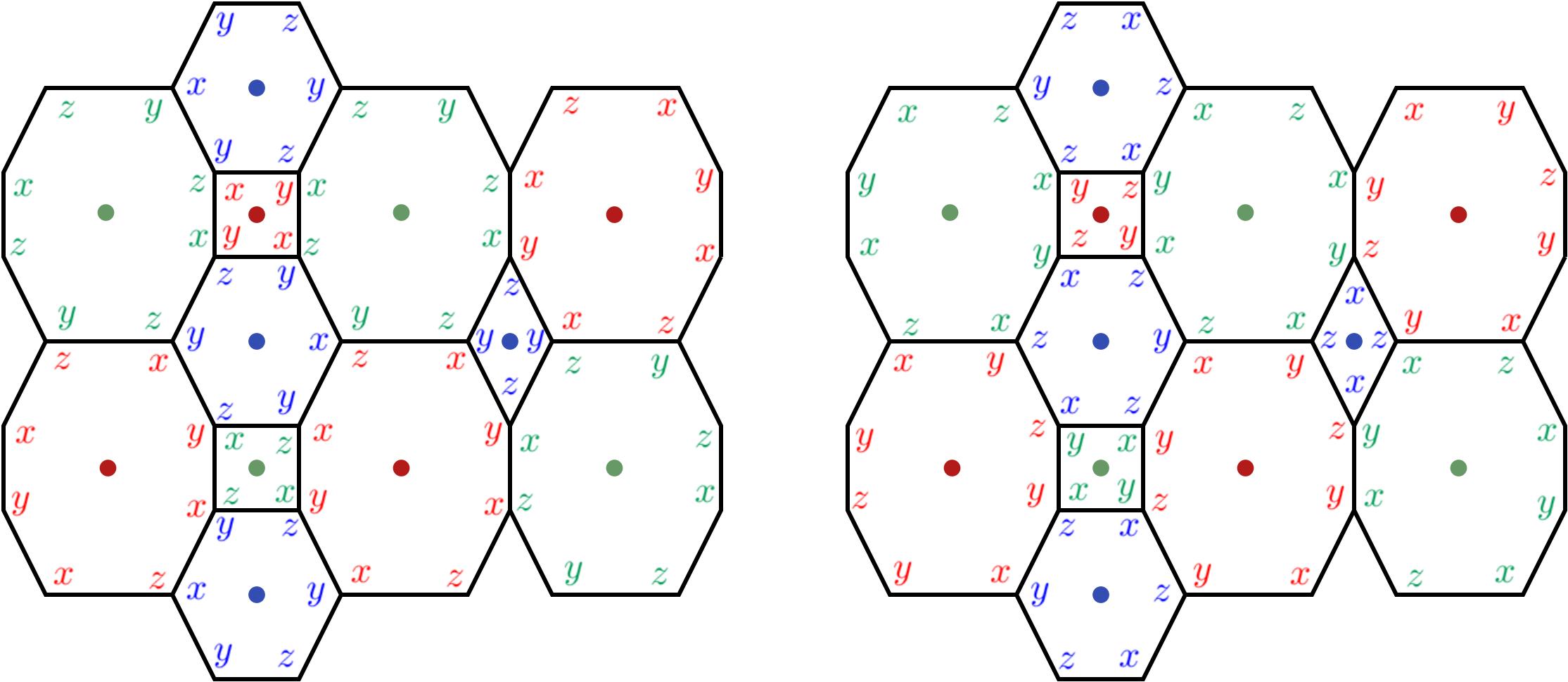}
    \includegraphics[width=9cm]{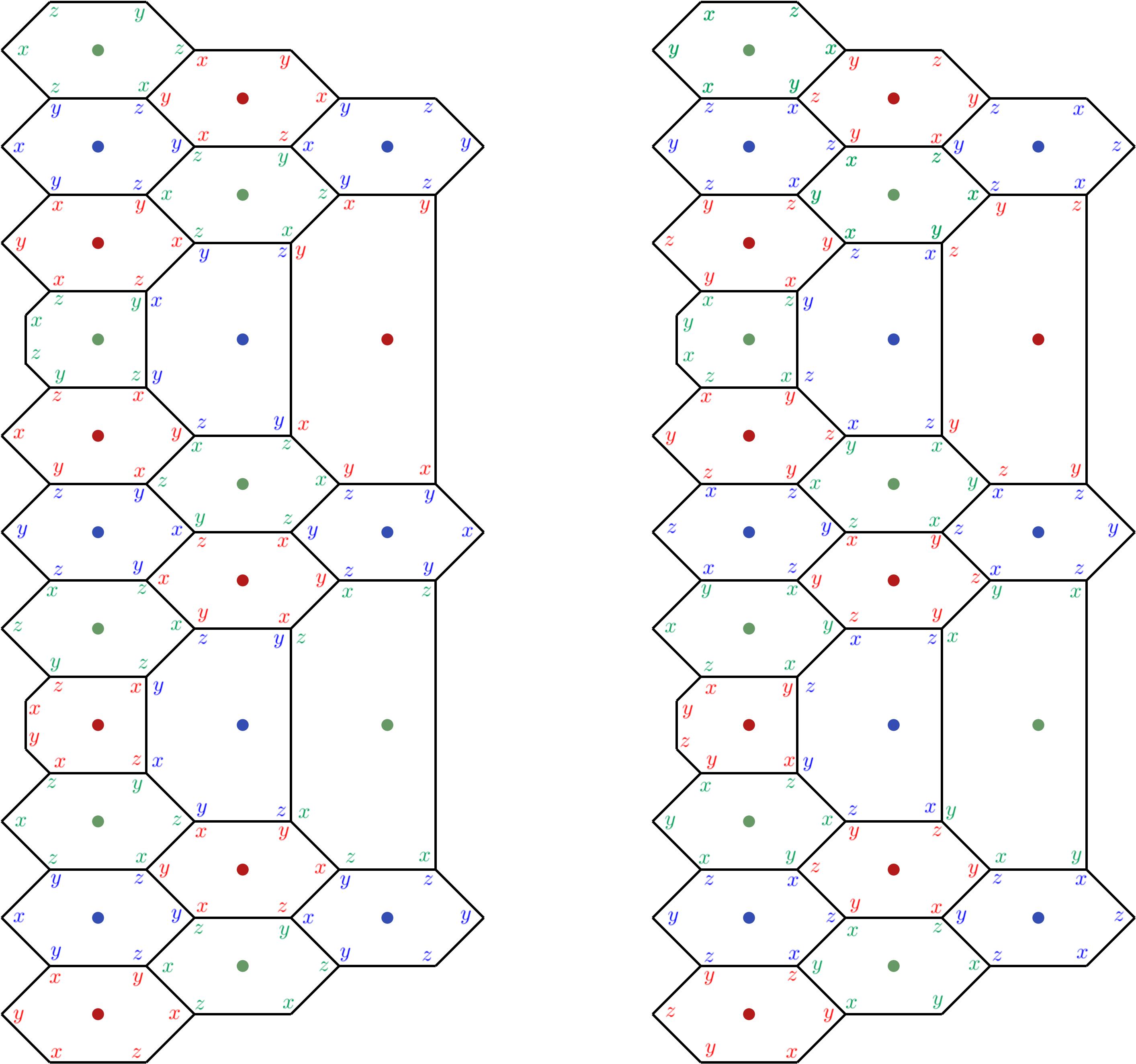}
\caption{A translationally invariant $[444]$-color code with fully mixed edge configurations on a higher-genus surface. The top panel shows the code on the tube connecting two tori and on octagons adjacent to the hole of each torus.
The bottom panel shows the code defined on a hexagonal lattice on a holed torus. The left end of the upper left (right) figure is identified with the right end of the lower left (right) figure. }
    \label{fig:(hg)[444]ccfullymixed}
\end{figure}

While the codes with homogeneous edge configurations are rather straightforward to write down, this is not the case where we have either fully mixed or partially mixed edge configurations. We show one such example for the $[444]$-color code with fully mixed edge configurations in Fig. \ref{fig:(hg)[444]ccfullymixed}.
%
%
Note that the translational invariance of the $[444]$-color code in Fig. \ref{fig:(hg)[444]ccfullymixed} occurs with the periodicity 2. This is a consequence of the way translational invariance for the $[444]$-color code with fully mixed edge configurations is implemented on the the square-octagon lattice (See the fully mixed case in Fig. \ref{fig:(488)[444]ccnopartialfullmixingwti}). The hexagonal lattice discretizing the torus is modified as shown in the bottom panel of Fig. \ref{fig:(hg)[444]ccfullymixed}. We have two sets of stabilizers, those that connect to the red octagon and those that connect to the green octagon. This also elaborates the translational invariance on the torus part. The translationally invariant $[444]$-color code with partially mixed configurations can be similarly written down but we do not show that here. The codes without translational invariance are easier to write down as they are less constraining than the previous case. This is also a result of the algorithm used in the construction of these codes in general.

\section{Outlook}
\label{sec:outlook}
We have presented a new approach to generalize the 2D color codes by using the short-range physics of these models. As a result we have discovered new versions, namely the $[444]$ and $[345]$-color codes, that are non-CSS by construction. While these models have the same long-range properties as the color code in its original version, they differ in the short-range physics. Apart from their potential applications in quantum error correction, especially the codes defined on finite lattices like the triangles of different sizes, there are several other interesting directions that one may to wish to pursue. 

\begin{itemize}
\item In this paper we have developed a systematic procedure to construct abelian stabilizer codes on trivalent lattices, with and without boundaries. The construction accommodates several other models, especially those that do not generate an abelian subgroup of the $n$-qubit Pauli group, $\mathcal{P}_n$, once we relax the anticommuting rule. This will then help us to define the model on more general trivalent lattices that include polygons with an odd number of sides.

\item This approach for building non-CSS stabilizer codes can be repeated for other kinds of lattices, like the quadrivalent ones that support surface codes like the toric code. However in this case we do not find any new model that is not equivalent to the original surface code by an LU. However this changes when we place two stabilizers on each face of the quadrivalent, bicolorable lattice. In this case we find codes with interesting properties. 

\item Another direction of generalization appears when we change the number of stabilizers defined on each face. In particular if we include just a single stabilizer on each face we obtain the recently studied {\it honeycomb code} \cite{Hastings2021DynamicallyGL}. The honeycomb defined in \cite{Hastings2021DynamicallyGL} can be thought of as the analog of the $[444]$-color codes presented in this work. Thus it is natural to write down the $[345]$ analogs of the honeycomb model and explore its properties. 

\item The toric code can be thought of as a discrete gauge theory in 2D realizing topological order. On the other hand the color code is not a lattice gauge theory in the usual sense, though it still realizes the $\mathbb{Z}_2\times\mathbb{Z}_2$ topolgical phase. In this context it is interesting to know what other quantum orders can be realized with models like the color code. In this direction the confined $[345]$-color codes seems to be promising as they have interesting excitations that are constructed using energetic string operators. They seem to resemble a fractonic phase and this direction is worth pursuing as well.

\item The codes on higher-genus surfaces are particularly interesting as they enlarge the code space. We have showed just a single model in this setting. There are more possibilities in this direction which we hope to visit in future publications. 

\item It would also be interesting to find appropriate generalizations of this construction on higher-dimensional lattices and hyperbolic spaces. 

\item In Apps. \ref{app:[345][336]mixedvertextypes}, \ref{app:[345][444]mixedvertextypes}, \ref{app:[444][345][336]mixedvertextypes} we have illustrated stabilizer codes with mixed vertex energetics. It would be interesting to explore their properties as well, though we expect the long range physics to coincide with that of the $\mathbb{Z}_2\times\mathbb{Z}_2$ topological phase.

\item Finally in Apps. \ref{app:[255]cc}, \ref{app:[345][255]cc} we give a glimpse of a code which does not satisfy the anticommuting rule. These models have {\it quasi-local} symmetries which warrants further study.
\end{itemize}

\begin{acknowledgements}
PP and AC would like to thank IIT-BBS where this work was carried out. AC acknowledges support from IIT-BBS seed grant (No. SP103). We thank Michael Vasmer and Paulo Teotonio-Sobrinho for useful discussions and references.
\end{acknowledgements}

\appendix
\section{Fusion and braiding rules for $[444]$-color codes with fully mixed edge configurations}
\label{app:fusionbraiding[444]fm}
Here we discuss the braiding and fusion rules of the anyons in the fully mixed $[444]$-color codes.

\paragraph{{\bf Braiding rules} -} It is easy to see that the electric charges and magnetic fluxes of the same color do not pick up a phase when they move around each other. This is due to the fact that two strings of the same color never cross each other at a vertex or the location of a physical qubit. 

\begin{figure}[h]
    \centering
    \includegraphics[width=9cm]{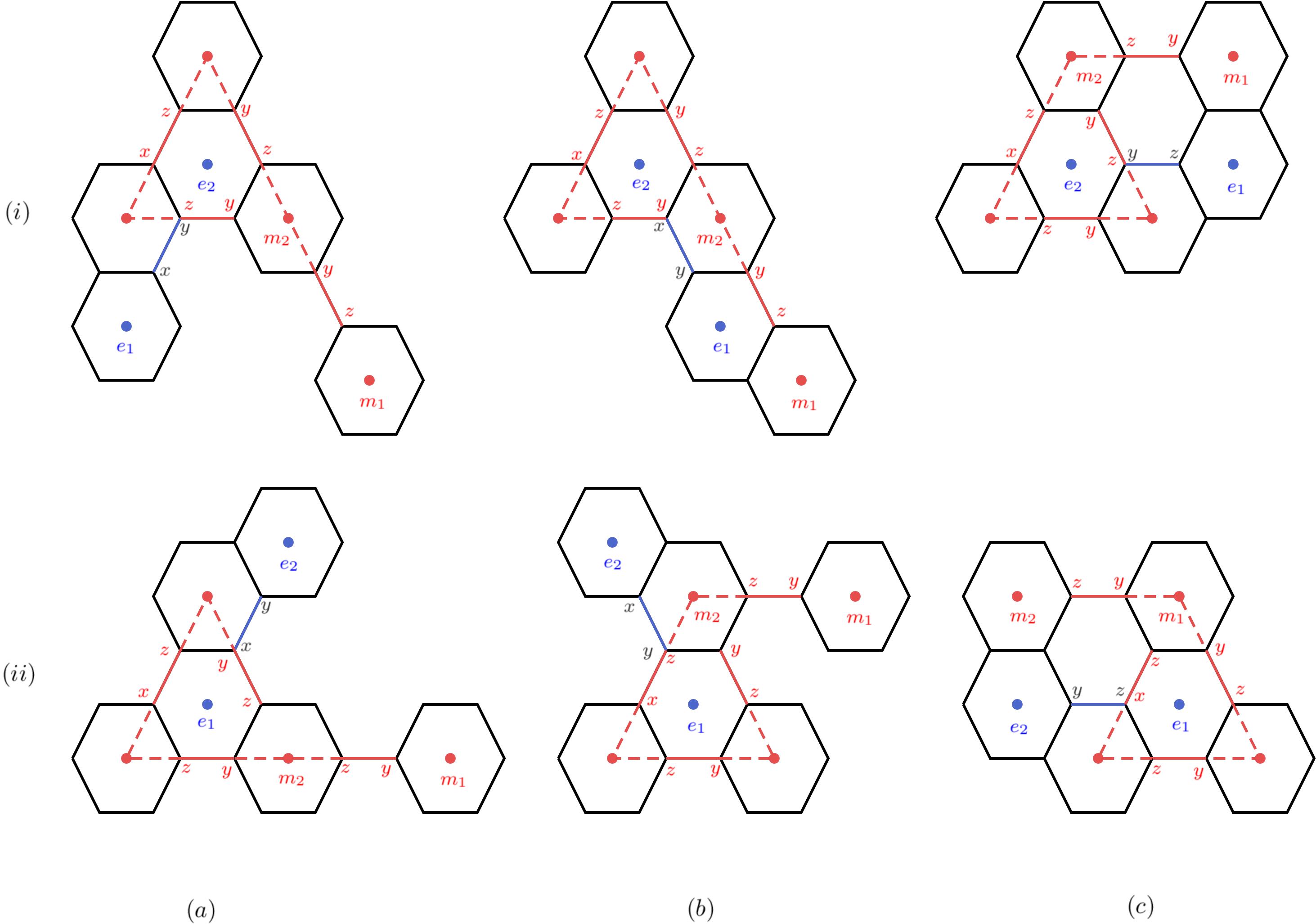}
 \caption{Exchanging ${\color{red}m}$ around ${\color{blue}e}$ shown for the three possible orientations of the pair of ${\color{blue}e}$ charges in (a), (b) and (c) respectively. The subfigures (i) and (ii) show the exchange around ${\color{blue}e_2}$ and ${\color{blue}e_1}$ respectively.}
 \label{fig:4ccanyonbraiding1}
\end{figure}

\begin{figure}[h]
    \centering
    \includegraphics[width=9cm]{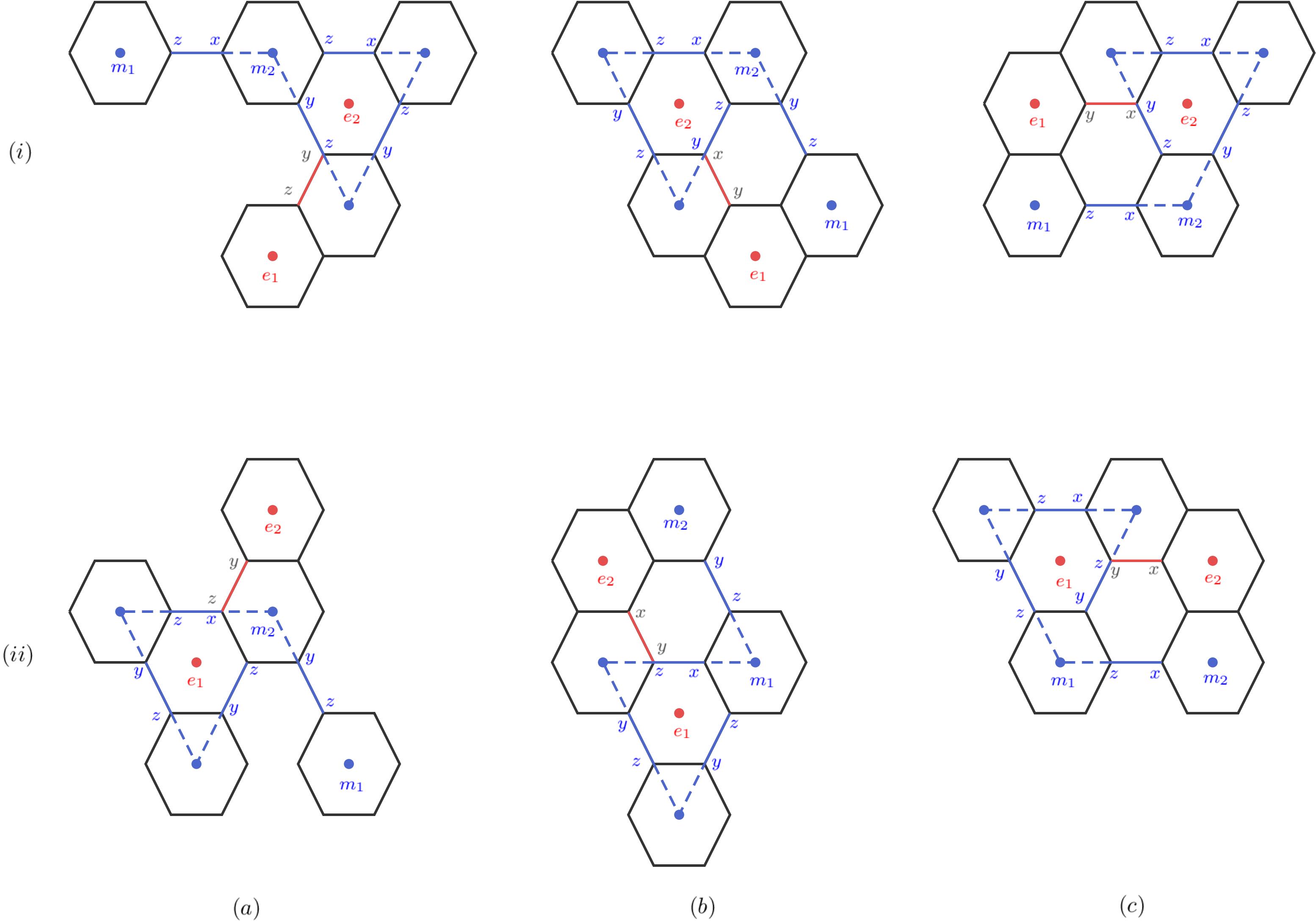}
 \caption{Exchanging ${\color{blue}m}$ around ${\color{red}e}$ shown for the three possible orientations of the pair of ${\color{red}e}$ charges in (a), (b) and (c) respectively. The subfigures (i) and (ii) show the exchange around ${\color{red}e_2}$ and ${\color{red}e_1}$ respectively.}
 \label{fig:4ccanyonbraiding2}
\end{figure}

\begin{figure}[h]
    \centering
    \includegraphics[width=9cm]{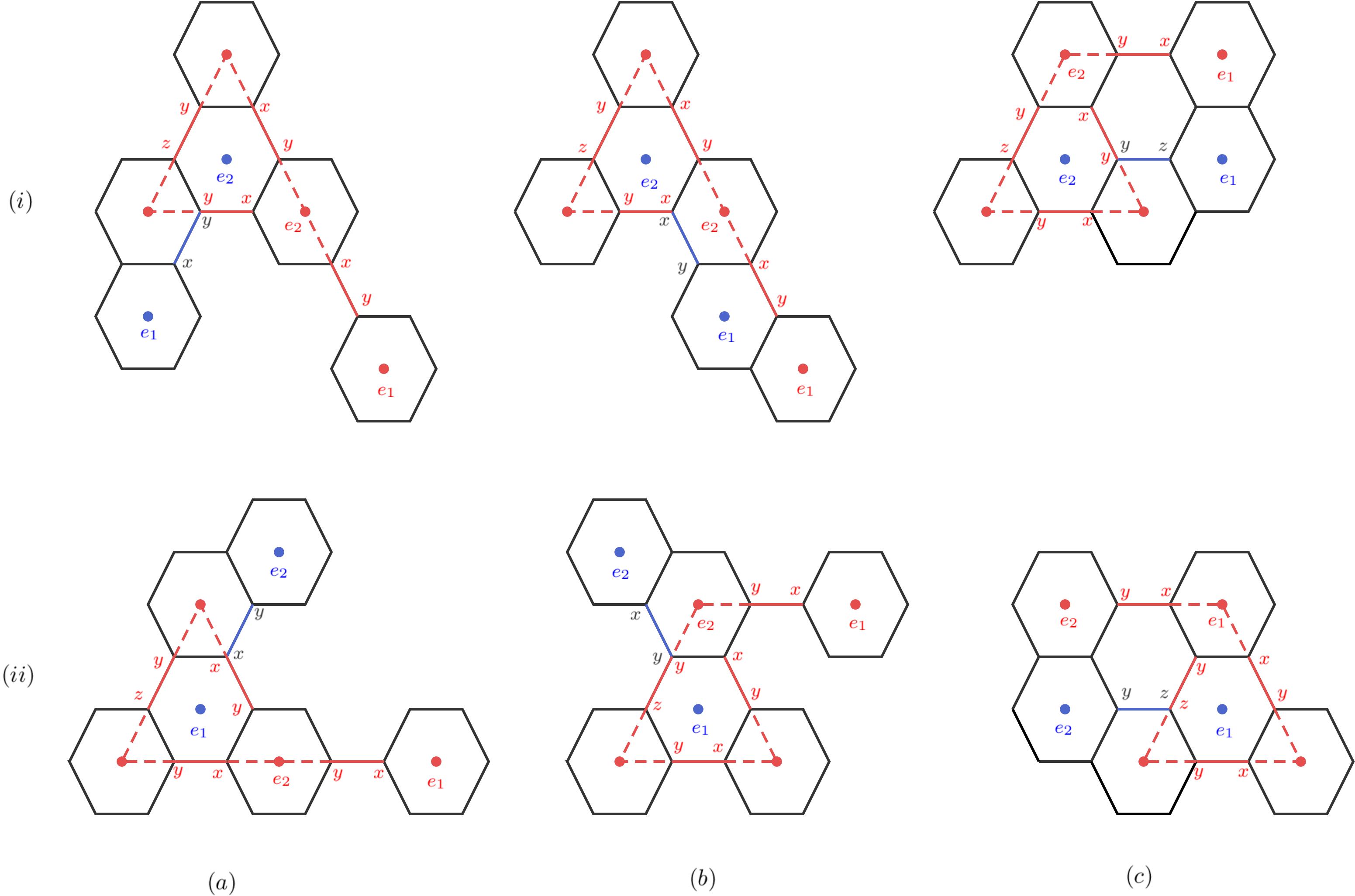}
 \caption{Exchanging ${\color{red}e}$ around ${\color{blue}e}$ shown for the three possible orientations of the pair of ${\color{blue}e}$ charges in (a), (b) and (c) respectively. The sub-figures (i) and (ii) show the exchange around ${\color{blue}e_2}$ and ${\color{blue}e_1}$ respectively.}
 \label{fig:4ccbraidingee}
\end{figure}

\begin{figure}[h]
    \centering
    \includegraphics[width=9cm]{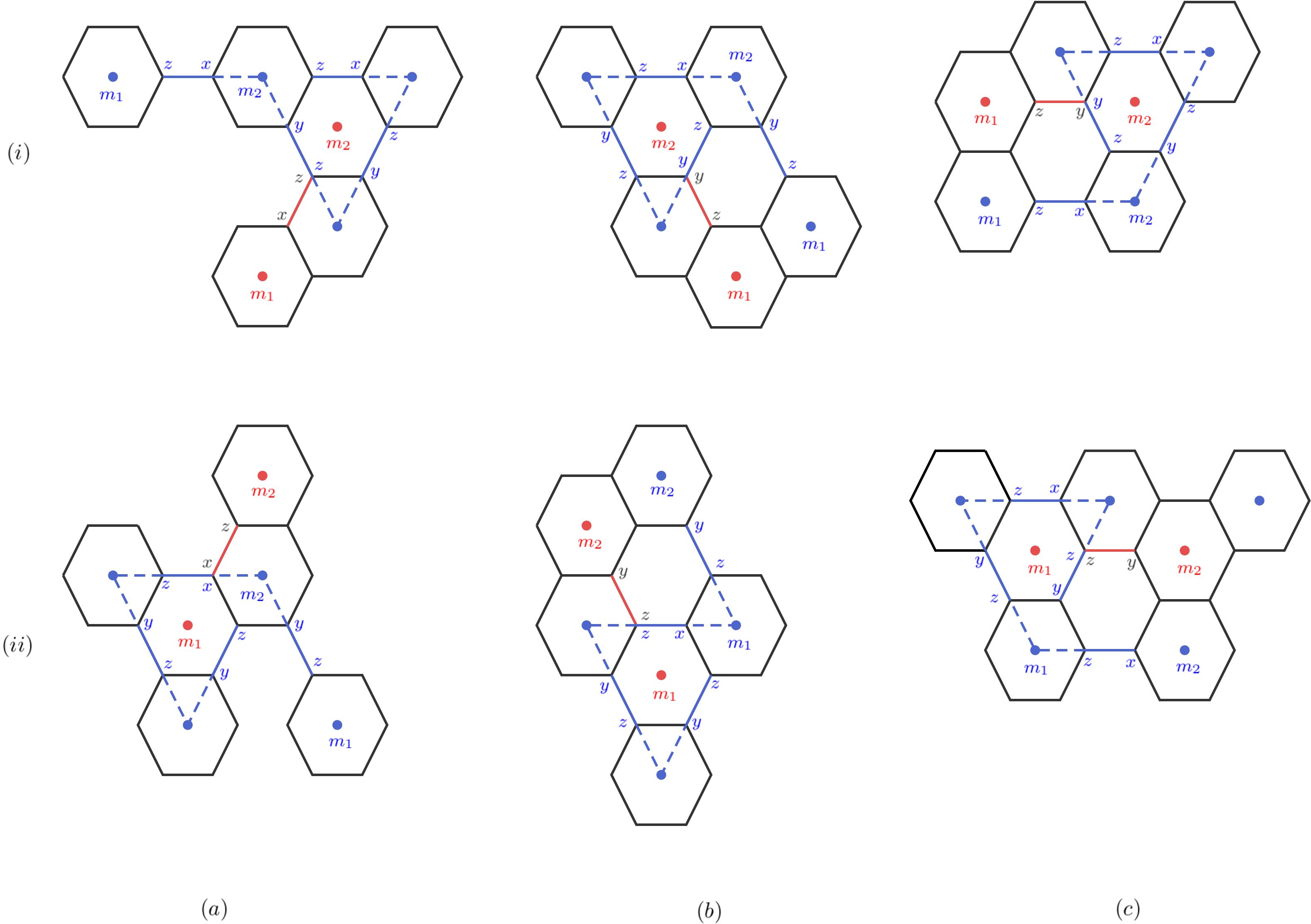}
 \caption{Exchanging ${\color{blue}m}$ around ${\color{red}m}$ shown for the three possible orientations of the pair of ${\color{red}m}$ fluxes in (a), (b) and (c) respectively. The subfigures (i) and (ii) show the exchange around ${\color{red}m_2}$ and ${\color{red}m_1}$ respectively.}
 \label{fig:4ccbraidingmm}
\end{figure}

Thus we should check the braiding of particles of different color around each other. We expect to see a phase of -1 when ${\color{red}e}$ goes around  ${\color{blue}m}$ and when ${\color{blue}e}$ goes around ${\color{red}m}$. However as the anyons can be oriented in three different directions we need to check that the -1 phase is obtained in all cases. This is verified in Figs. \ref{fig:4ccanyonbraiding1} and \ref{fig:4ccanyonbraiding2}.
Furthermore we need to ensure that ${\color{red}e}$ and ${\color{blue}e}$, and ${\color{red}m}$ and ${\color{blue}m}$ are mutually bosonic. This is shown in Figs. \ref{fig:4ccbraidingee} and \ref{fig:4ccbraidingmm}.

\begin{figure}[h]
    \centering
    \includegraphics[width=8cm]{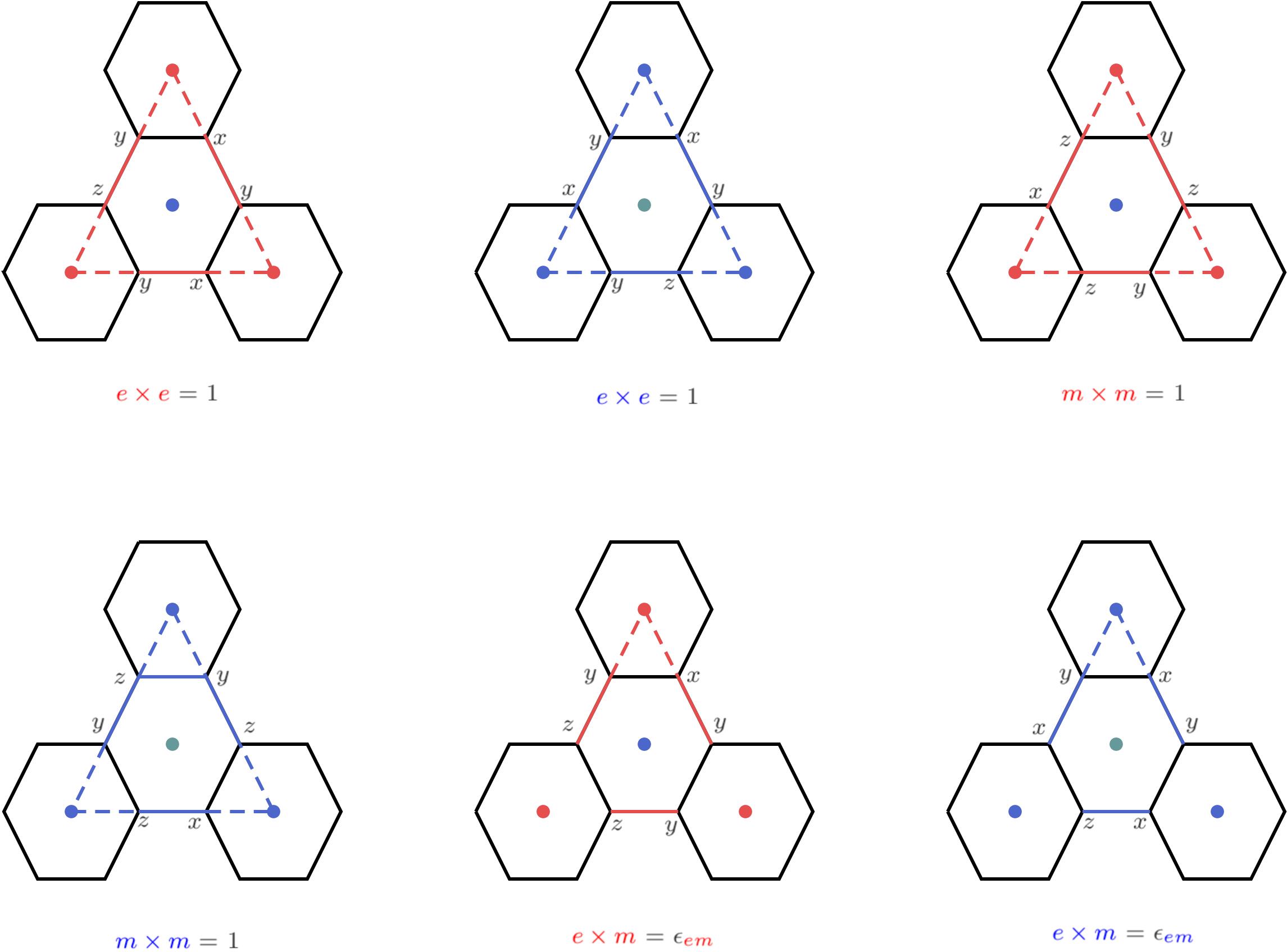}
 \caption{Fusion of the anyons of the $[444]$-color code defined by Fig. \ref{fig:4ccstabilisers}.}
 \label{fig:4ccanyonfusion}
\end{figure}

\paragraph{{\bf Fusion rules} -} The anyons $\{{\color{blue}e}, {\color{red}m}\},~ \{{\color{red}e}, {\color{blue}m}\}$ obey the fusion rules of the $\mathbb{Z}_2\times\mathbb{Z}_2$ toric code,
\begin{eqnarray}
{\color{red}e}\times {\color{red}e} & = & {\color{blue}e}\times {\color{blue}e} = {\color{red}m}\times {\color{red}m} = {\color{blue}m}\times {\color{blue}m} = 1, \nonumber \\
{\color{red}e}\times {\color{red}m} & = & \epsilon_{{\color{red}e}{\color{red}m}},~ {\color{blue}e}\times {\color{blue}m}  =  \epsilon_{{\color{blue}e}{\color{blue}m}}, \nonumber \\
{\color{red}e}\times {\color{blue}m} & = & \epsilon_{{\color{red}e}{\color{blue}m}},~
{\color{blue}e}\times {\color{red}m}  =  \epsilon_{{\color{blue}e}{\color{red}m}},
\end{eqnarray}
along with ${\color{blue}e}\times {\color{red}e}$ and ${\color{blue}m}\times {\color{red}m}$ which correspond to direct products of its components. There are four kinds of `dyonic' excitations in this case,$\epsilon_{{\color{blue}e}{\color{blue}m}}, \epsilon_{{\color{red}e}{\color{red}m}}, \epsilon_{{\color{red}e}{\color{blue}m}}, \epsilon_{{\color{blue}e}{\color{red}m}}$. In a diagrammatic way (See Fig. \ref{fig:4ccanyonfusion}) we ensure that we obtain the vacuum in case two particles of the same type fuse and that dyonic charges are obtained when particles of the same color fuse.

\section{Non-translationally invariant color code models}
\label{app:NTIccs}
Here we write down the non-translationally invariant non-CSS color codes on two dimensional trivalent and tricolorable lattices. We emphasize that the source for the loss of translational invariance is not the lattice geometry, i.e. we will continue to work on either the hexagonal or the square-octagonal lattice, but now the stabilizer pairs, on each face of a chosen color, are different as we move along the shrunk lattice of the same color.
\subsection{Non-translationally invariant $[444]$-color codes}
\label{app:NTIccs[444]}
\begin{figure}[h]
    \centering
    \includegraphics[width=8cm]{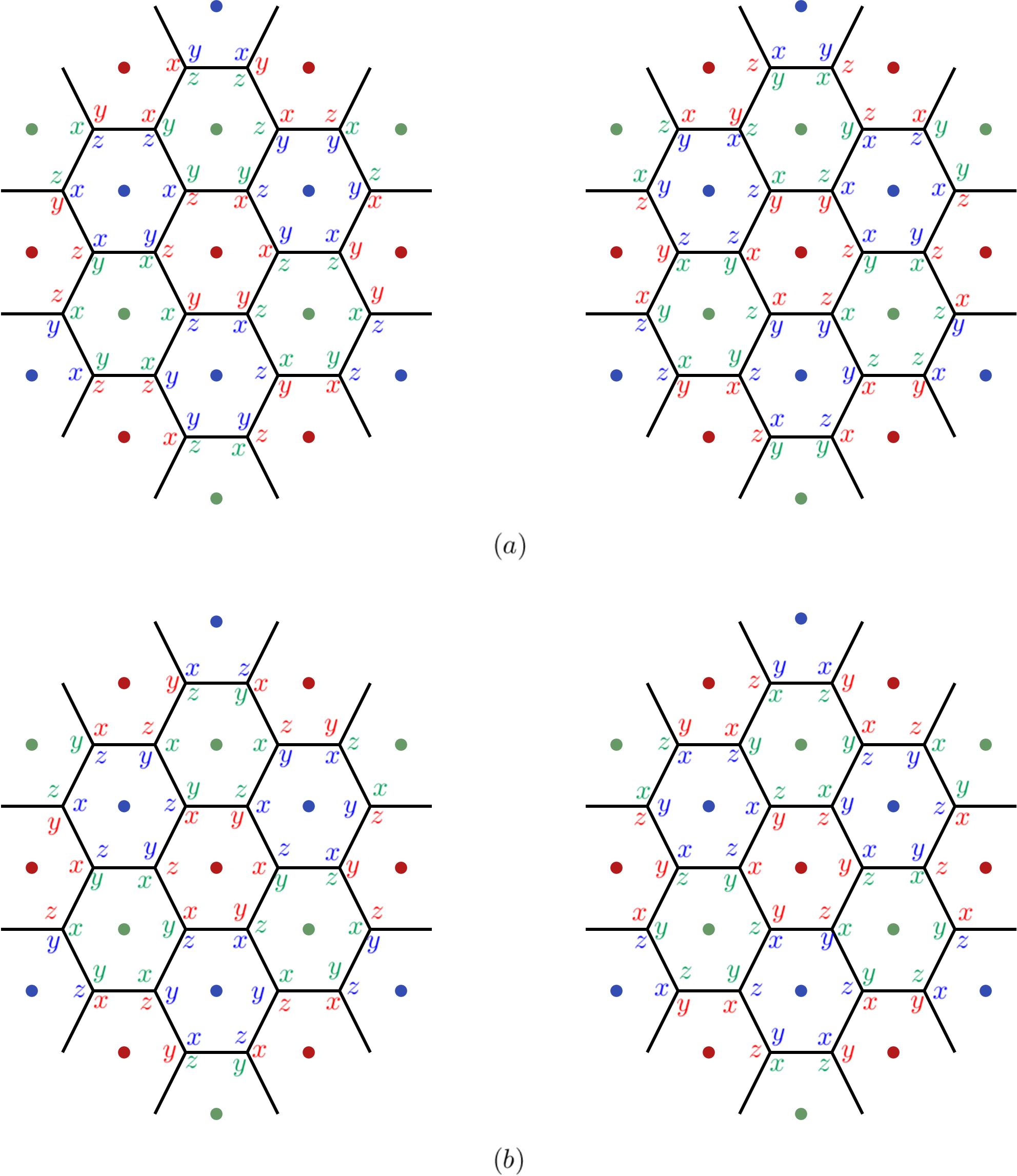}
 \caption{Non-translationally invariant $[444]$-color codes with (a) partially mixed and (b) fully mixed edge configurations. }
    \label{fig:[444]ccpartialfullmixingwoti}
\end{figure}

The non-translationally invariant $[444]$-color codes on the hexagonal lattice are shown in Fig. \ref{fig:[444]ccpartialfullmixingwoti}. Notice that is not possible to give up translational invariance and have homogeneous edge configurations.\\
\subsection{Non-translationally invariant and deconfined $[345]$-color codes}
\label{app:NTIccs[345]deconfined}
\begin{figure}[h]
    \centering
    \includegraphics[width=8cm]{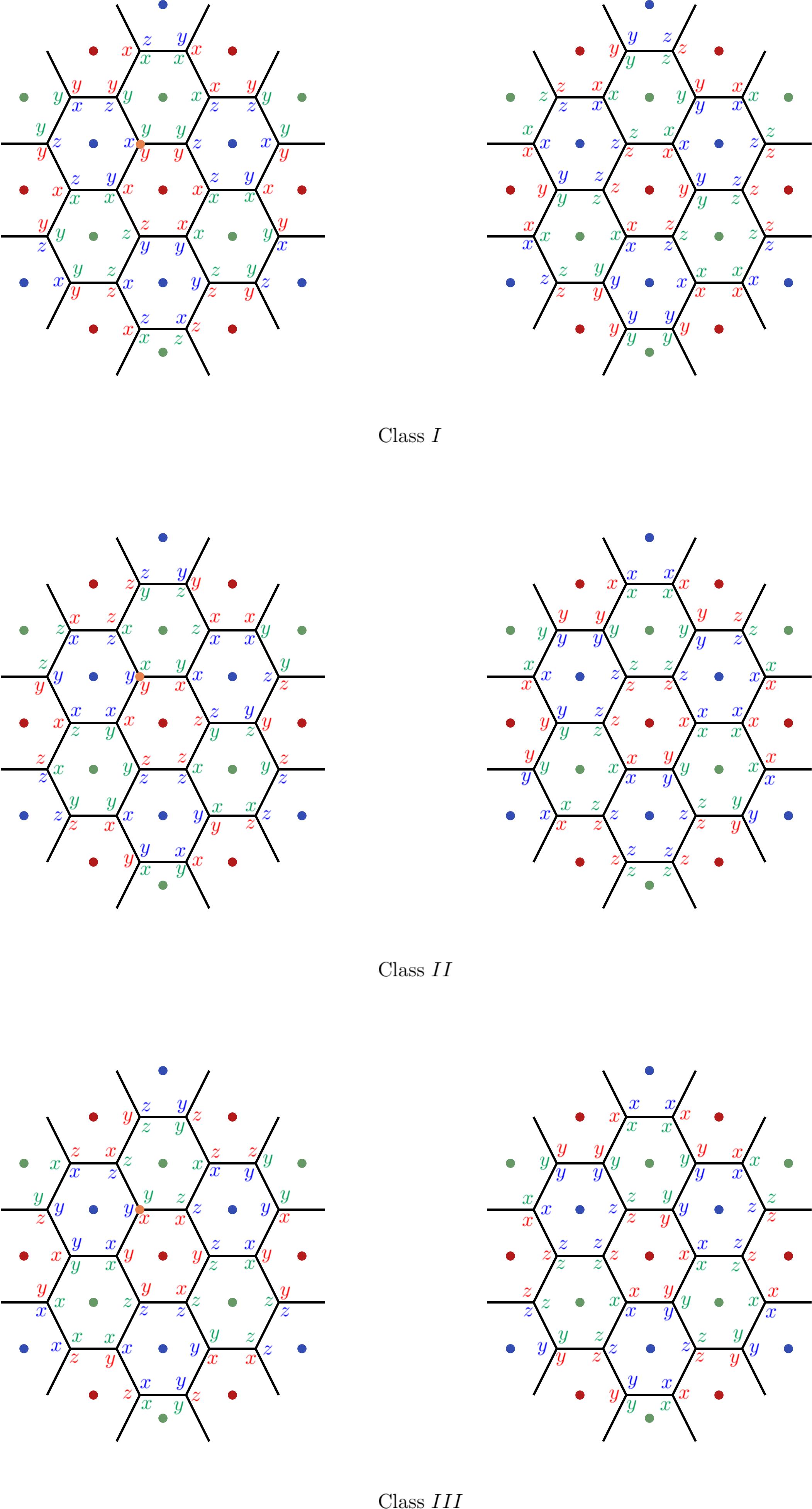}
 \caption{Non-translationally invariant, deconfined $[345]$-color codes with partially mixed edge configurations. The three inequivalent classes are distinguished by the bulk vertex configurations at the orange dot.}
    \label{fig:[345]ccdeconfinedpartialmixingwoti}
\end{figure}

\begin{figure}[h]
    \centering
    \includegraphics[width=8cm]{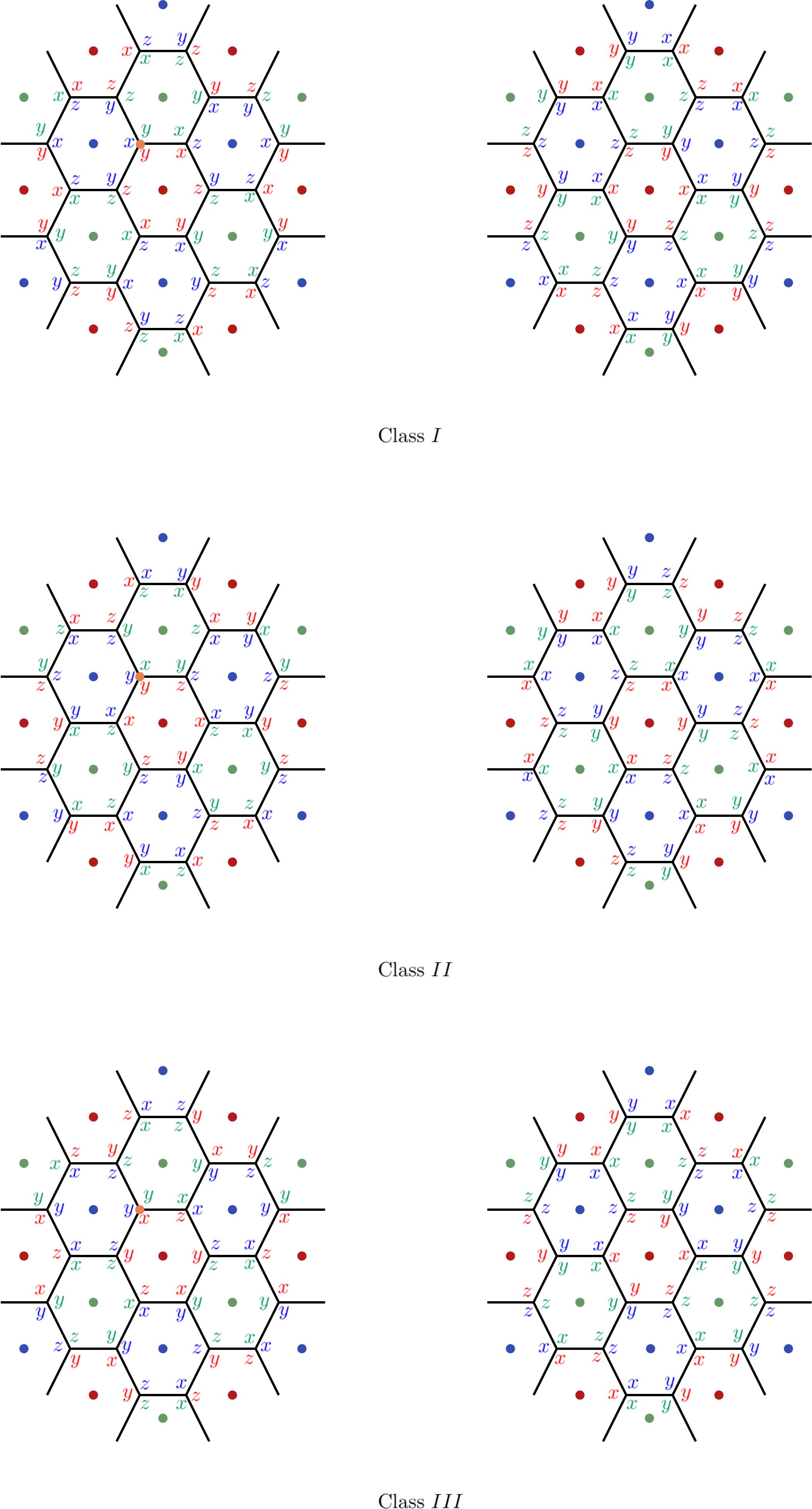}
 \caption{Non-translationally invariant, deconfined $[345]$-color codes with fully mixed edge configurations. The three inequivalent classes are distinguished by the bulk vertex configurations at the orange dot.}
    \label{fig:[345]ccdeconfinedfullmixingwoti}
\end{figure}

The non-translationally invariant $[345]$-color codes with partially and fully mixed edge configurations corresponding to the three inequivalent classes are shown in Figs. \ref{fig:[345]ccdeconfinedpartialmixingwoti} and  \ref{fig:[345]ccdeconfinedfullmixingwoti}, for the hexagonal lattice.

\begin{figure}[h]
    \centering
    \includegraphics[width=8cm]{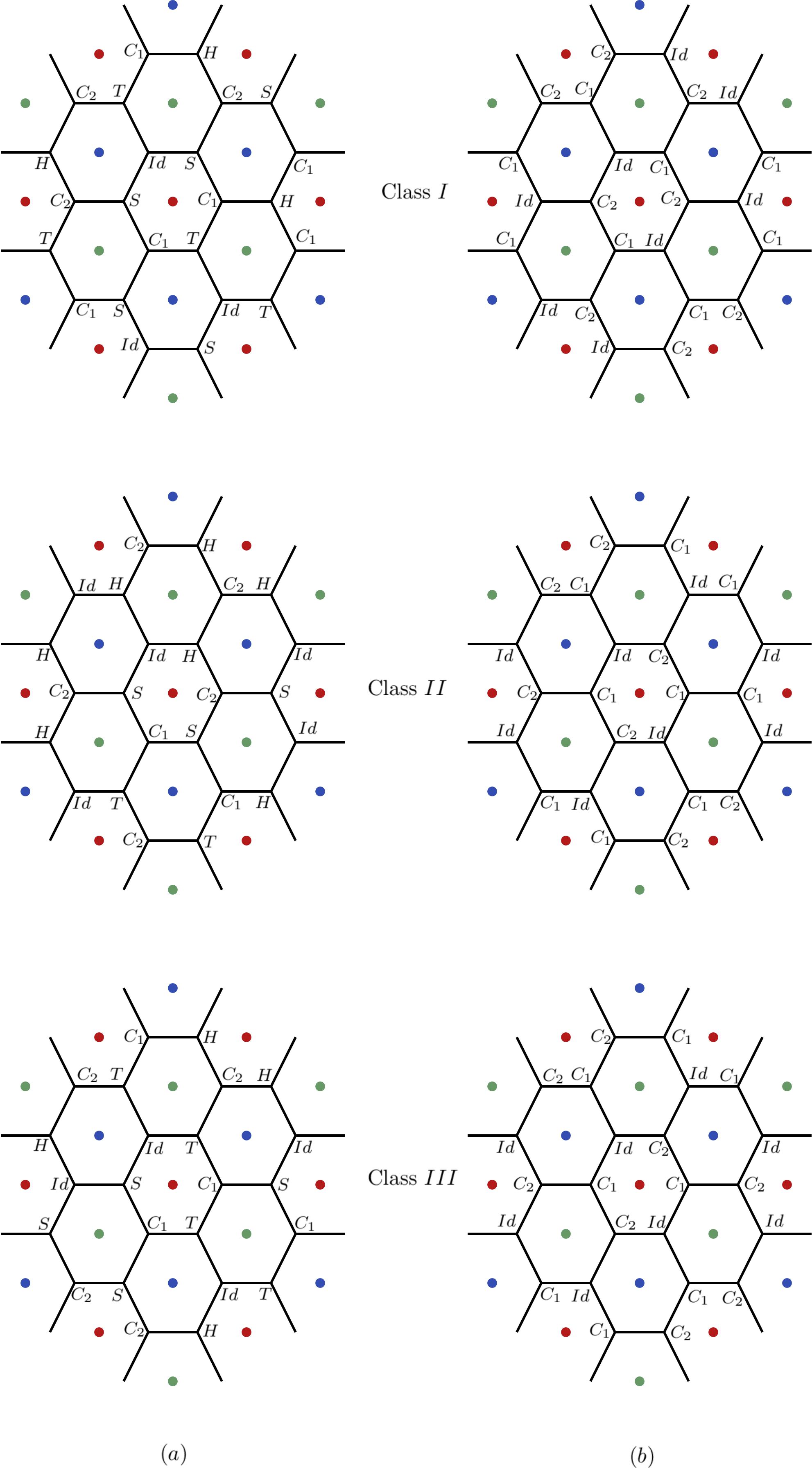}
 \caption{The LU's between the canonical deconfined $[345]$-color code on the hexagon lattice and the (a) partially mixed, (b) fully mixed non-translationally invariant deconfined $[345]$-color codes.}
    \label{fig:[345]ccLUtocanonicalwoti}
\end{figure}

The LU to the canonical deconfined [345]-color codes are shown in Fig. \ref{fig:[345]ccLUtocanonicalwoti}.

\subsection{Non-translationally invariant and confined $[345]$-color codes}
\label{app:NTIccs[345]confined}
\begin{figure}[h]
    \centering
    \includegraphics[width=8cm]{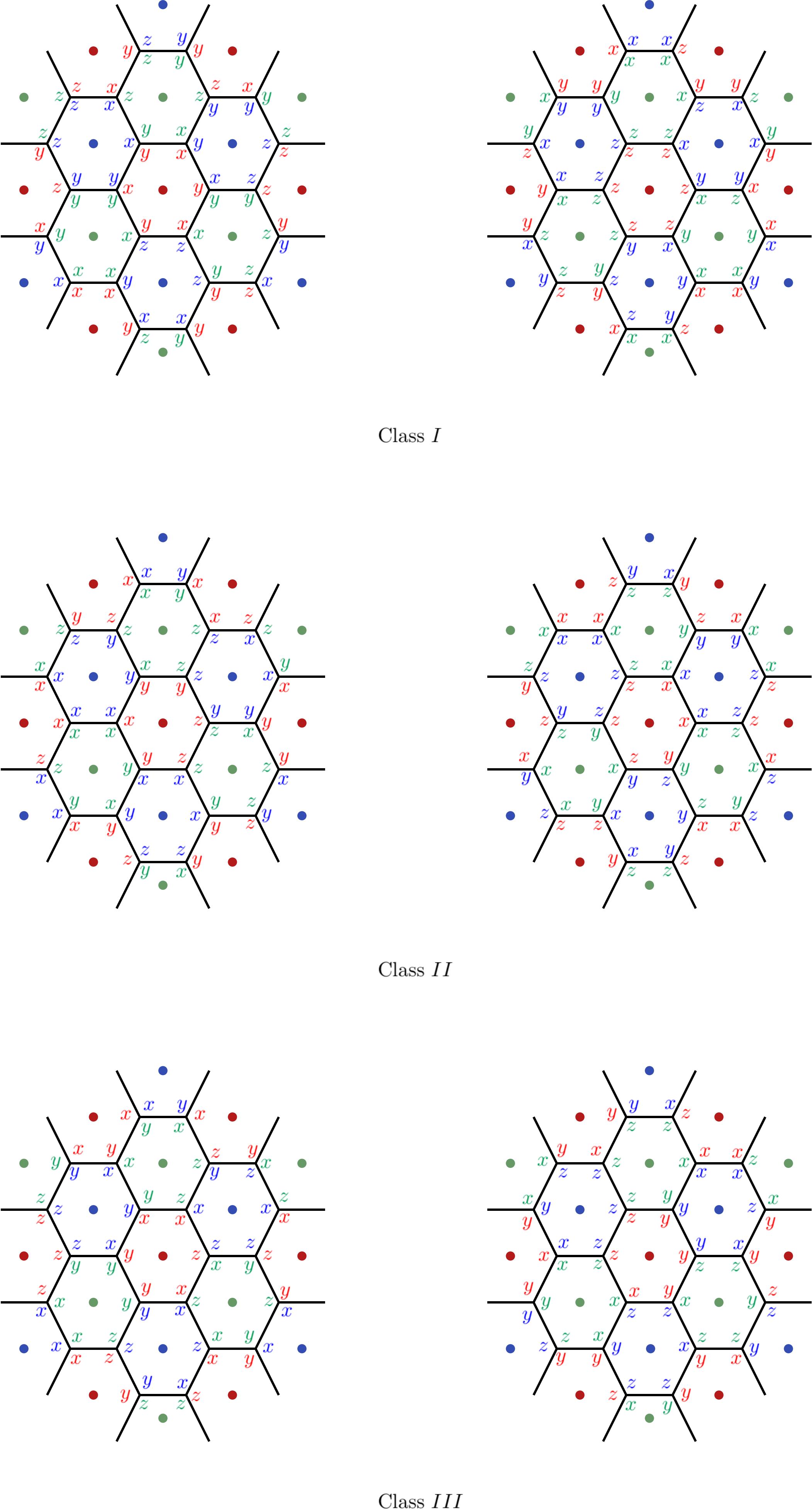}
    \caption{The confined $[345]$-color codes on the hexagonal lattice with partially mixed edge configurations and without translational invariance. This code has vertices belonging to the three inequivalent $[345]$ bulk vertices (See Fig. \ref{fig:[345]bulkvertexconfigs}).}
    \label{fig:[345]ccconfinedpartialmixingwoti}
\end{figure}

\begin{figure}[h]
    \centering
    \includegraphics[width=8cm]{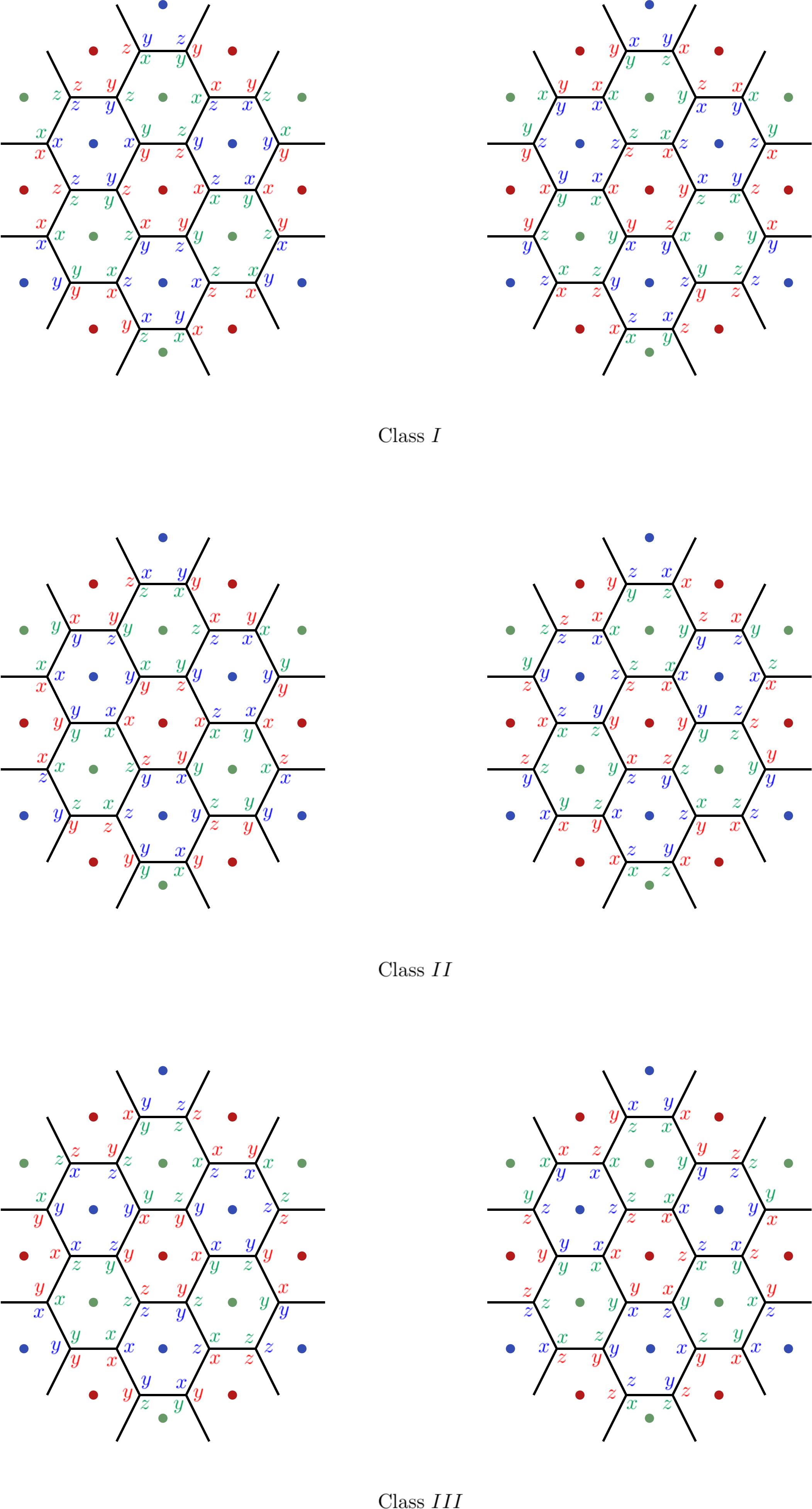}
    \caption{The confined $[345]$-color codes on the hexagonal lattice with fully mixed edge configurations and without translational invariance. This code has vertices belonging to the three inequivalent $[345]$ bulk vertices (See Fig. \ref{fig:[345]bulkvertexconfigs}).}
    \label{fig:[345]ccconfinedfullmixingwoti}
\end{figure}

The consistent models on the hexagonal lattice, with non-translational invariance, are shown in Figs.\,\ref{fig:[345]ccconfinedpartialmixingwoti} and \ref{fig:[345]ccconfinedfullmixingwoti}.

\subsection{Non-translationally invariant $[336]$-color codes}
\label{app:NTIccs[336]}
\begin{figure}[h]
    \centering
    \includegraphics[width=8cm]{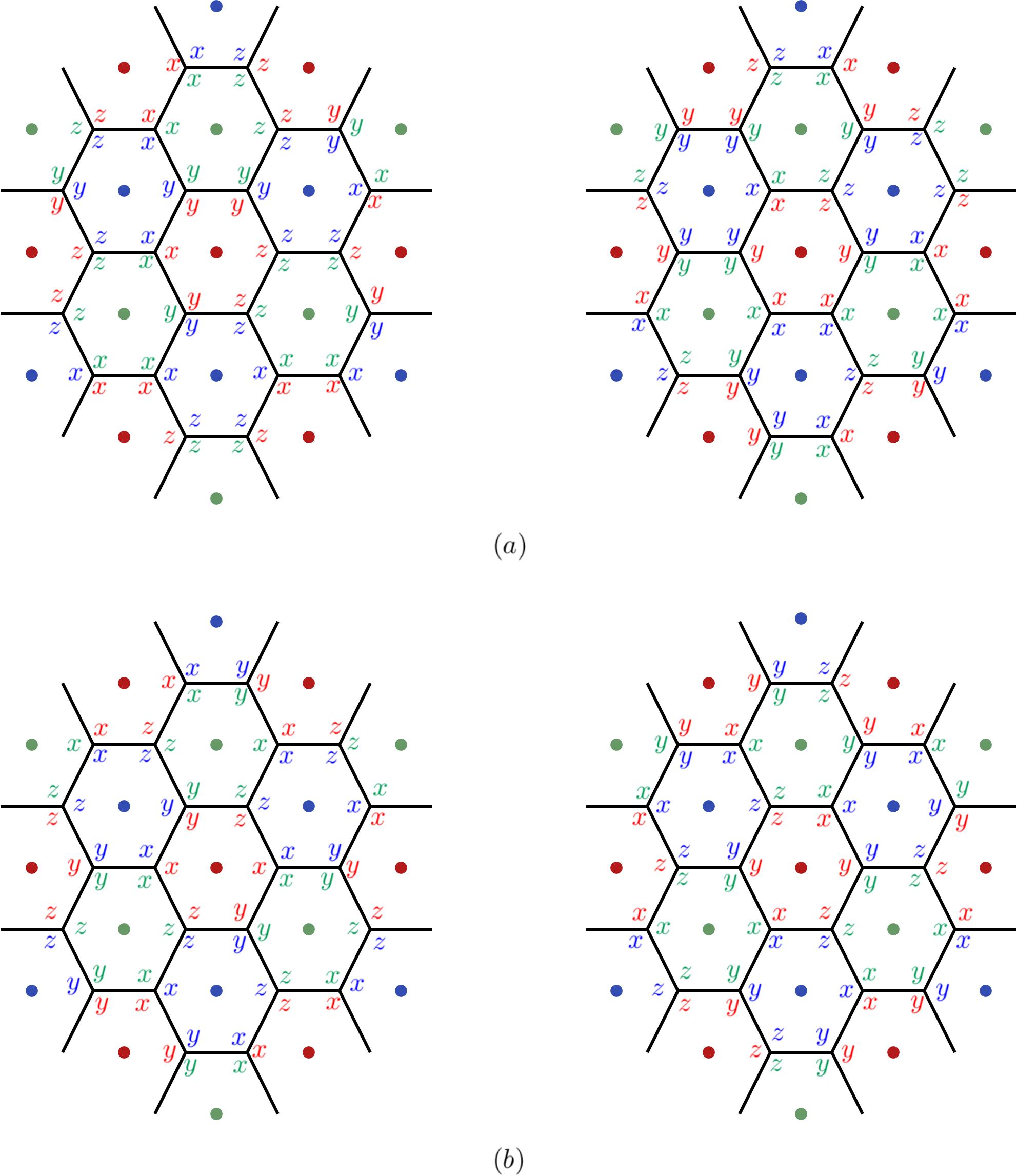}
    \caption{The $[336]$-color codes on the hexagonal lattice with (a) partially and (b) fully mixed edge configurations and without translational invariance. These codes are equivalent to the original $[336]$-color code by an LU.}
    \label{fig:[336]ccpartialfullmixingwoti}
\end{figure}

The non-CSS $[336]$-color codes on the hexagonal lattice for partially and fully mixed edge configurations, without  translational invariance is shown in Fig.\,\ref{fig:[336]ccpartialfullmixingwoti}.

\section{Steps for local code generation for deconfined and confined $[345]$-color codes and the $[336]$-color codes}
\label{app:localconstruction}
\begin{figure*}[h]
    \centering
    \includegraphics[width=15cm]{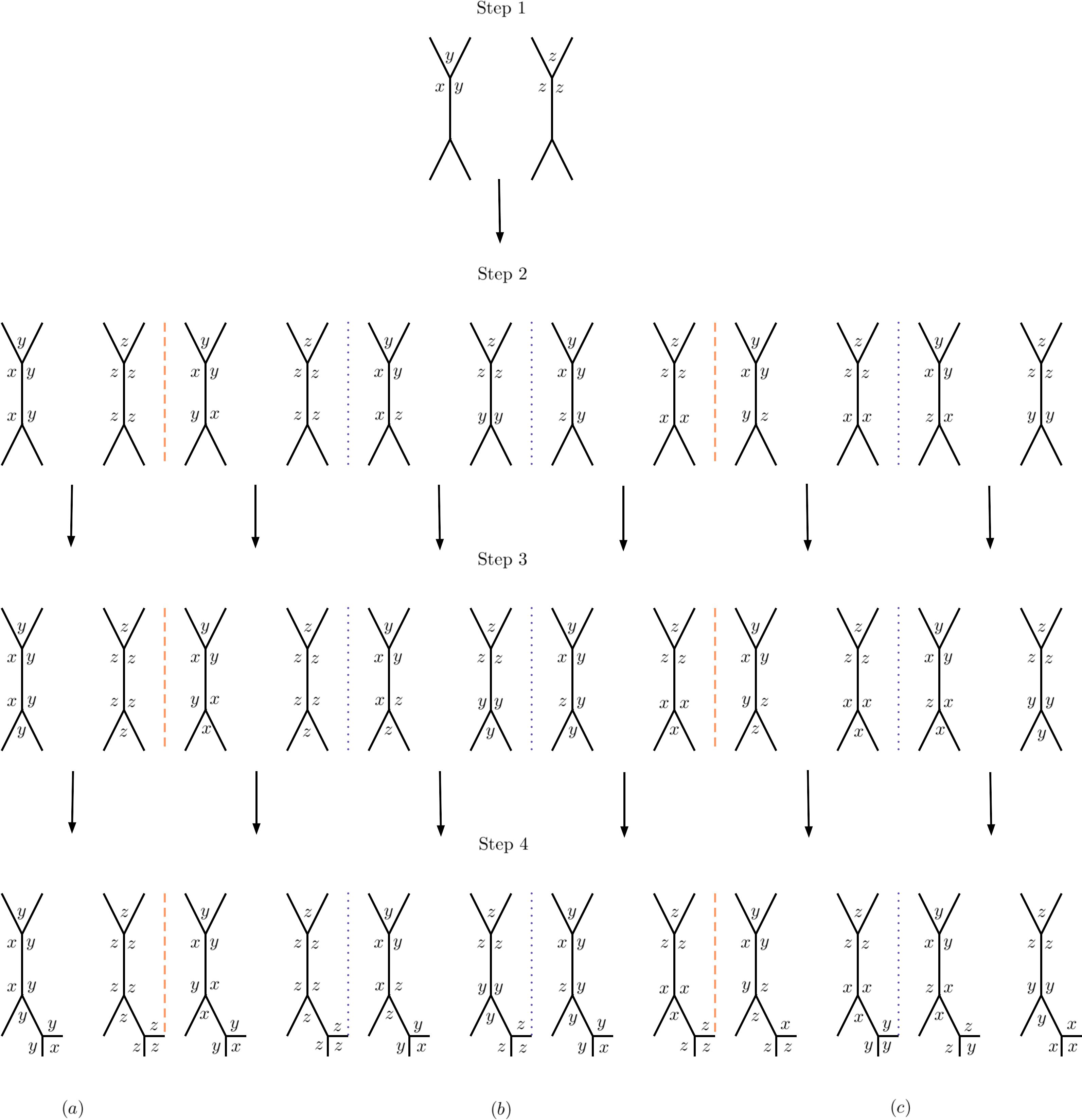}
 \caption{The steps for the local construction of a deconfined $[345]$-color code. The procedure can be continued to extend the code on a finite trivalent lattice just as in the case of the $[444]$-color code. In step 1 we have chosen the first class of the three inequivalent choices for a $[345]$ Y vertex (See Fig. \ref{fig:[345]bulkvertexconfigs}). While choosing the edge configurations in the second step we can go with either a (a) homogeneous (no mixing), (b) partially mixed or (c) fully mixed edge configuration. In the third step we choose the second vertex to be of the $[345]$ type and such that the vertex-edge combine leads to a deconfined stabilizer pair. In step 4 the procedure is continued for another edge where we continue to maintain the deconfined nature of the $[345]$-color code. We show just one among the many possibilities in this step. }
    \label{fig:[345]ccdeconfinedlocalconstruction}
\end{figure*}

The steps for the local construction in the deconfined $[345]$ setting are illustrated in Fig. \ref{fig:[345]ccdeconfinedlocalconstruction}.

\begin{figure*}[h]
    \centering
    \includegraphics[width=15cm]{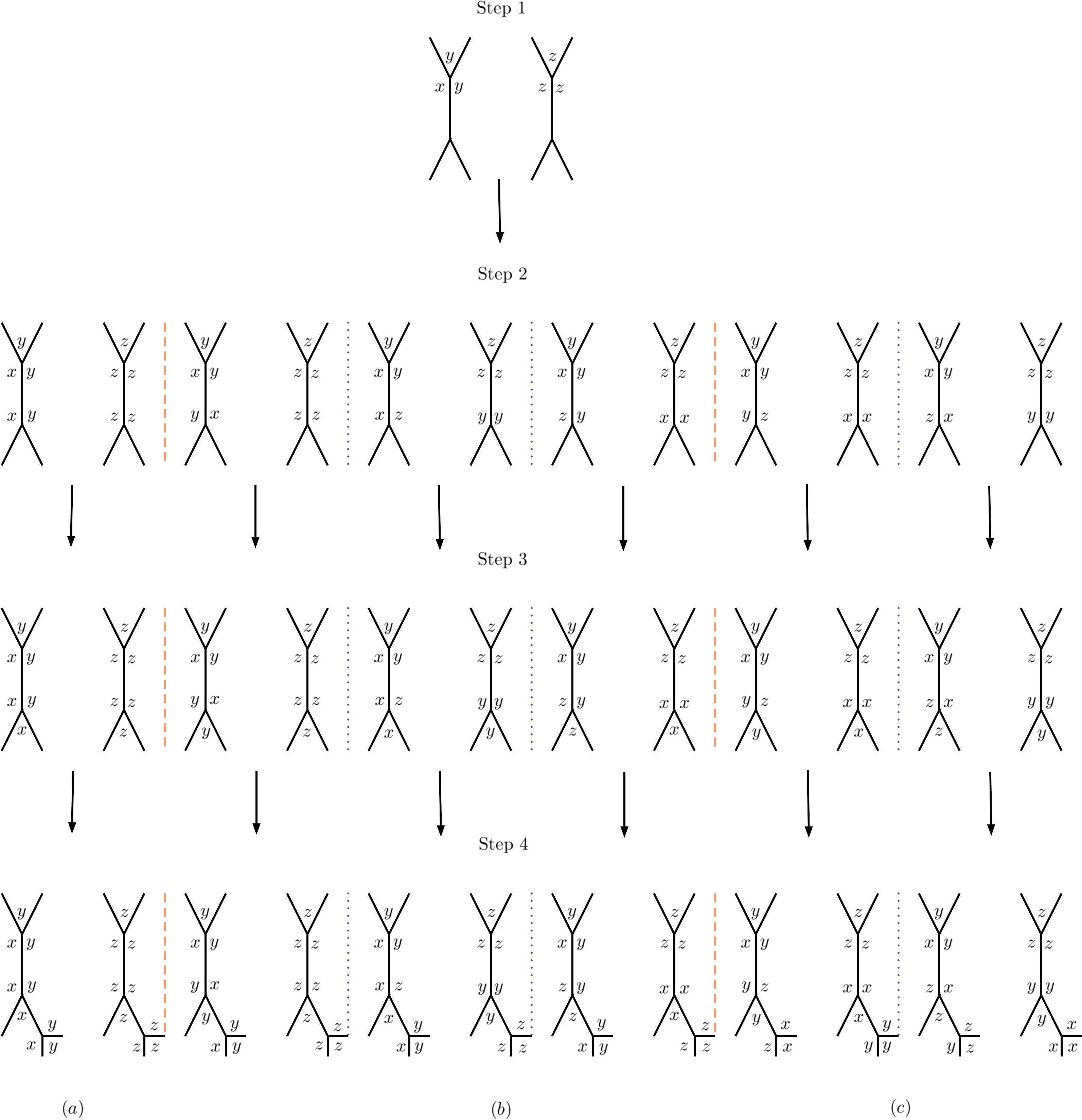}
 \caption{The steps to construct the stabilizer pair of the confined $[345]$-color code. A $[345]$ bulk vertex of type 1 (See Fig. \ref{fig:[345]bulkvertexconfigs}) has been chosen in step 1. The construction proceeds in a similar manner for the other two inequivalent $[345]$ bulk vertices. The possible edge configurations for the stabilizer pair are shown in step 2 with (a) homogeneous (no mixing), (b) partially and (c) fully mixed edge configurations. In step 3 the second vertex is chosen such that the vertex is of the $[345]$ type and the vertex-edge combine is of the confined type. In step 4 the procedure is extended to one more edge. There are more choices possible in step 4 though we just show one of them.}
    \label{fig:[345]ccconfinedlocalconstruction}
\end{figure*}

Fig. \ref{fig:[345]ccconfinedlocalconstruction}
shows the steps for the local construction of the confined $[345]$-color codes.

\begin{figure*}[h]
    \centering
    \includegraphics[width=15cm]{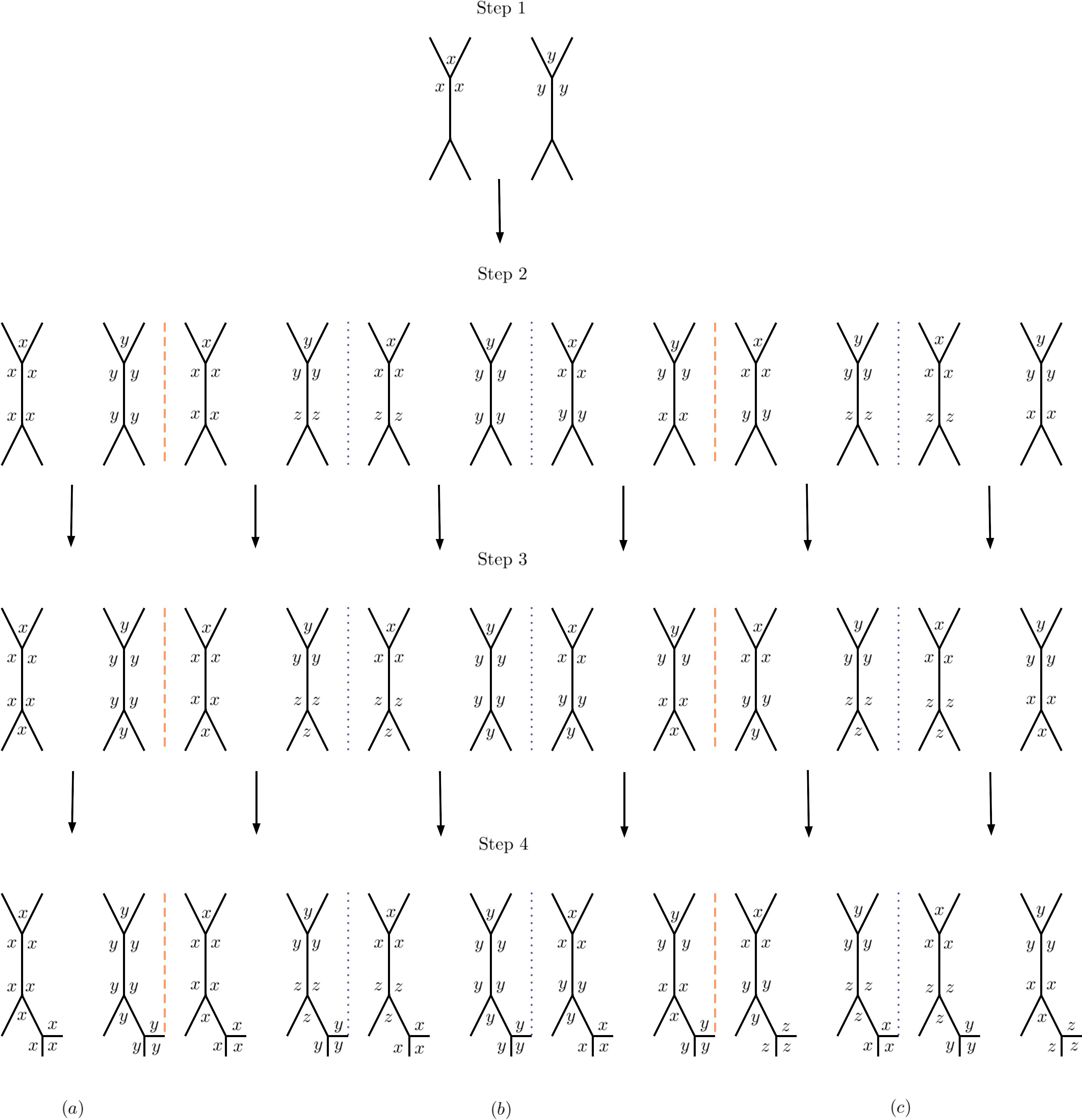}
    \caption{The steps in the construction of the $[336]$-color codes on a local part of the trivalent lattice. The choice of the bulk $[336]$ vertex is equivalent to the original color code by an LU. The edge configurations are shown in Step 2 - (a) homogeneous, (b) partially and (c) fully mixed edge configurations. The choice for the bulk vertex in Step 3 is determined by the choice of the edge configurations in Step 2. The construction is continued for another edge in Step 4. The resulting codes are equivalent to the original $[336]$-color code by an LU.}
    \label{fig:[336]cclocalconstruction}
\end{figure*}

The local construction for the $[336]$-color code is shown in Fig. \ref{fig:[336]cclocalconstruction}.

\section{The other inequivalent classes of the deconfined and confined $[345]$-color codes with translational invariance}
\label{app:[345]classIIclassIII}
There are three inequivalent classes for the bulk vertices of the $[345]$ type (See Fig. \ref{fig:[345]bulkvertexconfigs}). The class $I$ models for both the deconfined and confined varieties with all possible edge configurations were shown in the main text. Here we complete the classification by writing down the class $II$ and $III$ $[345]$-color codes in all its variations. 

\begin{figure}[h]
    \centering
    \includegraphics[width=8cm]{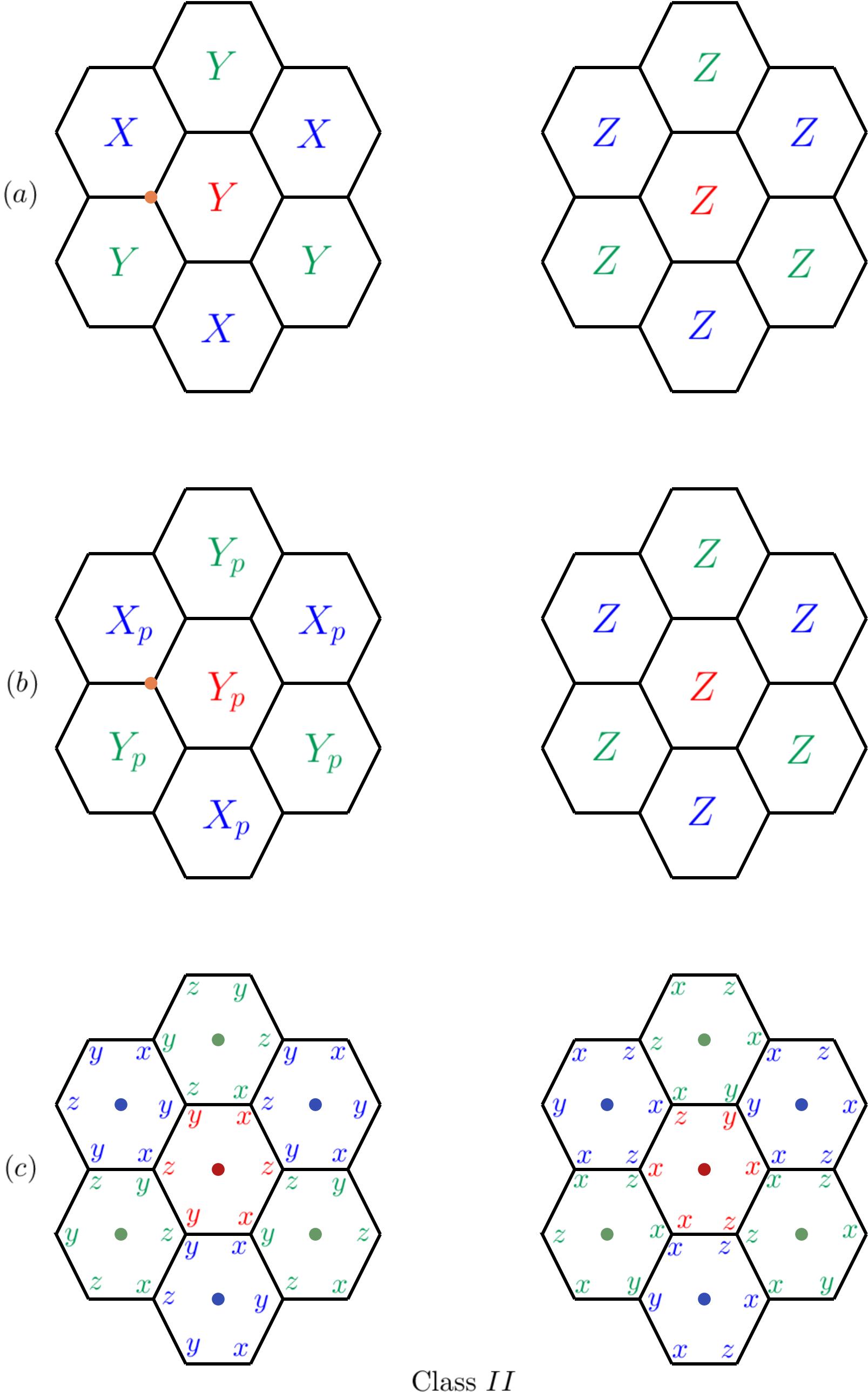}
 \caption{Translationally invariant, deconfined $[345]$-color codes with (a) homogeneous, (b) partially and (c) fully mixed edge configurations. }
    \label{fig:[345]ccdeconfinedClassII}
\end{figure}

\begin{figure}[h]
    \centering
    \includegraphics[width=8cm]{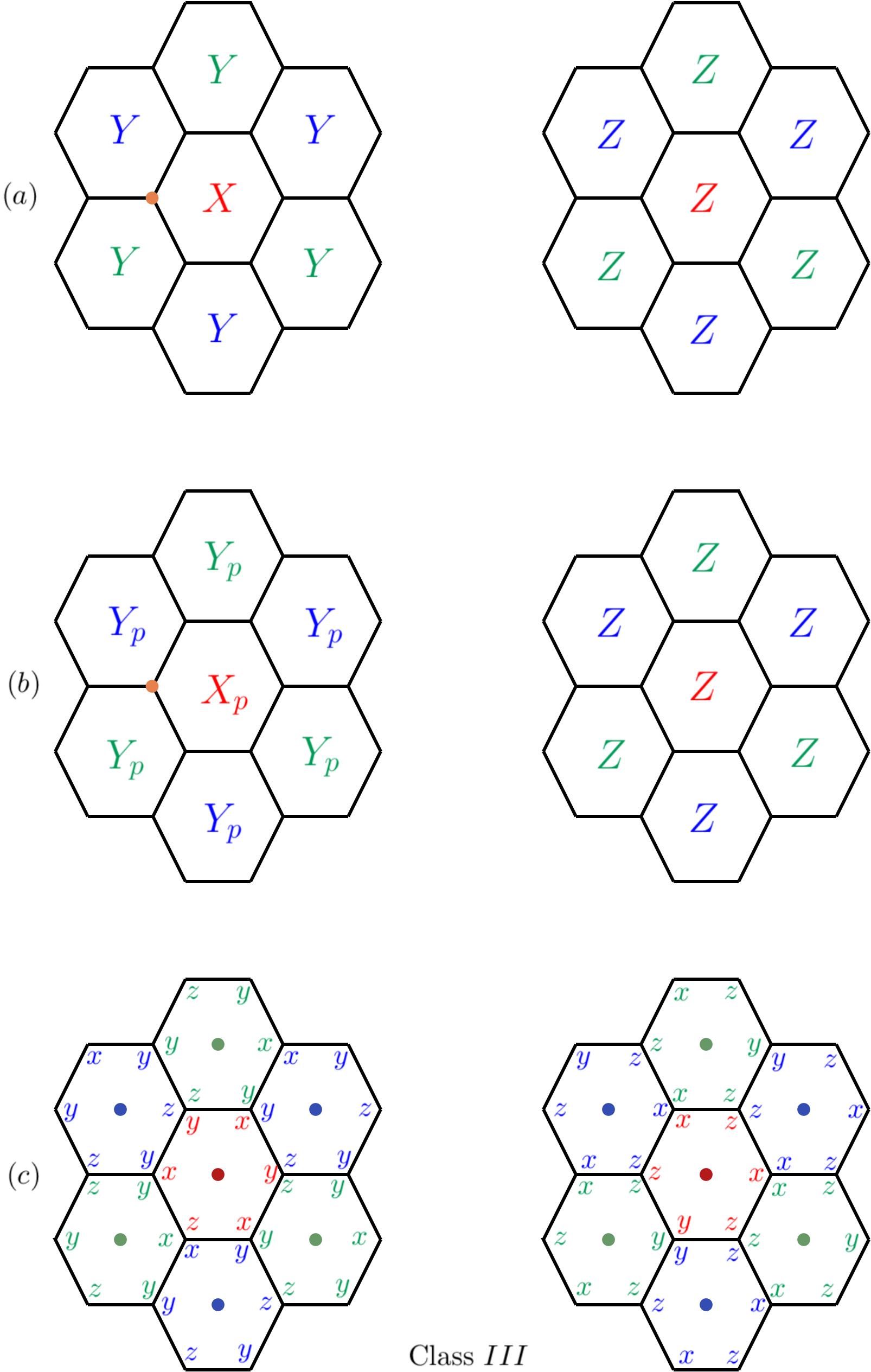}
 \caption{Translationally invariant, deconfined $[345]$-color codes with (a) homogeneous, (b) partially and (c) fully mixed edge configurations. }
    \label{fig:[345]ccdeconfinedClassIII}
\end{figure}

The translationally invariant $[345]$-color codes with partially and fully mixed edge configurations corresponding to classes II and III are shown in Figs. \ref{fig:[345]ccdeconfinedClassII} and \ref{fig:[345]ccdeconfinedClassIII}, for the hexagonal lattice.

\begin{figure}[h]
    \centering
    \includegraphics[width=8cm]{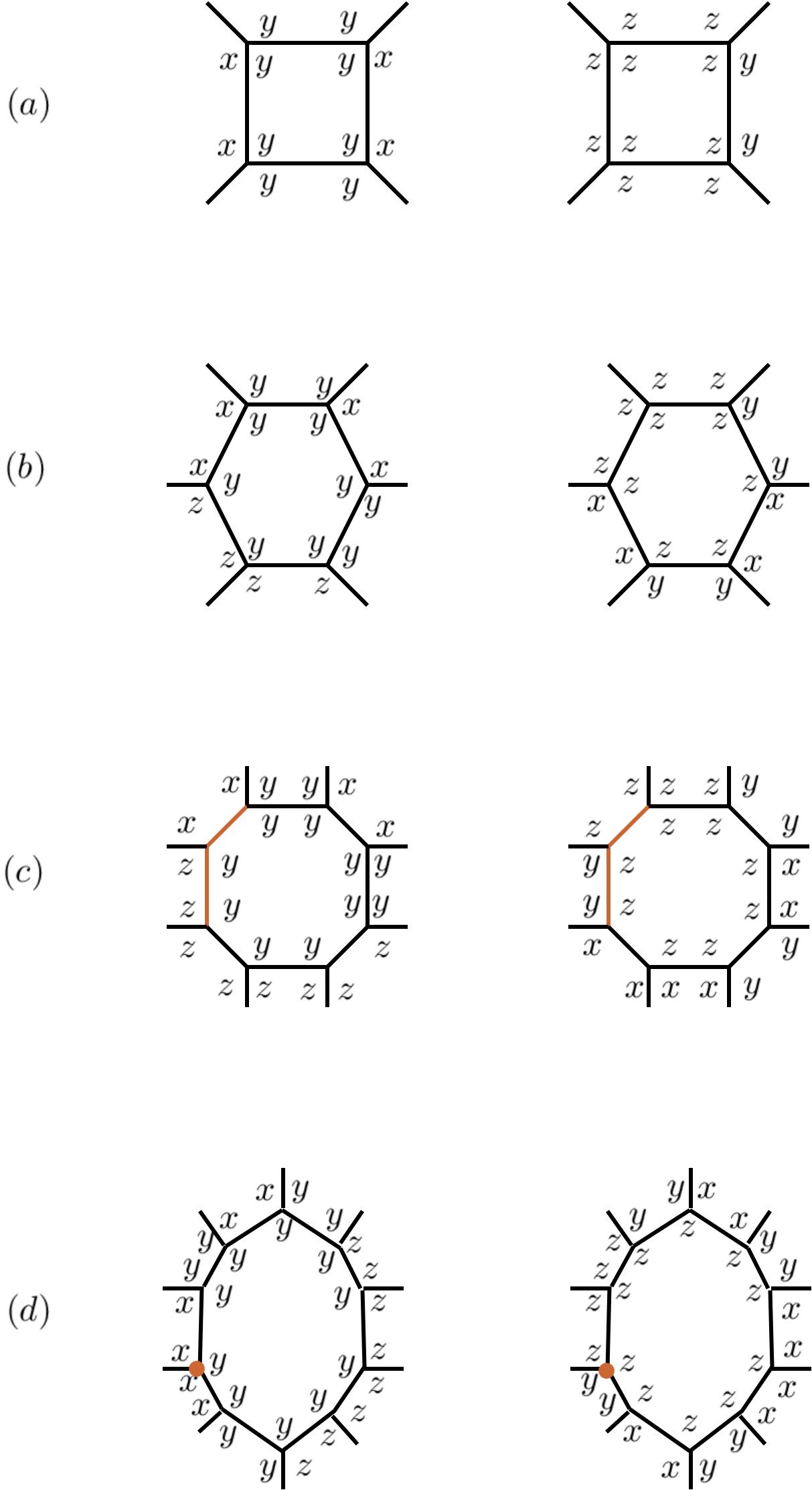}
 \caption{Polygons where the confined $[345]$-color code can be consistently defined are shown in, (a) a square and (b) a hexagon. On (c) the octagon the edges indicated in orange form deconfined vertex-edge combines and in (d) the decagon, the orange dot indicates a $[444]$ bulk vertex. }
    \label{fig:[345]ccconfinedpossiblepolygons}
\end{figure}

Fig. \ref{fig:[345]ccconfinedpossiblepolygons} explores the polygons where the confined $[345]$-color codes can be defined consistently.

\begin{figure}[h]
    \centering
    \includegraphics[width=8cm]{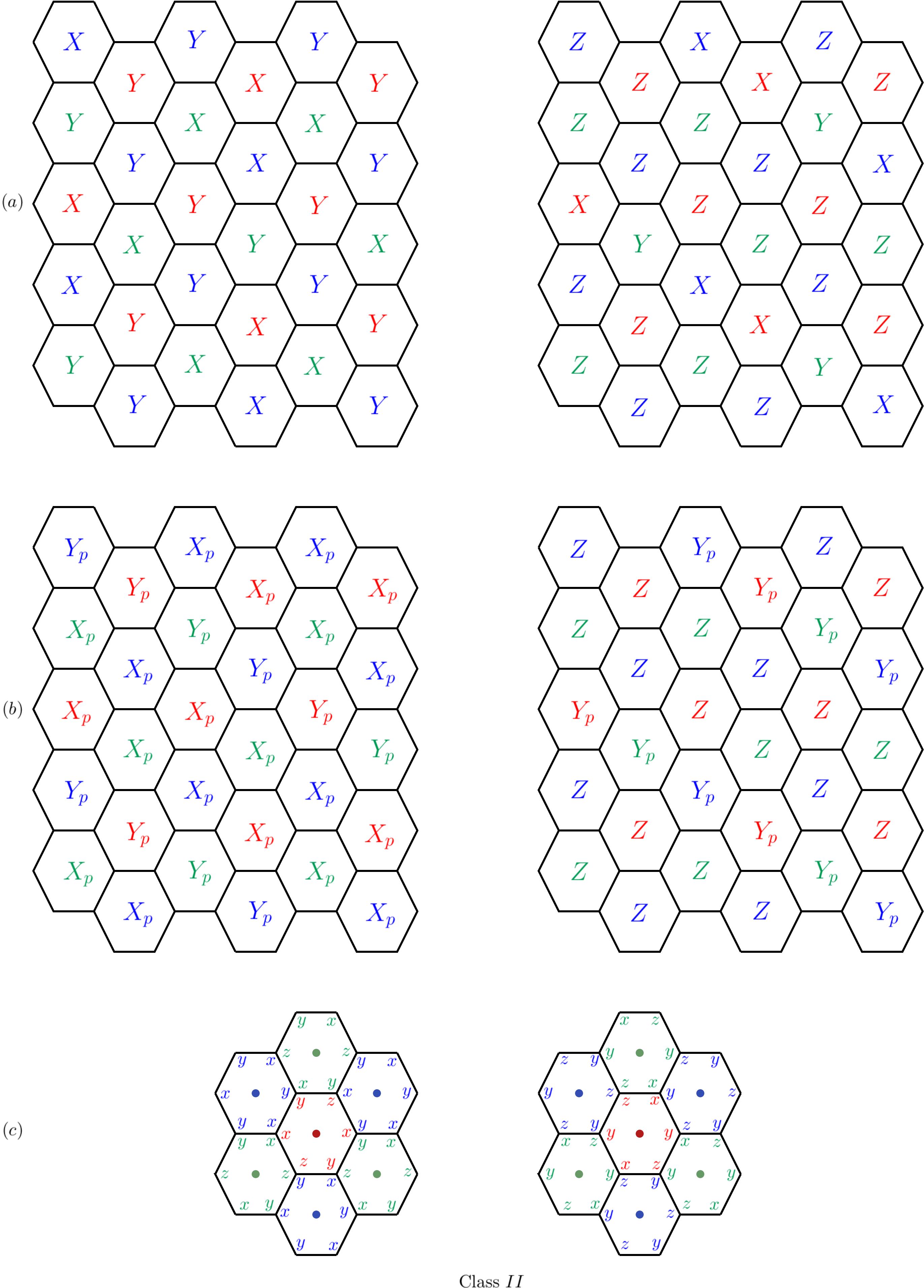}
 \caption{The Class II confined $[345]$-color codes on the hexagonal lattice with (a) homogeneous, (b) partially and (c) fully  mixed edge configurations. }
    \label{fig:[345]ccconfinedClassII}
\end{figure}

\begin{figure}[h]
    \centering
    \includegraphics[width=8cm]{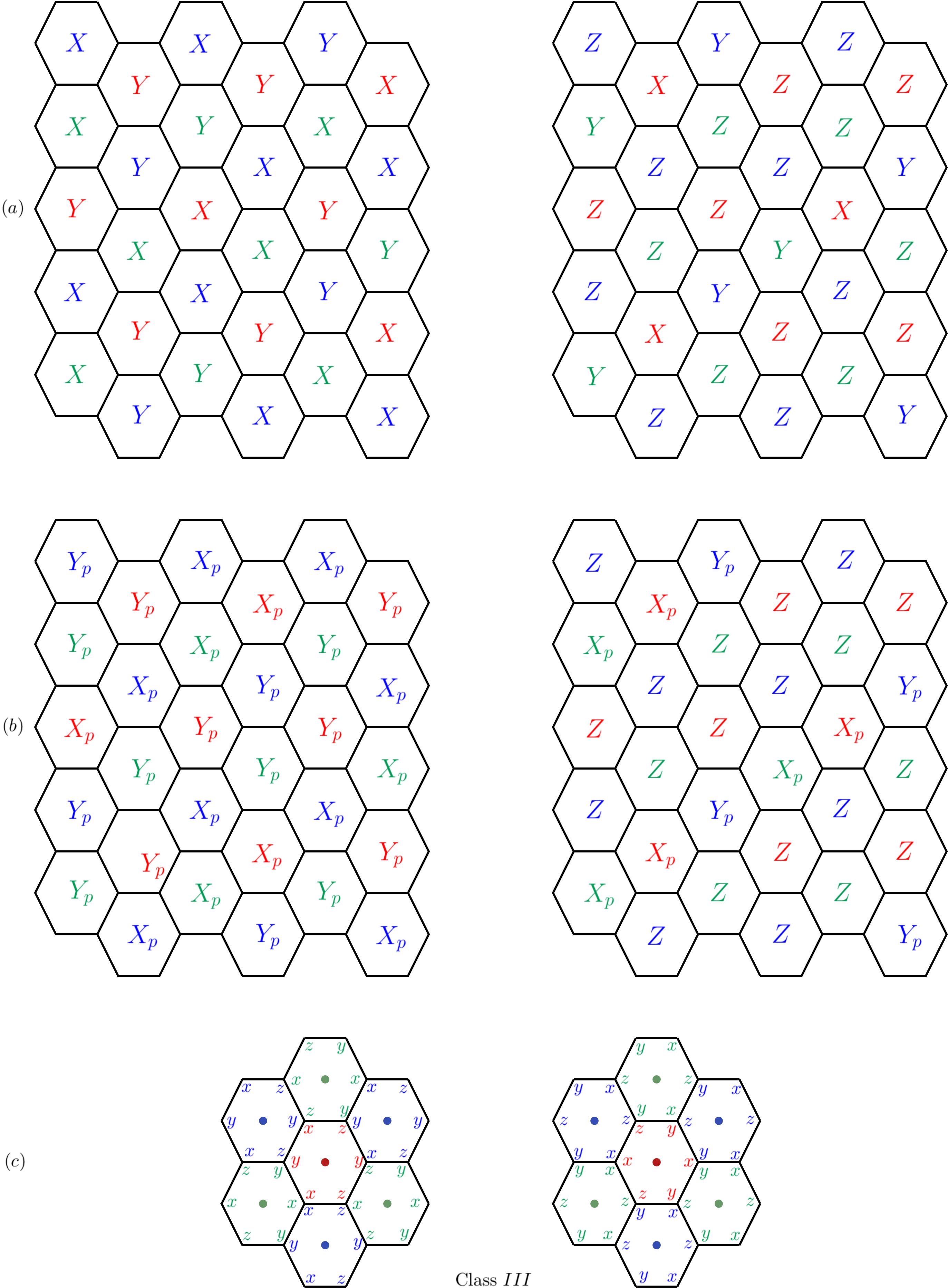}
 \caption{The Class III confined $[345]$-color codes on the hexagonal lattice with (a) homogeneous, (b) partially and (c) fully  mixed edge configurations. }
    \label{fig:[345]ccconfinedClassIII}
\end{figure}

The class II and III confined $[345]$-color codes are shown in Figs. \ref{fig:[345]ccconfinedClassII} and \ref{fig:[345]ccconfinedClassIII}

\section{Non-CSS color codes on square-octagon lattices on a torus}
\label{app:488latticemodels}

\subsection{$[444]$-color codes on the $(4,8,8)$ lattice}
\label{app:(488)[444]}
\begin{figure}[h]
    \centering
    \includegraphics[width=8cm]{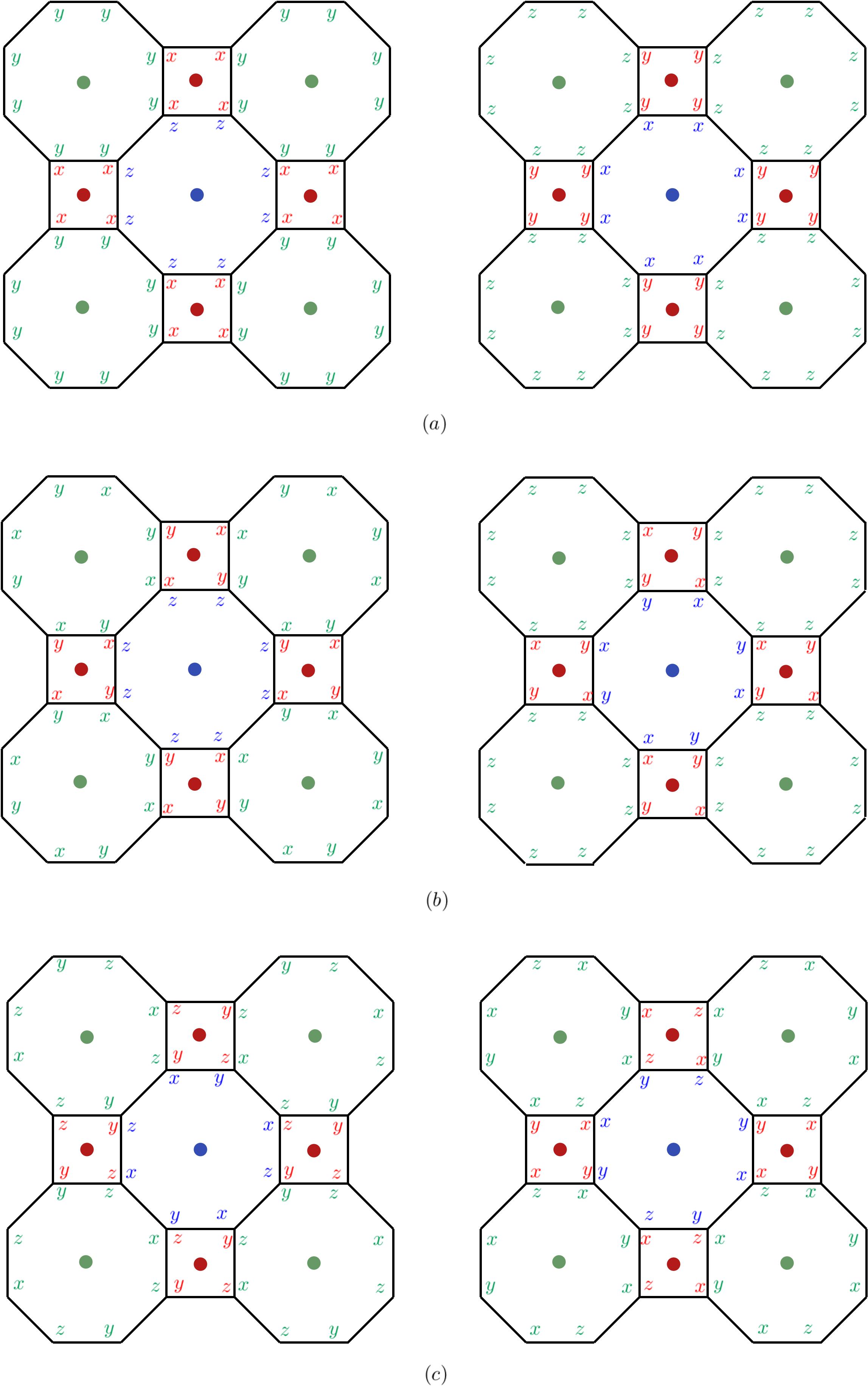}
 \caption{Translationally invariant $[444]$-color codes with (a) homogeneous, (b) partially and (c) fully mixed edge configurations on the square-octagon lattice. }
    \label{fig:(488)[444]ccnopartialfullmixingwti}
\end{figure}

\begin{figure}[h]
    \centering
    \includegraphics[width=9cm]{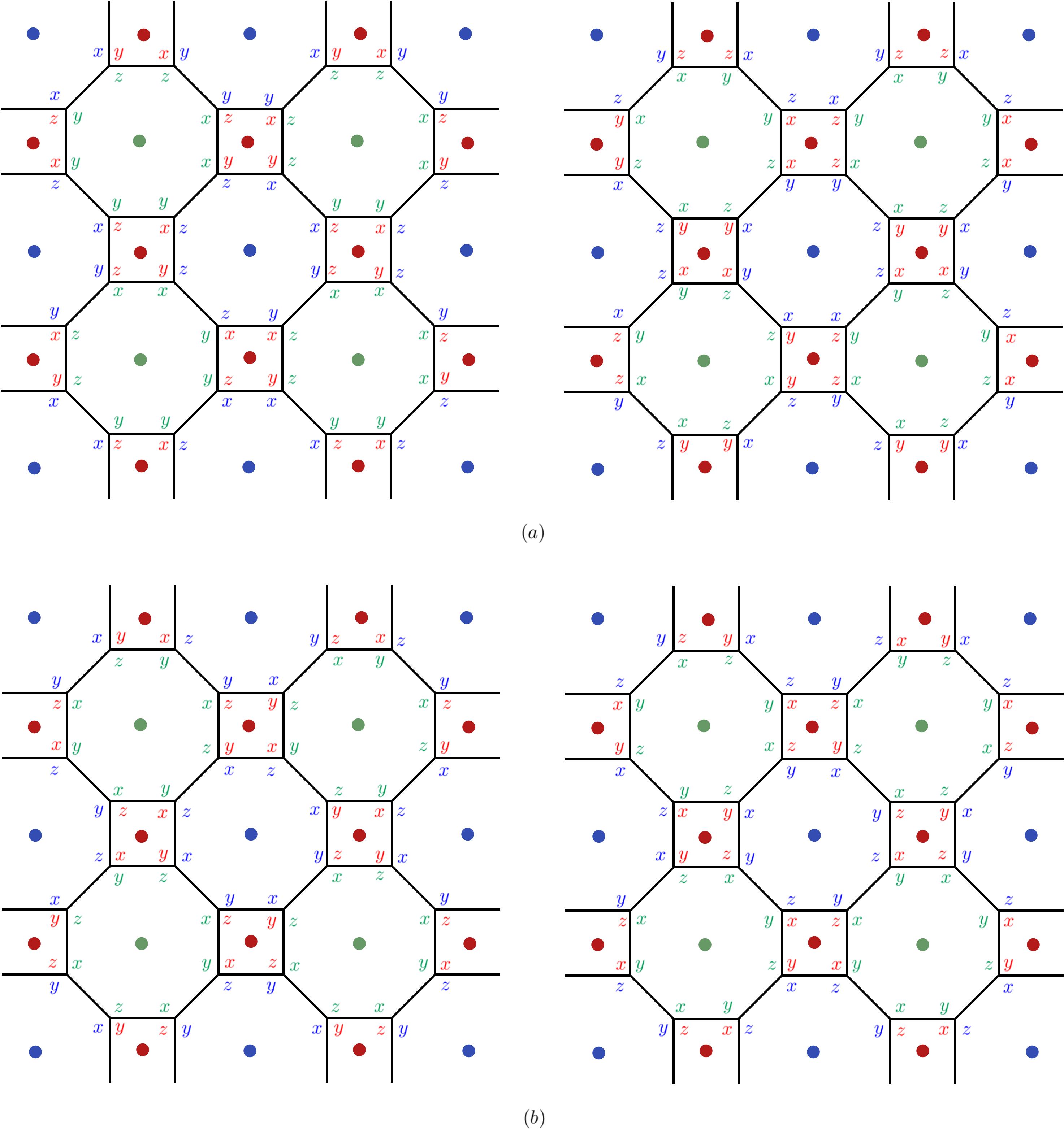}
 \caption{Non-translationally invariant $[444]$-color codes with (a) partially mixed and (b) fully mixed edge configurations.}
    \label{fig:(488)[444]ccpartialfullmixingwoti}
\end{figure}

The translationally invariant models are shown in Fig. \ref{fig:(488)[444]ccnopartialfullmixingwti}, that are LU related to each other.
We can also write down a model without translational invariance on the square-octagon lattice where the edges have partially (fully) mixed configurations, in Fig. \ref{fig:(488)[444]ccpartialfullmixingwoti}. 


\subsection{Deconfined $[345]$-color codes}
\label{app:(488)[345]}

\begin{figure}[h]
    \centering
    \includegraphics[width=8cm]{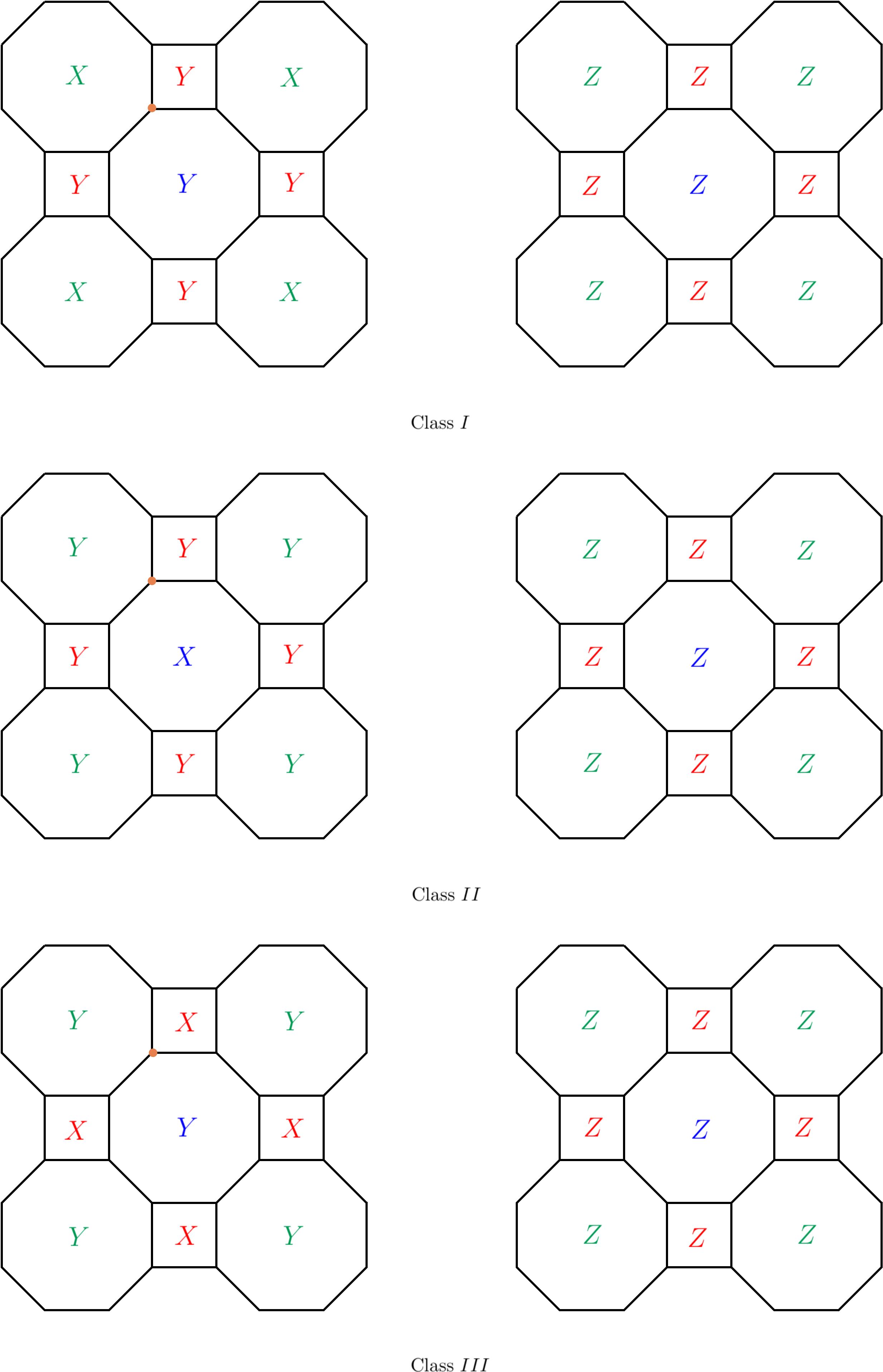}
 \caption{Translationally invariant, deconfined $[345]$-color codes on the square-octagon lattice with homogeneous edge configurations. The three inequivalent classes can be distinguished by the bulk vertex configurations at the orange dot.}
    \label{fig:(488)[345]ccdeconfinednomixing}
\end{figure}

\begin{figure}[h]
    \centering
    \includegraphics[width=8cm]{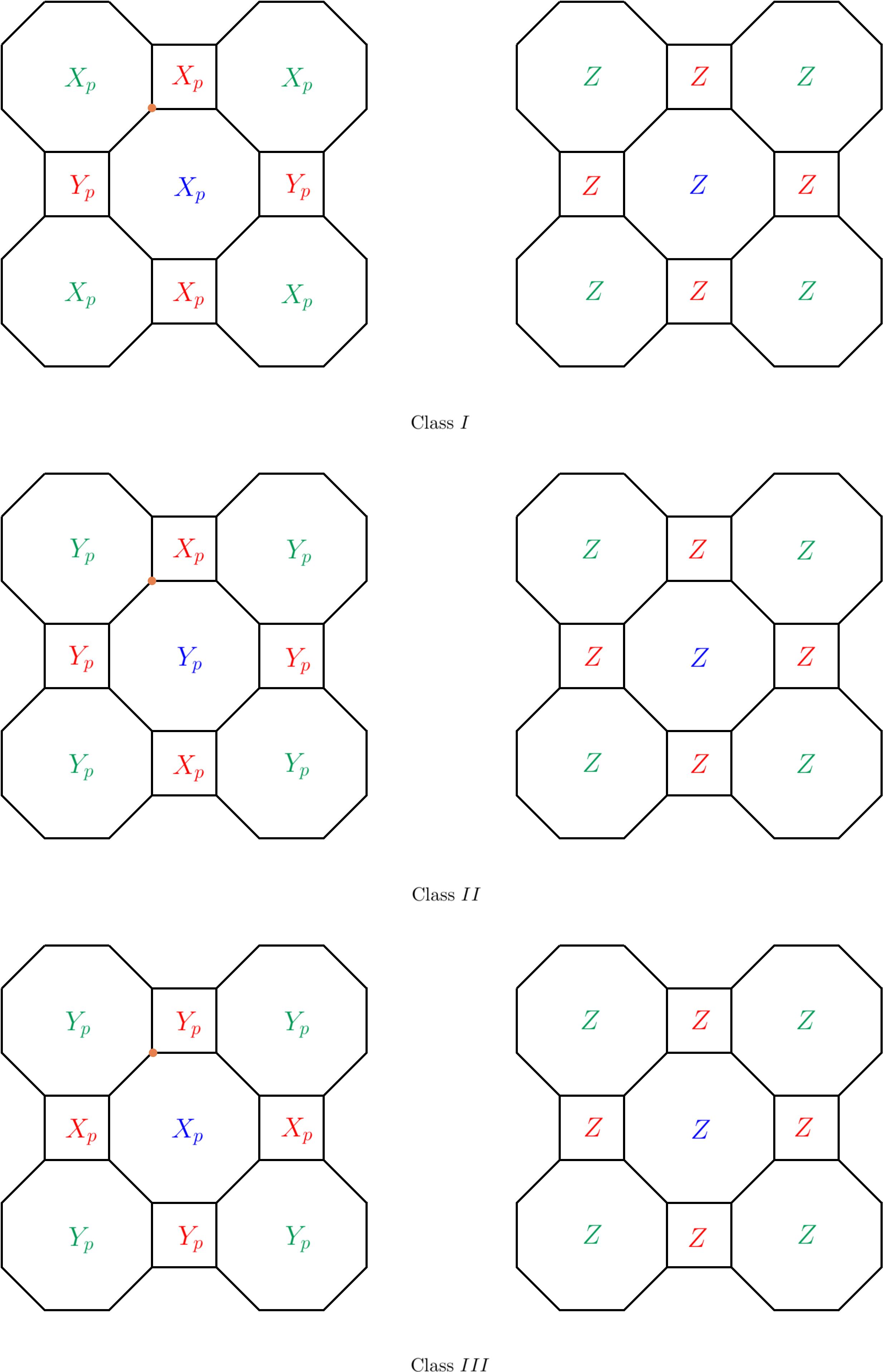}
 \caption{Translationally invariant, deconfined $[345]$-color codes on the square-octagon lattice with partially mixed edge configurations. The three inequivalent classes can be distinguished by the bulk vertex configurations at the orange dot. Notice that the red stabilizers in this code are translationally invariant with the period 2.}
    \label{fig:(488)[345]ccdeconfinedpartialmixingwti}
\end{figure}

\begin{figure}[h]
    \centering
    \includegraphics[width=8cm]{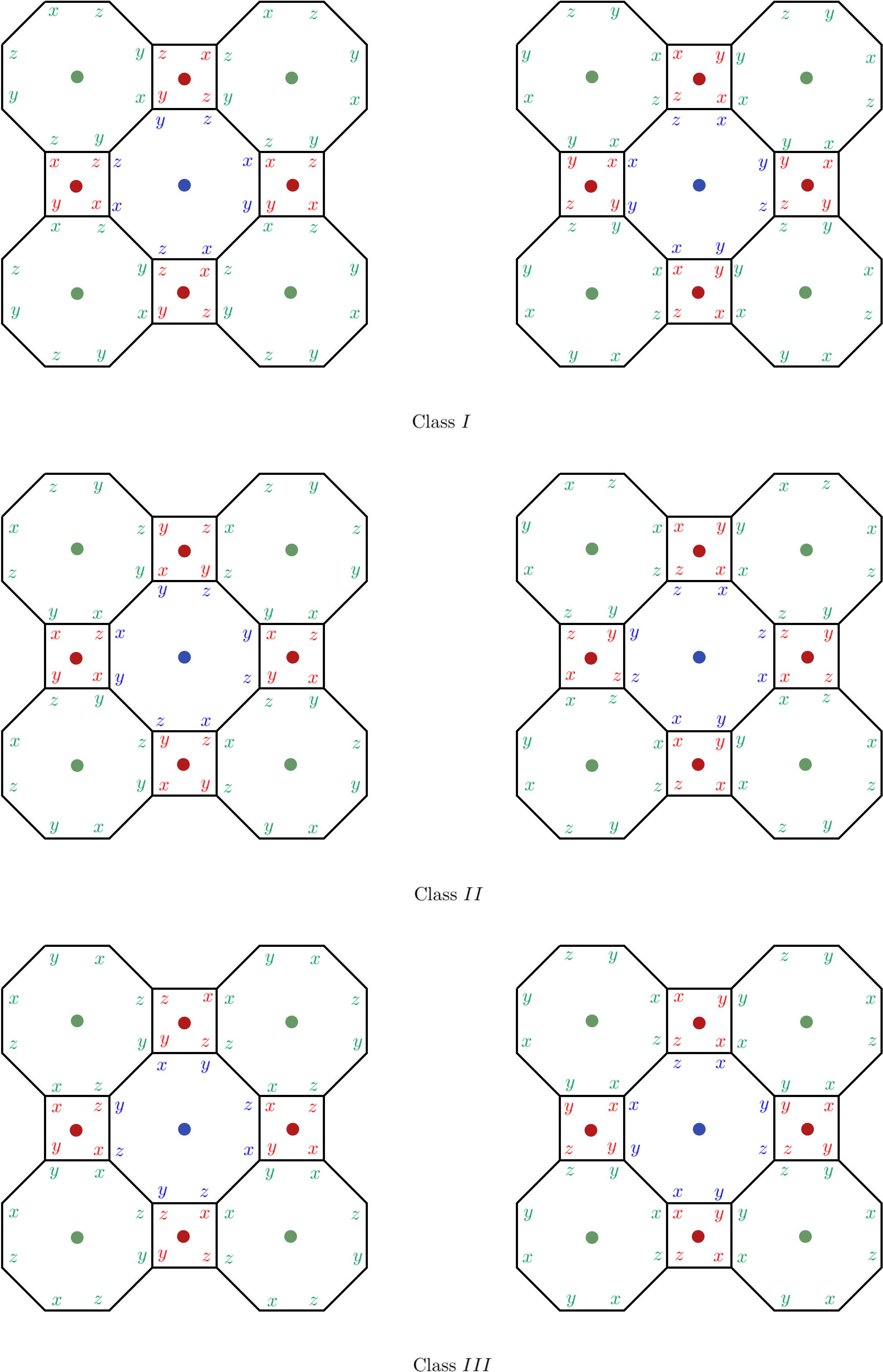}
 \caption{Translationally invariant, deconfined $[345]$-color codes on the square-octagon lattice with fully mixed edge configurations. The three inequivalent classes are distinguished by the bulk vertex configurations at the orange dot. Notice that the red stabilizers in this code are translationally invariant with the period 2.}
    \label{fig:(488)[345]ccdeconfinedfullmixingwti}
\end{figure}

\begin{figure}[h]
    \centering
    \includegraphics[width=9cm]{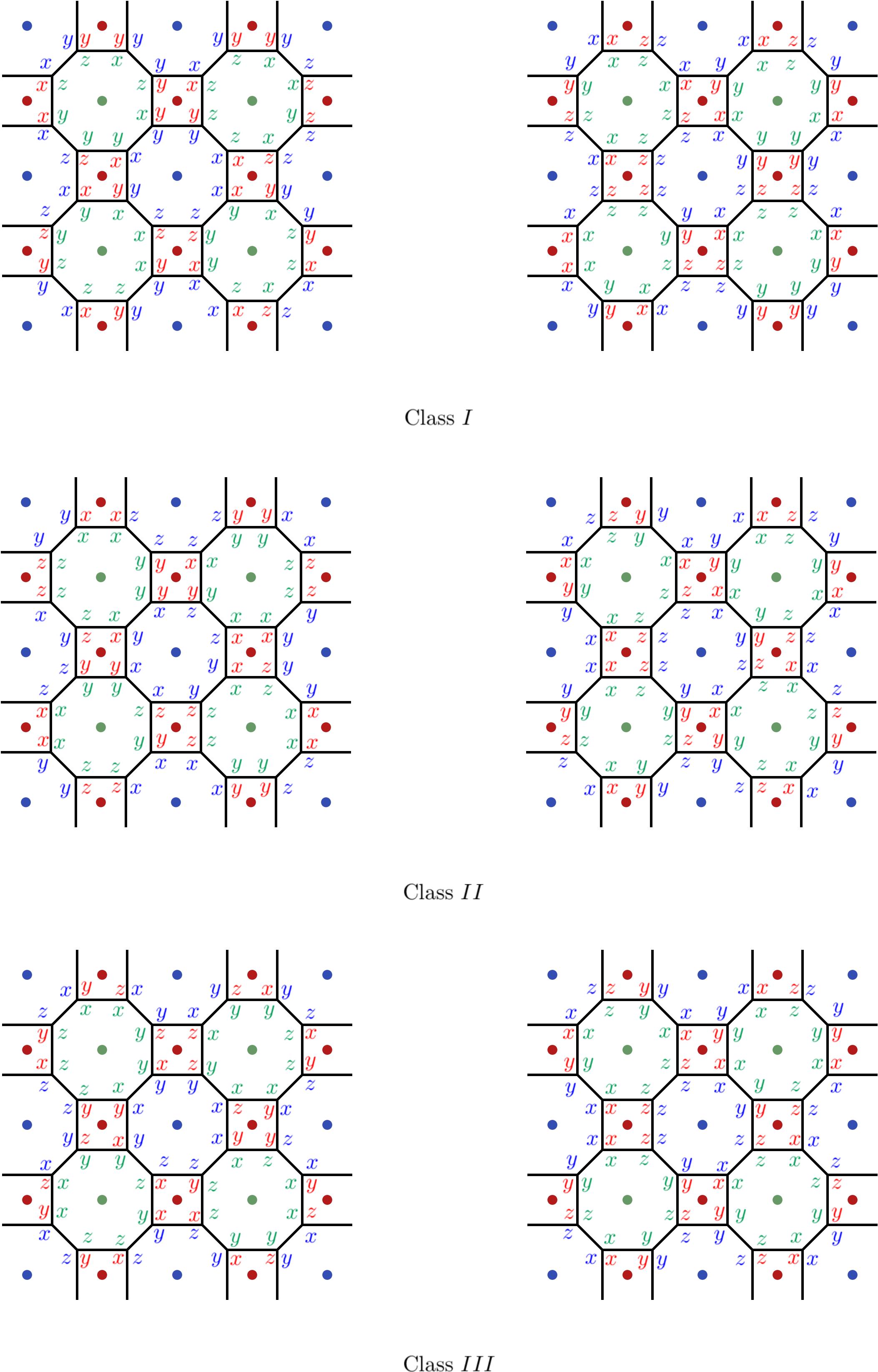}
 \caption{Non-translationally invariant, deconfined $[345]$-color codes on the square-octagon lattice with partially mixed edge configurations.}
    \label{fig:(488)[345]ccdeconfinedpartialmixingwoti}
\end{figure}

\begin{figure}[h]
    \centering
    \includegraphics[width=9cm]{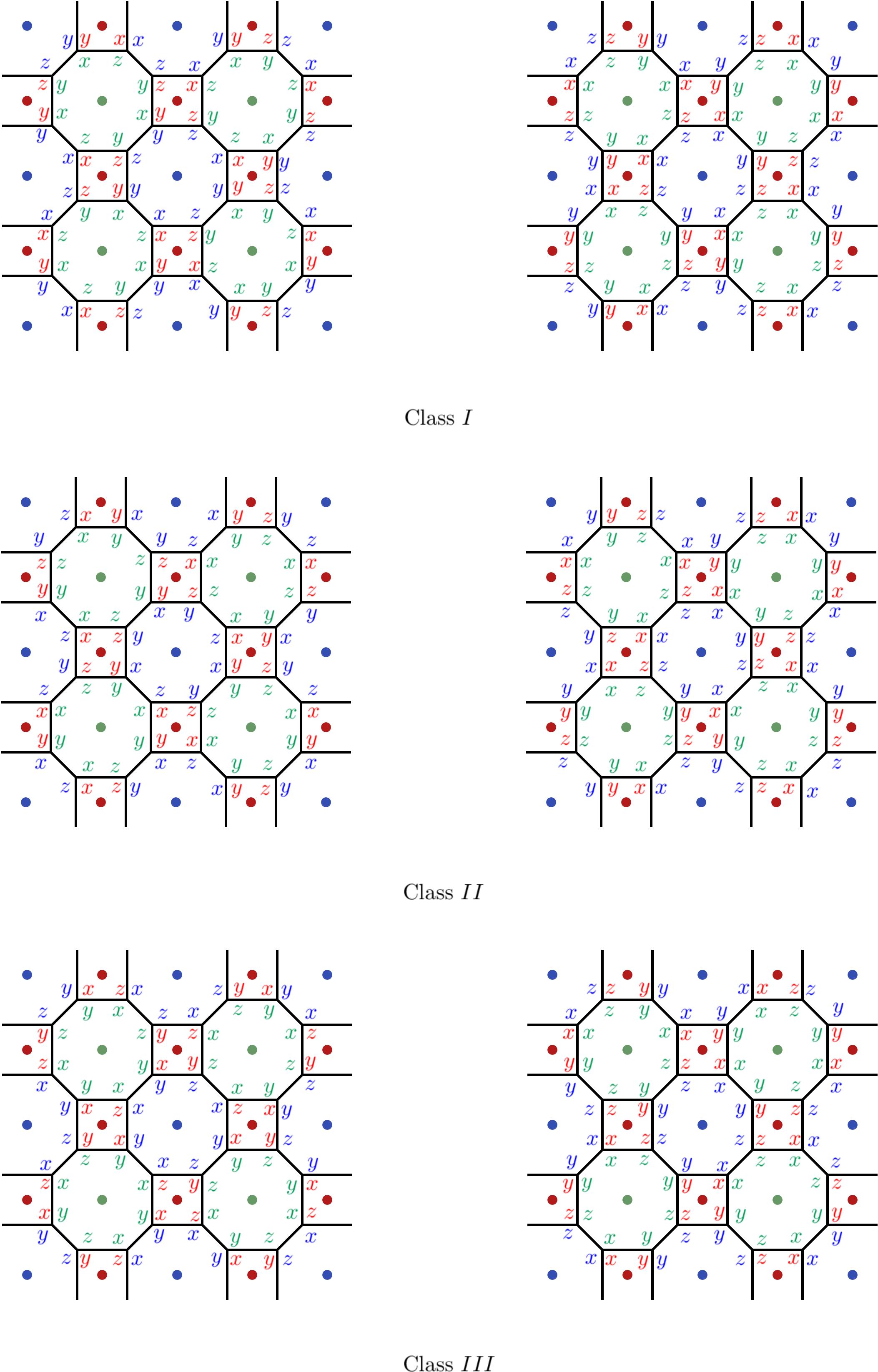}
 \caption{Non-translationally invariant, deconfined $[345]$-color codes on the square-octagon lattice with fully mixed edge configurations.}
    \label{fig:(488)[345]ccdeconfinedfullmixingwoti}
\end{figure}

 The deconfined $[345]$-color codes with and without translational invariance on the $(4,8,8)$-lattice are shown in Figs. \ref{fig:(488)[345]ccdeconfinednomixing}, \ref{fig:(488)[345]ccdeconfinedpartialmixingwti}, \ref{fig:(488)[345]ccdeconfinedfullmixingwti}, \ref{fig:(488)[345]ccdeconfinedpartialmixingwoti} and \ref{fig:(488)[345]ccdeconfinedfullmixingwoti}.

\section{Non-CSS color codes with mixed bulk vertex types}
\label{app:mixedvertextypes}
In this appendix we present further possibilities that open up with the algorithm for constructing two-dimensional color codes introduced in this paper. We focus our attention on color codes which have more than one type of vertex, namely a mixture of $[345]$ and $[444]$, a mixture of $[345]$ and $[336]$ or a mixture of all three types of vertices. It is clear that by construction these models are not LU equivalent to each other or the color codes in the main text which sport just a single vertex type. There are plenty of possibilities as a result, and we show just a few of them relegating a more thorough analysis to future papers. 

\subsection{$[345]-[444]$-color codes}
\label{app:[345][444]mixedvertextypes}
\begin{figure}[h]
    \centering
    \includegraphics[width=8cm]{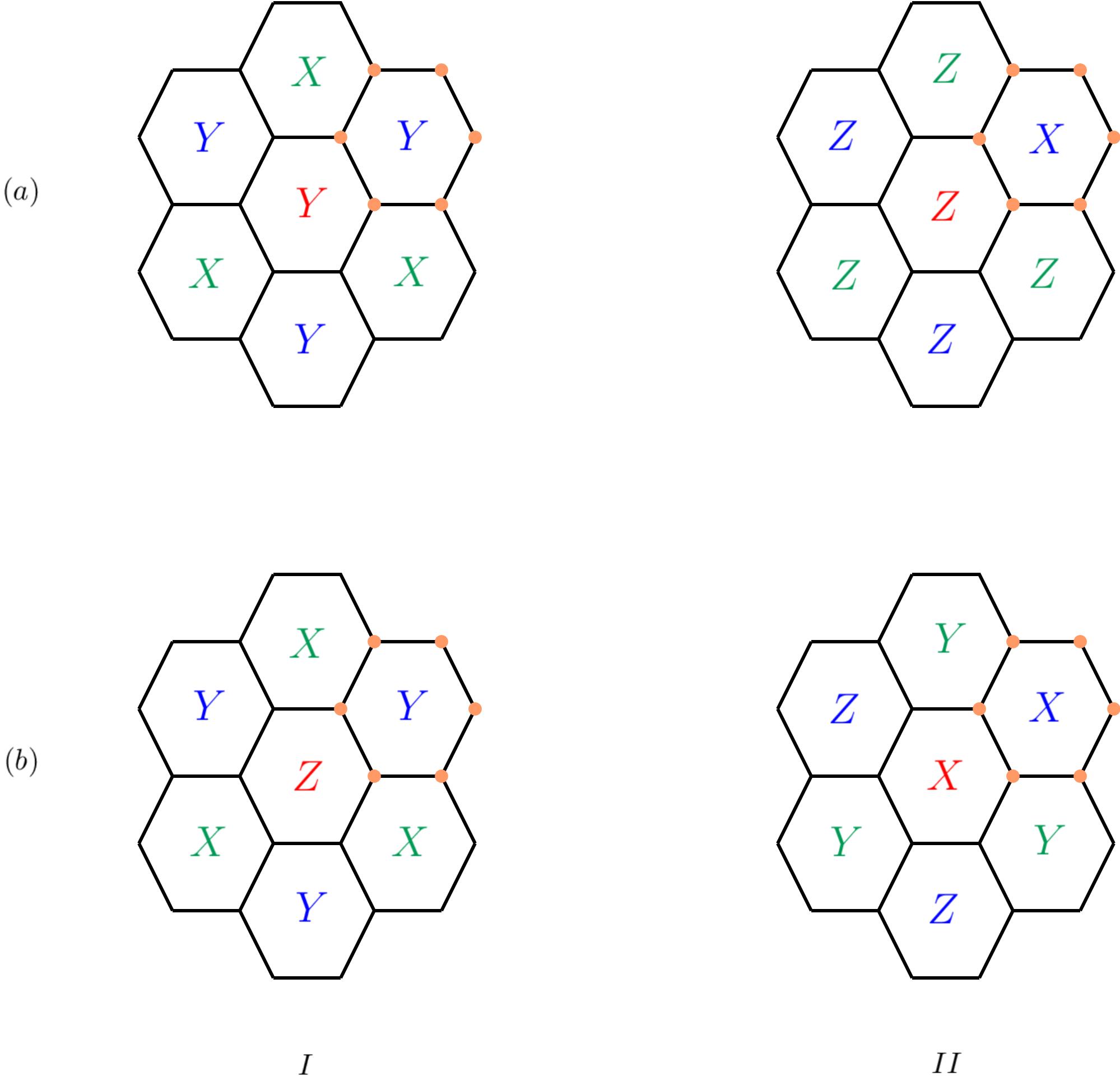}
 \caption{Example of color codes with mixed vertex types. (a) The $[345]-([444])$-color code. All the vertices are of the $[345]$ type except on the hexagon with orange vertices which are of the $[444]$ type. (b) The $[444]-([345])$-color code. Here the situation is reversed with all the vertices of the $[444]$ type except the orange ones which are of the $[345]$ type. In the former we can interpret the code as including an energetic defect of the $[444]$ type in the $[345]$-color code, whereas in the latter it is the opposite. These models are translationally invariant, with the stabilizer set $II$, containing $z$ Pauli operators on all the blue faces except the on the defect indicated with orange dots. }
 \label{fig:[345][444]ccnomixing}
\end{figure}

Here we consider color codes that have vertices of both the $[345]$ and the $[444]$ types. We show two of the simplest such examples on hexagonal lattices in Fig. \ref{fig:[345][444]ccnomixing} where the vertices of a single hexagon is of a different type compared to the rest of the vertices. In this setup the two possibilities are denoted $[345]-([444])$- and $[444]-([345])$-color codes where the defect hexagons contain the vertex types indicated in the $()$. We can consider more general color codes with an increased number of energetic defects such that they are placed in a random manner on the hexagonal lattice. 

\subsection{$[345]-[336]$-color codes}
\label{app:[345][336]mixedvertextypes}
\begin{figure}[h]
    \centering
    \includegraphics[width=8cm]{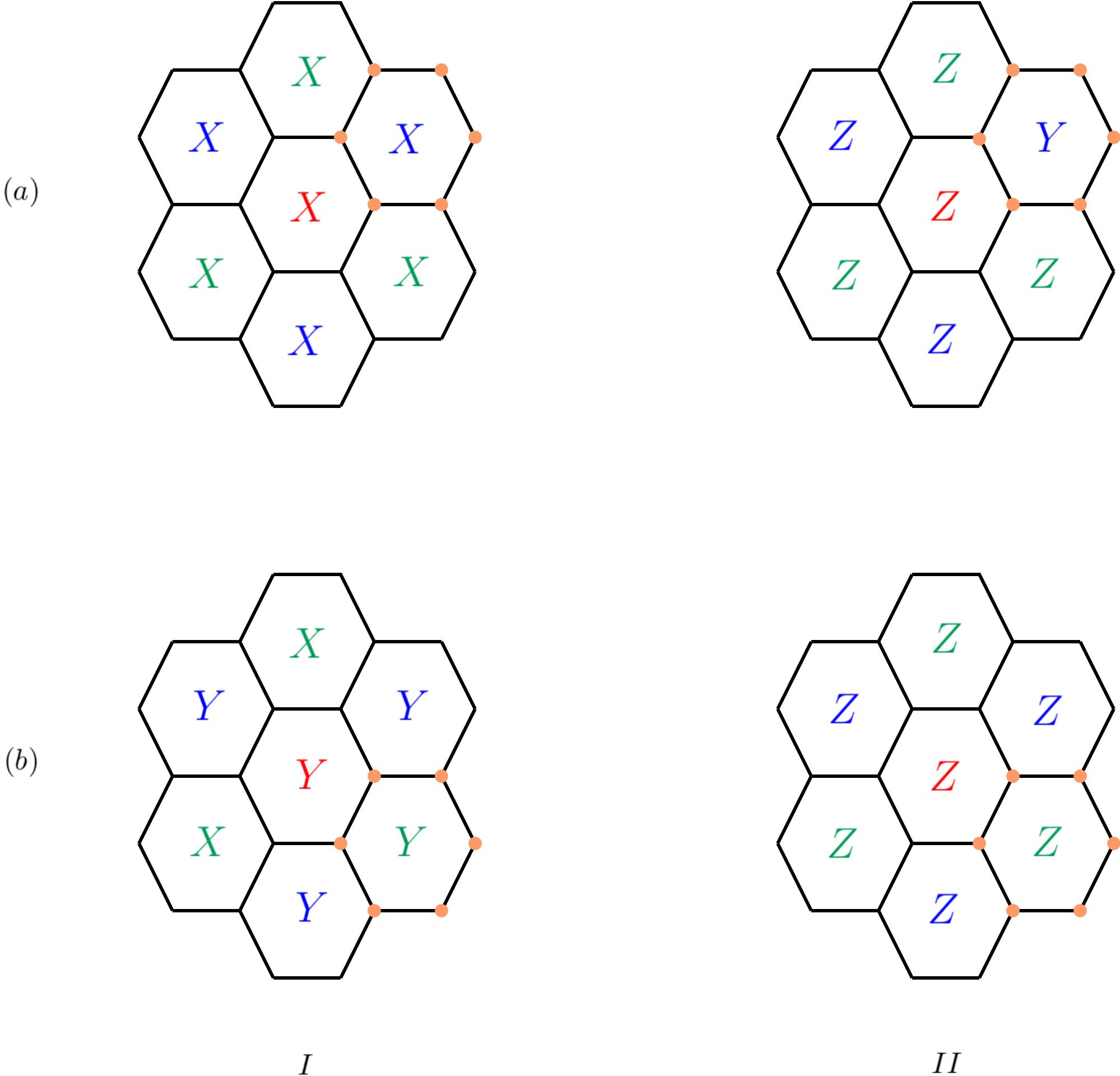}
 \caption{Example of color codes with mixed vertex types. (a) The $[336]-([345])$-color code. All the vertices are of the $[336]$ type except on the hexagon with orange vertices which are of the $[345]$ type. (b) The $[345]-([336])$-color code. Here the situation is reversed with all the vertices of the $[345]$ type except the orange ones which are of the $[336]$ type. In the former we can interpret the code as including an energetic defect of the $[345]$ type in the $[336]$-color code, whereas in the latter it is the opposite. These models are translationally invariant, with the stabilizer set $II$, containing (a) $z$ Pauli operators on all the blue faces except the on the defect indicated with orange dots and the stabilizer set $I$, containing (b) $x$ Pauli on all the green faces except on the one indicated by the orange dots. }
 \label{fig:[345][336]ccnomixing}
\end{figure}

The color codes here include vertices of types $[345]$ and $[336]$. The simplest examples where there is a single $[345]$ type hexagon in a sea of $[336]$ vertices or vice versa are shown in Fig. \ref{fig:[345][336]ccnomixing}. 

\subsection{$[444]-[345]-[336]$-color codes}
\label{app:[444][345][336]mixedvertextypes}
\begin{figure}[h]
    \centering
    \includegraphics[width=9cm]{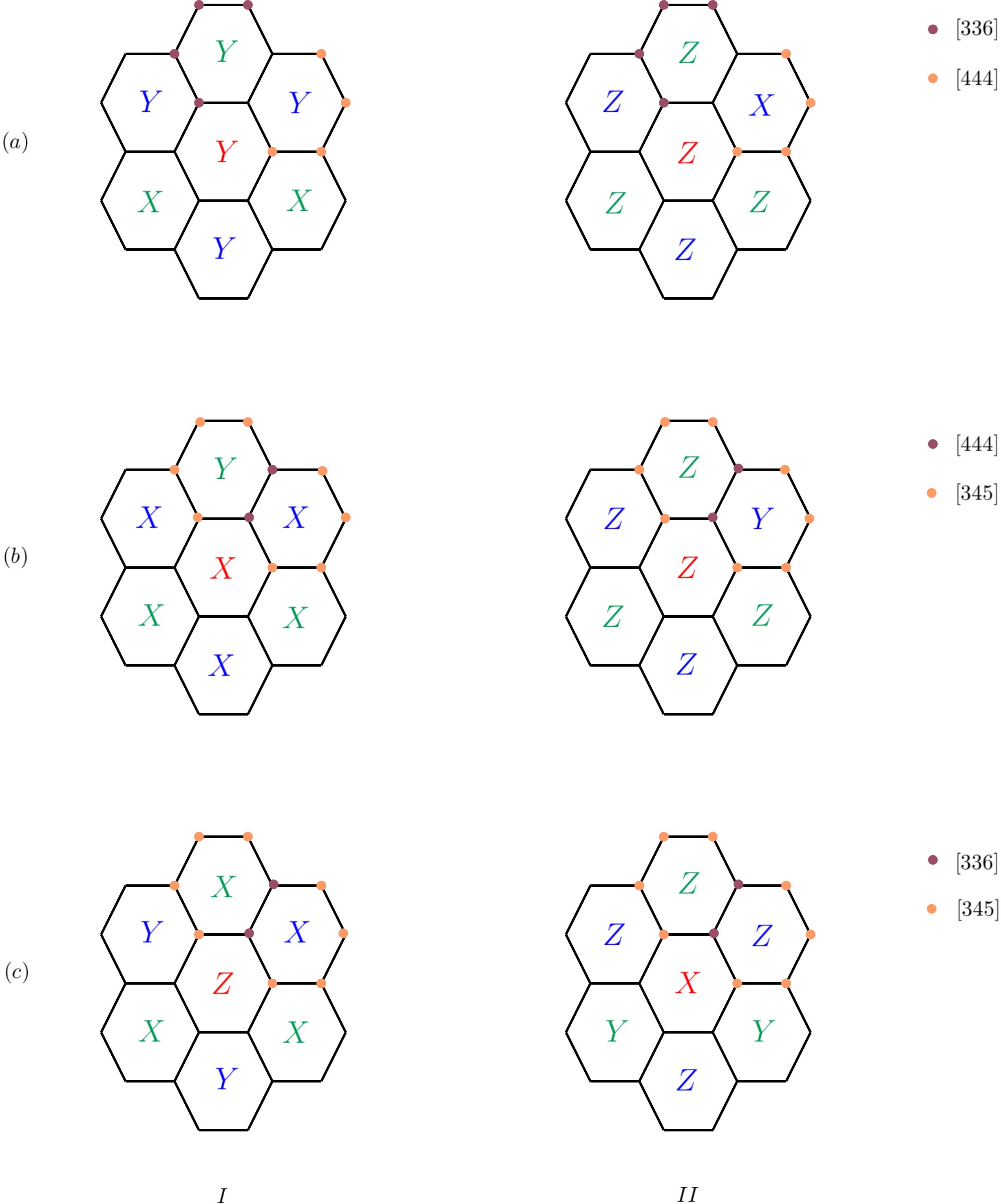}
 \caption{Example of color codes containing all three vertex types. (a) The $[345]-([336]-[444])$-color code. All the vertices are of the $[345]$ type except the vertices indicated by orange and purple dots which are of the $[444]$ and $[336]$ types respectively. (b) The $[336]-([345]-[444])$-color code. Here all the vertices of the $[336]$ type except the orange and purple ones which are of the $[345]$ and $[444]$ types respectively. (c) The $[444]-([345]-[336])$-color code. Here all the vertices of the $[444]$ type except the orange and purple ones which are of the $[345]$ and $[336]$ types respectively.  
 In this case each code has two energetic defects, placed next to each other. All these codes are translationally invariant, with the stabilizers modified at the location of the two hexagon defects indicated by the orange and the purple dots. }
 \label{fig:[444][345][336]ccnomixing}
\end{figure}

Finally we consider color codes where all the three bulk vertex types are present. In the examples shown in Fig. \ref{fig:[444][345][336]ccnomixing}, we pick the vertices on two of the hexagons to be of different vertex types with respect to the remaining vertices. 

\section{$[255]$-color codes}
\label{app:[255]cc}
\begin{figure}[h]
    \centering
    \includegraphics[width=8cm]{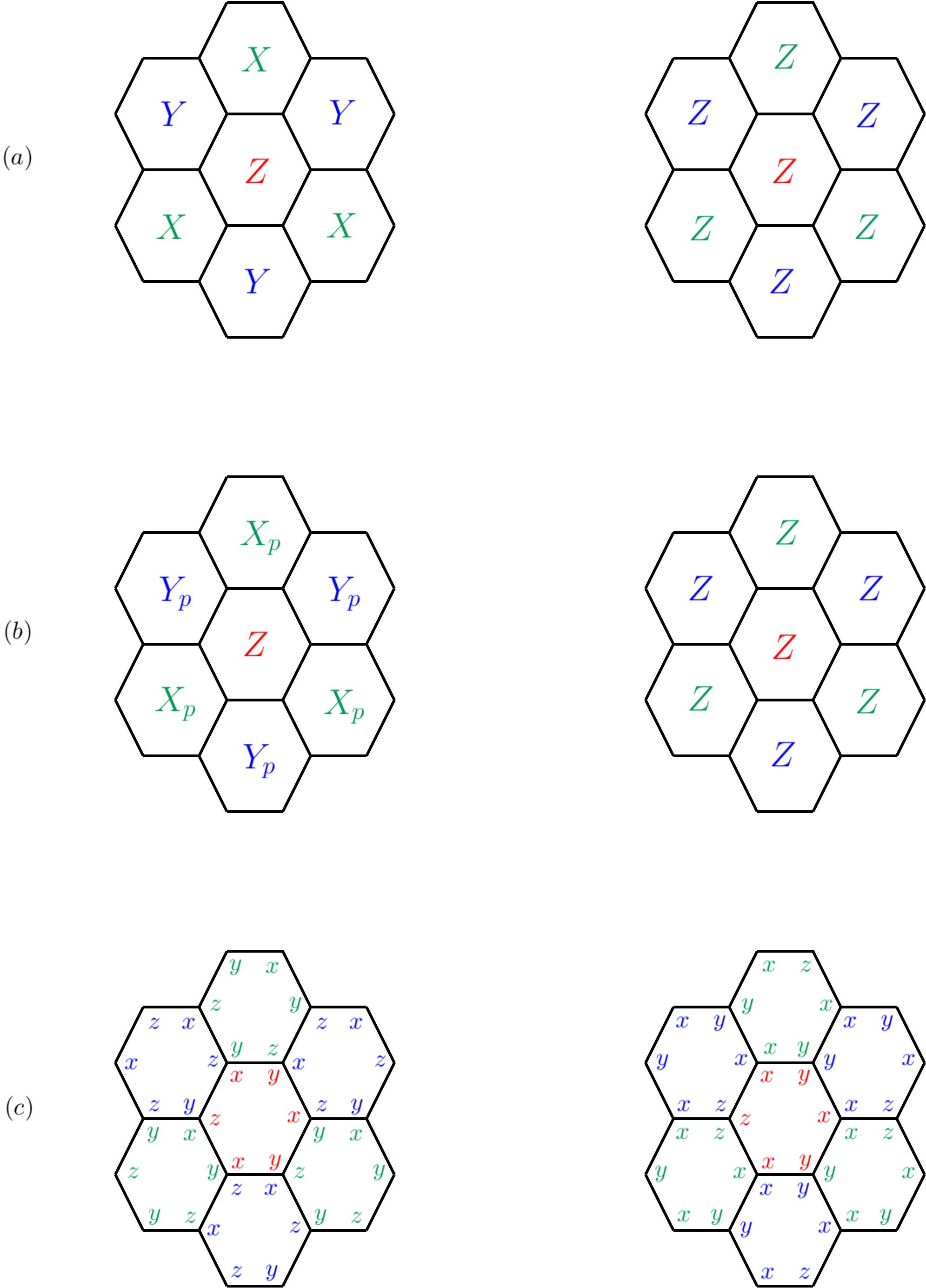}
 \caption{Translationally invariant $[255]$-color codes on the hexagonal lattice. Figure shows the different possible edge configurations, (a) homogeneous, (b) partially mixed and (c) fully mixed. Note that there is just one type of stabilizer on each of the red faces for the three codes. It is easy to verify that these models are equivalent to each other by LU's. }
 \label{fig:[255]ccwti}
\end{figure}

Here we briefly look at other possibilities that emerge from possible Y-vertex configurations listed in Table \ref{table:Yconfigs}, particularly the $[255]$ vertex type. These energy values are a result of the configurations containing, $\{xyzzzz\}$ and their LU equivalents. Stabilizer pairs created using these bulk vertices do not obey the anticommutation rule stated in Sec. \ref{sec:construction1}. As a consequence the color codes made up of these vertices have just a single type of stabilizer on one of the colored faces. In Fig. \ref{fig:[255]ccwti} we show translationally invariant $[255]$-color codes on the hexagonal lattice with homogeneous, partially and fully mixed edge configurations. Notice that while the blue and green faces have two stabilizers each, as it happens in the $[336]$-, $[345]$- and $[444]$-color codes, the red faces have just a single stabilizer. 
A crucial difference between these models and the $[336]$-, $[444]$- or $[345]$-color codes is that these models are not entirely topological. This is seen due to the presence of weight-two symmetries supported on the edges connecting red faces. $z_{v_1}z_{v_2}$ is a symmetry for the edge containing the vertices, $v_1$ and $v_2$ (See Fig. \ref{fig:[255]ccwti} (a)). Moreover there are a fewer number of independent constraints among the stabilizers in this case,
\begin{eqnarray}
C_1 & = & \prod\limits_{j=1}^{{\color{red}f}}~{\color{red}r}_j\prod\limits_{j=1}^{{\color{blue}f}}~{\color{blue}b_{II}}_j = 1,~ C_2=\prod\limits_{j=1}^{{\color{red}f}}~{\color{red}r}_j\prod\limits_{j=1}^{{\color{ForestGreen}f}}~{\color{ForestGreen}g_{II}}_j=1, \nonumber \\
C_3 & = & \prod\limits_{j=1}^{{\color{ForestGreen}f}}~{\color{ForestGreen}g_I}_j\prod\limits_{j=1}^{{\color{blue}f}}~{\color{blue}b_{I}}_j\prod\limits_{j=1}^{{\color{blue}f}}~{\color{blue}b_{II}}_j = 1,
\end{eqnarray}
three as opposed to four found for the $[336]$-, $[444]$- and $[345]$-color codes. Thus the number of independent stabilizers on the hexagonal lattice becomes, $v-\frac{5|F|}{3}+3=\frac{|F|}{3}+3$, making the code space dimension $2^{\frac{|F|}{3}+3}$, an extensive quantity.

\section{$[345]-([255])$-color codes}
\label{app:[345][255]cc}
\begin{figure}[h]
    \centering
    \includegraphics[width=8cm]{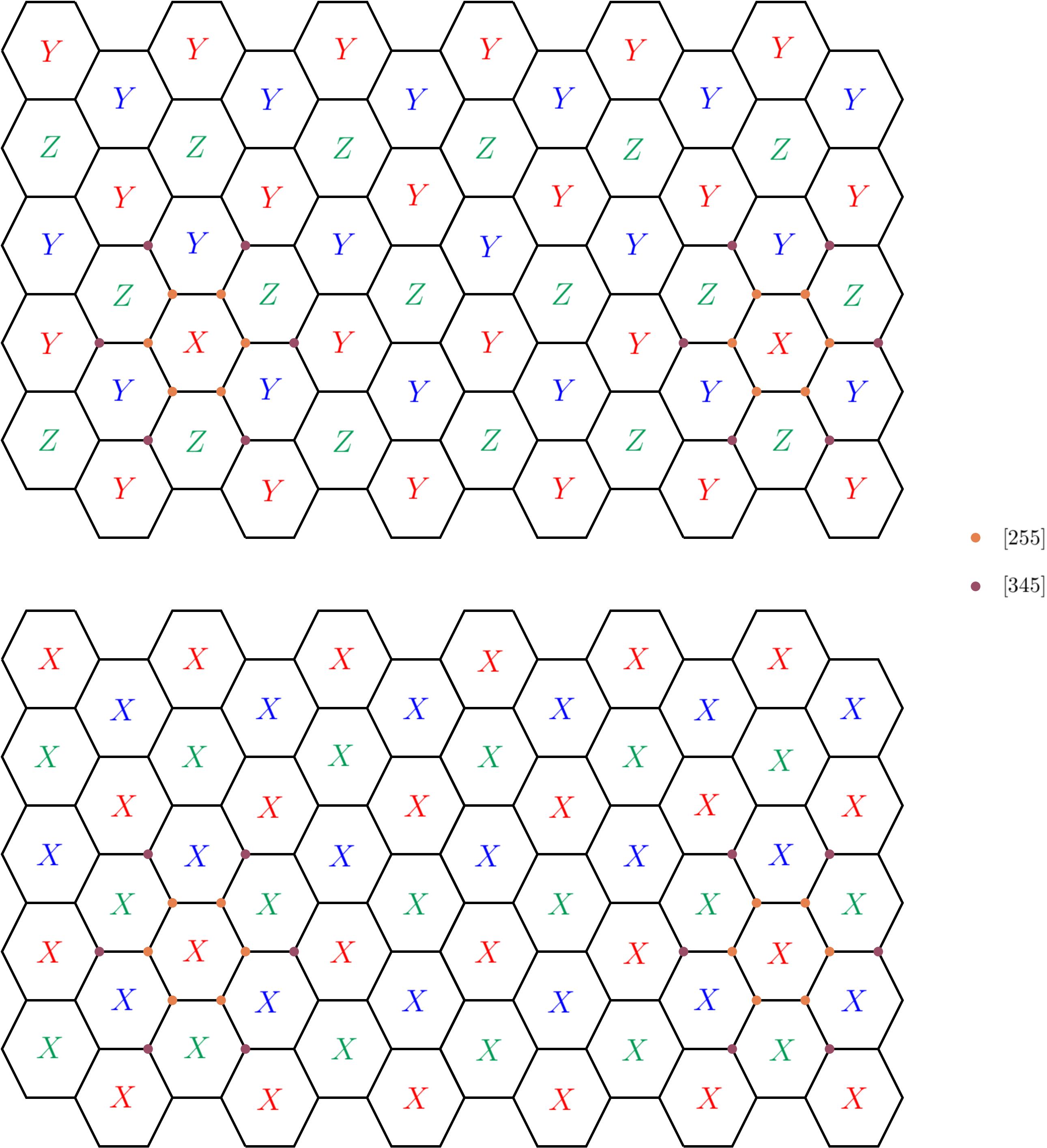}
 \caption{Translationally invariant $[345]$-color code on the hexagonal lattice with a $[255]$ `defect'. The vertex-edge combine around the $[255]$-defect hexagon are of the confined type. }
 \label{fig:345255ccnomixing}
\end{figure}

Fig. \ref{fig:345255ccnomixing} shows a translationally invariant deconfined $[345]$-color code with a `defect' hexagon made of $[255]$ vertices. Note that there is just one stabilizer on the defect hexagon. Any string operator made of the Pauli $x$'s along the red shrunk lattice connecting these two defects is a symmetry of this code. We can think of these as a quasi-local symmetry.

\clearpage

\bibliography{refs}
\end{document}